\documentclass{article}
\pdfoutput=1

\usepackage{natbib}
\usepackage{wasysym}
\usepackage{amssymb}
\usepackage{booktabs}
\usepackage{epubtk}
\usepackage{graphicx}
\usepackage{textcomp}
\usepackage{xspace}

\newcommand{\R}[1]{\textit{R\sub{#1}}\xspace}
\newcommand{\F}[1]{\textit{F}\sub{#1}\xspace}
\newcommand{\f}[1]{\textit{f}\sub{#1}\xspace}
\newcommand{\cms}{cm\super{-2}~s\super{-1}\xspace}

\begin{document}

\title{A History of Solar Activity over Millennia}

\author{
\epubtkAuthorData{Ilya G.\ Usoskin}
		 {Sodankyl{\"a} Geophysical Observatory (Oulu unit)\\
		  FIN-90014 University of Oulu, Finland}
		 {ilya.usoskin@oulu.fi}
		 {http://cc.oulu.fi/~usoskin/}
}

\date{}
\maketitle

\begin{abstract}
Presented here is a review of present knowledge of the long-term
 behavior of solar activity on a multi-millennial timescale, as
 reconstructed using the indirect proxy method.
The concept of solar activity is discussed along with an overview of
 the special indices used to quantify different aspects of variable
 solar activity, with special emphasis upon sunspot number.

Over long timescales, quantitative information about past solar
 activity can only be obtained using a method based upon indirect
 proxies, such as the cosmogenic isotopes \super{14}C and \super{10}Be in
 natural stratified archives (e.g., tree rings or ice cores).
We give an historical overview of the development of the proxy-based method
 for past solar-activity reconstruction over millennia, as well as a
 description of the modern state.
Special attention is paid to the verification and cross-calibration of reconstructions.
It is argued that this method of cosmogenic isotopes makes a solid basis for
 studies of solar variability in the past on a long timescale
 (centuries to millennia) during the Holocene.

A separate section is devoted to reconstructions of strong
 solar energetic-particle (SEP) events in the past, that suggest that
 the present-day average SEP flux is broadly consistent with estimates
 on longer timescales, and that the occurrence of extra-strong events
 is unlikely.

Finally, the main features of the long-term evolution of solar
 magnetic activity, including the statistics of grand minima and maxima
 occurrence, are summarized and their possible implications, especially
 for solar/stellar dynamo theory, are discussed.
\end{abstract}

\epubtkKeywords{solar activity, paleo-astrophysics, cosmogenic
  isotopes, solar-terrestrial relations, solar physics, long-term
  reconstructions, solar dynamo}

\newpage
\epubtkUpdate
    [Id=A,
     ApprovedBy=subjecteditor,
     AcceptDate={7 March 2013},
     PublishDate={21 March 2013},
     Type=major]{%
The review has been thoroughly revised and updated. Added
Sections~\ref{S:compo} and \ref{sec:SEP_ter} and 8 new figures (3 were
removed). 55 new references have been included (4 were removed).
}

\newpage
\tableofcontents

\newpage

\section{Introduction}
\label{section:introduction}

The concept of the perfectness and constancy of the sun, postulated by Aristotle, was a strong belief
 for centuries and an official doctrine of Christian and Muslim countries.
However, as people had noticed even before the time of Aristotle, some slight transient changes of the sun can
 be observed even with the naked eye.
Although scientists knew about the existence of ``imperfect'' spots on the sun since the early 17th century,
 it was only in the 19th century that the scientific community recognized that solar activity varies
 in the course of an 11-year solar cycle.
Solar variability was later found to have many different manifestations, including the fact that the ``solar constant'',
 or the total solar irradiance, TSI, (the amount of total incoming solar electromagnetic radiation in all wavelengths per
 unit area at the top of the atmosphere) is not a constant.
The sun appears much more complicated and active than a static hot plasma ball,
 with a great variety of nonstationary active processes going beyond the adiabatic equilibrium
 foreseen in the basic theory of sun-as-star.
Such transient nonstationary (often eruptive) processes can be broadly regarded as
 solar activity, in contrast to the so-called ``quiet'' sun.
Solar activity includes active transient and long-lived phenomena on the solar surface, such as spectacular solar flares,
 sunspots, prominences, coronal mass ejections (CMEs), etc.

The very fact of the existence of solar activity poses an enigma for solar physics, leading
 to the development of sophisticated models of an upper layer known as the convection zone and the solar corona.
The sun is the only star, which can be studied in great detail and thus can be considered as a proxy
 for cool stars.
Quite a number of dedicated ground-based and space-borne experiments are being carried out to
 learn more about solar variability.
The use of the sun as a paradigm for cool stars leads to a better understanding of the processes driving
 the broader population of cool sun-like stars.
Therefore, studying and modelling solar activity can increase the level of our understanding of nature.

On the other hand, the study of variable solar activity is not of purely academic interest,
 as it directly affects the terrestrial environment.
Although changes in the sun are barely visible without the aid of precise scientific instruments,
 these changes have great impact on many aspects of our lives.
In particular, the heliosphere (a spatial region of about 200\,--\,300 astronomical units across) is
 mainly controlled by the solar magnetic field.
This leads to the modulation of galactic cosmic rays (GCRs) by the solar magnetic activity.
Additionally, eruptive and transient phenomena in the sun/corona and in the interplanetary medium can
 lead to sporadic acceleration of energetic particles with greatly enhanced flux.
Such processes can modify the radiation environment on Earth and need to be taken
 into account for planning and maintaining space missions and even transpolar jet flights.
Solar activity can cause, through coupling of solar wind and the Earth's magnetosphere,
 strong geomagnetic storms in the magnetosphere and ionosphere, which may disturb radio-wave
 propagation and navigation-system stability, or induce dangerous spurious currents in
 long pipes or power lines.
Another important aspect is the link between solar-activity variations
 and the Earth's climate \citep[see, e.g., reviews by][]{haighLR,gray10}.

It is important to study solar variability on different timescales.
The primary basis for such studies is observational (or reconstructed) data.
The sun's activity is systematically explored in different ways (solar, heliospheric,
 interplanetary, magnetospheric, terrestrial), including ground-based and space-borne
 experiments and dedicated missions during the last few decades, thus covering 3\,--\,4 solar cycles.
However, it should be noted that the modern epoch is characterized by unusually-high solar activity
 dominated by an 11-year cyclicity, and it is not straightforward to extrapolate present
 knowledge (especially empirical and semi-empirical relationships and models) to a longer
 timescale.
The current cycle 24 indicates the return to the normal moderate level of solar activity,
 as manifested, e.g., via the extended and weak solar minimum in 2008\,--\,2009 and weak solar and heliospheric
 parameters, which are unusual for the space era but may be quite typical for the normal
 activity \citep[see, e.g.,][]{gibson11}.
 Thus, we may experience, in the near future, the interplanetary conditions quite different
  with respect to those we got used to during the last decades.

Therefore, the behavior of solar activity in the past, before the era of direct measurements,
 is of great importance for a variety of reasons.
For example, it allows an improved knowledge of the statistical behavior of the
 solar-dynamo process, which generates the cyclically-varying solar-magnetic field,
 making it possible to estimate the fractions of time the sun
 spends in states of very-low activity, what are called grand minima.
Such studies require a long time series of solar-activity data.
The longest direct series of solar activity is the 400-year-long
 sunspot-number series, which depicts the dramatic contrast between
 the (almost spotless) Maunder minimum and the modern period of very high activity.
Thanks to the recent development of precise technologies, including accelerator mass
 spectrometry, solar activity can be reconstructed over multiple millennia
 from concentrations of cosmogenic isotopes \super{14}C and \super{10}Be in terrestrial archives.
This allows one to study the temporal evolution of solar magnetic activity,
 and thus of the solar dynamo, on much longer timescales than are available from
 direct measurements.

This paper gives an overview of the present status of our knowledge of long-term
 solar activity, covering the period of Holocene (the last 11 millennia).
A description of the concept of solar activity and a discussion of observational methods and indices are presented in Section~\ref{sec:1}.
The proxy method of solar-activity reconstruction is described in some detail in Section~\ref{S:4}.
Section~\ref{sec:3} gives an overview of what is known about past solar activity.
The long-term averaged flux of solar energetic particles is discussed in Section~\ref{S:SEP}.
Finally, conclusions are summarized in Section~\ref{sec:conc}.

\newpage


\section{Solar Activity: Concept and Observations}
\label{sec:1}

\subsection{The concept of solar activity}

The sun is known to be far from a static state, the so-called ``quiet'' sun described
 by simple stellar-evolution theories, but instead goes through various nonstationary active processes.
Such nonstationary and nonequilibrium (often eruptive) processes can be broadly regarded as
 solar activity.
Whereas the concept of solar activity is quite a common term nowadays, it is
 neither straightforwardly interpreted nor unambiguously defined.
For instance, solar-surface magnetic variability, eruption phenomena, coronal
 activity, radiation of the sun as a star or even interplanetary transients and geomagnetic
 disturbances can be related to the concept of solar activity.
A variety of indices quantifying solar activity have been proposed in order to represent
 different observables and caused effects.
Most of the indices are highly correlated to each other due to the dominant 11-year
 cycle, but may differ in fine details and/or long-term trends.
In addition to the solar indices, indirect proxy data is often used to quantify
 solar activity via its presumably known effect on the magnetosphere or heliosphere.
The indices of solar activity that are often used for long-term studies are reviewed below.

\subsection{Indices of solar activity}

Solar (as well as other) indices can be divided into physical and synthetic according
 to the way they are obtained/calculated.
Physical indices quantify the directly-measurable values of a real physical observable, such as, e.g.,
 the radioflux, and thus have clear physical meaning as they quantify physical features of different
 aspects of solar activity and their effects.
Synthetic indices (the most common being sunspot number) are calculated
 (or synthesized) using a special algorithm from observed (often not measurable in physical units)
 data or phenomena.
Additionally, solar activity indices can be either direct (i.e., directly relating to the
 sun) or indirect (relating to indirect effects caused by solar activity), as discussed
 in subsequent Sections~\ref{S:dir} and \ref{S:ind}.

\subsubsection{Direct solar indices}
\label{S:dir}


The most commonly used index of solar activity is based on \textbf{sunspot number}.
Sunspots are dark areas on the solar disc (of size up to tens of thousands of km,
 lifetime up to half-a-year), characterized by a strong magnetic field, which
 leads to a lower temperature (about 4000~K compared to 5800~K in the photosphere)
 and observed as darkening.

Sunspot number is a synthetic, rather than a physical, index, but it has still become
 quite a useful parameter in quantifying the level of solar activity.
This index presents the weighted number of individual sunspots and/or sunspot groups,
 calculated in a prescribed manner from simple visual solar observations.
The use of the sunspot number makes it possible to combine together
 thousands and thousands of regular and fragmentary solar observations made by
 earlier professional and amateur astronomers.
The technique, initially developed by Rudolf Wolf,
 yielded the longest series of directly and regularly-observed scientific quantities.
Therefore, it is common to quantify solar magnetic activity via sunspot numbers.
For details see the review on sunspot numbers and solar cycles \citep{hathaway04, hathawayLR}.

\vspace{0.2cm}

\noindent\textbf{Wolf sunspot number (WSN) series}

\noindent
The concept of the sunspot number was developed by Rudolf Wolf of the Z\"urich
 observatory in the middle of the 19th century.
The sunspot series, initiated by him, is called the Z\"urich or Wolf sunspot number
 (WSN) series.
The relative sunspot number \R{z} is defined as
\begin{equation}
R_z = k\, (10\, G + N) \,,
\label{eq:Rz}
\end{equation}
where $G$ is the number of sunspot groups, $N$ is the number of individual sunspots
 in all groups visible on the solar disc and $k$ denotes the individual correction factor, which
 compensates for differences in observational techniques and instruments used by
 different observers, and is used to normalize different observations to each other.

\epubtkImage{LR_SN_long.png}{%
  \begin{figure}[htbp]
    \centerline{\includegraphics{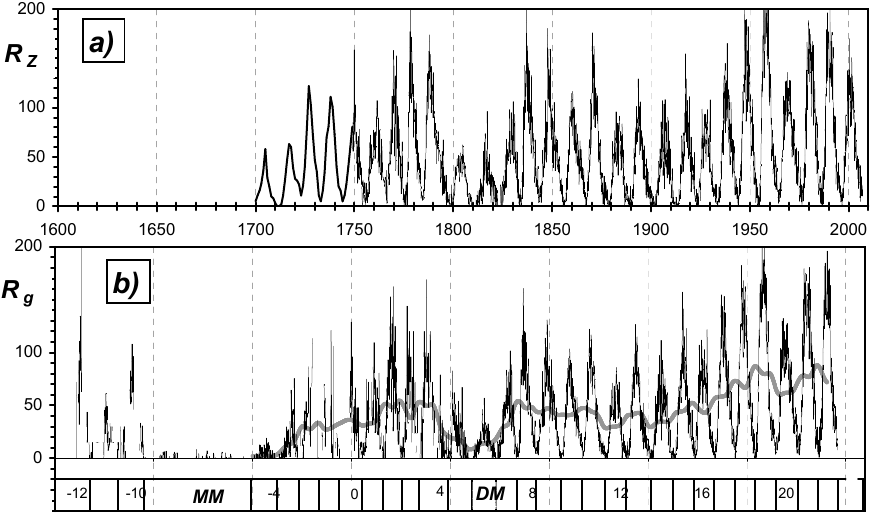}}
    \caption{Sunspot numbers since 1610. a) Monthly (since 1749) and yearly (1700\,--\,1749) Wolf sunspot number series.
    b) Monthly group sunspot number series.
    The grey line presents the 11-year running mean after the Maunder minimum.
    Standard (Z\"urich) cycle numbering as well as the Maunder minimum (MM) and Dalton minimum
    (DM) are shown in the lower panel.}
    \label{Fig:SA}
\end{figure}}

The value of \R{z} (see Figure~\ref{Fig:SA}a) is calculated for each
 day using only one observation made by the ``primary''
 observer (judged as the most reliable observer during a given time) for the day.
The primary observers were Staudacher (1749\,--\,1787), Flaugergues (1788\,--\,1825), Schwabe (1826\,--\,1847),
 Wolf (1848\,--\,1893), Wolfer (1893\,--\,1928), Brunner (1929\,--\,1944), Waldmeier (1945\,--\,1980) and
 Koeckelenbergh (since 1980).
If observations by the primary observer are not available for a certain day, the secondary, tertiary, etc.
 observers are used \citep[see the hierarchy of observers in][]{waldmeier61}.
The use of only one observer for each day aims to make \R{z} a homogeneous time series.
As a drawback, such an approach ignores all other observations available for the day, which constitute
 a large fraction of the existing information.
Moreover, possible errors of the primary observer cannot be caught or estimated.
The observational uncertainties in the monthly \R{z} can be up to 25\% \citep[e.g.,][]{vitinsky86}.
The WSN series is based on observations performed at the
 Z\"urich Observatory during 1849\,--\,1981 using almost the same technique.
This part of the series is fairly stable and homogeneous although an offset due to
 the change of the weighting procedure might have been introduced in 1945\,--\,1946 \citep{svalgaard12}.
However, prior to that there have been many gaps in the data that were interpolated.
If no sunspot observations are available for some period, the data gap is filled,
 without note in the final WSN series, using an interpolation between
 the available data and by employing some proxy data.
In addition, earlier parts of the sunspot series were ``corrected'' by Wolf using
 geomagnetic observation \citep[see details in ][]{svalgaard12}, which makes the series
 less homogeneous.
Therefore, the WSN series is a combination of direct observations and interpolations for the
 period before 1849, leading to possible errors and inhomogeneities as discussed, e.g, by \citet{vitinsky86, wilson98, letfus99, svalgaard12}.
The quality of the Wolf series before 1749 is rather poor and hardly reliable \citep{hoyt94, hoyt98, hathaway04}.

Note that the sun has been routinely photographed since 1876 so that full information on
 daily sunspot activity is available (the Greenwich series)
 with observational uncertainties being negligible for the last 140~years.

The routine production of the WSN series was terminated in Z\"urich in 1982.
Since then, the sunspot number series is routinely updated as
 the International sunspot number \R{i}, provided by the Solar Influences Data Analysis Center
 in Belgium \citep{clette07}.
The international sunspot number series is computed using the same definition (Equation~\ref{eq:Rz}) as WSN
 but it has a significant distinction from the WSN; it is based not on a single primary
 solar observation for each day but instead uses a weighted average of more than 20 approved
 observers.

In addition to the standard sunspot number \R{i}, there is also a series of hemispheric
 sunspot numbers $R_{\mathrm{N}}$ and $R_{\mathrm{S}}$, which account for spots only in
 the northern and southern solar hemispheres, respectively (note that $R_i=R_{\mathrm{N}}+R_{\mathrm{S}}$).
These series are used to study the N-S asymmetry of solar activity \citep{temmer02}.

\vspace{0.2cm}

\noindent\textbf{Group sunspot number (GSN) series}

\noindent
Since the WSN series is of lower quality before the 1850s and is hardly reliable
 before 1750, there was a need to re-evaluate early sunspot data.
This tremendous work has been done by \citet{hoyt96,hoyt98}, who performed
 an extensive archive search and nearly doubled the amount
 of original information compared to the Wolf series.
They have produced a new series of sunspot activity called
 the group sunspot numbers (GSN -- see Figure~\ref{Fig:SA}b),
 including all available archival records.
The daily group sunspot number \R{g} is defined as follows:
\begin{equation}
R_g={12.08\over n}\sum_i{k'_iG_i} \,,
\label{eq:Rg}
\end{equation}
where $G_i$ is the number of sunspot groups recorded by the $i$-th observer,
 $k'$ is the observer's individual correction factor, $n$ is the number of observers
 for the particular day, and 12.08 is a normalization number scaling
 \R{g} to \R{z} values for the period of 1874\,--\,1976.
\R{g} is more robust than \R{z} since it is based on more easily identified
 sunspot groups and does not include the number of individual spots.
The GSN series includes not only one ``primary'' observation, but all available observations, and
 covers the period since 1610, being, thus, 140~years longer than the original WSN series.
It is particularly interesting that the period of the Maunder minimum (1645\,--\,1715) was surprisingly
 well covered with daily observations \citep{ribes93, hoyt96} allowing for a detailed analysis of sunspot
 activity during this grand minimum (see also Section~\ref{sec:MM}).
Systematic uncertainties of the \R{g} values are estimated to be about 10\% before 1640,
 less than 5\% from 1640\,--\,1728 and from 1800\,--\,1849, 15\,--\,20\% from 1728\,--\,1799, and about 1\% since
 1849 \citep{hoyt98}.
The GSN series is more reliable and homogeneous than the WSN series before 1849.
The two series are nearly identical after the 1870s \citep{hoyt98, letfus99, hathaway04}.
However, the GSN series still contains some lacunas, uncertainties and possible inhomogeneities
 \citep[see, e.g.,][]{letfus00, usoskin_SP_daily03,vaquero12}.

The search for other lost or missing records of past solar instrumental observations has not
 ended even since the extensive work by Hoyt and Schatten.
Archival searches still give new interesting findings of forgotten
 sunspot observations, often outside major observatories -- see a detailed review book by \citet{vaquero09}
 and original papers by \citet{casas06, vaquero05, vaquero07, arlt08, arlt09}.
Interestingly, not only sunspot counts but also regular drawings, forgotten for centuries,
 are being restored nowadays in dusty archives.
A very interesting work has been done by Rainer Arlt \citep{arlt08, arlt09, arlt11, arlt13} on recovering, digitizing,
 and analyzing regular drawings by S.H.~Schwabe of 1825\,--\,1867 and J.C.~Staudacher of 1749\,--\,1796.
This work led to the extension of the Maunder butterfly diagram for several solar cycles backwards \citep{arlt09,usoskin_lost_09,arlt11,arlt13} --
 see a newly built diagram for solar cycles Nos.~7\,--\,10 shown in Figure~\ref{Fig:arlt}.
In particular, this data confirms that GSN series is more homogenous before 1874 that WSN. 
A recent finding of the lost data by G.~Marcgraf and correcting some earlier uncertain data for the period 1636\,--\,1642 by \citet{vaquero11}
 made it possible to revise the pattern of the beginning of the Maunder minimum.

\epubtkImage{LR_arlt_butterfly.png}{%
  \begin{figure}[htbp]
    \centerline{\includegraphics[width=\textwidth]{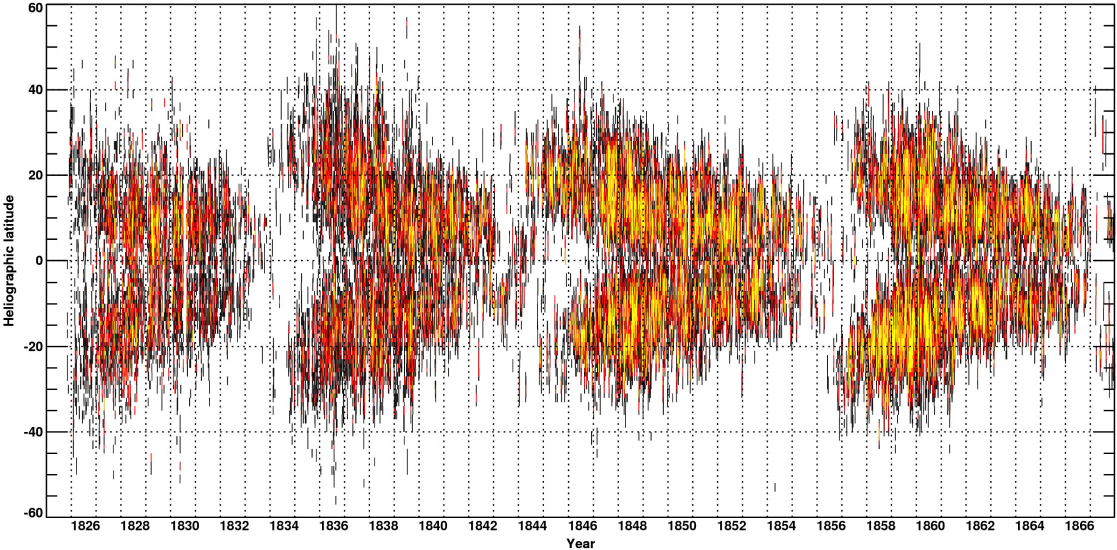}}
    \caption{Maunder butterfly diagram of sunspot occurrence reconstructed by \citet{arlt13}
     for 1825\,--\,1867 using recovered drawing of S.H.~Schwabe.}
    \label{Fig:arlt}
\end{figure}}


\vspace{0.2cm}
\noindent\textbf{Other indices}

\noindent
An example of a synthetic index of solar activity is the \emph{flare index}, representing solar flare activity
 \citep[e.g.,][]{ozguc03, kleczek52}.
The flare index quantifies daily flare activity in the following manner;
 it is computed as a product of the flare's relative importance $I$ in the $\mathrm{H}_{\alpha}$-range and
 duration $t$, $Q=I\, t$, thus being a rough measure of the total energy emitted by the flare.
The daily flare index is produced by Bogazici University \citep{ozguc03} and is
 available since 1936.


A traditional physical index of solar activity is related to the radioflux of
 the sun in the wavelength range of 10.7~cm and is called the \emph{F10.7 index} \citep[e.g.,][]{tapping94}.
This index represents the flux (in solar flux units, 1~sfu~=~10\super{-22}~Wm\super{-2}~Hz\super{-1}) of solar
 radio emission at a centimetric wavelength.
There are at least two sources of 10.7~cm flux -- free-free emission from hot coronal plasma and gyromagnetic
 emission from active regions \citep{tapping87}.
It is a good quantitative measure of the level of solar activity, which is not directly related
 to sunspots.
Close correlation between the F10.7 index and sunspot number indicates that the latter
 is a good index of general solar activity, including coronal activity.
The solar F10.7~cm record has been measured continuously since 1947.

Another physical index is the \emph{coronal index} \citep[e.g.,][]{rybansky05},
 which is a measure of the irradiance of the sun as a star in the coronal green line.
Computation of the coronal index is based on observations of green corona
 intensities (Fe~XIV emission line at 530.3~nm wavelength) from coronal stations all
 over the world, the data being transformed to the Lomnick\'y \v{S}tit photometric scale.
This index is considered a basic optical index of solar activity.
A synthesized homogeneous database of the Fe~XIV 530.3~nm coronal-emission line intensities has existed
 since 1943 and covers seven solar cycles.

Often \emph{sunspot area} is considered as a physical index representing
 solar activity \citep[e.g.,][]{baranyi01, balmaceda05}.
This index gives the total area of visible spots on the solar disc in units
 of millionths of the sun's visible hemisphere, corrected for apparent
 distortion due to the curvature of the solar surface.
The area of individual groups may vary between tens of millionths (for small groups) up to
 several thousands of millionths for huge groups.
This index has a physical meaning related to the solar magnetic flux emerging at sunspots.
Sunspot areas are available since 1874 in the Greenwich series obtained
 from daily photographic images of the sun.
In addition, some fragmentary data of sunspot areas, obtained from solar drawings,
 are available for earlier periods \citep[]{vaquero04, arlt08}.

An important quantity is solar irradiance, total and spectral \citep{frohlich12}.
Irradiance variations are physically related to solar magnetic variability \citep[e.g.,][]{solanki00},
 and are often considered manifestations of solar activity, which is of primary
 importance for solar-terrestrial relations.

Other physical indices include spectral sun-as-star observations, such as the
 \emph{Ca II-K index} \citep[e.g.,][]{donnelly94, foukal96},
 the space-based \emph{Mg II core-to-wing ratio} as an index of solar UVI \citep[e.g.,][]{donnelly94, viereck99, snow05}
 and many others.

All the above indices are closely correlated to sunspot numbers on the solar-cycle scale,
 but may depict quite different behavior on short or long timescales.

\subsubsection{Indirect indices}
\label{S:ind}

Sometimes quantitative measures of solar-variability effects are also considered
 as indices of solar activity.
These are related not to solar activity \emph{per se}, but rather to its effect on different environments.
Accordingly, such indices are called indirect, and can be roughly divided into
 terrestrial/geomagnetic and heliospheric/interplanetary.

Geomagnetic indices quantify different effects of geomagnetic activity ultimately
 caused by solar variability, mostly by variations of solar-wind properties and
 the interplanetary magnetic field.
For example, the \textit{aa}-index, which provides a global index of magnetic activity relative to a
 quiet-day curve for a pair of antipodal magnetic observatories (in England and Australia),
 is available from 1868 \citep{mayaud72}.
An extension of the geomagnetic series is available from the 1840s using the
 Helsinki \textit{Ak(H)} index \citep{nevanlinna04, nevanlinna04a}.
Although the homogeneity of the geomagnetic series is compromised \citep[e.g.,][]{lukianova09,love11},
 it still remains an important indirect index of solar activity.
A review of the geomagnetic effects of solar activity can be found, e.g., in \citet{pulkkinenLR}.
It is noteworthy that geomagnetic indices, in particular low-latitude aurorae \citep{silverman06},
 are associated with coronal/interplanetary activity (high-speed solar-wind streams,
 interplanetary transients, etc.) that may not be directly related to the sunspot-cycle phase
 and amplitude, and therefore serve only as an approximate index of solar activity.
One of the earliest instrumental geomagnetic indices is related to the daily magnetic declination range,
 the range of diurnal variation of magnetic needle readings at a fixed location, and is available from the 1780s
 \citep{nevanlinna95}.
However, this data exists as several fragmentary sets, which are difficult to combine into
 a homogeneous data series.

Heliospheric indices are related to features of the solar wind or the interplanetary
 magnetic field measured (or estimated) in the interplanetary space.
For example, the time evolution of the total (or open) solar magnetic flux is
 extensively debated \citep[e.g.,][]{lockwood99, wang05, krivova07}.

A special case of heliospheric indices is related to the galactic cosmic-ray intensity
 recorded in natural terrestrial archives.
Since this indirect proxy is based on data recorded naturally throughout the ages
 and revealed now, it makes possible the reconstruction of solar activity
 changes on long timescales, as discussed in Section~\ref{S:4}.

\subsection{Solar activity observations in the pre-telescopic epoch}
\label{sec:2}

Instrumental solar data is based on regular observation (drawings or
 counting of spots) of the sun using optical instruments, e.g.,
 the telescope used by Galileo in the early 17th century.
These observations have mostly been made by professional astronomers whose qualifications
 and scientific thoroughness were doubtless.
They form the basis of the Group sunspot-number series \citep{hoyt98}, which can be
 more-or-less reliably extended back to 1610 (see discussion in Section~\ref{S:dir}).
However, some fragmentary records of qualitative solar and geomagnetic observations
 exist even for earlier times, as discussed below (Sections~\ref{S:co}\,--\,\ref{S:ny}).

\subsubsection{Instrumental observations: Camera obscura}
\label{S:co}

The invention of the telescope revolutionized astronomy.
However, another solar astronomical instrument, the camera obscura, also made it possible to provide relatively
 good solar images and was still in use until the late 18th century.
Camera obscuras were known from early times, and they have been used in major cathedrals to define the
 sun's position \citep[see the review by][]{vaquero_rev07,vaquero09}.
The earliest known drawing of the solar disc was made by Frisius, who observed the solar eclipse in 1544
 using a camera obscura.
That observation was performed during the Sp\"orer minimum and no spots were observed on the sun.
The first known observation of a sunspot using a camera obscura was done by Kepler in May 1607,
 who erroneously ascribed the spot on the sun to a transit of Mercury.
Although such observations were sparse and related to other phenomena (solar eclipses or transits of planets),
 there were also regular solar observations by camera obscura.
For example, about 300 pages of logs of solar observations made in the cathedral of San
 Petronio in Bologna from 1655\,--\,1736 were published by Eustachio Manfredi
 in 1736 \citep[see the full story in][]{vaquero_rev07}.

Therefore, observations and drawings made using camera obscura can be regarded as instrumental observations.

\subsubsection{Naked-eye observations}
\label{S:ny}

Even before regular professional observations performed with the aid of specially-developed
 instruments (what we now regard as scientific observations) people were interested in unusual phenomena.
Several historical records exist based on naked-eye observations of transient phenomena on the sun or
 in the sky.


From even before the telescopic era, a large amount of evidence of spots being observed on the solar disc can be traced back
 as far as to the middle of the 4th century BC (Theophrastus of Athens).
The earliest known drawing of sunspots is dated to December~8, 1128~AD  as published
 in ``The Chronicle of John of Worcester'' \citep{willis01}.
However, such evidence from occidental and Moslem sources is scarce and mostly related to observations of
 transits of inner planets over the sun's disc, probably because of the dominance of the
 dogma on the perfectness of the sun's body, which dates back to Aristotle's doctrine
 \citep{bray64}.
Oriental sources are much richer for naked-eye sunspot records, but that data is also
 fragmentary and irregular \citep[see, e.g.,][]{clark78, wittman_xu87, yau_steph88}.
Spots on the sun are mentioned in official Chinese and Korean chronicles from 165~BC to 1918~AD.
While these chronicles are fairly reliable, the data is not
 straightforward to interpret since it can be influenced by meteorological
 phenomena, e.g., dust loading in the atmosphere due to dust storms \citep{willis80} or
 volcanic eruptions \citep{scuderi90} can facilitate sunspots observations.
Direct comparison of Oriental naked-eye sunspot observations and European
 telescopic data shows that naked-eye observations can serve only as a qualitative
 indicator of sunspot activity, but can hardly be quantitatively interpreted
 \citep[see, e.g.,][and references therein]{willis96}.
Moreover, as a modern experiment of naked-eye observations \citep{mossman89} shows,
 Oriental chronicles contain only a tiny ($^{1}/_{200}\mbox{\,--\,}^{1}/_{1000}$) fraction of the number
 of sunspots potentially visible with the naked eye \citep{eddy89}.
This indicates that records of sunspot observations in the official chronicles were highly
 irregular \citep{eddy83} and probably dependent on dominating traditions during specific historical periods
 \citep{clark78}.
Although naked-eye observations tend to qualitatively follow the general trend in solar activity
 according to \emph{a posteriori} information \citep[e.g.,][]{vaquero02},
 extraction of any independent quantitative information from these records seems impossible.


Visual observations of aurorae borealis at middle latitudes form another proxy for
 solar activity \citep[e.g.,][]{siscoe80, schove83, krivsky84, silverman92, schroder92, lee04, basurah04, vaquero10}.
Fragmentary records of aurorae can be found in both occidental and oriental sources since antiquity.
The first known dated notation of an aurora is from March~12, 567~BC from Babylon \citep{stephenson04}.
Aurorae may appear at middle latitudes as a result of enhanced geomagnetic
 activity due to transient interplanetary phenomena.
Although auroral activity reflects coronal and interplanetary features rather
 than magnetic fields on the solar surface, there is a strong correlation between long-term variations of
 sunspot numbers and the frequency of aurora occurrences.
Because of the phenomenon's short duration and low brightness, the probability of seeing aurora
 is severely affected by other factors such as the weather (sky overcast, heat lightnings),
 the Moon's phase, season, etc.
The fact that these observations were not systematic in early times (before the beginning of the 18th century)
 makes it difficult to produce a homogeneous data set.
Moreover, the geomagnetic latitude of the same geographical location may
 change quite dramatically over centuries, due to the migration of the geomagnetic axis, which also affects the probability of watching aurorae \citep{siscoe83, oguti95}.
For example, the geomagnetic latitude of Seoul (37.5\textdegree~N~127\textdegree~E), which is currently less than
 30\textdegree, was about 40\textdegree\ a millennium ago \citep{kovaltsov07}.
This dramatic change alone can explain the enhanced frequency of aurorae observations recorded in oriental
 chronicles.

\subsubsection{Mathematical/statistical extrapolations}
\label{sec:math}

Due to the lack of reliable information regarding solar activity in the pre-instrumental
 era, it seems natural to try to extend the sunspot series back in time, before 1610~AD,
 by means of extrapolating its statistical properties.
Indeed, numerous attempts of this kind have been made even recently
 \citep[e.g.,][]{nagovistyn97, demeyer98, rigozo01}.
Such models aim to find the main feature of the actually-observed sunspot series,
 e.g., a modulated carrier frequency or a multi-harmonic
 representation, which is then extrapolated backwards in time.
The main disadvantage of this approach is that it is not a reconstruction based upon
 measured or observed quantities, but rather a ``post-diction'' based on extrapolation.
This method is often used for short-term predictions, but it can hardly be
 used for the reliable long-term reconstruction of solar activity.
In particular, it assumes that the sunspot time series is stationary, i.e., a limited time
 realization contains full information on its future and past.
Clearly such models cannot include periods exceeding the time span of
 observations upon which the extrapolation is based.
Hence, the pre- or post-diction becomes increasingly unreliable with growing
 extrapolation time and its accuracy is hard to estimate.


Sometimes a combination of the above approaches is used, i.e.,
 a fit of the mathematical model to indirect qualitative proxy data.
In such models a mathematical extrapolation of the sunspot series is slightly tuned
 and fitted to some proxy data for earlier times.
For example, \citet{schove55, schove79} fitted the slightly variable but phase-locked carrier frequency
 (about 11~years) to fragmentary data from naked-eye sunspot observations and auroral sightings.
The phase locking is achieved by assuming exactly nine solar cycles per calendar century.
This series, known as \emph{Schove} series, reflects qualitative long-term variations
 of the solar activity, including some grand minima, but cannot pretend to be a quantitative representation
 in solar activity level.
The Schove series played an important historical role in the 1960s.
In particular, a comparison of the $\Delta$\super{14}C data with this series succeeded
 in convincing the scientific community that secular variations of \super{14}C in tree rings
 have solar and not climatic origins \citep{stuiver61}.
This formed a cornerstone of the precise method of solar-activity reconstruction, which uses
 cosmogenic isotopes from terrestrial archives.
However, attempts to reconstruct the phase and amplitude of the 11-year cycle, using
 this method, were unsuccessful.
For example, \citet{schove55} made predictions of forthcoming solar cycles up to 2005, which failed.
We note that all these works are not able to reproduce, for example, the Maunder minimum
 (which cannot be represented as a result of the superposition of different harmonic
 oscillations), yielding too high sunspot activity compared to that observed.
From the modern point of view, the Schove series can be regarded as archaic,
 but it is still in use in some studies.

\subsection{The solar cycle and its variations}
\label{sec:cycle}

\subsubsection{Quasi-periodicities}

The main feature of solar activity is its pronounced quasi-periodicity
 with a period of about 11~years, known as the Schwabe cycle.
However, the cycle varies in both amplitude and duration.
The first observation of a possible regular variability in sunspot numbers was made
 by the Danish astronomer Christian Horrebow in the 1770s on the basis
 of his sunspot observations from 1761\,--\,1769 \citep[see details in][]{gleissberg52, vitinsky65}, but
 the results were forgotten.
It took over 70~years before the amateur astronomer Schwabe announced in 1844
 that sunspot activity varies cyclically with a period of about 10~years.
This cycle, called the 11-year or Schwabe cycle, is the most prominent variability in the sunspot-number series.
It is recognized now as a fundamental feature of solar activity originating from the
 solar-dynamo process.
This 11-year cyclicity is prominent in many other parameters including solar, heliospheric, geomagnetic, space weather, climate and others.
The background for the 11-year Schwabe cycle is the 22-year Hale magnetic polarity cycle.
Hale found that the polarity of sunspot magnetic fields
 changes in both hemispheres when a new 11-year cycle starts \citep{hale1919}.
This relates to the reversal of the global magnetic field of the sun with the period of 22~years.
It is often considered that the 11-year Schwabe cycle is the modulo of the
 sign-alternating Hale cycle \citep[e.g.,][]{sonett83, bracewell86, kurths90, demeyer98, mininni01},
 but this is only a mathematical representation.
A detailed review of solar cyclic variability can be found in \citep{hathawayLR}.

Sometimes the regular time evolution of solar activity is broken up by
 periods of greatly depressed activity called grand minima.
The last grand minimum (and the only one covered by direct solar observations)
 was the famous Maunder minimum from 1645\,--\,1715 \citep{eddy76, eddy83}.
Other grand minima in the past, known from cosmogenic isotope data, include, e.g., the Sp\"orer minimum around 1450\,--\,1550 and the Wolf
 minimum around the 14th century (see the detailed discussion in Section~\ref{sec:MM}).
Sometimes the Dalton minimum (ca.\ 1790\,--\,1820) is also considered to be a grand minimum.
However, sunspot activity was not completely suppressed
 and still showed Schwabe cyclicity during the Dalton minimum.
As suggested by \citet{schussler97}, this can be a separate, intermediate
 state of the dynamo between the grand minimum and normal activity, or an
 unsuccessful attempt of the sun to switch to the grand minimum state \citep{frick97, sokoloff04}.
This is observed as the phase catastrophe of solar-activity evolution \citep[e.g.,][]{vitinsky86, kremliovsky94}.
A peculiarity in the phase evolution of sunspot activity around 1800 was also
 noted by \citet{sonett83}, who ascribed it to a possible error in Wolf sunspot data and by \citet{wilson88},
 who reported on a possible misplacement of sunspot minima for cycles 4\,--\,6 in the WSN series.
It has been also suggested that the phase catastrophe can be related to
 a tiny cycle, which might have been lost at the end of the 18th century because
 of very sparse observations \citep{usoskin_lost_AA_01, usoskin_lost_GRL_02, usoskin_lost_03, zolotova07}.
We note that a new independent evidence proving the existence of the lost cycle has been found recently
 in the reconstructed sunspot butterfly diagram for that period \citep{usoskin_lost_09}.

The long-term change (trend) in the Schwabe cycle amplitude
 is known as the secular Gleissberg cycle \citep{gleissberg39}
 with the mean period of about 90~years.
However, the Gleissberg cycle is not a cycle in the strict periodic sense
 but rather a modulation of the cycle envelope with a varying timescale of 60\,--\,120~years
 \citep[e.g.,][]{gleissberg71, kuklin76, ogurtsov02}.

Longer (super-secular) cycles cannot be studied using direct solar observations, but only indicatively
 by means of indirect proxies such as cosmogenic isotopes discussed in Section~\ref{S:4}.
Analysis of the proxy data also yields the Gleissberg secular cycle \citep{feynman90, peristykh03},
 but the question of its phase locking and persistency/intermittency still remains open.
Several longer cycles have been found in the cosmogenic isotope data.
A cycle with a period of 205\,--\,210~years, called the de~Vries or Suess cycle in different sources,
 is a prominent feature, observed in various cosmogenic data \citep[e.g.,][]{suess80, sonett90, zhentao90, usoskin_AA_04}.
Sometimes variations with a characteristic time of 600\,--\,700~years or 1000\,--\,1200 years are discussed
 \citep[e.g.,][]{vitinsky86, sonett90, vassiliev02,steinhilber12,abreu12}, but they are intermittent and can hardly be
 regarded as a typical feature of solar activity.
A 2000\,--\,2400-year cycle is also noticeable in radiocarbon data series \citep[see, e.g.,][]{vitinsky86, damon91, vassiliev02}.
However, the non-solar origin of these super-secular cycles (e.g., geomagnetic or climatic variability) cannot be excluded.

\subsubsection{Randomness vs.\ regularity}
\label{S:RvR}

The short-term (days - months) variability of sunspot numbers is greater than the
 observational uncertainties indicating the presence of random fluctuations (noise).
As typical for most real signals, this noise is not uniform (white), but rather
 red or correlated noise \citep[e.g.,][]{ostryakov90b, oliver96, frick97}, namely,
 its variance depends on the level of the signal.
While the existence of regularity and randomness in sunspot series is apparent,
 their relationship is not clear \citep[e.g.,][]{wilson94} --
 are they mutually independent or intrinsically tied together?
Moreover, the question of whether randomness in sunspot data
 is due to chaotic or stochastic processes is still open.

Earlier it was common to describe sunspot activity as a \emph{multi-harmonic
 process} with several basic harmonics \citep[e.g.,][]{vitinsky65, sonett83, vitinsky86}
 with an addition of random noise, which plays no role in the solar-cycle evolution.
However, it has been shown \citep[e.g.,][]{rozelot94, weiss00, charbonneau01, mininni02}
 that such an oversimplified approach depends on the chosen reference time
 interval and does not adequately describe the long-term evolution of solar activity.
A multi-harmonic representation is based on an assumption of the stationarity of the
 benchmark series, but this assumption is broadly invalid for solar activity
 \citep[e.g.,][]{kremliovsky94, sello00, polygiannakis03}.
Moreover, a multi-harmonic representation cannot, for an apparent reason, be
 extrapolated to a timescale larger than that covered by the benchmark series.
The fact that purely mathematical/statistical models cannot give
 good predictions of solar activity (as will be discussed later) implies
 that the nature of the solar cycle is not a multi-periodic or other purely
 deterministic process, but random (chaotic or stochastic) processes play an essential
 role in sunspot cycle formation \citep[e.g.,][]{moss08,kapyla12}.
An old idea of the possible planetary influence on the dynamo has received a new pulse recently
 with some unspecified torque effect on the assumed quasi-rigid non-axisymmetric
 tahocline \citep{abreu12}.
If confirmed this idea would imply a significant multi-harmonic driver of the solar activity,
 but the question is still open.
Different numeric tests, such as an analysis of the Lyapunov exponents \citep{ostryakov90, mundt91,
 kremliovsky95, sello00}, Kolmogorov entropy \citep{carbonell94, sello00} and Hurst exponent \citep{ruzmaikin94, oliver98},
 confirm the chaotic/stochastic nature of the solar-activity time evolution \citep[see, e.g., the recent review by][]{panchev07}.

It was suggested quite a while ago that the variability of the solar cycle may be
 a temporal realization of a \emph{low-dimensional chaotic system} \citep[e.g.,][]{ruzmaikin81}.
This concept became popular in the early 1990s, when many authors considered solar activity as
 an example of low-dimensional deterministic chaos, described
 by the strange attractor \citep[e.g.,][]{kurths90, ostryakov90, morfill91, mundt91, rozelot95, salakhutdinova99, serre00,hanslmeier13}.
Such a process naturally contains randomness, which is an intrinsic feature of the system
 rather than an independent additive or multiplicative noise.
However, although this approach easily produces features seemingly similar to those of solar activity,
 quantitative parameters of the low-dimensional attractor have varied greatly as obtained by
 different authors.
Later it was realized that the analyzed data set was too short \citep{carbonell93, carbonell94},
 and the results were strongly dependent on the choice of filtering methods \citep{price92}.
Developing this approach, \citet{mininni00, mininni01} suggest that one consider sunspot
 activity as an example of a 2D Van der Pol relaxation oscillator with an
 intrinsic stochastic component.

Such phenomenological or basic principles models, while succeeding in reproducing
 (to some extent) the observed features of solar-activity variability, do not
 provide insight into the nature of regular and random components of solar variability.
In this sense efforts to understand the nature of randomness in sunspot activity in the
 framework of dynamo theory are more advanced.
Corresponding theoretical dynamo models have been developed \citep[see reviews by][]{ossendrijver03, charbonneau10},
 which include stochastic processes
 \citep[e.g.,][]{weiss84, feynman90, schmaltz91, moss92, hoyng93, brooke94, lawrence95, schmitt96, charbonneau00, brandenburg02}.
For example, \citet{feynman90} suggest that the transition from a regular to
 a chaotic dynamo passes through bifurcation.
\citet{charbonneau00} studied stochastic fluctuations in a Babcock--Leighton dynamo
 model and succeeded in the qualitative reproduction of the anti-correlation
 between cycle amplitude and length (Waldmeier rule).
Their model also predicts a phase-lock of the Schwabe cycle, i.e., that
 the 11-year cycle is an internal ``clock'' of the sun.
Most often the idea of fluctuations is related to the $\alpha$-effect, which is the result of the
 electromotive force averaged over turbulent vortices, and thus can contain a fluctuating contribution
 \citep[e.g.,][]{hoyng93, ossendrijver96, brandenburg08, moss08}.
Note that a significant fluctuating component (with the amplitude more than
 100\% of the regular component) is essential in all these model.

\subsubsection{A note on solar activity predictions}

Randomness (see Section~\ref{S:RvR}) in the SN series is directly related to the predictability of solar activity.
Forecasting solar activity has been a subject of intense study for many years
 \citep[e.g.,][]{yule927, newton928, gleissberg48, vitinsky65} and has greatly intensified recently with a
 hundred of journal articles being published to predict the solar cycle No.~24 maximum \citep[see, e.g., the review by][]{pesnell12},
 following the boost of space-technology development and increasing debates on solar-terrestrial relations.
In fact, the situation has not been improved since the previous cycle, No. 23.
The predictions for the peak sunspot number of solar cycle No. 24 range by a factor of 5,
 between 40 and 200, reflecting the lack of a reliable consensus method
 \citep{tobias06}.
Detailed review of the solar activity prediction methods and results have been recently provided by
 \citep{hathaway09,petrovay10,pesnell12}.

A detailed classification of the prediction methods is given by \citet{pesnell12} who
 separates climatology, precursor, theoretical (dynamo model), spectral, neural network, and
 stock market prediction methods.
All prediction methods can be generically divided into precursor and statistical (including the
 majority of the above classifications) techniques or their combinations \citep{hathaway99}.
The fact that the prediction of solar cycle is not improved with adding more data (the new solar cycle)
 suggests that such methods are not able to give reliable prognoses.

The precursor methods are usually based on phenomenological, but sometimes physical, links between the
 poloidal solar-magnetic field, estimated, e.g., from geomagnetic activity
 in the declining phase of the preceding cycle or in the minimum time \cite[e.g.,][]{hathaway09},
  with the toroidal field responsible for sunspot formation.
These methods usually yield better short-term predictions of a forthcoming cycle maximum than the statistical
 methods, but cannot be applied to timescales longer than one solar cycle.

Statistical methods, including a low-dimensional
 solar-attractor representation \citep{kurths90}, are based solely on the statistical
 properties of sunspot activity and may give a reasonable result for short-term forecasting, but
 yield very poor results for long-term predictions
 \cite[see reviews by, e.g.,][]{conway98, hathaway99, li01, usoskin_SP03, kane07}
 because of chaotic/stochastic behavior (see Section~\ref{S:RvR}).

A new method based on sophisticated dynamo numerical simulations emerges \citep[e.g.,][]{dikpati06,dikpati08,choudhuri07,jiang07},
 but the results are contradictory with each other.
Prospectives of this approach are also not clear because of the stochastic component, which drives the
 dynamo out of the deterministic regime, and uncertainties in the input parameters \citep{tobias06,bushby07,karak12}.

Some models, mostly based on precursor method, succeed in reasonable predictions of a forthcoming
 solar cycle (i.e., several years ahead), but they do not pretend to extend further in time.
On the other hand, many claims of the solar activity forecast for 40\,--\,50~years ahead and even beyond
 have been made recently, often without sensible argumentation.
However, so far there is no evidence of any method giving a reasonable prediction of solar activity beyond
 the solar-cycle scale (see, e.g., Section~\ref{sec:math}), probably because of the intrinsic
 limit of solar-activity predictability due to its stochastic/chaotic nature \citep{kremliovsky95,tobias06}.
Accordingly, such attempts can be regarded as speculative, unless they are verified
 by the actual behavior of solar activity.
Note that even an exact prediction of the amplitude of one solar cycle can be just a random
 coincidence and cannot serve as a proof of the method's veracity.
Only a sequence of successful predictions can form a basis for confidence, which requires
 several decades.

Note that several ``predictions'' of the general decline of the coming solar activity have been
 made recently \citep{solanki_Nat_04,abreu08,lockwood11}, however, these are not really true predictions
 but rather the acknowledge of the fact that the Modern Grand maximum \citep{usoskin_PRL_03,solanki_Nat_04}
 must cease.
Similar caution can be made about predictions of a Grand minimum \citep[e.g.,][]{lockwood11,miyahara10} --
 a grand minimum should appear soon or later, but presently we are hardly able to predict its occurrence.

\subsection{Summary}

In this section, the concept of solar activity and quantifying indices is discussed,
 as well as the main features of solar-activity temporal behavior.

The concept of solar activity is quite broad and covers non-stationary and non-equilibrium
 (often eruptive) processes, in contrast to the ``quiet'' sun concept, and their effects
 upon the terrestrial and heliospheric environment.
Many indices are used to quantify different aspects of variable solar activity.
Quantitative indices include direct (i.e., related directly to solar variability)
 and indirect (i.e., related to terrestrial and interplanetary effects caused by solar activity),
 they can be physical or synthetic.
While all indices depict the dominant 11-year cyclic variability, their relationships on other
 timescales (short scale or long-term trends) may vary to a great extent.

The most common and the longest available index of solar activity is the sunspot number,
 which is a synthetic index and is very useful for the quantitative
 representation of overall solar activity outside the grand minimum.
During the grand Maunder minimum, however, it may give only a clue about solar activity
 whose level may drop below the sunspot formation threshold.
The sunspot number series is available for the period from 1610~AD, after the invention of the telescope,
 and covers, in particular, the Maunder minimum in the late 17th century.
Fragmentary non-instrumental observations of the sun before 1610, while giving a possible hint of
 relative changes in solar activity, cannot be interpreted in a quantitative manner.

Solar activity in all its manifestations is dominated by the 11-year Schwabe cycle, which
 has, in fact, a variable length of 9\,--\,14~years for individual cycles.
The amplitude of the Schwabe cycle varies greatly -- from the almost spotless Maunder minimum to
 the very high cycle~19, possibly in relation to the Gleissberg or secular cycle.
Longer super-secular characteristic times can also be found in various proxies of solar activity,
 as discussed in Section~\ref{sec:3}.

Solar activity contains essential chaotic/stochastic components, that lead to irregular
 variations and make the prediction of solar activity for a timescale exceeding one solar cycle impossible.

\newpage


\section{The Proxy Method of Past Solar-Activity Reconstruction}
\label{S:4}

In addition to direct solar observations, described in Section~\ref{S:dir},
 there are also indirect solar proxies, which are used to study solar activity in
 the pre-telescopic era.
Unfortunately, we do not have any reliable data that could give a direct index of
 solar variability before the beginning of the sunspot-number series.
Therefore, one must use indirect proxies, i.e., quantitative parameters, which can be measured nowadays
 but represent different effects of solar magnetic activity in the past.
It is common to use, for this purpose, signatures of terrestrial indirect effects
 induced by variable solar-magnetic activity, that is stored in natural archives.
Such traceable signatures can be related to nuclear (used in the cosmogenic-isotope method) or chemical
 (used, e.g., in the nitrate method) effects caused by cosmic rays (CRs) in the Earth's atmosphere, lunar
 rocks or meteorites.

The most common proxy of solar activity is formed by the data on cosmogenic radionuclides
 (e.g., \super{10}Be and \super{14}C), which are produced by cosmic rays in the Earth's atmosphere
 \cite[e.g,][]{stuiver80, beer90, bard97, beer00}.
Other cosmogenic nuclides, which are used in geological and paleomagnetic dating, are less suitable
 for studies of solar activity \citep[see e.g.,][]{beer00,beer12}.
Cosmic rays are the main source of cosmogenic nuclides in the atmosphere (excluding
 anthropogenic factors during the last decades) with the maximum production being in the
 upper troposphere/stratosphere.
After a complicated transport in the atmosphere, the cosmogenic isotopes are stored in natural
 archives such as polar ice, trees, marine sediments, etc.
This process is also affected by changes in the geomagnetic field and climate.
Cosmic rays experience heliospheric modulation due to solar wind and the frozen-in
 solar magnetic field.
The intensity of modulation depends on solar activity and, therefore,
 cosmic-ray flux and the ensuing cosmogenic isotope intensity depends inversely on solar activity.
An important advantage of the cosmogenic data is that primary archiving is done naturally in a similar
 manner throughout the ages, and these archives are measured nowadays in laboratories using modern techniques.
If necessary, all measurements can be repeated and improved, as has been done for some radiocarbon samples.
In contrast to fixed historical archival data (such as sunspot or auroral observations)
 this approach makes it possible to obtain homogeneous data sets of stable quality and to
 improve the quality of data with the invention of new methods (such as accelerator mass spectrometry).
Cosmogenic isotope data is the main regular indicator of solar activity on the
 very long-term scale but it cannot resolve the details of individual solar cycles.
The redistribution of nuclides in terrestrial reservoirs and archiving may be affected
 by local and global climate/circulation processes, which are, to a large extent, unknown for the past.
However, a combined study of different nuclides data, whose responses to terrestrial effects
 are very different, may allow for disentangling external and terrestrial signals.

\subsection{The physical basis of the method}

\subsubsection{Heliospheric modulation of cosmic rays}
\label{S:mod}

The flux of cosmic rays (highly energetic fully ionized nuclei) is considered roughly
 constant (at least at the time scales relevant for the present study) in the vicinity of the
 Solar system.
However, before reaching the vicinity of Earth, galactic cosmic rays experience
 complicated transport in the heliosphere that leads to modulation of their flux.
Heliospheric transport of GCR is described by Parker's theory \citep{parker65, toptygin85}
 and includes four basic processes:
the diffusion of particles due to their scattering on magnetic inhomogeneities,
the convection of particles by out-blowing solar wind,
adiabatic energy losses in expanding solar wind,
drifts of particles in the magnetic field, including the gradient-curvature drift
 in the regular heliospheric magnetic field, and the drift along the heliospheric current sheet,
 which is a thin magnetic interface between the two heliomagnetic hemispheres.
Because of variable solar-magnetic activity, CR flux in the vicinity of Earth is
 strongly modulated (see Figure~\ref{Fig:NM_SN}).
The most prominent feature in CR modulation is the 11-year cycle, which is in inverse relation to solar activity.
The 11-year cycle in CR is delayed (from a month up to two years) with respect to the sunspots \citep{usoskin_JGR_98}.
The time profile of cosmic-ray flux as measured by a neutron monitor (NM) is shown in Figure~\ref{Fig:NM_SN} (panel~b) together with the sunspot numbers (panel~a).
Besides the inverse relation between them, some other features can also be noted.
A 22-year cyclicity manifests itself in cosmic-ray modulation through the alteration of sharp and flat maxima in cosmic-ray data,
 originated from the charge-dependent drift mechanism.
One may also note short-term fluctuations, which are not directly related to sunspot numbers but are
 driven by interplanetary transients caused by solar eruptive events, e.g., flares or CMEs.
An interesting feature is the increase of CR flux in 2009, when it was the highest ever recorded by NMs
 \citep{moraal10}, as caused by the favorable heliospheric conditions (unusually weak heliospheric magnetic field and
 the flat heliospheric current sheet) \citep{mcdonald10}.
For the previous 50~years of high and roughly-stable solar activity, no trends have been observed in CR data;
 however, as will be discussed later, the overall level of CR has changed significantly on the centurial-millennial
 timescales.

\epubtkImage{LR_NM_SN.png}{%
  \begin{figure}[htbp]
    \centerline{\includegraphics[width=0.6\textwidth]{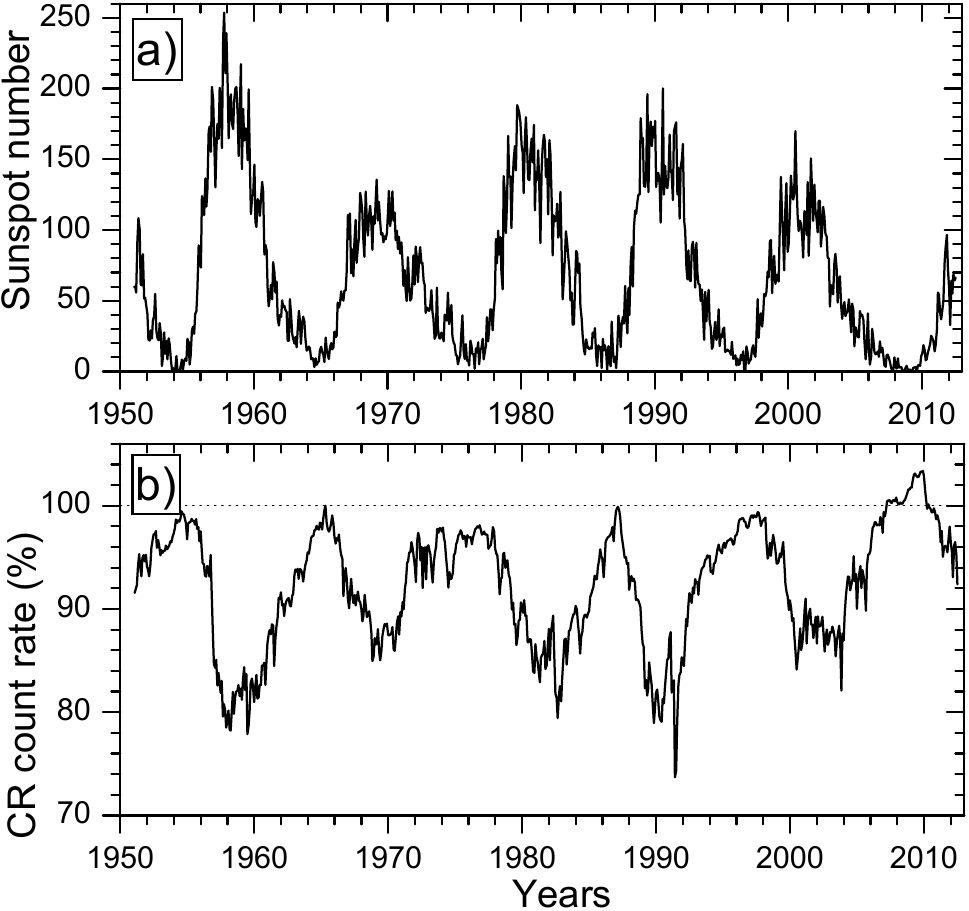}}
    \caption{Cyclic variations since 1951.
    \emph{Panel a:} Time profiles of sunspot numbers (\url{http://sidc.oma.be/sunspot-data/});
    \emph{Panel b:} Cosmic-ray flux as the count rate of a polar neutron monitor (Oulu NM \url{http://cosmicrays.oulu.fi}, Climax NM data used before 1964),
     100\% NM count rate corresponds to May 1965.}
    \label{Fig:NM_SN}
\end{figure}}

Full solution of the CR transport problems is a complicated task and requires sophisticated 3D time-dependent
 self-consistent modelling.
However, the problem can be essentially simplified for applications at a long-timescale.
An assumption on the azimuthal symmetry (requires times longer that
 the solar-rotation period) and quasi-steady changes reduces it to a 2D quasi-steady problem.
Further assumption of the spherical symmetry of the heliosphere reduces the problem to a 1D case.
This approximation can be used only for rough estimates, since it neglects the drift effect, but
 it is useful for long-term studies, when the heliospheric parameters cannot be evaluated independently.
Further, but still reasonable, assumptions (constant solar-wind speed, roughly power-law
 CR energy spectrum, slow spatial changes of the CR density) lead to the force-field approximation
 \citep{gleeson68}, which can be solved analytically.
The differential intensity $J_i$ of the cosmic-ray nuclei of type $i$ with kinetic energy $T$ at 1~AU is given
 in this case as
\begin{equation}
  J_i(T,\phi)=J_{{\mathrm{LIS},}i}(T+\Phi_i){(T)(T+2T_{\mathrm{r}})\over
  (T+\Phi_i)(T+\Phi_i+2T_{\mathrm{r}})} \,,
\label{Eq:ff_sol}
\end{equation}
where $\Phi_i=(Z_i e/A_i)\phi$ for a cosmic nuclei of $i$-th type (charge and mass numbers are $Z_i$
 and $A_i$), $T$ and $\phi$ are expressed in MeV/nucleon and in MV,
 respectively, $T_{\mathrm{r}}=938\mathrm{\ MeV}$.
$T$ is the CR particle's kinetic energy, and $\phi$ is the modulation potential.
The local interstellar spectrum (LIS) $J_{\mathrm{LIS}}$ forms the boundary condition for the heliospheric transport problem.
Since LIS is not measured directly, i.e., outside the heliosphere, it is not well known in the
 energy range affected by CR modulation (below 100~GeV).
Presently-used approximations for LIS \citep[e.g.,][]{garcia75, burger00, webber03, webber09} agree with each other
 for energies above 20~GeV but may contain uncertainties of up to a factor of 1.5 around 1~GeV.
These uncertainties in the boundary conditions make the results of the modulation theory slightly model-dependent
 \citep[see discussion in][]{usoskin_Phi_05,herbst10} and require the LIS model to be explicitly cited.
This approach gives results, which are at least dimensionally consistent with the full theory and can be
 used for long-term studies\epubtkFootnote{Note that the famous work by \citet{castagnoli80} contains
 an inconsistency in the force-field formula -- see details in \citet{usoskin_Phi_05},} \citep[]{usoskin_JGR_02, caballero04}.
Differential CR intensity is described by the only time-variable parameter, called the
 modulation potential $\phi$, which is mathematically interpreted as the averaged rigidity (i.e., the
 particle's momentum per unit of charge) loss of a CR particle in the heliosphere.
However, it is only a formal spectral index whose physical
 interpretation is not straightforward, especially on short timescales and during active periods
 of the sun \citep{caballero04}.
Despite its cloudy physical meaning, this force-field approach provides a very useful and simple single-parametric approximation
 for the differential spectrum of GCR, since the spectrum of different GCR species directly
 measured near the Earth can be perfectly fitted by Equation~\ref{Eq:ff_sol} using only the parameter $\phi$ in a wide range
 of solar activity levels \citep{usoskin_bazi_11}.
Therefore, changes in the whole energy spectrum (in the energy range from 100~MeV/nucleon to 100~GeV/nucleon) of cosmic rays
 due to the solar modulation can be described by this single number within the
 framework of the adopted LIS.
The concept of modulation potential is a key concept for the method of solar-activity
 reconstruction by cosmogenic isotope proxy
 as it makes it possible to parameterize the GCR with one single parameter.

\subsubsection{Geomagnetic shielding}
\label{S:geom}

Cosmic rays are charged particles and therefore are affected by the Earth's magnetic field.
Thus the geomagnetic field puts an additional shielding on the incoming flux of cosmic rays.
The shielding effect of the geomagnetic field is usually expressed in terms of the cutoff rigidity
 $P_{\mathrm{c}}$, which is the minimum rigidity a vertically incident CR particle
 must posses (on average) in order to reach the ground at a given location and time \citep{cooke91}.
Neglecting such effects as the East-West asymmetry, which is roughly averaged out for the isotropic particle
 flux, or nondipole magnetic momenta, which decay rapidly with distance, one can come to
 a simple approximation, called the St\"ormer's equation, that describes
 the vertical geomagnetic cutoff rigidity $P_{\mathrm{c}}$:
\begin{equation}
  P_{\mathrm{c}}\approx 1.9\,M\ \left({R_{\mathrm{o}}/ R}\right)^2\ \cos^4
  \lambda_G \ \mathrm{[GV]} \,,
\label{Eq:Pc}
\end{equation}
where $M$ is the geomagnetic dipole moment (in 10\super{25}~G~cm\super{3}),
 $R_{\mathrm{o}}$ is the Earth's mean radius, $R$ is the distance from the given location
 to the dipole center, and $\lambda_G$ is the geomagnetic latitude.
The cutoff concept works like a Heaviside step-function so that all cosmic rays whose rigidity is
 below the cutoff are not allowed to enter the atmosphere while all particles with higher rigidity
 can penetrate.
This approximation provides a good compromise between simplicity and reality,
 especially when using the eccentric dipole description of the geomagnetic field \citep{fraser87}.
The eccentric dipole has the same dipole moment and orientation as the centered dipole,
 but the dipole's center and consequently the poles, defined
 as crossings of the axis with the surface, are shifted with respect to geographical ones.

The shielding effect is the strongest at the geomagnetic equator, where the present-day
 value of $P_{\mathrm{c}}$ may reach up to 17~GV in the region of India.
There is almost no cutoff in the geomagnetic polar regions ($\lambda_G\ge 60^\circ$).
However, even in the latter case the atmospheric cutoff becomes important, i.e., particles must
 have rigidity above 0.5~GV in order to initiate the atmospheric cascade which
 can reach ground (see Section~\ref{S:cas}).

The geomagnetic field is seemingly stable on the short-term scale, but it changes essentially
 on centurial-to-millennial timescales \citep[e.g.,][]{korte06}.
Such past changes can be evaluated based on measurements of the residual magnetization of
 independently-dated samples.
These can be paleo- (i.e., natural stratified archives such as lake or
 marine sediments or volcanic lava)
 or archaeological (e.g., clay bricks that preserve magnetization upon baking) samples.
Most paleo-magnetic data preserve not only the magnetic field intensity but also the direction of
 the local field, while archeo-magnetic samples provide information on the intensity only.
Using a large database of such samples, it is possible to reconstruct (under reasonable assumptions)
 the large-scale magnetic field of the Earth.
Data available provides good global coverage for the last 3 millennia, allowing for a reliable paleomagnetic reconstruction
 of the true dipole moment (DM) or virtual dipole moment\epubtkFootnote{The concept of VDM assumes that the geomagnetic dipole
  is centered at the planet's center and its axis is aligned with the true magnetic axis.} (VDM)
   and its orientation \citep[the ArcheoInt collaboration --][]{genevey08}.
Less precise, but still reliable reconstructions of the DM and its orientation are possible
 for the last seven millennia \citep[the CALS7K.2 model by][]{korte05}, however
 they may slightly underestimate the dipole moment, especially in the earlier part of the period \citep{korte08}.
Directional paleomagnetic reconstruction are less reliable on a longer timescale,
 because of the spatial sparseness of the paleo/archeo-magnetic samples in the earlier part of the Holocene \citep{korte11}.
Some paleomagnetic reconstructions are shown in Figure~\ref{Fig:geom}.
All paleomagnetic models depict a similar long-term trend -- an enhanced intensity during the
 period between 1500~BC and 500~AD and a significantly lower field before that.

Changes in the dipole moment $M$ inversely modulate the flux of CR at Earth, with strong effects
 in tropical regions and globally.
The migration of the geomagnetic axis, which changes the geomagnetic latitude $\lambda_G$
 of a given geographical location is also important;
 while not affecting the global flux of CR, it can dramatically change the CR effect regionally,
 especially at middle and high latitude.
These changes affect the flux of CR impinging on the Earth's atmosphere both locally and globally
 and must be taken into account when reconstructing solar activity from terrestrial proxy data
 \citep{usoskin_CRII_08,usoskin_Geo_10}.
Accounting for these effects is quite straightforward provided the geomagnetic changes in the past
 are known independently, e.g., from archeo and paleo-magnetic studies \citep{donadini10}.
However, because of progressively increasing uncertainties of paleomagnetic reconstructions back in time,
 it presently forms the main difficulty for the proxy method on the long-term scale \citep{snowball07},
 especially in the early part of the Holocene.
On the other hand, the geomagnetic field variations are relatively well
 known for the last few millennia \citep{genevey08, korte08}.

\epubtkImage{LR_Geom.png}{%
  \begin{figure}[htbp]
  \centerline{\includegraphics[width=0.8\textwidth]{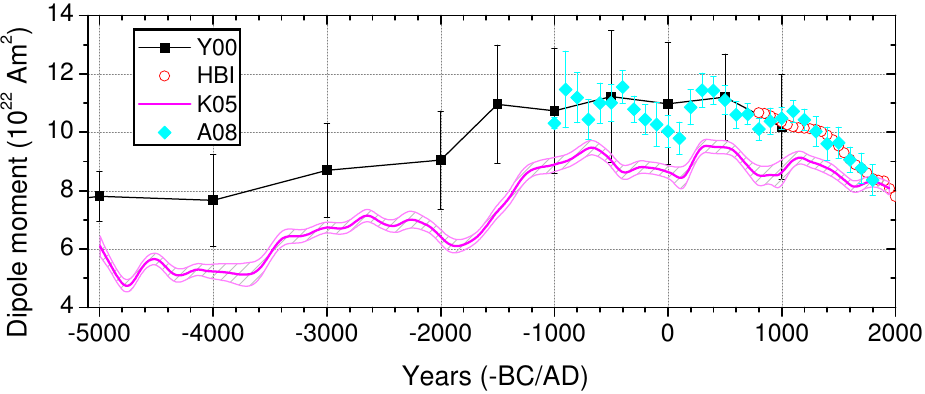}}
  \caption{Geomagnetic field intensity over millennia: VADM compilation by
  \citet[][-- Y00 curve with $1\sigma$ statistical errors of the
  sample distribution]{yang00}; dipole moment according to \citet[][--
  HBI red dots]{hongre98} since 800~AD, dipole moment according to
  CALS7K.2 model \citep[][-- K05 magenta curve with $1\sigma$ error
  band]{korte05} as well as a recent ArcheoInt compilations of VADM/VDM
  \citep[][-- A08 azure diamonds]{genevey08}.}
  \label{Fig:geom}
\end{figure}}


\subsubsection{Cosmic-ray--induced atmospheric cascade}
\label{S:cas}

When an energetic CR particle enters the atmosphere, it first moves straight in the upper layers,
 suffering mostly from ionization energy losses that lead to the ionization of the ambient rarefied air
 and gradual deceleration of the particles.
However, after traversing some amount of matter (the nuclear interaction mean-free path
 is on the order of 100~g/cm\super{2} for a proton in the air)
 the CR particle may collide with a nucleus in the atmosphere, producing
 a number of secondaries.
These secondaries have their own fate in the atmosphere, in particular they
 may suffer further collisions and interactions forming an atmospheric cascade \citep[e.g.,][]{dorman04}.
Because of the thickness of the Earth's atmosphere (1033~g/cm\super{2} at sea level) the number
 of subsequent interactions can be large, leading to a fully-developed cascade (also called an air shower)
 consisting of secondary rather than primary particles.
A schematic view of the atmospheric cascade is shown in Figure~\ref{Fig:cascade}.
\epubtkImage{LR_cascade.png}{%
  \begin{figure}[htb]
  \centerline{\includegraphics[width=9cm]{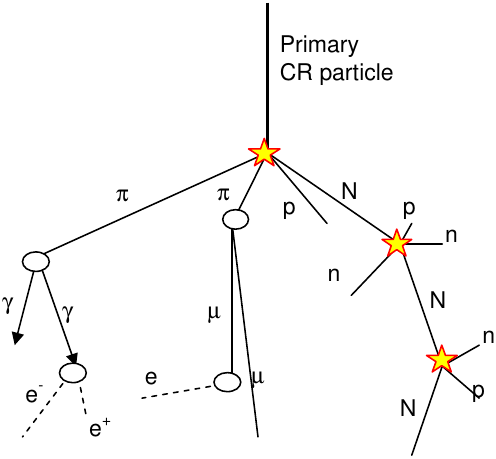}}
  \caption{Schematic view of an atmospheric cascade caused by energetic cosmic rays in the atmosphere.
Left-to-right are denoted, respectively, the soft, muon and hadronic components of the cascade.
Symbols ``N, p, n, $\mu$, $\pi$, e\super{-}, e\super{+}, and $\gamma$'' denote nuclei, protons, neutrons, muons, pions,
 electrons, positrons, and photons, respectively.
Stars denote nuclear collisions, ovals -- decay processes.
This sketch does not represent the full development of the cascade and serves solely as
an illustration for the processes discussed in the text.
Image reproduced by permission from~\cite{usoskin_MSI_11}, copyright by SAIt.}
  \label{Fig:cascade}
\end{figure}}
Three main components can be separated in the cascade:

\begin{itemize}

\item The \emph{``hadronic'' nucleonic} component is formed by the
  products of nuclear collisions of primary cosmic rays and their
  secondaries with the atmospheric nuclei, and consists mostly of
  superthermal protons and neutrons.

\item The \emph{``soft''} or electromagnetic component consists of
  electrons, positrons and photons.

\item The \emph{``hard''} or muon component consists mostly of muons;
  pions are short lived and decay almost immediately upon production,
  feeding muons and the ``soft'' component.

\end{itemize}

The development of the cascade depends mostly on the amount of matter traversed and is usually linked to residual
 atmospheric depth, which is very close to the static barometric pressure, rather than to the actual altitude,
 that may vary depending on the exact atmospheric density profile.

Cosmogenic isotopes are a by-product of the hadronic branch of the cascade (details are given below).
Accordingly, in order to evaluate cosmic-ray flux from the cosmogenic isotope data, one
 needs to know the physics of cascade development.
Several models have been developed for this cascade, in particular its hadronic branch
 with emphasis on the generation of cosmogenic isotope production.
The first models were simplified quasi-analytical \citep[e.g.,][]{lingenfelter63, obrien73}
 or semi-empirical models \citep[e.g.,][]{castagnoli80}.
With the fast advance of computing facilities it became possible to exploit the best
 numerical method suitable for such problems -- Monte-Carlo
  \citep[e.g.,][]{masarik99, masarik09, webber03, webber07, usoskin_7Be_08, kovaltsov10,kovaltsov12,argento12}.
The fact that models, based on different independent Monte-Carlo packages, namely, a general
 GEANT tool and a specific CORSIKA code, yield similar results provides additional
 verification of the approach.


\subsubsection{Transport and deposition}

A scheme for the transport and redistribution of the two most useful cosmogenic isotopes,
 \super{14}C and \super{10}Be, is shown in Figure~\ref{Fig:scheme}.
After a more-or-less similar production, the two isotopes follow quite different
 fates, as discussed in detail in Sections~\ref{S:14C_cycle} and \ref{S:10Be_tran}.
Therefore, expected terrestrial effects are quite different for the isotopes
 and comparing them with each other can help in disentangling solar and climatic effects
 (see Section~\ref{S:comp}).
A reader can find great detail also in a book by \citet{beer12}.

\epubtkImage{LR_scheme.png}{%
  \begin{figure}[htbp]
    \centerline{\includegraphics[width=0.8\textwidth]{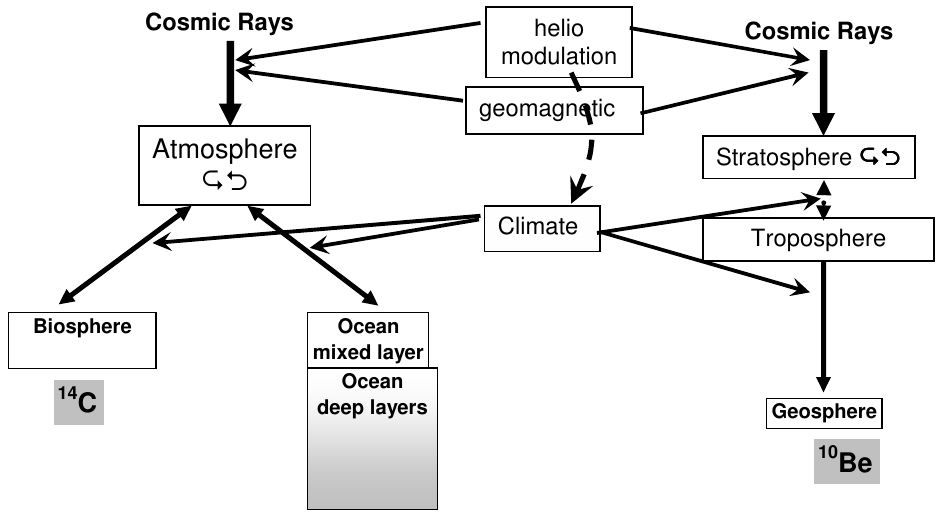}}
    \caption{Schematic representation of \super{14}C (left) and \super{10}Be
    (right) production chains. The flux of cosmic rays impinging on
    the Earth is affected by both heliospheric modulation and
    geomagnetic field changes. The climate may affect the
    redistribution of the isotopes between different
    reservoirs. Dashed line denotes a possible influence of solar
    activity on climate.}
    \label{Fig:scheme}
\end{figure}}

\subsection{Radioisotope \super{14}C}
\label{sec:14C}

The most commonly used cosmogenic isotope is radiocarbon \super{14}C.
This radionuclide is an unstable isotope of carbon with a
 half-life $\left(T_{1/2}\right)$ of about 5730~years.
Since the radiocarbon method is extensively used in other science disciplines where accurate dating
 is a key issue (e.g., archeology, paleoclimatology, quaternary geology), it was developed primarily for this task.
The solar-activity--reconstruction method, based on radiocarbon data, was initially
 developed as a by-product of the dating techniques used in archeology
 and Quaternary geology, in an effort to improve the quality of the
 dating by means of better information on the \super{14}C variable source function.
The present-day radiocarbon calibration curve, based on a dendrochronological scale, uninterruptedly covers
 the whole Holocene \citep[and extending to 50,000 BP --][]{reimer09} and provides a solid quantitative basis for studying solar
 activity variations on the multi-millennial time scale.

\subsubsection{Measurements}

Radiocarbon is usually measured in tree rings, which allows an
absolute dating of the samples by means of dendrochronology. Using a
complicated technique, the \super{14}C activity%
\epubtkFootnote{Isotope's activity quantifies (a) in the radiometric
  \super{14} technique its decay rate, and is usually given in terms
  of disintegrations per minute per gram of carbon, and (b) in the AMS
  technique, the \super{14}C/\super{12}C ratio, all normalized to the
  standard.} 
$A$ is measured in an independently dated sample, which is then
corrected for age as
\begin{equation}
A^* = A\cdot\exp{\left({0.693\, t\over T_{1/2}}\right)},
\end{equation}
where $t$ and $T_{1/2}$ are the age of the sample and the half-life of the isotope, respectively.
Then the relative deviation from the standard activity $A_o$ of oxalic acid (the National Bureau of Standards)
 is calculated:
\begin{equation}
\delta^{14}\mathrm{C} = \left({A^*-A_o\over A_o}\right)\cdot 1000 \,.
\label{Eq:delta}
\end{equation}
After correction for the carbon isotope fractionating (account for the \super{13}C isotope)
 of the sample, the radiocarbon value of $\Delta$\super{14}C is calculated \citep[see details in][]{stuiver77}.
\begin{equation}
\Delta^{14}\mathrm{C} = \delta^{14}\mathrm{C} -
(2\cdot\delta^{13}\mathrm{C} +50)\cdot(1+\delta^{14}\mathrm{C}/1000) \,,
\end{equation}
where $\delta$\super{13}C is the per mille deviation of the \super{13}C content in the sample from that
 in the standard belemnite sample calculated similarly to Equation~\ref{Eq:delta}.
The value of $\Delta$\super{14}C (measured in per mille $\permil$) is further
 used as the index of radiocarbon relative activity.
The series of $\Delta$\super{14}C for the Holocene is presented in Figure~\ref{Fig:Q}A
 as published by the INTCAL04 collaboration of 21 dating laboratories as a result of systematic precise measurements
 of dated samples from around the world \citep{reimer04}.
The most recent INTCAL09 dataset is available at \url{http://www.radiocarbon.org/IntCal09.htm}.

\epubtkImage{LR_14C_Q.png}{%
  \begin{figure}[htbp]
    \centerline{\includegraphics[width=10cm]{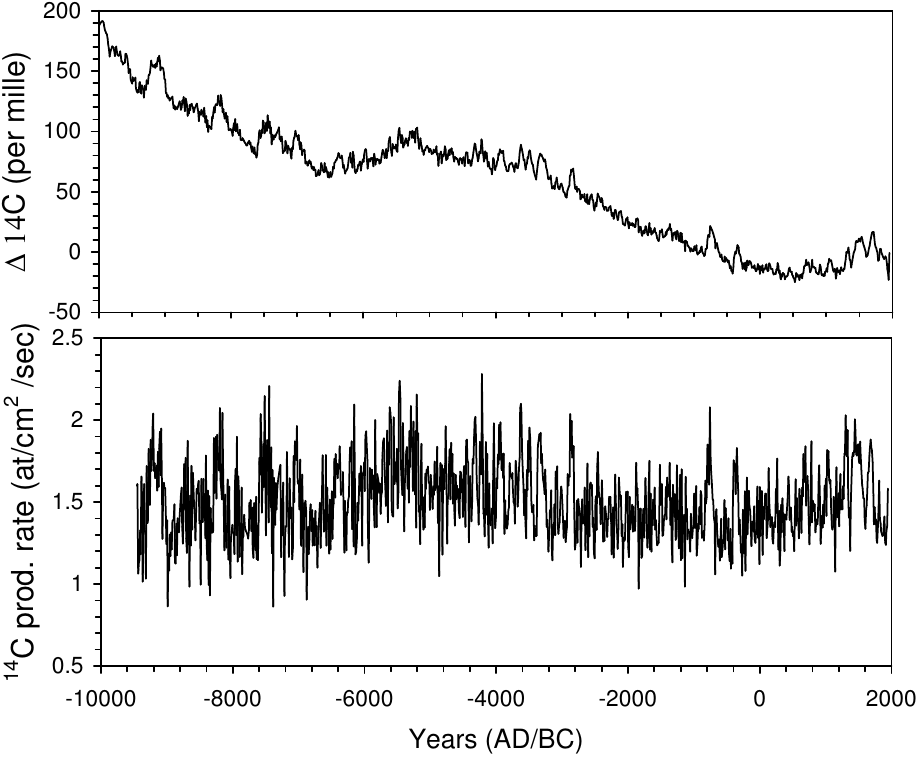}}
    \caption{Radiocarbon series for the Holocene. \emph{Upper panel:}
    Measured content of $\Delta$\super{14}C in tree rings by INTCAL-98/04
    collaboration \citep{stuiver98, reimer04}. The long-term trend is
    caused by the geomagnetic field variations and the slow response
    of the oceans. \emph{Lower panel:} Production rate of \super{14}C in the
    atmosphere, reconstructed from the measured $\Delta$\super{14}C
    \citep{usoskin_14C_05}.}
    \label{Fig:Q}
\end{figure}}

A potentially interesting approach has been made by \citet{lal05}, who
 measured the amount of \super{14}C directly produced by CR in polar ice.
Although this method is free of the carbon-cycle influence, the first results,
 while being in general agreement with other methods, are not precise.

\subsubsection{Production}
\label{S:14C_pro}

The main source of radioisotope \super{14}C (except anthropogenic sources during the last decades)
 is cosmic rays in the atmosphere.
It is produced as a result of the capture of a thermal neutron by atmospheric nitrogen
\begin{equation}
^{14}\mathrm{N} + n \rightarrow\, \, ^{14}\mathrm{C} + p.\,
\end{equation}
Neutrons are always present in the atmosphere as a product of the cosmic-ray--induced
 cascade (see Section~\ref{S:cas}) but their flux varies in time along with the modulation
 of cosmic-ray flux.
This provides continuous source of the isotope in the atmosphere, while the sinks are
 isotope decay and transport into other reservoirs as described below (the carbon cycle).

The connection between the cosmogenic-isotope--production rate, $Q$, at a given location
 (quantified via the geomagnetic latitude $\lambda_{\mathrm{G}}$) and the cosmic-ray flux is given by
\begin{equation}
  Q = \int^\infty_{P_{\mathrm{c}}(\lambda_{\mathrm{G}})}
  S(P,\phi)\ Y(P)\ dP \,,
\label{Eq:R}
\end{equation}
 where $P_{\mathrm{c}}$ is the local cosmic-ray--rigidity cutoff (see Section~\ref{S:geom}), $S(P,\phi)$ is
 the differential energy spectrum of CR
 (see Section~\ref{S:mod}) and $Y(P)$ is the differential yield function of cosmogenic
 isotope production, calculated using a Monte-Carlo simulation of the cosmic-ray--induced
 atmospheric cascade \citep{kovaltsov12}.
Because of the global nature of the carbon cycle and its long attenuation time, the
 radiocarbon is globally mixed before the final deposition, and Equation~\ref{Eq:R} should be
 integrated over the globe.
The yield function $Y(P)$ of the \super{14}C production is shown in Figure~\ref{Fig:yield}A together
 with  those for \super{10}Be (see Section~\ref{S:10Be_prod}) and for a ground-based neutron monitor
 (NM), which is the main instrument for studying cosmic-ray variability during the modern epoch.
One can see that the yield function increases with the energy of CR.
On the other hand, the energy spectrum of CR decreases with energy.
Accordingly, the differential production rate (i.e., the product of the yield
 function and the spectrum, $F=Y\cdot S$ -- the integrand of Equation~\ref{Eq:R}), shown in Figure~\ref{Fig:yield}B, is more informative.
The differential production rate reflects the sensitivity to cosmic rays, and the total production
 rate is simply an integral of $F$ over energy above the geomagnetic threshold.

\epubtkImage{LR_Yield.png}{%
  \begin{figure}[htb]
    \centerline{\includegraphics[width=10cm]{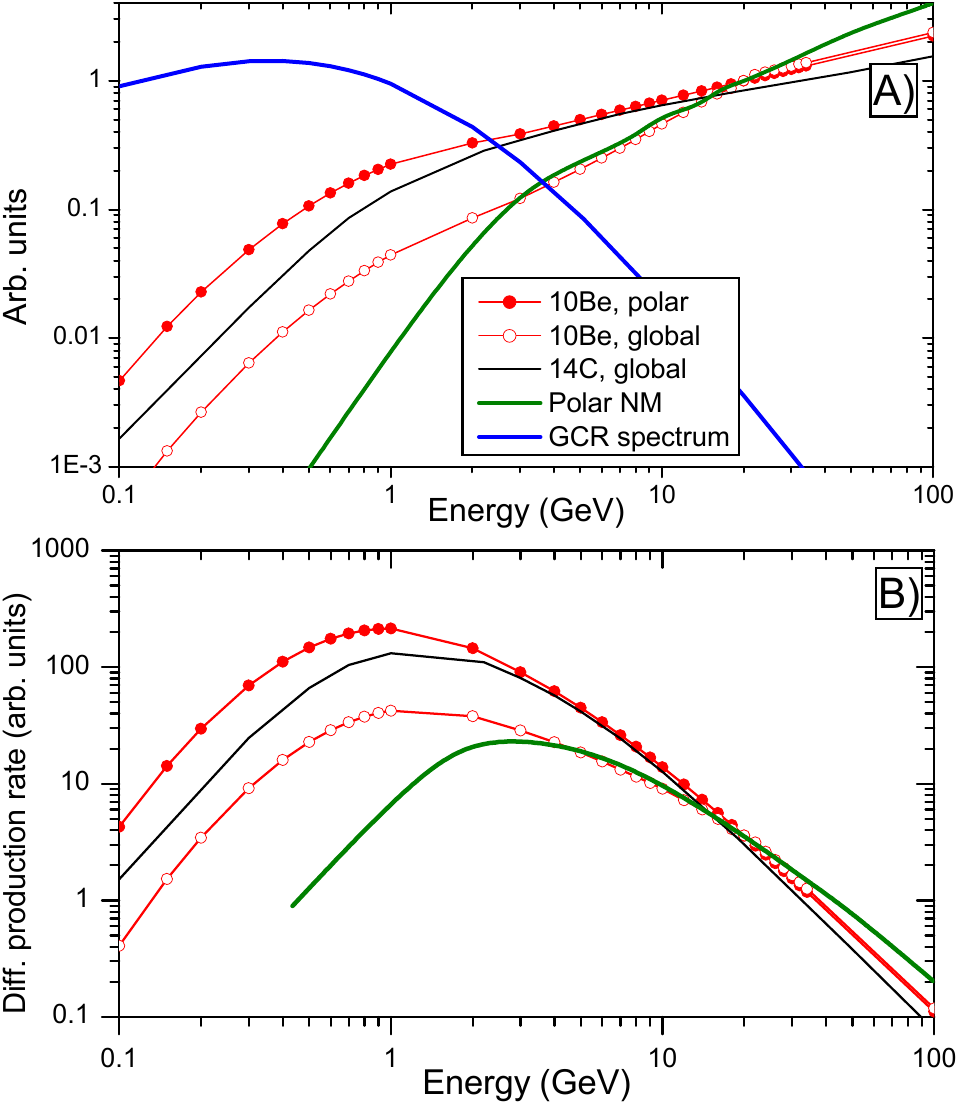}}
    \caption{Differential production rate for cosmogenic isotopes and
    ground-based neutron monitors as a function of cosmic-ray energy.
    \emph{Panel~A:} Yield functions of the globally averaged and polar \super{10}Be production \citep{webber03}, global \super{14}C
    production \citep{kovaltsov12}, polar neutron monitor \citep{clem00} as well as the energy spectrum of galactic cosmic
    protons for medium modulation ($\phi$~=~550~MV).
    \emph{Panel~B:} The differential production rate for global and polar \super{10}Be
    production, global \super{14}C production, and polar neutron
    monitor.}
    \label{Fig:yield}
\end{figure}}

Thanks to the development of atmospheric cascade models (Section~\ref{S:cas}), there are
 numerical models that allow one to compute the radiocarbon production rate as a function of the
 modulation potential $\phi$ and the geomagnetic dipole moment $M$.
The overall production of \super{14}C is shown in Figure~\ref{Fig:14C_prost}.

The production rate of radiocarbon, $Q_{^{14}\mathrm{C}}$, can vary as
 affected by different factors \citep[see, e.g.,][]{damon91}:

\begin{itemize}

\item Variations of the cosmic-ray flux on a geological timescale due
  to the changing galactic background (e.g., a nearby supernova
  explosion or crossing the dense galactic arm).

\item Secular-to-millennial variations are caused by the slowly-changing geomagnetic field.
 This is an important component of the variability, which needs to be independently evaluated from paleo and
 archeo-magnetic studies.

\item Modulation of cosmic rays in the heliosphere by solar magnetic
  activity. This variation is the primary aim of the present method.

\item Short-term variability of CR on a daily scale (suppression due to interplanetary transients or
 enhancement due to solar energetic-particle events) can be hardly resolved in radiocarbon data.

\end{itemize}

\epubtkImage{LR_14C-prost.png}{%
  \begin{figure}[htbp]
    \centerline{\includegraphics[width=10cm]{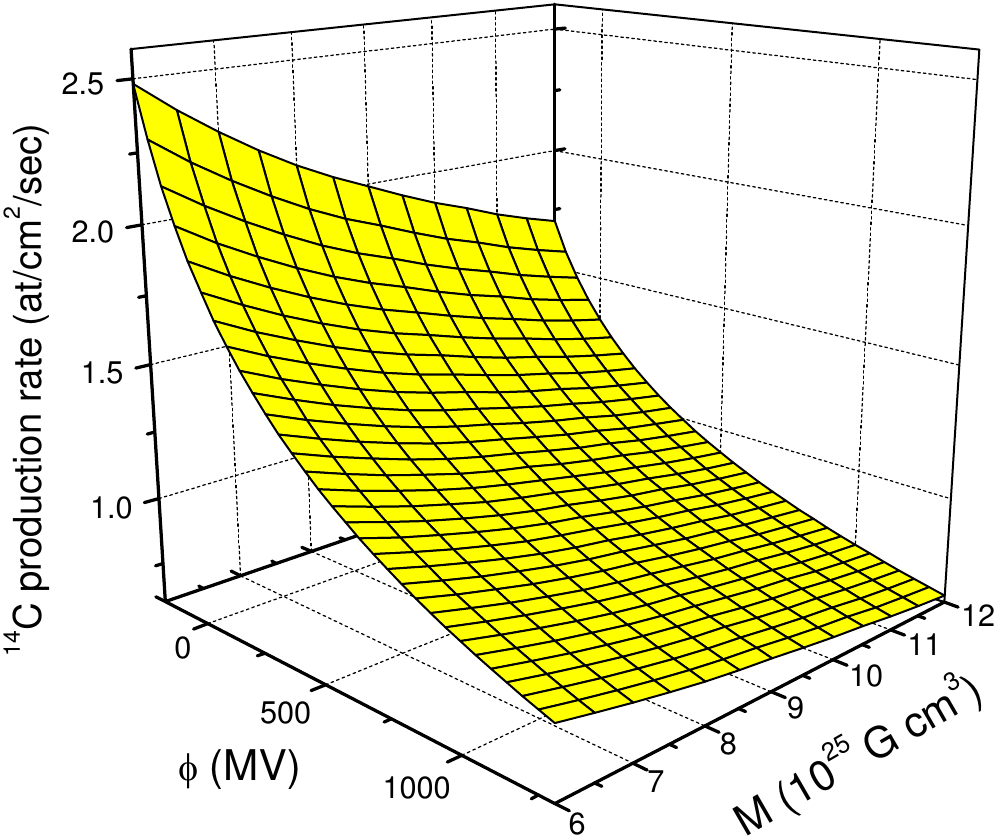}}
    \caption{Globally-averaged production rate of \super{14}C as a function
    of the modulation potential $\phi$ and geomagnetic dipole moment
    $M$, computed using the yield function by \citet{kovaltsov12},
    LIS by \citet{burger00} and cosmic-ray--modulation model by
    \citet{usoskin_Phi_05}. Another often used model \citep{masarik09}
    yields a similar result.}
    \label{Fig:14C_prost}
\end{figure}}

Therefore, the production rate of \super{14}C in the atmosphere can be modelled for a given time (namely,
 the modulation potential and geomagnetic dipole moment) and location.
The global production rate $Q$ is then obtained as a result of global averaging.

There is still a small unresolved discrepancy in the absolute value of the modeled \super{14}C production rate.
Different models yield the global-average production rate of 1.7\,--\,2.3~atoms~\cms
 \citep[see, e.g.,][and references therein]{obrien79, masarik99, goslar01, usoskin_GRL_SCR06,kovaltsov12},
 which is consistent with a direct estimate of the radiocarbon reservoir, based on analyses of the
 specific \super{14}C activity on the ground, 1.76\,--\,2.0 \citep{suess65, damon78, obrien79}.
On the other hand, the steady-state production calculated from the \super{14}C inventory in the
 carbon-cycle model (see Section~\ref{S:14C_cycle}) typically yields 1.6\,--\,1.7~atoms~\cms for the
  pre-industrial period \citep[e.g.,][and references therein]{goslar01}.
Thus, results obtained from the carbon cycle models and production models agree
 only marginally in the absolute values, and this needs further detailed studies.
The situation has been improved recently, when the newest numerical model
 \citep{kovaltsov12} yields the global average production of 1.64 and 1.88~atoms~\cms for the modern
  and pre-industrial periods, respectively.
In \super{14}C-based reconstructions, the pre-industrial steady-state production is commonly used.

\subsubsection{Transport and deposition}
\label{S:14C_cycle}

Upon production cosmogenic radiocarbon gets quickly oxidized to carbon dioxide CO\sub{2} and
 takes part in the regular carbon cycle of interrelated systems: atmosphere-biosphere-ocean
 (Figure~\ref{Fig:scheme}).
Because of the long residence time, radiocarbon becomes globally mixed in the atmosphere and
 involved in an exchange with the ocean.
It is common to distinguish between an upper layer of the ocean, which can directly exchange CO\sub{2}
 with the air and deeper layers.
The measured $\Delta$\super{14}C comes from the biosphere (trees), which receives radiocarbon from the
 atmosphere.
Therefore, the processes involved in the carbon cycle are quite complicated.
The carbon cycle is usually described using a box model \citep{oeschger74, siegenthaler80}, where it is
 represented by fluxes between different carbon reservoirs and mixing within the ocean reservoir(s),
 as shown in Figure~\ref{Fig:pandora}.
Production and radioactive decay are also included in box models.
Free parameters in a typical box model are the \super{14}C production rate $Q$, the air-sea
 exchange rate (expressed as turnover rate $\kappa$), and the vertical--eddy-diffusion coefficient $K$,
 which quantifies ocean ventilation.
Starting from the original representation \citep{oeschger74}, a variety of box models have been
 developed, which take into account subdivisions of the ocean reservoir and direct exchange between
 the deep ocean and the atmosphere at high latitudes.
More complex models, including a diffusive approach, are able to simulate more realistic scenarios,
 but they require knowledge of a large number of model parameters.
These parameters can be evaluated for the present time using the bomb test -- studying the
 transport and distribution of the radiocarbon produced during the atmospheric nuclear tests.
However, for long-term studies, only the production rate is considered variable, while the
 gas-exchange rate and ocean mixing are kept constant.
Under such assumptions, there is no sense in subdividing reservoirs or processes, and a simple
 carbon box model is sufficient.

\epubtkImage{LR_pandora.png}{%
  \begin{figure}[htbp]
    \centerline{\includegraphics[width=10cm]{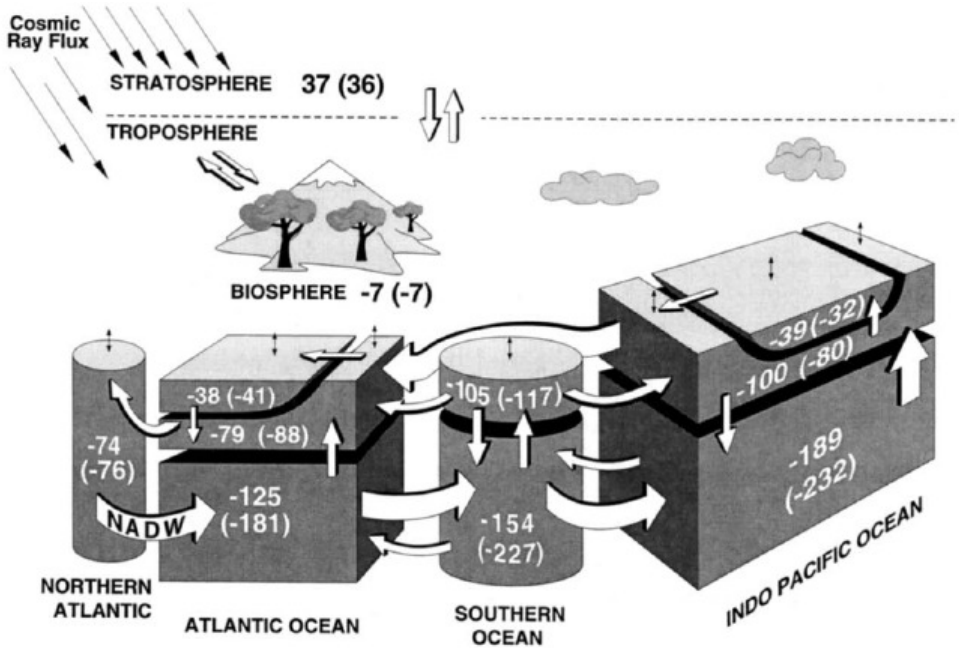}}
    \caption{A 12-box model of the carbon cycle \citep{broeker86, siegenthaler80}.
    The number on each individual box is the
    steady-state $\Delta$\super{14}C of this particular reservoir expressed
    in per mil. Image reproduced by permission from~\cite{bard97},
    copyright by Elsevier.}
    \label{Fig:pandora}
\end{figure}}

Using the carbon cycle model and assuming that all its parameters are constant
 in time, one can evaluate the production rate $Q$ from the measured $\Delta$\super{14}C data.
This assumption is well validated for the the Holocene \citep{damon78, stuiver91} as there is no
 evidence of considerable oceanic change or other natural variability of the carbon cycle
 \citep{gerber03}, and accordingly all variations of
 $\Delta$\super{14}C predominantly reflect the production rate.
This is supported by the strong similarity of the fluctuations of the \super{10}Be data in polar ice cores
 (Section~\ref{S:10Be}) compared to \super{14}C, despite their completely different geochemical fate \citep{bard97,steinhilber12}.
However, the changes in the carbon cycle during the last glaciation and deglaciation were dramatic,
 especially regarding ocean ventilation; this and the lack of independent information about the
 carbon cycle parameters, make it hardly possible to qualitatively estimate solar activity
 from \super{14}C before the Holocene.

First attempts to extract information on production-rate variations from measured
 $\Delta$\super{14}C were based on simple frequency separations of the signals.
All slow changes were ascribed to climatic and geomagnetic variations, while short-term
 fluctuations were believed to be of solar origin.
This was done by removing the long-term trend from the $\Delta$\super{14}C series and claiming the
 residual as being a series of solar variability \citep[e.g.,][]{peristykh03}.
This oversimplified approach was natural at earlier times, before the development of carbon cycle models,
 but later it was replaced by the inversion of the carbon cycle (i.e., the reconstruction of the production rate from the measured
 \super{14}C concentration).
Although mathematically this problem can be solved correctly as a system of linear differential equations,
 the presence of fluctuating noise with large magnitude makes it not straightforward, since the time derivative
 cannot be reliably identified leading thus to possible amplification of the high-frequency noise in $\Delta$\super{14}C data.
One traditional approach \citep[e.g.,][]{stuiver80} is based on an iterative procedure, first assuming a constant
 production rate, and then fitting the calculated $\Delta$\super{14}C variations
  to the actual measurements using a feedback scheme.
A concurrent approach based on the presentation of the carbon cycle as a Fourier filter \citep{usoskin_14C_05}
 produces similar results.
Roughly speaking, the carbon cycle acts as an attenuating and delaying filter for the \super{14}C signal
 (see Figure~\ref{Fig:phase}).
The higher the frequency is, the greater the signal is attenuated.
In particular, the large 11-year solar cycle expected in the \super{14}C is attenuated by a factor of hundred in the
 measured $\Delta$\super{14}C data, making it hardly detectable.
Because of the slow oceanic response, the \super{14}C data is also delayed with respect to the production signal.
The production rate $Q_{^{14}\mathrm{C}}$ for the Holocene is shown in Figure~\ref{Fig:Q}
 and depicts both short-term fluctuations as well as slower variations, mostly due to geomagnetic field changes (see Section~\ref{S:geoma}).

\epubtkImage{LR_14C_phase.png}{%
  \begin{figure}[htbp]
    \centerline{\includegraphics[width=0.8\textwidth]{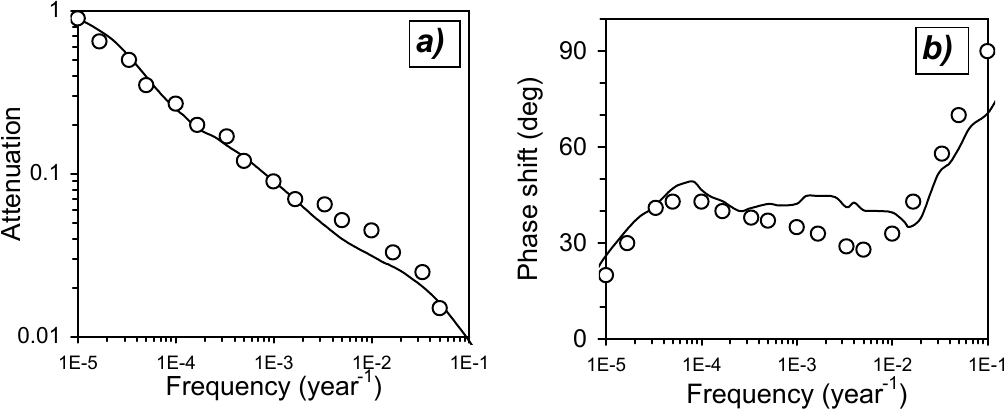}}
    \caption{The frequency characteristics of the carbon cycle:
      attenuation (left-hand panel) and phase shift (right-hand panel)
      as a function of the frequency of the \super{14}C production
      signal. Lines stand for a classical Oeschger--Siegenthaler box
      model \citep{siegenthaler80}, and open circles for a
      sophisticated PANDORA model \citep{bard97}.}
    \label{Fig:phase}
\end{figure}}

\subsubsection{The Suess effect and nuclear bomb tests}

Unfortunately, cosmogenic \super{14}C data cannot be easily used for the last century,
 primarily because of the extensive burning of fossil fuels.
Since fossil fuels do not contain \super{14}C, the produced CO\sub{2} dilutes the atmospheric \super{14}CO\sub{2}
 concentration with respect to the pre-industrial epoch.
Therefore, the measured $\Delta$\super{14}C cannot be straightforwardly translated into the production
 rate $Q$ after the late 19th century, and a special correction for fossil fuel burning is needed.
This effect, known as the Suess effect \citep[e.g.,][]{suess55}, can be up to
 $-25\permil$ in $\Delta$\super{14}C in 1950 \citep{tans79}, which is an order of magnitude
 larger than the amplitude of the 11-year cycle of a few $\permil$.
Moreover, while the cosmogenic production of \super{14}C is roughly homogeneous
 over the globe and time, the use of fossil fuels is highly nonuniform \citep[e.g.,][]{dejong82} both spatially
 (developed countries, in the northern hemisphere) and temporarily (World Wars, Great Depression, industrialization, etc.).
This makes it very difficult to perform
 an absolute normalization of the radiocarbon production to the direct measurements.
Sophisticated numerical models \citep[e.g.,][]{sabine04, mikaloff06} aim to account for
 the Suess effect and make good progress.
However, the results obtained indicate that the determination of the Suess effect does not yet
 reach the accuracy required for the precise modelling and reconstruction of the \super{14}C
 production for the industrial epoch.
As noted by \citet{matsumoto04},
``\ldots not all is well with the current generation of ocean carbon cycle models.
At the same time, this highlights the danger in simply using the available models to
 represent state-of-the-art modeling without considering the credibility of each model.''
Note that the atmospheric concentration of another carbon isotope \super{13}C is partly affected by land use, which has
 also been modified during the last century.

Another anthropogenic activity greatly disturbing the natural variability of \super{14}C is related to the
 atmospheric nuclear bomb tests actively performed in the 1960s.
For example, the radiocarbon concentration nearly doubled in the early 1960s in the northern hemisphere
 after nuclear tests performed by the USSR and the USA in 1961 \citep{damon78}.
On one hand, such sources of momentary spot injections of radioactive tracers (including \super{14}C)
 provide a good opportunity to verify and calibrate the exchange parameters for different carbon-cycle
 reservoirs and circulation models \citep[e.g.,][]{bard87, sweeney07}.
Thus, the present-day carbon cycle is more-or-less known.
On the other hand, the extensive additional production of isotopes during nuclear tests
 makes it hardly possible to use the \super{14}C as a proxy for solar activity after the 1950s \citep{joos94}.

These anthropogenic effects do not allow one to make a straightforward link between pre-industrial
 data and direct experiments performed during more recent decades.
Therefore, the question of the absolute normalization of \super{14}C model is still open
 \citep[see, e.g., the discussion in][]{solanki_Nat_04, solanki_Nat_05, muscheler_Nat_05}.

\subsubsection{The effect of the geomagnetic field}
\label{S:geoma}

As discussed in Section~\ref{S:geom}, knowledge of geomagnetic shielding is an
 important aspect of the cosmogenic isotope method.
Since radiocarbon is globally mixed in the atmosphere before deposition, its production is affected by changes in
 the geomagnetic dipole moment $M$, while magnetic-axis migration plays hardly any role in \super{14}C data.

The crucial role of paleomagnetic reconstructions has long been known \citep[e.g.,][]{elsasser56, kigoshi66}.
Many earlier corrections for possible geomagnetic-field changes were performed by detrending
 the measured $\Delta$\super{14}C abundance or production rate $Q$ \citep{stuiver80, voss96, peristykh03},
  under the assumption that geomagnetic and solar signals can be disentangled from the
  production in the frequency domain.
Accordingly, the temporal series of either measured $\Delta$\super{14}C or its production
 rate $Q$ is decomposed into the slow changing trend and faster oscillations.
The trend is supposed to be entirely due to geomagnetic changes, while the
 oscillations are ascribed to solar variability.
Such a method, however, obliterates all information on possible long-term variations
 of solar activity.
Simplified empirical correction factors were also often used \citep[e.g.,][]{stuiver80, stuiver91}.
The modern approach is based on a physics-based model \citep[e.g.,][]{solanki_Nat_04, vonmoos06}
 and allows the quantitative reconstruction of solar activity, explicitly using independent
 reconstructions of the geomagnetic field.
In this case the major source of errors in solar activity reconstructions
 is related to uncertainties in the paleomagnetic data \citep{snowball07}.
These errors are insignificant for the last millennium \citep{usoskin_GRL_06},
 but become increasingly important for earlier times.

\subsection{Cosmogenic isotope \super{10}Be}
\label{S:10Be}

\subsubsection{Measurements}

The cosmogenic isotope \super{10}Be is useful for long-term studies of solar activity because of
 its long half-life of around 1.5~\texttimes~10\super{6}~years.
Its concentration is usually measured in stratified ice cores allowing for independent dating.
Because of its long life, the beryllium concentration is difficult to measure by the decay rate \citep{beer00}.
Accordingly, the \super{10}Be/\super{9}Be ratio needs to be precisely measured at an accuracy better than 10\super{-13}.
This can be done using AMS (Accelerator Mass Spectrometry) technique, which makes
 the measurements complicated and expensive.
Correction for the decay is straightforward and does not include isotope fractionating.
From the measured samples, first the \super{10}Be concentration is defined, usually in
 units of 10\super{4}~atoms/g.
Sometimes, a correction for the snow precipitation amount is considered leading
 to the observable \super{10}Be flux, which is the number of atoms, precipitating to the surface per cm\super{2} per second.

There exist different \super{10}Be series suitable for studies of long-term solar activity,
 coming from ice cores in Greenland and Antarctica.
They have been obtained from different cores with different resolutions, and include
 data from Milcent, Greenland \citep{beer83}, Camp Century, Greenland \citep{beer88},
 Dye~3, Greenland \citep{beer90}, Dome Concordia and South Pole, Antarctica \citep{raisbeck90},
 GRIP, Greenland \citep{yiou97}, GISP2, Greenland \citep{finkel97}, Dome Fuji, Antarctica
 \citep{horiuchi07, horiuchi08}, Dronning Maud Land, Antarctica \citep{ruth07}, etc.
We note that data on \super{10}Be in other archives, e.g., lake sediments, is usually
 more complicated to interpret because of the potential influence of the climate \citep{horiuchi99, belmaker08}.

Details of the \super{10}Be series and their comparison with each
other can be found in \citet{beer00}, \citet{muscheler07}, and \citet{beer12}.

\subsubsection{Production}
\label{S:10Be_prod}

The isotope \super{10}Be is produced as a result of spallation of atmospheric nitrogen and
 oxygen (carbon is less abundant by far in the atmosphere and makes a negligible contribution)
 by the nucleonic component of the cosmic-ray--induced atmospheric cascade (Section~\ref{S:cas}).

A small contribution may also exist from photo-nuclear reactions \citep{bezuglov12}.
The cross section (a few mb) of the spallation reactions is almost independent
 of the energy of impacting particles and has a threshold of about 15~MeV.
Thus, the production of \super{10}Be is defined mostly by the multiplicity of the nucleonic
 component, which increases with the energy of primary cosmic rays (see Figure~\ref{Fig:yield}).
Maximum production occurs at an altitude of 10\,--\,15 km due to a balance
 between the total energy of the cascade (which increases with altitude) and the number
 of secondaries (decreasing with altitude).
Most of the global \super{10}Be is produced in the stratosphere (55\,--\,70\%) and the rest in the troposphere
 \citep{lal67, masarik99, masarik09, usoskin_7Be_08, kovaltsov10}.

Computation of \super{10}Be isotope production is straightforward, provided a model of the
 atmospheric cascade is available.
The first consistent model was developed by D.~Lal \textit{et al.} \citep{bhandari66, lal67, lal68},
 using an empirical approach based on fitting simplified model calculations to measurements of
 the isotope concentrations and ``star'' (inelastic nuclear collisions) formations in the atmosphere.
Next was an analytical model by \citet{obrien79}, who
 solved the problem of the GCR-induced cascade in the atmosphere using an analytical
 stationary approximation in the form of the Boltzmann equation.
Those models were based on calculating the rate of inelastic collisions or ``stars''
 and then applying the mean spallation yield per ``star''.
A new step in the modelling of isotope production was made by \citet{masarik99},
 who performed a full Monte-Carlo simulation of a GCR-initiated cascade in the atmosphere
 and used cross sections of spallation reactions directly instead of the average ``star'' efficiency.
Modern models \citep{webber03,webber07,usoskin_7Be_08,kovaltsov10} are based on
 a full Monte-Carlo simulation of the atmospheric cascade, using improved cross sections.
The global production rate of \super{10}Be is about 0.02\,--\,0.03~atoms~\cms
 \citep{masarik99, webber07, kovaltsov10}, which is lower than that for
 \super{14}C by two orders of magnitude (about 2~atoms~\cms; see Section~\ref{S:14C_pro}).
The yield function of \super{10}Be production is shown in Figure~\ref{Fig:yield}A and
 the differential production rate in Figure~\ref{Fig:yield}B.
One can see that the peak of \super{10}Be sensitivity, especially in polar regions, is
 shifted towards lower energies (below 1~GeV) compared with a neutron monitor.
This implies that the \super{10}Be isotope is relatively more sensitive to less energetic CR and
 is, therefore, more affected by solar energetic particles \citep{usoskin_GRL_SCR06}.
Comparison of model computations with direct beryllium production experiments \citep{usoskin_7Be_08, kovaltsov10}, and also
 the results of modelling of the short-living \super{7}Be isotope \citep{usoskin_7Be_09} suggest that some numerical models
 \citep{masarik99, webber03, webber07} tend to underestimate the production.


Although the production of \super{10}Be can be more or less precisely modelled, a simple normalization
 ``surface'', similar to that shown in Figure~\ref{Fig:14C_prost} for \super{14}C, is not easy to produce
 because of partial mixing in the atmosphere (see Section~\ref{S:10Be_tran}).
Simplified models, assuming either only global \citep[e.g.,][]{beer00} or polar
 production \citep{bard97, usoskin_AA_04}, have been used until recently.
However, it has been recognized that a more realistic model of the limited atmospheric mixing
 should be used.
Without detailed knowledge of \super{10}Be transport in the atmosphere, it is impossible
 to relate the quantitatively-measured concentration to the production (as done for \super{14}C using
 the carbon cycle), and one has to assume that the measured abundance is proportional (with an unknown
 coefficient) to the production rate in a specific geographical region (see Section~\ref{S:10Be_tran}).

\subsubsection{Atmospheric transport}
\label{S:10Be_tran}

After production, the \super{10}Be isotope has a seemingly simple (Figure~\ref{Fig:scheme}) but difficult-to-account-for
 fate in the atmosphere.
Its atmospheric residence time depends on scavenging, stratosphere-troposphere exchange
 and inter-tropospheric mixing \citep[e.g.,][]{mchargue91}.
Soon after production, the isotope becomes attached to atmospheric aerosols and follows their fate.
In addition, it may be removed from the lower troposphere by wet deposition (rain and snow).
The mean residence time of the aerosol-bound radionuclide in the atmosphere is quite different
 for the troposphere, being a few weeks, and stratosphere, where it is one to two years \citep{raisbeck81}.
Accordingly, \super{10}Be produced in the troposphere is deposited mostly locally,
 i.e., in the polar regions, while stratospheric \super{10}Be can be partly or totally mixed.
In addition, because of the seasonal (usually Spring) intrusion of stratospheric air
 into the troposphere at mid-latitudes, there is an additional contribution of stratospheric \super{10}Be.
Therefore, the measured \super{10}Be concentration (or flux) in polar ice is modulated not
 only by production but also by climate/precipitation effects \cite[e.g.,][]{steig96, bard97}.
This led \citet{lal87} to the extreme conclusion that variations of polar \super{10}Be reflect a meteorological,
 rather than solar, signal.
However, comparison between Greenland and Antarctic \super{10}Be series and between \super{10}Be
 and \super{14}C data \citep[e.g.,][]{bard97, horiuchi08, beer12} suggests that the beryllium data
 mostly depicts production variations (i.e., solar signal) on top of which some
 meteorological effects can be superposed (see also Section~\ref{S:comp}).

Since both assumptions of the global and purely-local polar production of \super{10}Be archived
 in polar ice are over-simplified, several attempts have been made to overcome this problem.
For instance, \citet{mccracken_JGR_04} proposed several simple mathematical models of partial atmospheric
 mixing (without division in the troposphere and stratosphere) and compared them with observed data.
From this semi-empirical approach McCracken concluded that M2 (full mixing above 60\textdegree\ latitude
 and a limited mixing between 40\textdegree\ and 60\textdegree\ latitude) is a reasonable model for Antarctica.
\citet{vonmoos06} assumed that the production of \super{10}Be recorded in Greenland
 is related to the entire hemisphere in the stratosphere (i.e, global stratospheric mixing) but
 is limited to latitudes above 40\textdegree\ latitude in the troposphere (partial tropospheric mixing).
This approach uses either semi-empirical or indirect arguments in choosing the unknown degree of mixing.

Recent efforts in employing modern atmospheric 3D circulation models for simulations of \super{10}Be transport
 and deposition, including realistic air-mass transport and dry-vs-wet deposition \citep{field06, heikkila07,heikkila09}, look more promising.
An example of \super{10}Be deposition computed on the world grid using the NASA GISS model \citep{field06}
 is shown in Figure~\ref{Fig:atmo}.
Precision of the models allows one to distinguish local effects, e.g., for Greenland \citep{heikkila07}.
A simulation performed by combining a detailed \super{10}Be-production
 model with an air-dynamics model can result in an
 absolute model relating production and deposition of the radionuclide.
We may expect this breakthrough to occur in the near future.
The validity and usefulness of this approach has been recently demonstrated by \citet{usoskin_7Be_09},
 who directly modeled production \citep[using the CRAC model --][]{usoskin_7Be_08} and
 transport \citep[using the GISS ModelE --][]{koch06} of a short-living beryllium isotope \super{7}Be
 and showed that such a combined model is able to correctly reproduce both the absolute level and
 temporal variations of the \super{7}Be concentration measured in near ground air around the globe.
Keeping in mind the similarity between production and transport of the two beryllium isotopes,
 \super{7}Be and \super{10}Be, this serves as support for the advanced modelling of \super{10}Be transport.
A similar general agreement between measured and modelled seasonal variability has been
 recently found for \super{10}Be in an Antarctic ice core \citep{pedro10}.


\epubtkImage{LR_atm.png}{%
  \begin{figure}[htbp]
    \centerline{\includegraphics[width=10cm]{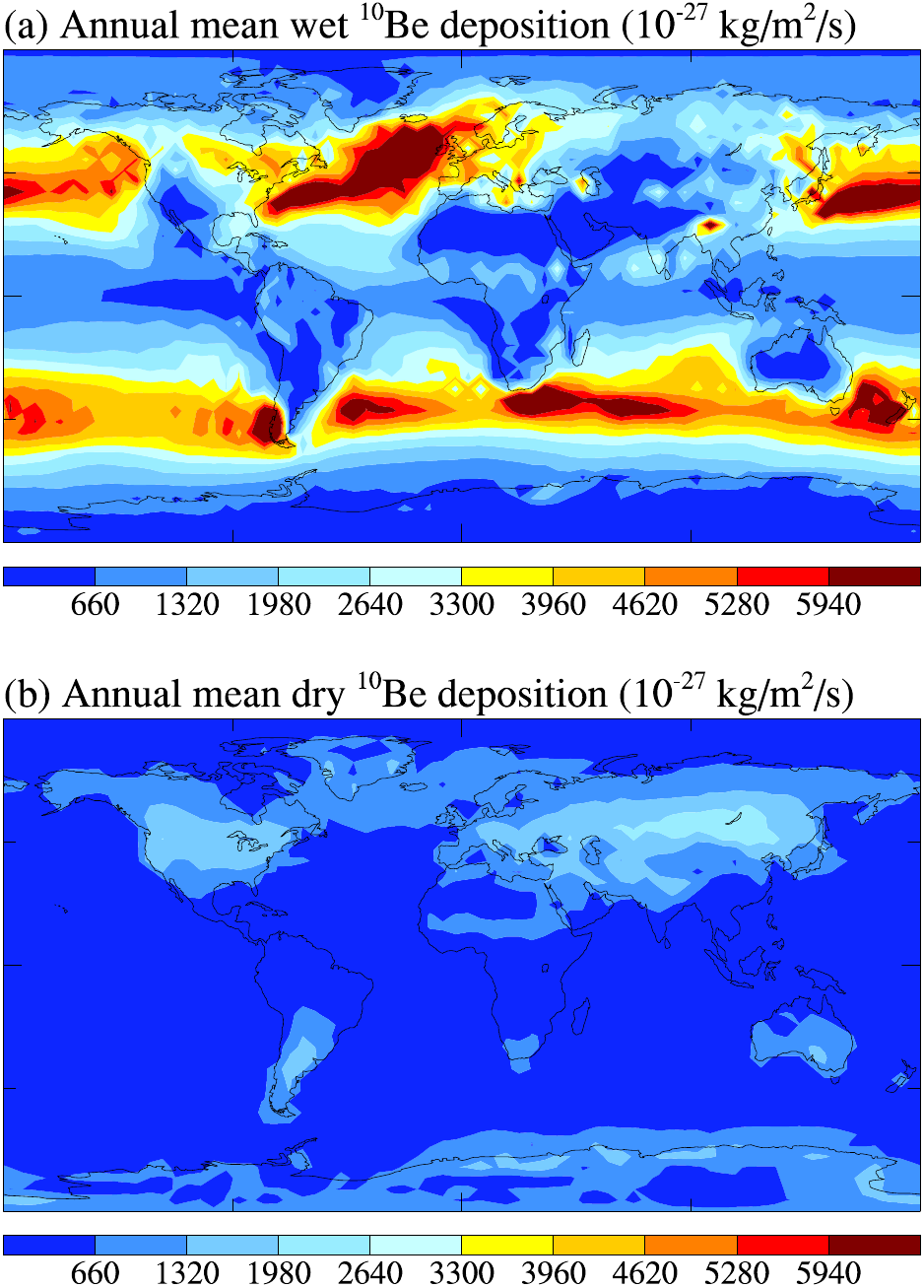}}
    \caption{Wet (panel a) and dry (panel b) deposition of \super{10}Be,
    computed using the NASA GISS model \citep{field06} for a fixed
    sea-surface temperature.}
    \label{Fig:atmo}
\end{figure}}

\subsubsection{Effect of the geomagnetic field}

In order to properly account for geomagnetic changes (Section~\ref{S:geom}), one needs to know
 the effective region in which the radionuclide is produced before being
 stored in the archive analyzed.
For instance, if the concentration of \super{10}Be measured in polar ice reflects mainly the
 isotope's production in the polar atmosphere \citep[as, e.g., assumed by][]{usoskin_PRL_03},
 no strong geomagnetic signal is expected to be observed, since the geographical
 poles are mostly related to high geomagnetic latitudes.
On the other hand, assuming global mixing of atmospheric \super{10}Be before deposition
 in polar ice \citep[e.g.,][]{masarik99}, one expects that only changes in the geomagnetic dipole moment
 affect will the signal.
However, because of partial mixing, which can be different in the stratosphere and troposphere, taking into
 account migration and displacement of the geomagnetic dipole axis may be essential for a reliable
 reconstruction of solar variability from \super{10}Be data \citep{mccracken_JGR_04}.
Therefore, only a full combination of the transport and production models, the latter
 explicitly including geomagnetic effects estimated from paleomagnetic reconstructions,
 can adequately account for geomagnetic changes and separate the solar signal.
These will form the next generation of physics-based models for the cosmogenic-isotope proxy method.
We note that paleomagnetic data should ideally not only provide the dipole moment (VADM or VDM)
 but should also provide estimates of the geomagnetic axis attitude and displacement of the dipole center \citep{korte11}.

\subsection{Other potential proxy}
\label{sec:nit_GCR}

An interesting new potential proxy for solar activity (or cosmic ray) variability on the
 long-term centennial-to-millennial time scale has been proposed recently by \citet{traversi12}.
This is the nitrate content in a polar ice core Talos Dome in Antarctica, which has a favorite location
 in the sense of snow accumulation and conservation of such volatile specie as nitrate.
Nitrate-related species are partly produced in the stratosphere/troposphere as a result of
 the ionization of the atmospheric air by cosmic rays and, partly, via terrestrial sources
 (e.g., lightnings) and are subject to air transport \citep{rozanov12}.
As shown by \citet{traversi12}, the nitrate concentration/flux measured in the Talos Dome
 ice core for the Holocene period agrees well with the cosmogenic data of \super{14}C in
 tree rings and \super{10}Be in both Antarctic and Greenland ice cores, on the time scales
 from centennia to millennia.
Due to the large errors of the ice core dating, 200\,--\,300 years \citep{schupbach11}, shorter time scales cannot be considered.
The level of the nitrate variability is generally consistent with that predicted by theoretical models assuming
 its production by GCR in the atmosphere \citep{semeniuk11,rozanov12}.
Thus, the nitrate in an ice core provides a potential to become a new proxy of long-term
 solar activity, with independent atmospheric fate, which would strengthen the robustness
 of the reconstructions.
However, an independent confirmation of the result and a more detailed model are needed
 before it can serve as a new quantitative proxy.
Note that the mechanism of the nitrate production and transport is not related to the
 possible nitrate peaks claimed to be caused by strong solar energetic-particle events
 (see Section~\ref{sec:nit}).

\subsection{Towards a quantitative physical model}

Several methods have been developed historically to convert measured
 cosmogenic-isotope data into a solar activity index, ranging from
 very simple regressions to physics-based models.
A new step in long-term solar-activity reconstruction has been made recently, which is
 the development of the proxy method in which physics-based models are used, instead of
 a phenomenological regression, to link SN with cosmogenic-isotope
 production \citep{usoskin_PRL_03, usoskin_AA_07, solanki_Nat_04, vonmoos06, muscheler07, steinhilber08}.
Due to recent theoretical developments, it is now possible to construct a chain
 of physical models to model the entire relationship between solar activity and cosmogenic data.


The physics-based reconstruction of solar activity (in terms of sunspot numbers) from
 cosmogenic proxy data includes several steps:

\begin{itemize}

\item Computation of the isotope's production rate in the atmosphere
  from the measured concentration in the archive (Sections~\ref{S:14C_pro}
  and \ref{S:10Be_prod});

\item Computation, considering independently-known secular geomagnetic
  changes (see Section \ref{S:geoma}) and a model of the CR-induced
  atmospheric cascade, of the GCR spectrum parameter quantified via
  the modulation potential $\phi$ (Section~\ref{sec:phii}), some
  reconstructions being terminated at this point;

\item Computation of a heliospheric index, whether of the open solar
  magnetic flux or of the average HMF intensity at the Earth's orbit
  (Section~\ref{sec:phii})

\item Computation of a solar index (sunspot number series),
  corresponding to the above-derived heliospheric parameter
  (Section~\ref{sec:sun}).

\end{itemize}

Presently, all these steps can be completed using appropriate models.
Some models stop after computations of the modulation potential as
 its translation into the solar index may include additional uncertainties.
Although the uncertainties of the models may be considerable, the models allow a full basic
 quantitative reconstruction of solar activity in the past.
However, much needs to be done, both theoretically and experimentally, to obtain an
 improved reconstruction.

\subsubsection{Regression models}
\label{S:regr}

\emph{Mathematical regression} is the most apparent and often used (even recently)
 method of solar-activity reconstruction from proxy data \citep[see, e.g.,][]{stuiver80, ogurtsov04}.
The reconstruction of solar activity is performed in two consecutive steps.
First, a phenomenological regression (either linear or nonlinear) is built between
 a proxy data set and a direct solar-activity index for the available
 ``training'' period (e.g., since 1750 for WSN or since 1610 for GSN).
Then this regression is extrapolated backwards to evaluate SN from the proxy data.
The main shortcoming of the regression method is that it depends on the time
 resolution and choice of the ``training'' period.
The former is illustrated by Figure~\ref{Fig_regr}, which shows the scatter
 plot of the \super{10}Be concentration vs.\ GSN for the annual and 11-year
 smoothed data.
\epubtkImage{LR_regr.png}{%
  \begin{figure}[htbp]
    \centerline{\includegraphics[width=\textwidth]{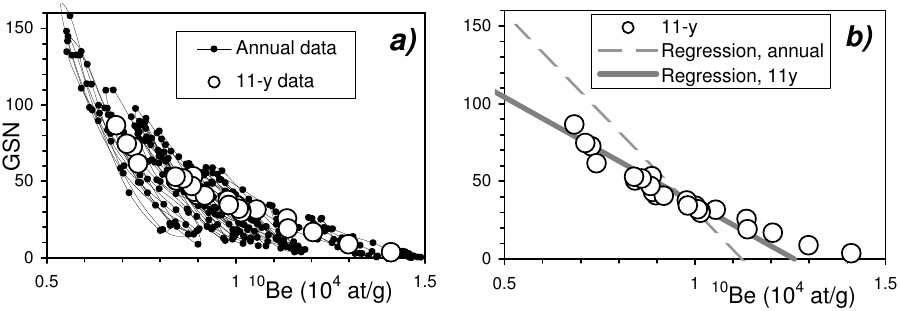}}
    \caption{Scatter plot of smoothed group sunspot numbers vs.\
    (2-year delayed) \super{10}Be concentration. a) Annual (connected
    small dots) and 11-year averaged (big open dots) values. b)
    Best-fit linear regressions between the annual (dashed line) and
    11-year averaged values (solid line). The dots are the same as in
    panel (a). \citep[After][]{usoskin_Rev_04}.}
    \label{Fig_regr}
\end{figure}}
One can see that the slope of the \super{10}Be-vs-GSN relation (about -500~g/atom) within individual cycles is significantly different from
 the slope of the long-term relation (about -100~g/atom), i.e., individual
 cycles do not lie on the line of the 11-year averaged cycles.
Moreover, the slope of the regression for individual 11-year cycles
 varies essentially depending on the solar activity level.
Therefore, a formal regression built using the annual data for 1610\,--\,1985 yields
 a much stronger GSN-vs-\super{10}Be dependence than for the cycle-averaged data (see
 Figure~\ref{Fig_regr}b), leading to a potentially-erroneous evaluation of the sunspot number
 from the \super{10}Be proxy data.

It is equally dangerous to evaluate other solar/heliospheric/terrestrial indices from sunspot numbers,
 by extrapolating an empirical relation obtained for the last few decades back in time.
This is because the last decades (after the 1950s), which are well covered by direct
 observations of solar, terrestrial and heliospheric parameters, correspond to a very
 high level of solar activity.
After a steep rise in activity level between the late 19th and mid 20th centuries,
 the activity remained at a roughly constant high level, being totally dominated by
 the 11-year cycle without a long-term trends.
Accordingly, all empirical relations built based on data for this period are
 focused on the 11-year variability and can overlook possible long-term trends \citep{mursula03}.
This may affect all regression-based reconstructions, whose results cannot be independently
 (directly or indirectly) tested.
In particular, this may be related to solar irradiance reconstructions, which are
 often based on regression-like models, built and verified using data for the last
 three solar cycles, when there was no strong trend in solar activity.

\epubtkImage{LR_NM_belov.png}{%
  \begin{figure}[htbp]
    \centerline{\includegraphics[width=0.8\textwidth]{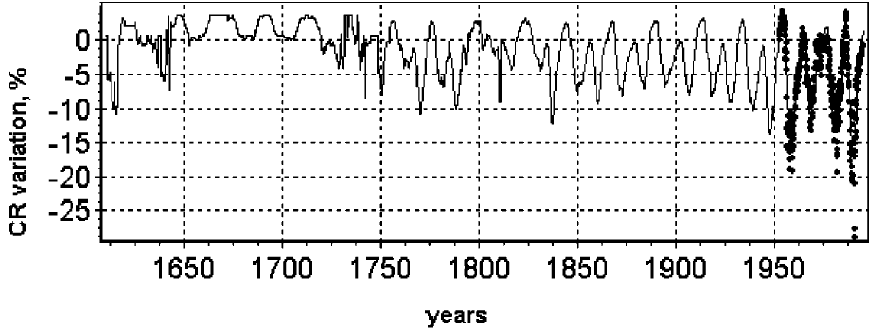}}
    \caption{An unsuccessful attempt of the reconstruction of
      cosmic-ray intensity in the past using a regression with sunspot
      numbers. Dots represent the observed cosmic-ray intensity since
      1951. Note the absence of a long-term trend. Image reproduced by
      permission from~\cite{belov06}, copyright by Elsevier.}
    \label{Fig:belov}
\end{figure}}

As an example let us consider an attempt \citep{belov06} to reconstruct cosmic-ray intensity since
 1610 from sunspot numbers using a (nonlinear) regression.
The regression between the count rate of a neutron monitor and sunspot numbers, established
 for the last 30~years, yields an agreement at a 95\% confidence level for the period 1976\,--\,2003
 (see Figure~\ref{Fig:NM_SN}).
Based on that, \citet{belov06} extrapolated the regression back in time to produce
 a reconstruction of cosmic-ray intensity (quantified in NM count rate) to 1560
 (see Figure~\ref{Fig:belov}).
One can see that there is no notable long-term trend in the reconstruction, and
 the fact that all CR maxima essentially lie at the same level, from the Maunder
 minimum to modern times, is noteworthy.
It would be difficult to dispute such a result if there was no direct test for CR
 levels in the past.
Independent reconstructions based on cosmogenic isotopes
 or theoretical considerations \citep[e.g.,][]{usoskin_JGR_02, scherer04, scherer_AA04}
 provide clear evidence that cosmic-ray intensity was essentially higher
 during the Maunder minimum than nowadays.
This example shows how easy it is to overlook an essential feature in a reconstruction
 based on a regression extrapolated far beyond the period it is based on.
Fortunately, for this particular case we do have independent information that can
 prevent us from making big errors.
In many other cases, however, such information does not exist (e.g., for total or
 spectral solar irradiance), and those who make such unverifiable reconstructions
 should be careful about the validity of their models beyond the range of
 the established relations.

\subsubsection{Reconstruction of heliospheric parameters}
\label{sec:phii}

The modulation potential $\phi$ (see Section~\ref{S:mod}) is directly related to cosmogenic isotope production in the atmosphere.
It is a parameter describing the spectrum of galactic
 cosmic rays \citep[see the definition and full description of this index in][]{usoskin_Phi_05}
 and is sometimes used as a stand-alone index of solar (or, actually, heliospheric) activity.
We note that, provided the isotope production rate $Q$ is estimated and geomagnetic changes
 can be properly accounted for, it is straightforward to obtain a time series of the
 modulation potential, using, e.g., the relation shown in Figure~\ref{Fig:14C_prost}.
Several reconstructions of modulation potential for the last few centuries are
 shown in Figure~\ref{Fig:phi}.
While being quite consistent in the relative changes, they differ in the absolute level
 and fine details.
Reconstructions of solar activity often end at this point,
 representing solar activity by the modulation potential, as some authors \citep[e.g.,][]{beer03, vonmoos06, muscheler07}
 believe that further steps (see Section~\ref{sec:sun}) may introduce additional uncertainties.
However, since $\phi$ is a heliospheric, rather than solar, index, the same uncertainties
 remain when using it as an index of solar activity.
Moreover, the modulation potential is a model-dependent quantity (see discussion in Section~\ref{S:mod})
 and therefore does not provide an unambiguous measure of heliospheric activity.
In addition, the modulation potential is not a physical index but rather a formal
 fitting parameter to describe the GCR spectrum near Earth \citep{usoskin_Phi_05}
 and, thus, is not a universal solar-activity index.

\epubtkImage{LR_phi.png}{%
  \begin{figure}[htbp]
    \centerline{\includegraphics[width=0.8\textwidth]{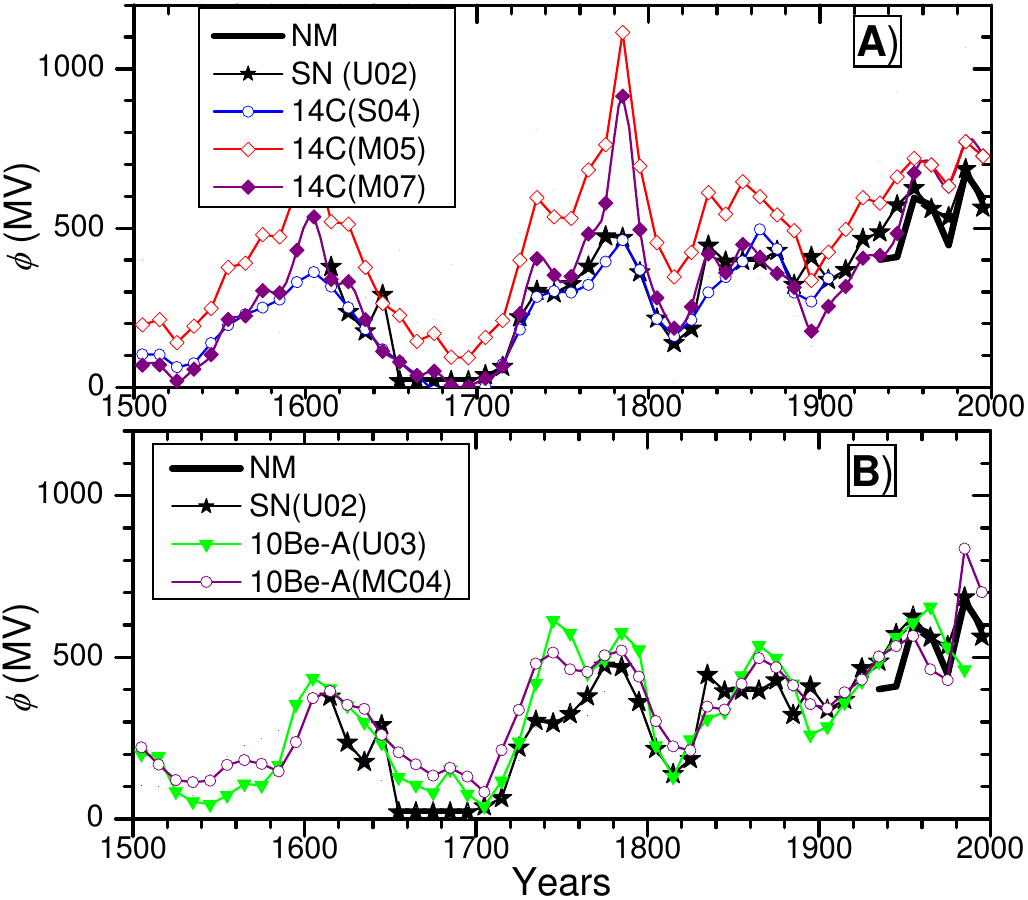}}
    \caption{Several reconstructions of the decade-averaged
    modulation potential $\phi$ for the last few centuries: from sunspot
    numbers \citep[SN(U02) --][]{usoskin_JGR_02}, from \super{14}C data
    \citep[14C(S04), 14C(M05), 14C(M07) --][respectively]{solanki_Nat_04, muscheler_Nat_05, muscheler07},
    from Antarctic \super{10}Be data \citep[10Be(U03), 10Be(MC04)
    --][respectively]{usoskin_PRL_03, mccracken04}. The thick black NM
    curve is based on direct cosmic-ray measurements by neutron
    monitors since 1951 \citep{usoskin_bazi_11} and ionization chambers
    since 1936 \citep{mccracken_beer_07}.}
    \label{Fig:phi}
\end{figure}}

Modulation of GCR in the heliosphere (see Section~\ref{S:mod}) is mostly defined
 by the turbulent heliospheric magnetic field (HMF), which ultimately
 originates from the sun and is thus related to solar activity.
It has been shown, using a theoretical model of the heliospheric transport of cosmic rays \citep[e.g.,][]{usoskin_JGR_02},
 that on the long-term scale (beyond the 11-year solar cycle) the modulation potential $\phi$ is
 closely related to the open solar magnetic flux \F{o}, which is
 a physical quantity describing the solar magnetic variability \citep[e.g.,][]{solanki00, krivova07}.

Sometimes, instead of the open magnetic flux, the mean HMF intensity at Earth orbit, \textit{B},
 is used as a heliospheric index \citep{caballero04, mccracken07, steinhilber10}.
Note that \textit{B} is linearly related to \F{o} assuming constant solar-wind speed,
 which is valid on long-term scales.
An example of HMF reconstruction for the last 600~years is shown in Figure~\ref{Fig:mccr}.
In addition, the count rate of a ``pseudo'' neutron monitor (i.e., a count rate of a neutron monitor if
 it was operated in the past) is considered as a solar/heliospheric index \citep[e.g.,][]{beer00, mccracken_beer_07}.

\epubtkImage{LR_mccr07.png}{%
  \begin{figure}[htbp]
    \centerline{\includegraphics[width=0.8\textwidth]{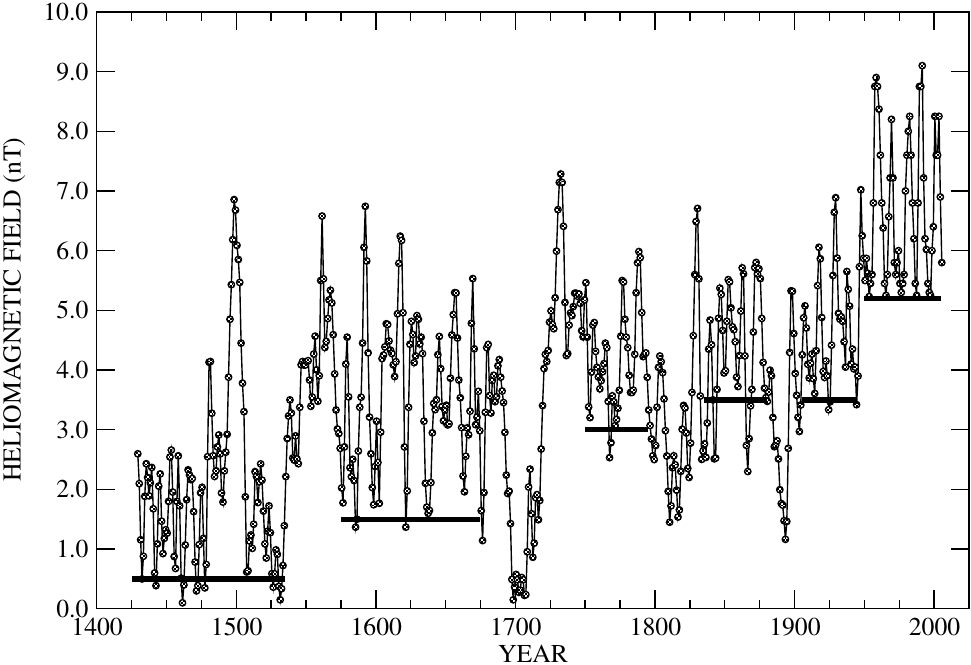}}
    \caption{An example of reconstruction of the heliospheric magnetic
    field at Earth orbit for the last 600~years from \super{10}Be
    data. Image reproduced by permission from~\cite{mccracken07},
    copyright by AGU.}
    \label{Fig:mccr}
\end{figure}}

\subsubsection{A link to sunspot numbers}
\label{sec:sun}

The open solar magnetic flux \F{o} described above is related to
 the solar surface magnetic phenomena such as sunspots or faculae.
Modern physics-based models allow one to calculate the
 open solar magnetic flux from data of solar observation, in
 particular sunspots \citep{solanki00, solanki02, krivova07}.
Besides the solar active regions, the model includes ephemeral regions.
Although this model is based on physical principals, it contains some unknowns like
 the decay time of the open flux, which cannot be measured or theoretically
 calculated and has to be found by means of fitting the model to data.
This free parameter has been determined by requiring the model output to
 reproduce the best available data sets for the last 30~years with the help of a genetic algorithm.
Inversion of the model, i.e., the computation of sunspot numbers for given
 \F{o} values is formally a straightforward solution of a system of linear differential
 equations, however, the presence of noise in the real data makes it only possible in a
 numerical-statistical way \citep[see, e.g.,][]{usoskin_AA_04, usoskin_AA_07}.
By inverting this model one can compute the sunspot-number series corresponding to
 the reconstructed open flux, thus forging the final link in a chain
 quantitatively connecting solar activity to the measured cosmogenic isotope abundance.
A sunspot-number series reconstructed for the Holocene using \super{14}C isotope
 data is shown in Figure~\ref{Fig:long}.
While the definition of the grand minima (Section~\ref{sec:MM}) is virtually insensitive to the uncertainties
 of paleomagnetic data, the definition of grand maxima depends on the paleomagnetic model used \citep{usoskin_AA_07}.
Since the Y00 paleomagnetic model forms an upper bound for the true geomagnetic strength (Section~\ref{S:geom}),
 the corresponding solar-activity reconstructions may underestimate the solar-activity level.
Accordingly, the grand maxima defined using the Y00 model are robust and can be
 regarded as ``maximum maximorum'' (see Section~\ref{S:Max}).

\epubtkImage{LR_SA_long.png}{%
  \begin{figure}[htbp]
    \centerline{\includegraphics[width=0.8\textwidth]{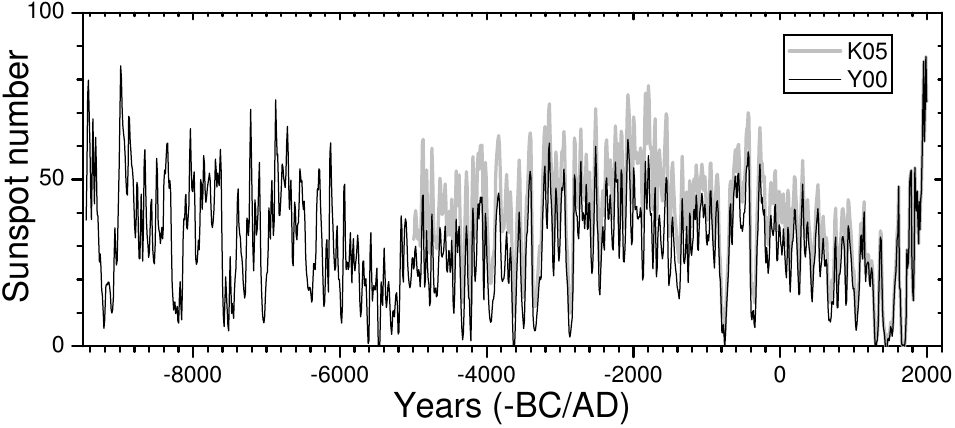}}
    \caption{Long-term sunspot-number reconstruction from \super{14}C
    data \citep[after][]{usoskin_AA_07}. All data are decade
    averages. Solid (denoted as `Y00') and grey (`K05') curves are
    based on the paleo-geomagnetic reconstructions of \citet{yang00}
    and \citet{korte05}, respectively. Observed group sunspot numbers
    \citep{hoyt98} are shown after 1610.}
    \label{Fig:long}
\end{figure}}

As very important for the climate research, the variations of the total solar irradiance (TSI)
 are sometimes reconstructed from the solar proxy data \citep{steinhilber09,vieira11}.
However, the absolute range of the TSI variability on the centennial-millennial time scales
 still remains unknown \citep{schmidt12}.

\subsection{Solar activity reconstructions}

Detailed computational models of cosmogenic isotope production in the atmosphere \citep[e.g.,][]{masarik99}
 have opened up a new possibility for long-term solar-activity reconstruction \citep[e.g.,][]{beer00}.
The first quantitative reconstructions of solar activity from cosmogenic proxy appeared in the early 2000s
 based on \super{10}Be deposited in polar ice \citep{beer03, usoskin_PRL_03}.

\citet{beer03} reconstructed the modulation potential on a multi-millennial timescale using the
 model computations by \citet{masarik99} and the \super{10}Be data from the GISP2 core in Greenland.
This result has been extended, even including the \super{14}C data set, and presently covers the
 whole Holocene \citep{vonmoos06, steinhilber08, steinhilber10}.
\citet{usoskin_PRL_03} presented the reconstruction of sunspot activity over the last millennium,
 based on \super{10}Be data from both Greenland and Antarctica, using a physics-based model described
 in detail in \citet{usoskin_AA_04}.
This result reproduces the four known grand minima of solar activity -- Maunder,
 Sp\"orer, Wolf and Oort minima (see Section~\ref{sec:MM}).
Later \citet{solanki_Nat_04} reconstructed 10-year--averaged sunspot numbers from
 the \super{14}C content in tree rings throughout the Holocene and estimated its
 uncertainties (see Figure~\ref{Fig_10Be}).
This result was disputed by \citet{muscheler_Nat_05}, whose concurrent model, however, rested on an erroneous
 normalization, as argued in \cite{solanki_Nat_05}.
The reconstruction of \citet{solanki_Nat_04} has been recently updated by \citet{usoskin_GRL_06}, using a newer paleomagnetic
 reconstruction by \citet{korte05}, and was later slightly revised \citep{usoskin_AA_07}, considering
 an updated model of the solar open magnetic flux by \citet{krivova07}.
Reconstruction of the HMF from \super{10}Be data has been performed by \citet{caballero04}, using
 a model of CR modulation in the heliosphere and a \super{10}Be production model by \citet{webber03}.
Recently it was revised \citep{mccracken07} to present a detailed reconstruction of HMF intensity
 since 1428.
The most recent reconstruction of the heliospheric modulation potential was done by
 \citet{steinhilber12} using the combined principal component analysis of several data sets.

The obtained results are discussed in Section~\ref{sec:3}.

\epubtkImage{LR_10Be_reco.png}{%
  \begin{figure}[htbp]
    \centerline{\includegraphics[width=0.8\textwidth]{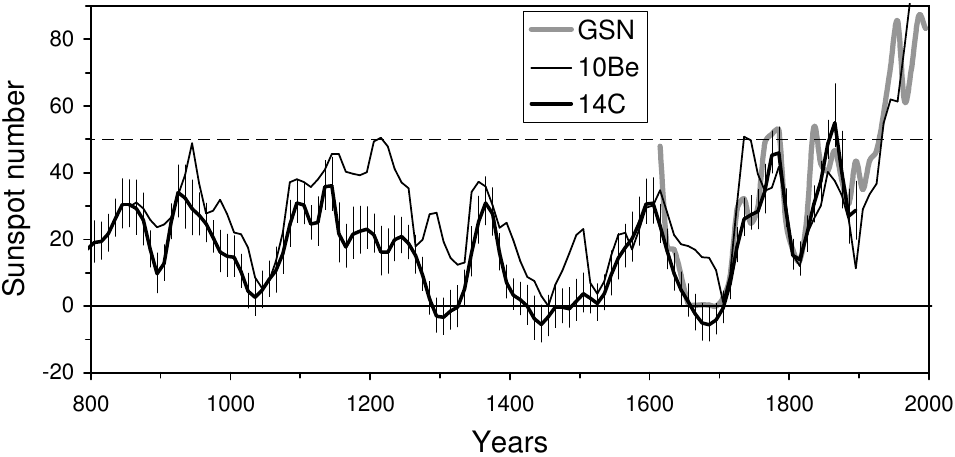}}
    \caption{10-year averaged sunspot numbers: Actual group sunspot
    numbers (thick grey line) and the reconstructions based on
    \super{10}Be \citep[thin curve,][]{usoskin_PRL_03} and on \super{14}C
    \citep[thick curve with error bars,][]{solanki_Nat_04}. The
    horizontal dotted line depicts the high activity threshold.}
    \label{Fig_10Be}
\end{figure}}

\subsection{Verification of reconstructions}
\label{sec:veri}

Because of the diversity of the methods and results of solar-activity reconstruction,
 it is vitally important to verify them.
Even though a full verification is not possible, there are different means of indirect
 or partial verification, as discussed below.
Several solar-activity reconstructions on the
 millennium timescale, which differ from each other to some degree and are based on terrestrial
  cosmogenic isotope data, have been published recently by various groups.
Also, they may suffer from systematic effects.
Therefore, there is a need for an independent method to verify/calibrate
 these results in order to provide a reliable quantitative estimate of
 the level of solar activity in the past, prior to the era of direct observations.

\subsubsection{Comparison with direct data}

The most direct verification of solar-activity reconstruction is a comparison
 with the actual GSN sunspot data for the last few centuries.
However, regression-based models (see Section~\ref{S:regr}) cannot be tested in this way, since it
 would require a long set of independent direct data outside the ``training'' interval.
It is usual to include all available data into the ``training'' period to increase
 the statistics of the regression, which rules out the possibility of testing the model.
On the other hand, such a comparison to the actual
 GSN since 1610 can be regarded as a direct test for a physics-based model since
 it does not include phenomenological links over the same time interval.
The period of the last four centuries is pretty good for testing purposes since it includes
 the whole range of solar activity levels from the nearly spotless Maunder minimum to the modern period
 of a very active sun.
As an example, a comparison between the observed GSN and the \super{14}C-based \citep{solanki_Nat_04}
 and \super{10}Be-based \citep{usoskin_PRL_03} reconstructions is shown in Figure~\ref{Fig_10Be}.
The agreement between the actual and reconstructed sunspot numbers is quite good, the correlation
 coefficient for the \super{14}C-based series is {\it r}~=~0.93 with the RMS deviation between
 the two series being six for the period of 1610\,--\,1900 \citep{solanki_Nat_04}.
We want to stress that this reconstruction is fully physics based and does not include any
 fitting to the whole GSN data series;
 thus this comparison verifies the approach in both absolute level and relative variations.
The agreement between GSN and \super{10}Be-based reconstructions \citep{usoskin_PRL_03} is
 also good ({\it r}~=~0.78, RMS~=~10 for 1700\,--\,1985).
In this case, however, the comparison can only test the relative variation because
 of the unknown proportionality coefficient between the measured concentration of \super{10}Be
 and the production rate (Section~\ref{S:10Be_tran}), which is fitted to match the overall
 level of the reconstructed solar activity.
One can see that the reconstructed sunspot series generally follows
 the real GSN series, depicting the same main features, namely, the Maunder
 minimum, the tiny Dalton minimum, a slight decrease of activity around 1900 (sometimes called
 the modern minimum) as well as a steep rise in the first half of the 20th
 century.
This validates the reliability of the physics-based reconstruction of sunspot numbers.
Note, however, that individual 11-year cycles are poorly resolved in these reconstructions.

Models focused on the reconstruction of heliospheric parameters (HMF or the modulation
 potential $\phi$) cannot be verified in this manner since no heliospheric data exists
 before the middle of the 20th century.
Comparison to direct cosmic-ray data after the 1950s \citep[or, with caveats, after the 1930s --][]{mccracken_beer_07}
 is less conclusive, since the latter are of shorter length and correspond to a period
 of high solar activity, leading to larger uncertainties during grand minima.

It is important that some (semi)empirical relations forming the basis for the proxy method
 are established for the recent decades of high solar activity.
The end of the Modern grand maximum of activity and the current low level of activity,
 characterized by the highest ever observed cosmic ray flux as recorded by ground-based
 neutron monitors, the very low level of the HMF and geomagnetic activity, should help to
 verify the connections between solar activity, cosmic ray fluxes, geomagnetic activity, the heliospheric magnetic field, and open field.
Since some of these connections are somewhat controversial, these extreme conditions should help to
 quantify them better.

\subsubsection{Meteorites and lunar rocks: A direct probe of the
  galactic cosmic-ray flux}

Another more-or-less direct test of solar/heliospheric activity in the past
 comes from cosmogenic isotopes measured in lunar rock or meteorites.
Cosmogenic isotopes, produced in meteoritic or lunar
 rocks during their exposure to CR in interplanetary space, provide
 a direct measure of cosmic-ray flux.
Uncertainties due to imprecisely known terrestrial processes, including the geomagnetic
 shielding and redistribution process, are naturally avoided in this case,
 since the nuclides are directly produced by cosmic rays in the body of the rock, where they
 remain until they are measured, without any transport or redistribution.
The activity of a cosmogenic isotope in meteorite/lunar rock represents an integral
 of the balance between the isotope's production and decay, thus
 representing the time-integrated CR flux over a period
 determined by the mean life of the radioisotope.
The results of different analyses of measurements of cosmogenic isotopes in meteoritic and lunar rocks show
 that the average GCR flux remained roughly constant -- within 10\% over the last
 million years and within a factor of 1.5 for longer periods of up to 10\super{9}~years \citep[e.g.,][]{vogt90, grieder01}.

By means of measuring the abundance of relatively short-lived cosmogenic isotopes
 in meteorites, which fell through the ages, one can evaluate the variability of the CR flux,
 since the production of cosmogenic isotopes ceases after the fall of the meteorite.
A nearly ideal isotope for studying centurial-scale variability is
 \super{44}Ti with a half-life of 59.2~\textpm~0.6~yr (a lifetime of about 85~years).
The isotope is produced in nuclear interactions of energetic CR with
 nuclei of iron and nickel in the body of a meteorite \citep{bonino95, taricco06}.
Because of its mean life, \super{44}Ti is relatively insensitive
 to variations in cosmic-ray flux on decade (11-year Schwabe cycle)
 or shorter timescales, but is very sensitive to the level of CR
 flux and its variations on a centurial scale.
Using a full model of \super{44}Ti production in a stony meteorite \citep{michel98} and
 data on the measured activity of cosmogenic isotope \super{44}Ti in meteorites, which fell during the past 235~years
 \citep{taricco06}, \citet{usoskin_Ti_06} tested, in a straightforward manner, several recent
 reconstructions of heliospheric activity after the Maunder minimum.
First, the expected \super{44}Ti activity has been calculated from the reconstructed
 series of the modulation potential, and then compared with the results of actual measurements
 (see Figure~\ref{Fig:Ti}).
It has been shown that \super{44}Ti data can distinguish between various
 reconstructions of past solar activity, allowing unrealistic models to be ruled out.
Since the life-time of the \super{44}Ti is much longer than the 11-year cycle, this method does not allow
 for the reconstruction of solar/heliospheric activity, but it serves as a direct way to test
 existing reconstructions independently.
Most of the reconstructions appear consistent with the measured \super{44}Ti activity
 in meteorites, including the last decades, thus validating their veracity.
The only apparently-inconsistent model is the one by \citet{muscheler_Nat_05}, which
 is based on erroneous normalization \citep[as discussed in][]{solanki_Nat_05}.
In particular, the \super{44}Ti data confirms significant secular
 variations of the solar magnetic flux during the last century \citep[cf.][]{lockwood99, solanki00, wang05}.

\epubtkImage{LR_Ti.png}{%
  \begin{figure}[htbp]
    \centerline{\includegraphics[width=0.8\textwidth]{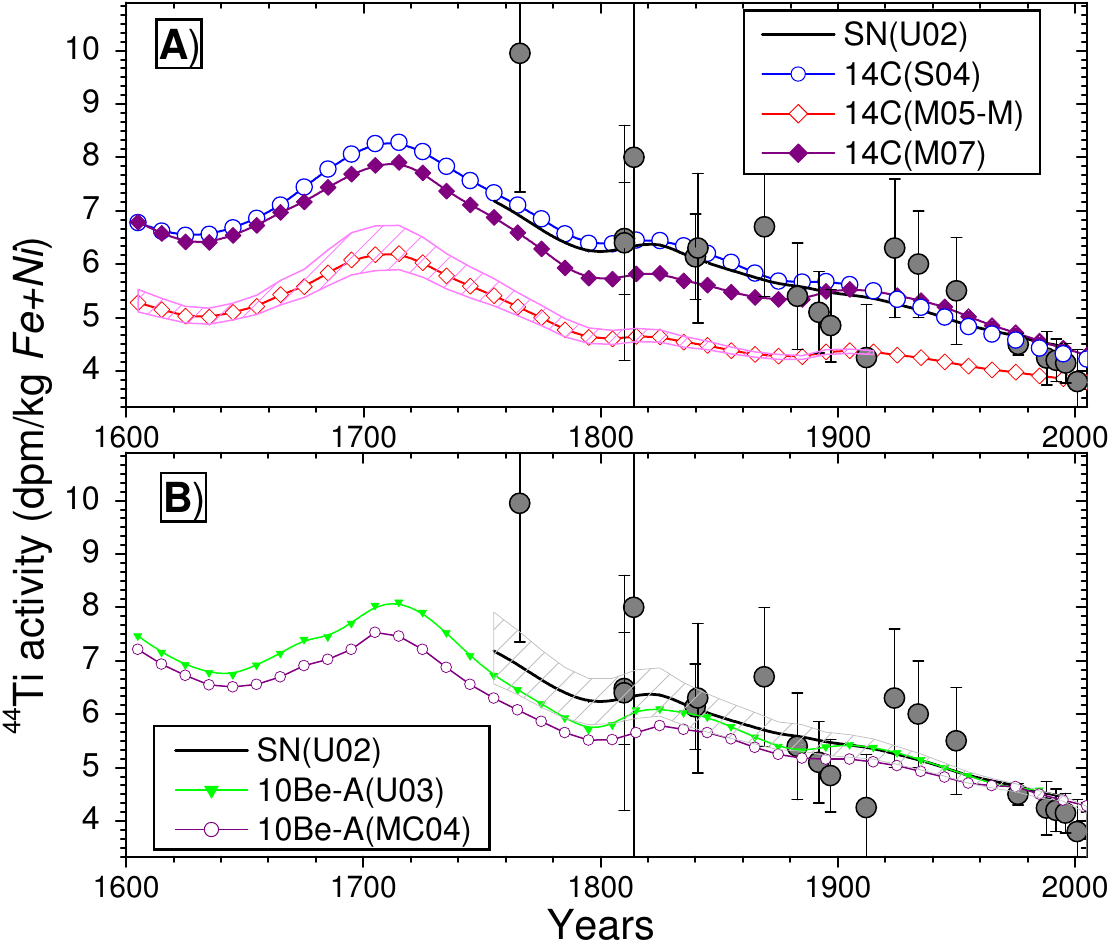}}
    \caption{Immediate \super{44}Ti activity in stony meteorites as a
    function of time of fall.
    Dots with error bars correspond to measured values \citep{taricco06}.
    Curves correspond to the theoretically expected \super{44}Ti activity, computed using the
    method of \citet{usoskin_Ti_06} and different reconstructions of
    $\phi$ shown in Figure~\ref{Fig:phi}.}
    \label{Fig:Ti}
\end{figure}}

\subsubsection{Comparison between isotopes}
\label{S:comp}

As an indirect test of the solar-activity reconstruction, one can compare different isotopes.
The idea behind this test is that two isotopes, \super{14}C and \super{10}Be, have essentially
 different terrestrial fates, so that only the production signal, namely, solar modulation of
 cosmic rays, can be regarded as common in the two series.
Processes of transport/deposition are completely different (moreover, the \super{14}C
 series is obtained as an average of the world-wide--distributed samples).
The effect of changing geomagnetic fields is also different (although not
 completely) for the two isotopes, since radiocarbon is globally mixed, while
 \super{10}Be is only partly mixed before being stored in an archive.
Even comparison between data of the same \super{10}Be isotope, but measured in far-spaced ice cores
 (e.g., Greenland and Antarctica), may help in separating climatic and extraterrestrial factors, since
 meteorology in the two opposite polar areas is quite different.

The first thorough consistent comparison between \super{10}Be and \super{14}C records for the last millennium
 was performed by \citet{bard97}.
They assumed that the measured \super{10}Be concentration in Antarctica is directly related
 to CR variations.
Accordingly, \super{14}C production was considered as proportional to \super{10}Be data.
Then, applying a 12-box carbon-cycle model, \citet{bard97} computed the expected $\Delta$\super{14}C
 synthetic record.
Finally, these \super{10}Be-based $\Delta$\super{14}C variations were compared with the actual measurements
 of $\Delta$\super{14}C in tree rings, which depicted a close agreement in the profile of temporal variation
 (coefficient of linear correlation {\it r}~=~0.81 with exact phasing).
Despite some fine discrepancies, which can indicate periods of climatic influence in either (or both)
 of the series, that result has clearly proven the dominance of solar modulation of cosmogenic nuclide
 production variations during the last millennium.
This conclusion has been confirmed \citep[e.g.,][]{usoskin_PRL_03, muscheler07} in the sense
 that quantitative solar-activity reconstructions, based on \super{10}Be and \super{14}C data series
 for the last millennium, yield very similar results, which differ only in small details.
However, a longer comparison over the entire Holocene timescale suggests that, while
 centennial variations of solar activity reconstructed from the two isotopes are very close
 to each other, there might be a discrepancy in the very long-term trend \citep{vonmoos06, muscheler07},
 whose nature is not clear (climate changes, geomagnetic effects or model uncertainties).

Recently, \citet{usoskin_10Be_09} studied the dominance of the solar signal in different cosmogenic
 isotope data on different time scales.
They compared the expected \super{10}Be variations computed from \super{14}C-based reconstruction of cosmic
 ray intensity with the actually measured \super{10}Be abundance at the sites and found that:
 (1) There is good agreement between the \super{14}C and \super{10}Be data sets,
  on different timescales and at different locations, confirming the existence of a common solar signal in both isotope data;
 (2) The \super{10}Be data are driven by the solar signal on timescales from about centennial to millennial time scales;
 (3) The synchronization is lost on short (\textless~100~years) timescales, either due to local climate or chronological
   uncertainties \citep{delaygue11} but the solar signal becomes important even at short scales during periods of Grand minima of solar activity,
 (4) There is an indication of a possible systematic uncertainty in the early Holocene \citep[cf.,][]{vonmoos06}, likely due to a
  not-perfectly-stable thermohaline circulation.
Overall, both \super{14}C- and \super{10}Be-based records are consistent with each other over a wide range of timescales and time intervals.


Thus, comparison of the results obtained from different sources implies that the
 variations of cosmogenic nuclides on the long-term scale (centuries to millennia) during
 the Holocene are primarily defined by the solar modulation of CR.

\subsection{Composite reconstruction}
\label{S:compo}

Most of the earlier solar activity reconstructions are based on single proxy records, either \super{14}C or
 \super{10}Be.
Although they are dominated by the same production signal, viz.\ solar activity, (see Section~\ref{S:comp}),
 they still contain essential fractions of noise.

A promising first step in the direction of extracting the common solar signal from different proxy records
 was made recently by \citet{steinhilber12}.
They combined, in a composite reconstruction, different \super{10}Be ice core records from Greenland
 and Antarctica with the global \super{14}C tree ring record.
The composite was made in a mathematical way, using the principal component analysis as a numerical tool.
This analysis formally finds the common variability in different series, that is assumed to be
 the solar signal.
The resultant reconstruction is shown in Figure~\ref{Fig:PNAS} and is the most consistent reconstruction
 of the cosmic ray (and thus solar magnetic activity) variability over the Holocene, available up-to-date.
However, since the used mathematical tool can only work with the relative variability, the reconstruction
 also yields the relative values rather than absolute values, and it is not available in the terms of sunspot numbers.
\epubtkImage{LR_PNAS.png}{%
  \begin{figure}[htbp]
    \centerline{\includegraphics[width=0.8\textwidth]{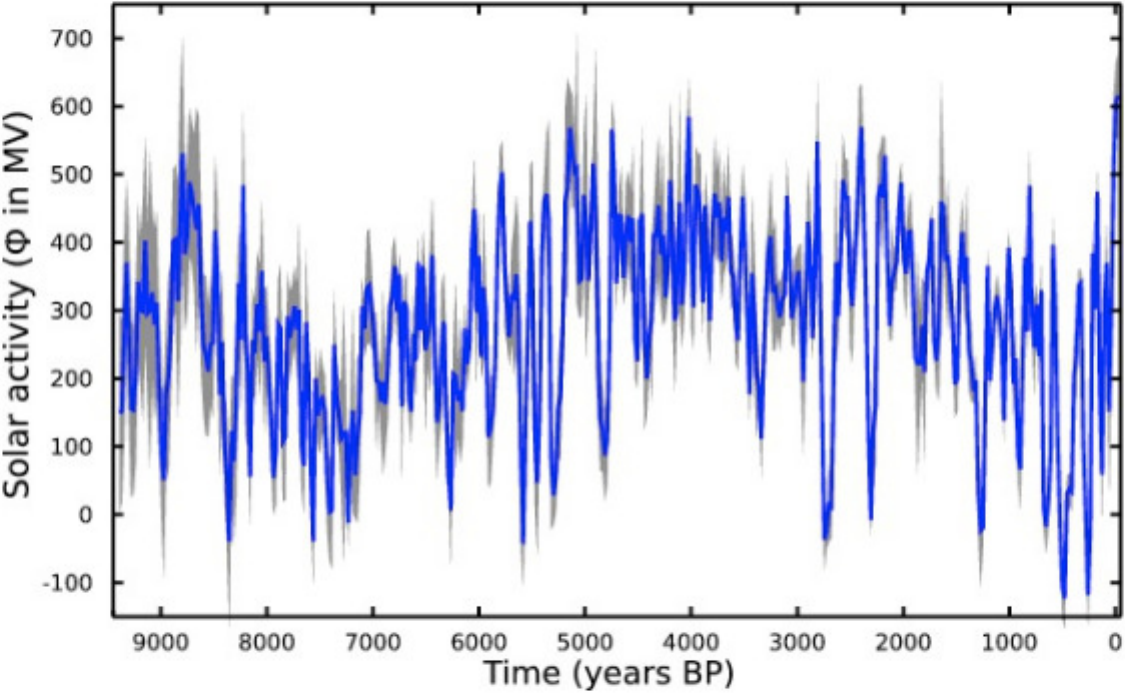}}
    \caption{Reconstruction of the solar modulation potential over the Holocene based on a
     composite record (Section~\ref{S:compo}), along with $1\sigma$ uncertainties.
     Time is given in year BP.
     Image reproduced by permission from~\cite{steinhilber12}.}
    \label{Fig:PNAS}
\end{figure}}

A full physics-based multi-proxy composite reconstruction of the solar activity on the millennial time scale
 is still pending.

\clearpage
\subsection{Summary}

In this section, a proxy method of past--solar-activity reconstruction is
 described in detail.

This method is based on the use of indirect proxies of solar activity, i.e., quantitative parameters, which can be
 measured now, but represent signatures, stored in natural archives, of the different effects of
 solar magnetic activity in the past.
Such traceable signatures can be related to nuclear or chemical effects caused by cosmic rays
 in the Earth's atmosphere, lunar rocks or meteorites.
This approach allows one to obtain homogeneous data sets with stable quality and to improve
 the quality of data when new measurement techniques become available.
It provides the only possible regular indicator of solar activity on a very long-term scale.

The most common proxy of solar activity is formed by data of the cosmogenic radionuclides,
 \super{10}Be and \super{14}C, produced by cosmic rays in the Earth's atmosphere.
After a complicated transport in the atmosphere, these cosmogenic isotopes are stored in natural archives
 such as polar ice, trees, marine sediments, from where they can now be measured.
This process is also affected by changes in the geomagnetic field and the climate.

Radioisotope \super{14}C, measured in independently-dated tree rings,
 forms a very useful proxy for long-term solar-activity variability.
It participates in the complicated carbon cycle, which smoothes out spatial and short-term
 variability of isotope production.
For the Holocene period, with its stable climate, it provides a useful tool
 for studying solar activity in the past.
Existing models allow the quantitative conversion between the measured relative abundance of
 \super{14}C and the production rate in the atmosphere.
The use of radiocarbon for earlier periods, the glacial and deglaciation epochs, is limited by
 severe climate and ocean ventilation changes.
Radiocarbon data cannot be used after the end of the 19th century because of the Suess effect and
 atmospheric nuclear tests.

Another solar activity proxy is the cosmogenic \super{10}Be isotope measured
 in stratified polar ice cores.
Atmospheric transport of \super{10}Be is relatively straightforward, but its details
 are as of yet unresolved, leading to the lack of a reliable quantitative model relating
 the measured isotope concentration in ice to the atmospheric production.
Presently, it is common to assume that the production rate is proportional,
 with an unknown coefficient, to the measured concentration.
However, a newly-developed generation of models, which include 3D atmospheric-circulation models,
 will hopefully solve this problem soon.

Recently a new proxy, nitrate concentration measured in an Antarctic ice core,
 has been proposed for long-term solar activity reconstructions, but it still needs
 verification and model support.

Modern physics-based models make it possible to build a chain, which
 quantitatively connects isotope production rate and sunspot activity,
 including subsequently the GCR flux quantified via the modulation potential,
 the heliospheric index, quantified via the open solar magnetic flux
 or the average HMF intensity at the Earth's orbit, and finally the
 sunspot-number series.
Presently, all these steps can be made using appropriate models allowing for
 a full basic quantitative reconstruction of solar activity in the past.
The main uncertainties in the solar-activity reconstruction arise from
 paleo-magnetic models and the overall normalization.

An independent verification of the reconstructions, including direct comparison with sunspot numbers,
 cosmogenic isotopes in meteorites and the comparison of different models with each other, confirms
 their veracity in both relative variations and absolute level.
It also implies that the variations in cosmogenic nuclides on the long-term scale (centuries to millennia) during
 the Holocene are primarily defined by the solar modulation of CR.

\clearpage


\section{Variability of Solar Activity Over Millennia}
\label{sec:3}

Several reconstructions of solar activity on multi-millennial timescales have been performed recently
 using physics-based models (see Section~\ref{S:4}) from measurements of \super{14}C in tree rings
 and \super{10}Be in polar ice.
The validity of these models for the last few centuries was discussed in Section~\ref{sec:veri}.
In this section we discuss the temporal variability of thus-reconstructed solar activity on
 a longer scale.


Here we consider the \super{14}C-based decade reconstruction of sunspot numbers (shown in Figure~\ref{Fig:SN_S04}).
It is identical to that shown in Figure~\ref{Fig:long}, but includes also a Gleissberg
 1-2-2-2-1 filter in order to suppress noise and short-term fluctuations.
This series forms the basis for the forthcoming analysis, while differences related to the
 use of other reconstructions are discussed.

\epubtkImage{LR_min.png}{%
  \begin{figure}[htbp]
    \centerline{\includegraphics[width=0.8\textwidth]{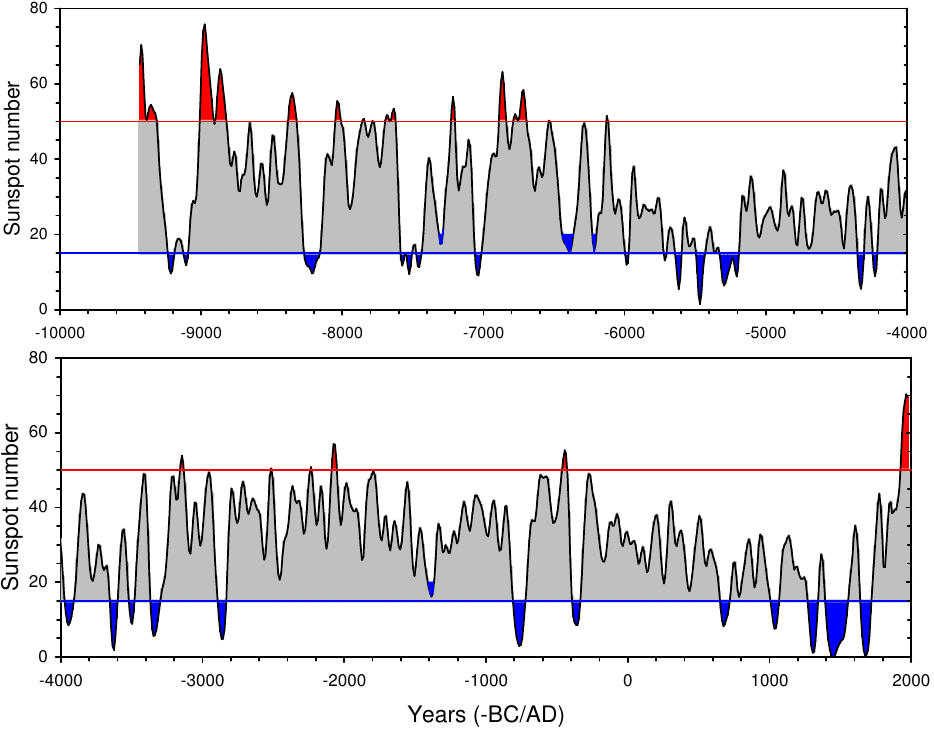}}
    \caption{Sunspot activity (over decades, smoothed with a 12221 filter)
    throughout the Holocene, reconstructed from \super{14}C by
    \citet{usoskin_AA_07} using geomagnetic data by
    \citet{yang00}. Blue and red areas denote grand minima and maxima,
    respectively.}
    \label{Fig:SN_S04}
\end{figure}}

\subsection{Quasi-periodicities and characteristic times}
\label{S:period}

In order to discuss spectral features of long-term solar-activity dynamics, we show in Figure~\ref{Fig:wv} a
 wavelet spectral decomposition of the sunspot number reconstruction throughout the
 Holocene shown in Figure~\ref{Fig:SN_S04}.
The left-hand panels show the conventional wavelet decomposition in the time-frequency domain,
 while the right-hand panels depict the global spectrum, namely, an integral over the time domain,
 which is comparable to a Fourier spectrum.
The peak in the global spectrum at about an 80-year period corresponds to the Gleissberg periodicity,
 known from a simple Fourier analysis of the $\Delta$\super{14}C series \citep{peristykh03}.
The peak at an approximately 150~year period does not correspond to a persistent periodicity,
 but is formed by a few time intervals (mostly 6000\,--\,4000~BC) and can be related to another
 ``branch'' of the secular cycle, according to \citet{ogurtsov02}.
The de~Vries/Suess cycle, with a period of about 210~years, is prominent in the global spectrum,
 but it is intermittent and tends to become strong with around 2400 clustering time \citep{usoskin_Rev_04}.
Another variation with a period of around 350~years can be observed after 6000~BC \citep[cf.][]{steinhilber12}.
Variations with a characteristic time of 600\,--\,700~years are intermittent and can be hardly
 regarded as a typical feature of solar activity.
There is also a weak millennial periodicity \citep{steinhilber12}.
Of special interest is the 2000\,--\,2400 year Hallstatt cycle \citep[see, e.g.,][]{vitinsky86, damon91, vassiliev02},
 which is relatively stable and mostly manifests itself as a modulation of long-term solar activity,
 leading to the clustering of grand minima \citep{usoskin_AA_07}.

On the other hand, an analysis of the occurrence of grand minima (see Section~\ref{sec:MM}) shows no clear
 periodicity except for a marginal 2400 year clustering, implying that the occurrence of grand minima and maxima is not
 a result of long-term cyclic variability but is defined by stochastic/chaotic processes.

\epubtkImage{LR_SN_wv.png}{%
  \begin{figure}[htbp]
    \centerline{\includegraphics[width=\textwidth]{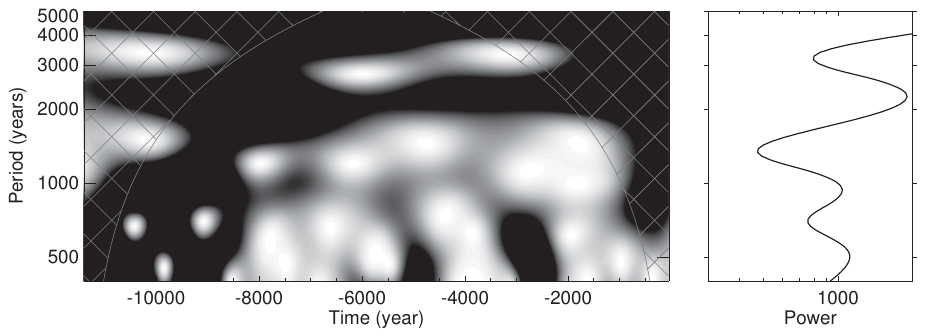}}
    \centerline{\includegraphics[width=\textwidth]{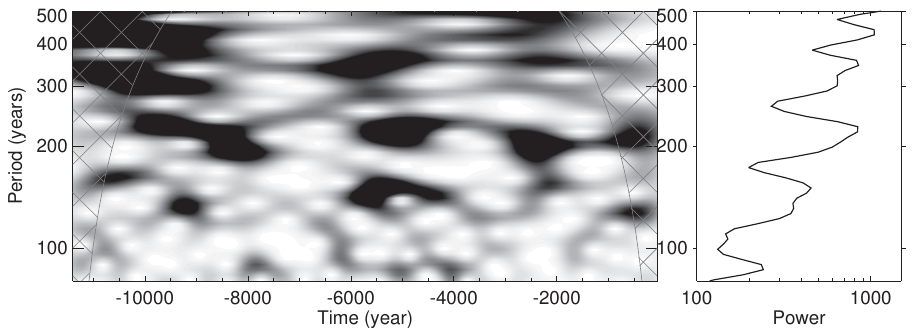}}
    \caption{Wavelet (Morlet basis) spectrum of the sunspot-number
    reconstruction shown in Figure~\ref{Fig:SN_S04}.
    Left and right-hand panels depict 2D and global wavelet spectra, respectively.
    Upper and lower panels correspond to period ranges of
    500\,--\,5000 years and 80\,--\,500~years, respectively.
    Dark/light shading denotes high/low power.
    Hatched areas depict the cone of influence where the result is not
     fully reliable because of the proximity of the edges of the time series.}
    \label{Fig:wv}
\end{figure}}

\subsection{Grand minima of solar activity}
\label{sec:MM}

A very particular type of solar activity is the grand minimum, when solar activity
 is greatly reduced.
The most famous is the Maunder minimum in the late 17th century, which is discussed below in some detail
 \citep[for a detailed review see the book by][]{soon03}.
Grand minima are believed to correspond to a special state of the dynamo
 \citep{sokoloff04, miyahara06}, and its very existence poses a challenge for the solar-dynamo theory.
It is noteworthy that dynamo models do not agree on how often such episodes occur in the sun's history
 and whether their appearance is regular or random.
For example, the commonly used mean-field dynamo yields a fairly-regular 11-year cycle \citep{charbonneau10},
 while dynamo models including a stochastic driver predict the intermittency of solar magnetic activity
 \citep{choudhuri92, schussler94, schmitt96, ossendrijver00, weiss00, mininni01, charbonneau01}.
Most of the models predict purely random occurrence of the grand minima, without any intrinsic long-term
 memory \citep{moss08}.
Although cosmogenic isotope data suggest the {possible} existence of such memory \citep{usoskin_AA_07}, statistics
 is not sufficient to distinguish between the two cases \citep{usoskin_sok_09}.


\subsubsection{The Maunder minimum}

The Maunder minimum is a representative of grand minima in solar activity \citep[e.g.,][]{eddy76},
 when sunspots have almost completely vanished from the solar surface, while the solar wind
 kept blowing, although at a reduced pace \citep{cliver98, usoskin_JGR_MM_01}.
There is some uncertainty in the definition of its duration; the ``formal'' duration is 1645\,--\,1715 \citep{eddy76},
 while its deep phase with the absence of apparent sunspot cyclic activity is often
 considered as 1645\,--\,1700, with the low, but very clear, solar cycle of 1700\,--\,1712 being ascribed to a recovery
 or transition phase \citep{usoskin_MM_AA_00}.
The Maunder minimum was amazingly well covered (more than 95\% of days) by direct sunspot
 observations \citep{hoyt96}, especially in its late phase \citep{ribes93}.
On the other hand, sunspots appeared rarely (during $\sim$~2\% of the days)
 and seemingly sporadically, without an indication of the 11-year cycle \citep{usoskin_SP03}.
This makes it almost impossible to apply standard methods of time-series
 analysis to sunspot data during the Maunder minimum \citep[e.g.,][]{frick97}).
Therefore, special methods such as the distribution of spotless days vs. days with
 sunspots \citep[e.g.,][]{harvey99,kovaltsov04} or an analysis of sparsely-occurring
 events \citep{usoskin_MM_AA_00} should be applied in this case.
Using these methods, \citet{usoskin_JGR_MM_01} have shown that sunspot occurrence during the Maunder minimum
 was gathered into two large clusters (1652\,--\,1662 and 1672\,--\,1689),
 with the mass centers of these clusters being in 1658 and 1679\,--\,1680.
Together with the sunspot maxima before (1640) and after (1705) the deep
 Maunder minimum, this implies a dominant 22-year periodicity in sunspot
 activity throughout the Maunder minimum \citep{mursula01},
 with a subdominant 11-year cycle emerging towards the end of the Maunder minimum
 \citep{ribes93, mendoza97, usoskin_MM_AA_00} and becoming dominant again after 1700.
Similar behavior of a dominant 22-year cycle and a weak subdominant Schwabe
 cycle during the Maunder minimum has been found in other indirect solar
 proxy data: auroral occurrence \citep{krivsky88, schlamminger90, silverman92} and
 \super{14}C data \citep{stuiver93, kocharov95, peristykh98, miyahara06}.
This is in general agreement with the concept of ``immersion'' of 11-year
 cycles during the Maunder minimum \citep[][and references therein]{vitinsky86}.
This concept means that full cycles cannot be resolved and sunspot activity only appears
 as pulses around cycle-maximum times.

An analysis of \super{10}Be data \citep{beer98} implied that the 11-year
 cycle was weak but fairly regular during the Maunder minimum, but its phase
 was inverted \citep{usoskin_JGR_MM_01}.
A recent theoretical study \citep{owens12,wang13} confirms that such a phase change between
 cosmic rays and solar activity can appear for very weak cycles.

The time behavior of sunspot activity during the Maunder minimum yielded the following
 general scenario \citep{vitinsky86, ribes93, sokoloff94, usoskin_MM_AA_00, usoskin_JGR_MM_01, miyahara06}.
Transition from the normal high activity to the deep minimum did not have any apparent precursor.
 On the other hand, newly recovered data suggest that the start of the Maunder minimum might had been not
 very sudden but via a regular cycle of reduced height \citep{vaquero11}.
A 22-year cycle was dominant in sunspot occurrence during the deep minimum (1645\,--\,1700), with
 the subdominant 11-year cycle, which became visible only in the late phase of the Maunder minimum.
There is an indication that the length of solar cycle may slightly extend during
 and already slightly before a grand minimum \citep{miyahara04,nagaya12}.
The 11-year Schwabe cycle started dominating solar activity after 1700.
Recovery of sunspot activity from the deep minimum to normal activity was
 gradual, passing through a period of nearly-linear amplification of the 11-year cycle.
It is interesting to note that such a qualitative evolution of a grand minimum is
 consistent with predictions of the stochastically-forced return map \citep{charbonneau01}.

Although the Maunder minimum is the only one with available direct sunspot observations,
 its predecessor, the Sp\"orer minimum from 1450\,--\,1550, is covered by precise bi-annual measurements of
 \super{14}C \citep{miyahara_JGR_06}.
An analysis of this data \citep{miyahara_JGR_06, miyahara06} reveals a similar pattern with the
 dominant 22-year cycle and suppressed 11-year cycle, thus supporting the
 idea that the above general scenario may be typical for a grand minimum.
A similar pattern has been recently also for an un-named grand minima in
 the 4-th century BC \citep{nagaya12}.

A very important feature of sunspot activity during the Maunder minimum was its strong
 north-south asymmetry, as sunspots were only observed in the southern solar hemisphere
 during the end of the Maunder minimum \citep{ribes93, sokoloff94}.
This observational fact has led to intensive theoretical efforts to explain a
 significant asymmetry of the sun's surface magnetic field in the framework of the
 dynamo concept \citep[see the review by][and references therein]{sokoloff04}.
Note that a recent discovery \citep{arlt08, arlt09} of the Staudacher's original drawings of
 sunspots in late 18th century shows that similarly asymmetric sunspot occurrence existed
 also in the beginning of the Dalton minimum in 1790s \citep{usoskin_lost_09}.
However, the northern hemisphere dominated at that period contrary to the situation
 during the Maunder minimum.


\subsubsection{Grand minima on a multi-millennial timescale}
\label{MM:long}

\enlargethispage{\baselineskip}
The presence of grand minima in solar activity on the long-term scale has been mentioned numerously
 \citep[e.g.,][]{eddy77, solanki_Nat_04}, using the radioisotope \super{14}C data in tree rings.
For example, \citet{eddy_climate_77} identified major excursions in the detrended \super{14}C record as
 grand minima and maxima of solar activity and presented a list of six grand minima
 and five grand maxima for the last 5000~years (see Table~\ref{Tab:min}).
\citet{stuiver89} and \citet{stuiver91} also studied grand minima as systematic excesses of the high-pass
 filtered \super{14}C data and suggested that the minima are generally of two distinct types:
 short minima of duration 50\,--\,80~years (called Maunder-type) and longer minima
 collectively called Sp\"orer-like minima.
Using the same method of identifying grand minima as significant peaks in
 high-pass filtered $\Delta$\super{14}C series, \citet{voss96} provided a list of 29
 such events for the past 8000~years.
A similar analysis of bumps in the \super{14}C production rate was presented recently by \citet{goslar03}.
However, such studies retained a qualitative element, since they are based on high-pass--filtered \super{14}C
 data and thus implicitly assume that \super{14}C variability can be
 divided into short-term solar variations and long-term changes attributed solely to
 the slowly-changing geomagnetic field.
This method ignores any possible long-term changes in solar activity
 on timescales longer than 500~years \citep{voss96}.
The modern approach, based on physics-based modelling (Section~\ref{S:4}), allows for the quantitative
 reconstruction of the solar activity level in the past, and thus, for a more realistic definition of
 the periods of grand minima or maxima.

\begin{table}[htb]
\caption{Approximate dates (in --BC/AD) of grand minima in reconstructed solar activity.}
\label{Tab:min}
\begin{center}
{\small
\begin{tabular}{rrrl}
\toprule
No.& center & duration & comment\\
\midrule
1  & 1680 & 80 & Maunder \\
2  & 1470 & 160 & Sp\"orer \\
3  & 1305 & 70 & Wolf \\
4  & 1040 & 60 &  a, d) \\
5  & 685 & 70 & b, d) \\
6  & --360 & 60 & a, b, c, d) \\
7  & --765 & 90 & a, b, c, d) \\
8  & --1390 & 40 & b, d) \\
9  & --2860 & 60 & a, c, d) \\
10 & --3335 & 70 & a, b, c, d) \\
11 & --3500 & 40 & a, b, c, d) \\
12 & --3625 & 50 & a, b, d) \\
13 & --3940 & 60 & a, c, d) \\
14 & --4225 & 30 & c, d) \\
15 & --4325 & 50 & a, c, d) \\
16 & --5260 & 140 & a, b, d) \\
17 & --5460 & 60 & c, d) \\
18 & --5620 & 40 & d) \\
19 & --5710 & 20 & c, d) \\
20 & --5985 & 30 & a, c, d) \\
21 & --6215 & 30 & c, d, e) \\
22 & --6400 & 80 & a, c, d, e) \\
23 & --7035 & 50 & a, c, d) \\
24 & --7305 & 30 & c, d) \\
25 & --7515 & 150 & a, c, d) \\
26 & --8215 & 110 & d) \\
27 & --9165 & 150 & d) \\
\bottomrule
\end{tabular}}
\end{center}
{\small
a) According to \citet{stuiver80, stuiver89}.\\
b) According to \citet{eddy77, eddy_climate_77}.\\
c) According to \citet{goslar03}.\\
d) According to \citet{usoskin_AA_07}.\\
e) Exact duration is uncertain.
}
\end{table}

A list of 27 grand minima, identified in the quantitative solar-activity reconstruction of the last 11,000~years,
 shown in Figure~\ref{Fig:SN_S04}, is presented in Table~\ref{Tab:min} \citep[after][]{usoskin_AA_07}.
The cumulative duration of the grand minima is about 1900~years, indicating
 that the sun in its present evolutionary stage spends $\sim {^1/_6}$ (17\%) of its time
 in a quiet state, corresponding to grand minima.
Note that the definition of grand minima is quite robust.

The question of whether the occurrence of grand minima in solar activity is a regular
 or chaotic process is important for understanding the {action of the} solar-dynamo machine.
Even a simple deterministic numerical dynamo model can produce events comparable with  grand
 minima \citep[]{brandenburg89}.
Such models can also simulate a sequence of grand minima occurrences, which are irregular
 and seemingly chaotic \citep[e.g.,][]{jennings91, tobias95, covas98}.
The presence of long-term dynamics in the dynamo process is often explained in terms
 of the $\alpha$-effect, which, being a result of the electromotive force averaged over
  turbulent vortices, can contain a fluctuating part \citep[e.g.,][]{hoyng93, ossendrijver96}
  leading to irregularly occurring grand minima \citep[e.g.,][]{brandenburg08}.
The present dynamo models can reproduce almost all the observed features of the solar cycle
  under \textit{ad hoc} assumptions \citep[e.g.,][]{pipin12}, although it is still unclear what
  leads to the observed variability.
Most of these models predict that the occurrence of grand minima is a purely random ``memoryless'' Poisson-like
 process, with the probability of a grand minimum occurring being constant at any given time.
This unambiguously leads to the exponential shape of the waiting-time distribution (waiting time
 is the time interval between subsequent events) for grand minima.

\citet{usoskin_AA_07} performed a statistical analysis of grand minima occurrence time
 (Table~\ref{Tab:min}) and concluded that their occurrence is not a result
 of long-term cyclic variations, but is defined by stochastic/chaotic processes.
Moreover, waiting-time distribution deviates from the exponential law.
This implies that the event occurrence is still random, but the probability
 is nonuniform in time and depends on the previous history.
In the time series it is observed as a tendency of the events to cluster together
 with a relatively-short waiting time, while the clusters are separated by
 long event-free intervals (cf.\ Section~\ref{S:period}).
Such behavior can be interpreted in different ways,
 e.g., self-organized criticality or processes related to accumulation and release of energy.
This poses a strong observational constraint on theoretical models
 aiming to explain the long-term evolution of solar activity (Section~\ref{sec:theor}).
However, as discussed by \citet{moss08} and \citet{usoskin_sok_09}, the observed feature
 can be an artefact of the small statistics (only 27 grand minima are identified during the Holocene),
 making this result only indicative and waiting for a more detailed investigation.


A histogram of the duration of grand minima from Table~\ref{Tab:min} is shown in Figure~\ref{Fig:MM_dur}.
The mean duration is 70~year but the distribution is bimodal.
The minima tend to be either of a short (30\,--\,90~years) duration similar to the Maunder minimum,
 or rather long (\textgreater~100~years), similar to the Sp\"orer minimum, in agreement with earlier
 conclusions \citep{stuiver89}.
This suggests that grand minima correspond to a special state of the dynamo.
Once falling into a grand minimum as a result of a stochastic/chaotic, but
 non-Poisson process, the dynamo is ``trapped'' in this state and its behavior is
 driven by deterministic intrinsic features.

\epubtkImage{LR_MM_duration.png}{%
  \begin{figure}[htbp]
    \centerline{\includegraphics[width=8cm]{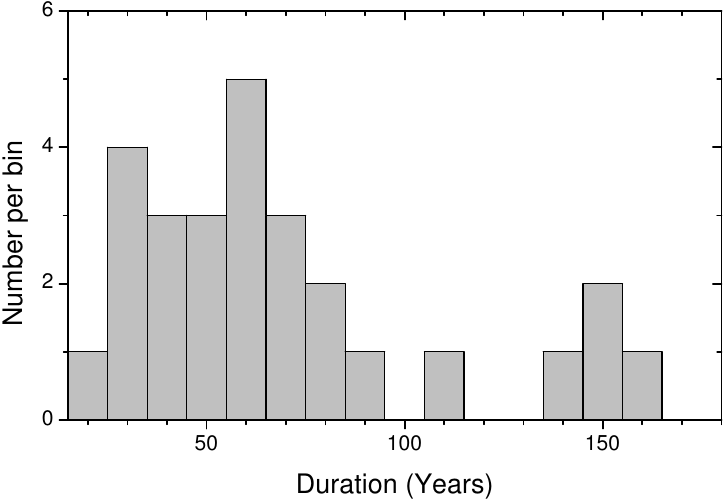}}
    \caption{Histogram of the duration of grand minima from
    Table~\ref{Tab:min}.}
    \label{Fig:MM_dur}
\end{figure}}

\clearpage
\subsection{Grand maxima of solar activity}
\label{S:Max}

\subsubsection{The modern episode of active sun}

{In the last decades we were living in a period of a very active sun} with a level of
 activity that is unprecedentedly high for the last few centuries covered by direct
 solar observation.
The sunspot number was growing rapidly between 1900 and 1940, with more than a doubling
 average group sunspot number, and has remained at that high level until recently (see Figure~\ref{Fig:SA}).
Note that growth comes mostly from raising the cycle maximum amplitude, while
 sunspot activity always returns to a very low level around solar cycle minima.
While the average group sunspot number for the period 1750\,--\,1900 was 35~\textpm~9
 (39~\textpm~6, if the Dalton minimum in 1797\,--\,1828 is not counted), it stands high at the level of
 75~\textpm~3 for 1950\,--\,2000.
Therefore, the modern active sun episode, which started in the 1940s, can be regarded as
 the modern grand maximum of solar activity, as opposed to a grand minimum \citep{wilson88a}.
As first shown by \citet{usoskin_PRL_03} and \citet{solanki_Nat_04}, such high activity
 episodes occur quite seldom.

However, as we can securely say now, after the very weak solar minimum in 2008\,--\,2009 \citep[e.g.,][]{gibson11},
 solar activity returns to its normal moderate level, or perhaps even to a low-activity stage,
 comparable to the Dalton minimum in the turn of 18\,--\,19th centuries \citep[e.g.,][]{lockwood11}.
Thus, the high activity episode known as the Modern grand maximum is over.

Is such high solar activity typical or is it something extraordinary?
While it is broadly agreed that the modern active sun episode is a special phenomenon,
 the question of how (a)typical such upward bumps are
 from ``normal'' activity is a topic of hot debate.

\subsubsection{Grand maxima on a multi-millennial timescale}
\label{sec:max}

The question of how often grand maxima occur and how strong they are, cannot be
 studied using the 400-year-long series of direct observations.
An increase in solar activity around 1200~AD, also related to the
 Medieval temperature optimum, is sometimes qualitatively regarded as a grand maximum
 \citep{wilson88a, demeyer98}, but its magnitude is lower than the modern maximum
 \citep{usoskin_PRL_03}.
Accordingly, it was not included in a list of grand maxima by \citet{eddy77, eddy_climate_77}.

\begin{table}[htb]
\caption[Approximate dates (in --BC/AD) of grand maxima in the SN-L
  series.]{Approximate dates (in --BC/AD) of grand maxima in the SN-L
  series \citep[after][]{usoskin_AA_07}.}
\label{Tab:max}
\begin{center}
{\small
\begin{tabular}{rrr}
\toprule
No. & center & duration \\
\midrule
1\,\textdagger   &   1960    &   80  \\
2   &   --445    &   40   \\
3   &   --1790   &   20   \\
4   &   --2070   &   40   \\
5   &   --2240   &   20   \\
6   &   --2520   &   20   \\
7   &   --3145   &   30   \\
8   &   --6125   &   20   \\
9   &   --6530   &   20   \\
10  &   --6740   &   100  \\
11  &   --6865   &   50    \\
12  &   --7215   &   30    \\
13  &   --7660   &   80    \\
14  &   --7780   &   20    \\
15  &   --7850   &   20    \\
16  &   --8030   &   50    \\
17  &   --8350   &   70    \\
18  &   --8915   &   190   \\
19  &   --9375   &   130   \\
\bottomrule
\end{tabular}}
\end{center}
{\footnotesize \textdagger~Center and duration of the modern maximum are preliminary
since it is still ongoing.}
\end{table}

A quantitative analysis is only possible using proxy data, especially cosmogenic isotope records.
Using a physics-based analysis of solar-activity series reconstructed from \super{10}Be data
 from polar (Greenland and Antarctica) archives,
 \citet{usoskin_PRL_03, usoskin_AA_04} stated that the modern maximum is unique in the last millennium.
Then, using a similar analysis of the \super{14}C calibrated series, \citet{solanki_Nat_04}
 found that the modern activity burst is not unique, but a very rare event, with the previous burst
 occurring about 8 millennia ago.
An update \citep{usoskin_GRL_06} of this result, using a more precise paleo-magnetic reconstruction by \citet{korte05}
 since 5000~BC, suggests that an increase of solar activity comparable with the modern episode might have
 taken place around 2000~BC, i.e., around 4 millennia ago.
This result is confirmed by the most recent composite reconstruction by \citet{steinhilber12}.
The result by \citet{solanki_Nat_04} has been disputed by \citet{muscheler_Nat_05} who claimed
 that equally high (or even higher) solar-activity bursts occurred several times during
 the last millennium, circa 1200~AD, 1600~AD and at the end of the 19th century.
We note that the latter claimed peak (ca. 1860) is not confirmed by direct solar
 or geomagnetic data.
However, as argued by \citet{solanki_Nat_05}, the level of solar activity reconstructed by \citet{muscheler_Nat_05}
 was overestimated because of an erroneous normalization to the data of ground-based ionization chambers
 \citep[see also][]{mccracken_beer_07}.
This indicates that the definition of grand maxima is less robust than grand minima and
 is sensitive to other parameters such as geomagnetic field data or overall normalization.

Keeping possible uncertainties in mind, let us consider a list of the largest grand maxima
 (the 50 year smoothed sunspot number stably exceeding 50), identified for the last 11,400~years using \super{14}C data,
 as shown in Table~\ref{Tab:max} \citep[after][]{usoskin_AA_07}.
A total of 19 grand maxima have been identified with a total duration of around 1030~years, suggesting that the sun spends
 around 10\% of its time in an active state.
A statistical analysis of grand-maxima--occurrence time
 suggests that they do not follow long-term cyclic variations, but like grand minima, are defined by stochastic/chaotic processes.
The distribution of the waiting time between consecutive grand maxima
 is not unambiguously clear, but also hints at a deviation from exponential law.
The duration of grand maxima has a smooth distribution, which nearly exponentially
 decreases towards longer intervals.
Most of the reconstructed grand maxima (about 75\%) were not longer than 50~years, and
 only four grand minima (including the modern one) have been longer than 70~years \citep[cf.][]{barnard11}.
This suggest that the probability of the modern active-sun episode continuing
 is low\epubtkFootnote{This is not a prediction of future solar activity,
 but only a statistical estimate.} \citep[cf.][]{solanki_Nat_04, abreu08}.

\subsection{Related implications}

Reconstructions of long-term solar activity have different implications
 in related areas of science.
The results, discussed in this overview, can be used in such diverse research
 disciplines as theoretical astrophysics, solar-terrestrial studies,
 paleo-climatology, and even archeology and geology.
We will not discuss all possible implications of long-term solar activity
 in great detail but only briefly mention them here.

\subsubsection{Theoretical constrains}
\label{sec:theor}
The basic principles of the occurrence of the 11-year Schwabe cycle are more-or-less understood in terms of
 the solar dynamo, which acts, in its classical form \citep[e.g.,][]{parker55}, as follows \citep[see detail in][]{charbonneau10}.
Differential rotation $\Omega$ produces a toroidal magnetic field from a poloidal one, while the
 ``$\alpha$-effect'', associated with the helicity of the velocity field or Joy's Law tilt of active
 regions, produces a poloidal magnetic field from a toroidal one.
This classical model results in a periodic process in the form of propagation of a toroidal
 field pattern in the latitudinal direction (the ``butterfly diagram'').
As evident from observation, the solar cycle is far from being a strictly periodic phenomenon,
 with essential variations in the cycle length and especially in the amplitude,
 varying dramatically between nearly spotless grand minima and very large values during
 grand maxima.
The mere fact of such great variability, known from sunspot data, forced solar
 physicists to develop dynamo models further.
Simple deterministic numerical dynamo models, developed on the basis of Parker's
 migratory dynamo, can simulate events, which are seemingly comparable with grand minima/maxima
 occurrence \citep[e.g.,][]{brandenburg89}.
However, since variations in the solar-activity level, as deduced from cosmogenic isotopes,
 appear essentially nonperiodic and irregular, appropriate models have been developed to reproduce
 irregularly-occurring grand minima \citep[e.g.,][]{jennings91, tobias95, covas98}.
Models, including an \emph{ad hoc} stochastic driver
 \citep{choudhuri92, schmitt96, ossendrijver00, weiss00, mininni01, charbonneau01, charbonneau04},
 are able to reproduce the great variability and intermittency found in the solar cycle \citep[see the review by][]{charbonneau10}.
A recent statistical result of grand minima occurrence \citep[][Section~\ref{sec:max}]{usoskin_AA_07} shows
 disagreement between observational data, depicting a degree of self-organization or ``memory'', and
 the above dynamo model, which predicts a pure Poisson occurrence rate for grand minima (see Section~\ref{sec:MM}).
This poses a new constraint on the dynamo theory, responsible for long-term
 solar-activity variations \citep{sokoloff04,moss08}.

In general, the following additional constraints can be posed on dynamo models
 aiming to describe the long-term (during the past 11,000~years) evolution of solar magnetic activity.

\begin{itemize}

\item The sun spends about $^3/_4$ of its time at moderate magnetic-activity levels, about $^1/_6$ of its time in a grand
  minimum and about $^1/_5\mbox{\,--\,}^1/_{10}$ in a grand maximum.
  Recent solar activity corresponds to a grand maximum, which has ceased after solar cycle 23.

\item Occurrence of grand minima and maxima is not a result of
  long-term cyclic variations but is defined by stochastic/chaotic
  processes.

\item Observed statistics of the occurrence of grand minima and
 maxima display deviation from a ``memory-less'' Poisson-like
 process, but tend to either cluster events together or produce long
 event-free periods.
 This can be interpreted in different ways, such
 as self-organized criticality \citep[e.g.,][]{carvalho00},
 a time-dependent Poisson process \citep[e.g.,][]{wheatland03}, or some
 memory in the driving process \citep[e.g.,][]{mega03}.

\item Grand minima tend to be of two different types: short minima
  of Maunder type and long minima of Sp\"orer type.
  This suggests that a grand minimum is a special state of the dynamo.

\item Duration of grand maxima resemble a random Possion-like
  process, in contrast to grand minima.

\end{itemize}

\subsubsection{Solar-terrestrial relations}

The sun ultimately defines the climate on Earth supplying it with energy via radiation
 received by the terrestrial system, but the role of solar variability
 in climate variations is far from being clear.
Solar variability can affect the Earth's environment and climate in different ways
 \citep[see, e.g.,  reviews by][]{haighLR,gray10}.
Variability of total solar irradiance (TSI) measured during recent decades is known
 to be too small to explain observed climate variations \cite[e.g.,][]{foukal06, frohlich06}.
On the other hand, there are other ways solar variability may affect the climate, e.g.,
 an unknown long-term trend in TSI \citep{solanki04a, wang05} or a terrestrial amplifier of spectral
 irradiance variations \citep{shindell99,haigh10}.
Uncertainties in the TSI/SSI reconstructions remain large \citep{shapiro11,schmidt12}, making it
 difficult to assess climate models on the long-term scale.
Alternatively, an indirect mechanism also driven by solar activity, such as ionization of the
 atmosphere by CR \citep{usoskin_JGR_06} or the global terrestrial current system \citep{tinsley06}
 can modify atmospheric properties, in particular cloud cover \citep{ney59, svensmark98,usoskin_CR_08}.
Even a small change in cloud cover modifies the transparency/absorption/reflectance of the atmosphere and
 affects the amount of absorbed solar radiation, even without changes in the solar irradiance.
However, the direct role of this effect is estimated to be small \citep{usoskin_CRII_08, gray10}.

Accordingly, improved knowledge of the solar driver's variability may help
 in disentangling various effects in the very complicated system that is the terrestrial climate \citep[e.g.,][]{dejager05, versteegh05,gray10}.
It is of particular importance to know the driving forces in the pre-industrial era, when
 all climate changes were natural.
Knowledge of the natural variability can lead to an improved understanding of
 anthropogenic effects upon the Earth's climate.

Studies of the long-term solar-terrestrial relations are mostly phenomenological, lacking
 a clear quantitative physical mechanism.
Even phenomenological and empirical studies suffer from large uncertainties, related
 to the quantitative interpretation of proxy data, temporal and spatial resolution \citep{versteegh05}.
Therefore, more precise knowledge of past solar activity, especially since it is accompanied by
 continuous efforts of the paleo-climatic community on improving climatic data sets,
 is crucial for improved understanding of the natural (including solar) variability
 of the terrestrial environment.

\subsubsection{Other issues}

The proxy method of solar-activity reconstruction, based on cosmogenic isotopes, was developed from
 the radiocarbon dating method, when it was recognized that the production rate of \super{14}C
 is not constant and may vary in time due to solar variability and geomagnetic field changes.
Neglect of these effects can lead to inaccurate radiocarbon (or more generally, cosmogenic
 nuclide) dating, which is a key for, e.g., archeology and Quaternary geology.
Thus, knowledge of past solar activity and geomagnetic changes allows for the improvement of the
 quality of calibration curves, such as the IntCal \citep{stuiver98, reimer04, reimer09} for
 radiocarbon, eventually leading to more precise dating.

Long-term variations in the geomagnetic field are often evaluated using cosmogenic
 isotope data.
Knowledge of source variability due to solar modulation is important for better results.

\subsection{Summary}

In this section, solar activity on a longer scale is discussed, based on recent reconstructions.

According to these reconstructions, the sun has spent about 70\% of its time during the Holocene, which is ongoing,
 in a normal state characterized by medium solar activity.
About 15\,--\,20\% of the time the sun has experienced a grand minimum, while
 10\,--\,15\% of the time has been taken up by periods of very high activity.

One of the main features of long-term solar activity is its irregular behavior,
 which cannot be described by a combination of quasi-periodic processes as it includes
 an essentially random component.

Grand minima, whose typical representative is the Maunder minimum of the late 17th century,
 are typical solar phenomena.
A total of 27 grand minima have been identified in reconstructions of the Holocene period.
Their occurrence suggests that they appear not periodically, but rather as the result of a chaotic process
 within clusters separated by 2000\,--\,2500~years.
Grand minima tend to be of two distinct types: short (Maunder-like) and longer (Sp\"orer-like).
The appearance of grand minima can be reproduced by modern stochastic-driven
 dynamo models to some extent, but some problems still remain to be resolved.

The modern level of solar activity (after the 1940s) was very high,
 corresponding to a grand maximum, which are typical but rare
 and irregularly-spaced events in solar behavior.
However, this grand maximum has ceased after solar cycle 23.
The duration of grand maxima resembles a random Possion-like process, in contrast to grand minima.

These observational features of the long-term behavior of solar activity
 have important implications, especially for the development of theoretical
 solar-dynamo models and for solar-terrestrial studies.

\newpage


\section{Solar Energetic Particles in the Past}
\label{S:SEP}

In addition to galactic cosmic rays, which are always present in the Earth's vicinity,
 sometimes sporadic solar energetic-particle (SEP) events with
 a greatly enhanced flux of less energetic particles in the interplanetary medium also occur
 \citep[e.g.,][]{klecker06}.
Strong SEP events mostly originate from CME-related shocks propagating in the solar corona
 and interplanetary medium, that lead to effective bulk acceleration of charged particles
 \citep[e.g.,][]{cane06}.
Although these particles are significantly less energetic than GCRs, they can occasionally be
 accelerated to an energy reaching up to several GeV, which is enough to initiate
 the atmospheric cascade.
Peak intensity of SEP flux can be very high, up to 10\super{4} particles
 (with energy \textgreater~30~MeV) per cm\super{2} per second.
In fact, the long-term average flux (or fluence) of SEP is mostly defined by rare
 major events, which occur a few times per solar cycle, with only minor contributions from a large number
 of weak events \citep{shea90, shea02}.
As an example, energy spectra of GCR and SEP are shown in Figure~\ref{Fig:2005} for
 the day of January~20, 2005, when an extreme SEP event took place.
Such SEPs dominate the low-energy section of cosmic rays (below hundreds of MeV of a particle's
 kinetic energy), which is crucial for the radiation
 environment, and play an important role in solar-terrestrial relations.
For many reasons it is important to know the variations of SEPs on
 long-term scales.

\epubtkImage{LR_Spec05.png}{%
  \begin{figure}[htbp]
    \centerline{\includegraphics[width=8cm]{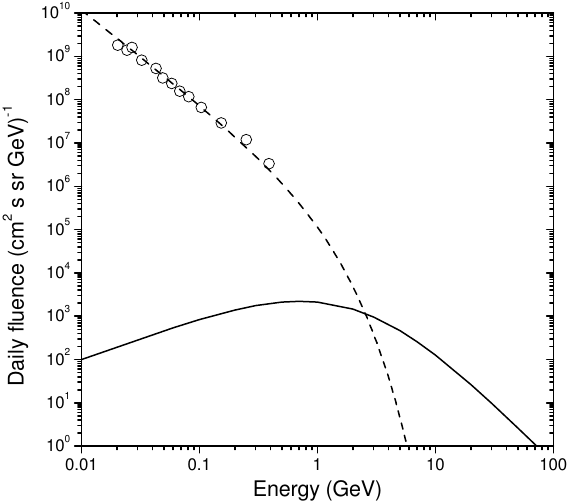}}
    \caption{Daily fluence of solar energetic particles \citep[dashed curve --][]{tylka09}
    and galactic cosmic rays (solid curve) for the day of January~20, 2005. Open
    circles represent space-borne measurements \citep{mewaldt06,mewaldt12}.}
    \label{Fig:2005}
\end{figure}}

It is not straightforward to evaluate the average SEP flux even for the modern instrumental epoch
 of direct space-borne measurements \citep[e.g.,][]{mewaldt07}.
For example, estimates for the average flux of SEPs with an energy above 30~MeV (called \f{30} henceforth) for
 individual cycles may vary by an order of magnitude, from 10~\cms for cycle 21 up to 70~\cms for cycle~19
 \citep{reedy12}.
Moreover, estimates of the SEP flux were quite uncertain during the earlier years of space-borne measurements
 because of two effects, which are hard to account for \citep[e.g.,][]{reeves92, tylka97}.
One is related to the very high flux intensities of SEPs during the peak phase of events,
 when a detector can be saturated because of the dead-time effect (the maximum trigger rate
 of the detector is exceeded).
The other is related to events with high energy solar particles, which can
 penetrate into the detector through the walls of the collimator or the detector,
 leading to an enhanced effective acceptance cone with respect to the ``expected'' one.
Since the SEP fluence is defined by major events, these effects may lead to an
 underestimate of the average flux of SEPs.
The modern generation of detectors are better suited for measuring high fluxes.
The average \f{30} flux for the last five solar cycles (1954\,--\,2006)
 is estimated at about 35~\cms \citep{smart02, shea06}.

\subsection{Cosmogenic isotopes}
\label{sec:SEP_ter}

The development of the method of cosmogenic isotopes makes it possible to estimate occurrence
 of extreme SEP events in the past.
Some earlier attempts were inconclusive.
For example, \citet{usoskin_GRL_SCR06} found that a typical strong SEP event leaves no distinguishable
 signature in \super{14}C but may be observed from ice core \super{10}Be records.
However, the question of the possible rare occurrence of extreme SEP events on the millennial time scale
 is important not only from the theoretical point of view, but also for assessment of radiation risks
 for space-borne missions, especially manned ones.
What can be the strongest SEP event originated from the sun, how often they can occur?
These questions need to be answered.
Several attempts have been made to evaluate that from the cosmogenic isotope data
 \citep{lingenfelter80,usoskin_GRL_SCR06,webber07}, but the result was grossly uncertain \citep{hudson10,schrijver12},
 mostly because of the large model uncertainties of the radionuclide production.

A new step forward has been done recently by \citet{usoskin_SEP_12}, who analyzed two \super{14}C
 and five \super{10}Be records over the last millennia and searched for possible signatures of extreme SEP events.

\epubtkImage{LR_14C_785.png}{%
  \begin{figure}[htbp]
    \centerline{\includegraphics[width=9cm]{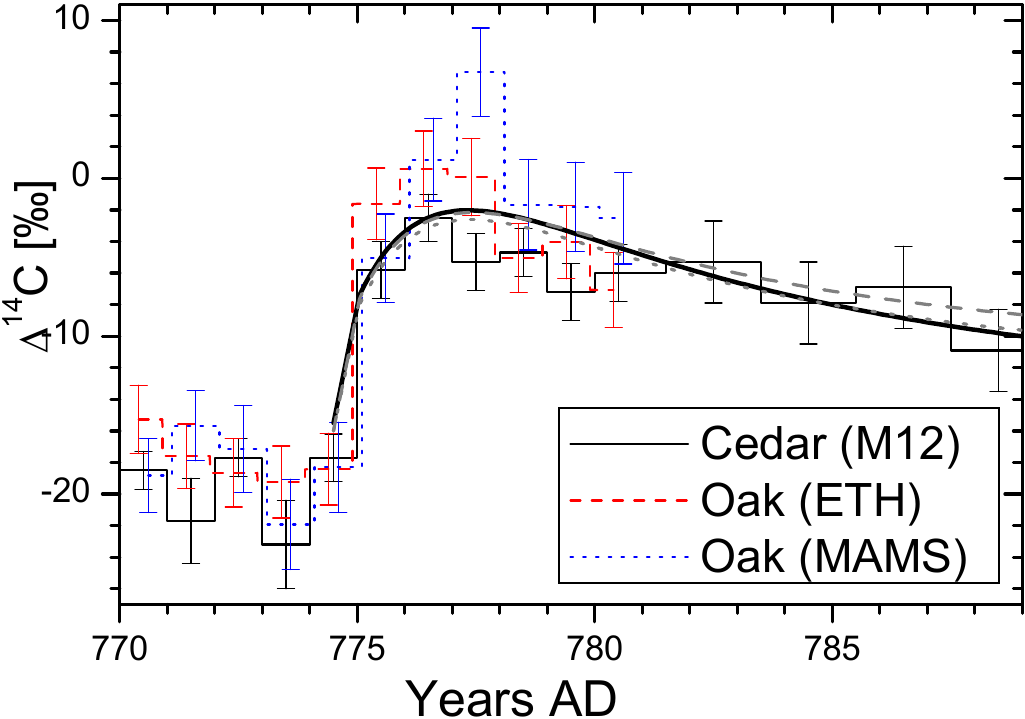}}
    \caption{Time profiles of the measured $\Delta$\super{14}C content in
      Japanese cedar \citep[M12 --][]{miyake12} and German oak
      \citep[ETH Z\"urich \& Mannheim AMS --][]{usoskin_775_13} trees for
      the period around 775~AD. Smooth black and grey lines depict a
      family of best fit $\Delta$\super{14}C profiles, calculated
      using a family of realistic carbon cycle models for an
      instantaneous injection of \super{14}C into the stratosphere
      \citep{usoskin_SEP_12}. Image after \citet{usoskin_775_13}.}
    \label{Fig:14C_e}
\end{figure}}

While the response of \super{10}Be to an SEP event is simply a 1\,--\,2-yr long peak, because of the simple
 atmospheric transport/deposition (see Section~\ref{S:10Be_tran}), the response of \super{14}C has a typical shape
  shown in Figure~\ref{Fig:14C_e} -- with a sharp peak and exponential decay of the length of several decades,
  due to the carbon cycle (see Section~\ref{S:14C_cycle}).
\citet{usoskin_SEP_12} checked all the available cosmogenic isotope data through the entire Holocene looking for
 a potential SEP signatures, and came up with a list of candidates of extreme SEP events and assessments of their strength
 (Table~\ref{Tab:SPE}).

\begin{table}[htb]
\caption[A list of candidates for extreme SEP events found in
  different cosmogenic isotope records throughout the Holocene.]{A
  list of candidates for extreme SEP events found in different
  cosmogenic isotope records throughout the Holocene: approximate
  year, dataset used (Dye3 -- \citet{mccracken04}; NGRIP --
  \citet{berggren09}; IntCal09 -- \citet{reimer09}; GRIP --
  \citet{yiou97}; Dome Fuji -- \citet{horiuchi08}; South Pole --
  \citet{raisbeck90}; M12 -- \citet{miyake12}), and the \F{30} fluence
        [cm\super{-2}]. Table after \citet{usoskin_SEP_12}.}
\label{Tab:SPE}
\begin{center}
{\small
\begin{tabular}{ccc}
\toprule
SPE year & Series & \F{30} \\
\midrule
1460\,--\,1462 AD & NGRIP(1460) & 1.5~\texttimes~10\super{10} \\
~             & Dye3 (1462) & 9.7~\texttimes~10\super{9}  \\
1505 AD       & Dye3        & 1.3~\texttimes~10\super{10} \\
1719 AD &     NGRIP  & 1~\texttimes~10\super{10}  \\
1810 AD &     NGRIP  & 1~\texttimes~10\super{10} \\
\midrule
 8910 BC & IntCal09   & 2.0~\texttimes~10\super{10} \\
 8155 BC & IntCal09   & 1.3~\texttimes~10\super{10} \\
 8085 BC & IntCal09   & 1.5~\texttimes~10\super{10} \\
 7930 BC & IntCal09   & 1.3~\texttimes~10\super{10} \\
 7570 BC & IntCal09   & 2.0~\texttimes~10\super{10} \\
 7455 BC & IntCal09   & 1.5~\texttimes~10\super{10} \\
 6940 BC & IntCal09   & 1.1~\texttimes~10\super{10} \\
 6585 BC & IntCal09   & 1.7~\texttimes~10\super{10} \\
 5835 BC & IntCal09   & 1.5~\texttimes~10\super{10} \\
 5165 BC & GRIP       & 2.4~\texttimes~10\super{10} \\
 4680 BC & IntCal09   & 1.6~\texttimes~10\super{10} \\
 3260 BC & IntCal09   & 2.4~\texttimes~10\super{10} \\
 2615 BC & IntCal09   & 1.2~\texttimes~10\super{10} \\
 2225 BC & IntCal09   & 1.2~\texttimes~10\super{10} \\
 1485 BC & IntCal09   & 2.0~\texttimes~10\super{10} \\
   95 AD & GRIP       & 2.6~\texttimes~10\super{10} \\
  265 AD & IntCal09   & 2.0~\texttimes~10\super{10} \\
  785 AD & IntCal09   & 2.4~\texttimes~10\super{10} \\
 ~       & Dome Fuji  & 5.3~\texttimes~10\super{10}\,\textdagger \\
 ~       & M12        & 4~\texttimes~10\super{10}\,\textdagger \\
 1455 AD & South Pole & 7.0~\texttimes~10\super{10}\,\textdagger \\
\bottomrule
\multicolumn{3}{l}{
{\footnotesize
\textdagger~Upper bound.
}}
\end{tabular}}
\end{center}
\end{table}

The list includes 23 candidates for extreme SEP events with the fluence \F{30}
 exceeding 10\super{10}~cm\super{-2}, viz.\ the greatest fluence observed for the space era in 1960 \citep{shea90}.
Note that only two of these candidates appear in more than one series -- the events of ca.\ 1460~AD and ca.\ 780~AD.
The former had signatures in two annual \super{10}Be series, NGRIP and Dye3.
The later was observed in two \super{14}C series, biennial M12 and 5-yr IntCal09, and in quasi-decadal Dome Fuji
 \super{10}Be series.
The quasi-decadal South Pole \super{10}Be series does not show an increase ca. 780 AD placing an upper limit on the
 strength of the event.

We note that the event of ca.\ 775~AD was analyzed using biennial \super{14}C data by \citet{miyake12}, who suggested
 that the event was probably caused by $\gamma$-rays from an unknown nearby supernova.
This event is confirmed by annual \super{14}C data from a German oak tree \citep{usoskin_775_13}.
However, because of the use of an inappropriate carbon cycle model, \citet{miyake12} grossly (by a factor of 5) overestimated the
 corresponding \super{14}C production, leading to the need of a supernova.
Moreover, this leads to a strong disagreement between \super{14}C and \super{10}Be data sets, since this event is
 not observed in the South Pole record and is not exceptionally strong in the Dome Fuji record.
However, if an appropriate model of the carbon cycle is used, the production of \super{14}C appears in a reasonable range,
 being consistent with \super{10}Be \citep{usoskin_775_13}.
Therefore, there is no need to involve such an exotic object as a nearby supernova whose remnants are unknown for us --
 the event of ca. 775\,--\,780~AD can be consistently explained by a extreme but not exceptional SEP event.

The integral probability distribution of the occurrence of strong SEP events, as revealed from the
 cosmogenic isotope data, is shown in Figure~\ref{Fig:SEP}.

\epubtkImage{LR_SEP.png}{%
  \begin{figure}[htbp]
    \centerline{\includegraphics[width=10cm]{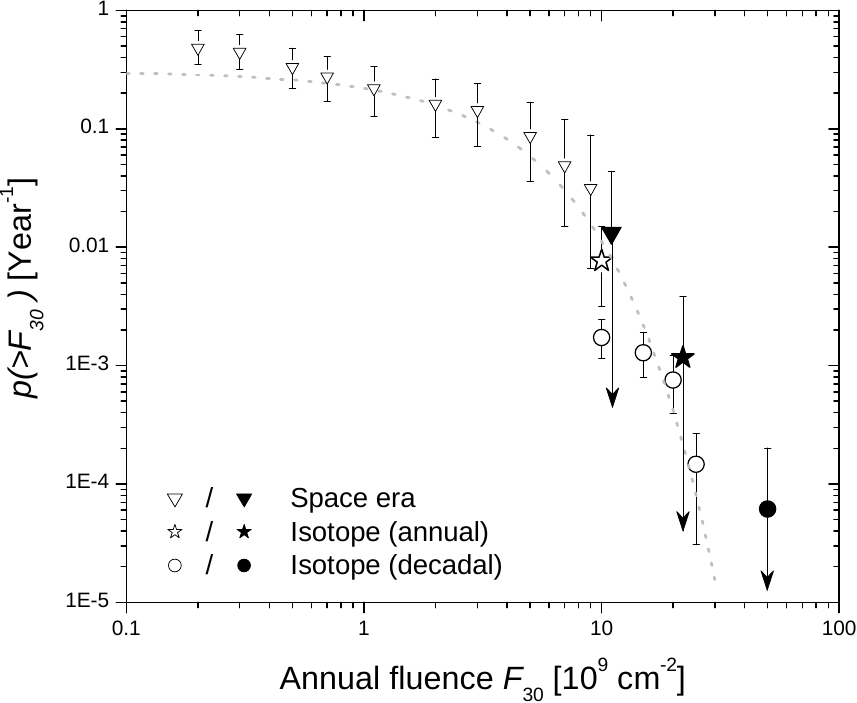}}
    \caption{Cumulative probability (with the 90\% confidence
      interval) of occurrence of a SEP event with fluence
      (\textgreater~30~MeV) exceeding the given value \F{30}, as
      assessed from the data for the space era 1956\,--\,2008
      (triangles), cosmogenic isotope annual data (stars), and
      cosmogenic isotope decadal data (circles). Gray dotted curve
      depicts the best-fit exponent. Image reproduced by permission
      from~\cite{usoskin_SEP_12}, copyright by AAS.}
    \label{Fig:SEP}
\end{figure}}

One can see that the break in the distribution marginally hinted in the directly observed
 SEP events at around \F{30}~=~(5\,--\,7)~\texttimes10\super{9}~cm\super{-2} (nonproportionally fewer strong events
 observed) is confidently confirmed by the cosmogenic isotope data.
In particular, no event with \F{30}~\textgreater~2~\texttimes~10\super{10}~cm\super{-2} was found over the last 600 years using annually resolved \super{10}Be data.
It is noteworthy that the idea of an possible extreme Carrington SPE of 1859 AD \citep{mccracken01} is discarded \citep[see also][]{wolff12}.
On the longer time scale of 11 millennia, no event with \F{30}~\textgreater~5~\texttimes~10\super{10}~cm\super{-2} has been found.
This gives a new strict observational constraint on the occurrence probability of extreme SPEs.

According to \citet{usoskin_SEP_12} practical limits can be set as \F{30}~$\approx$~1, 2\,--\,3 and 5~\texttimes~10\super{10}~cm\super{-2}
 (10, 20\,--\,30 and 50 times greater than the SEP event of February 23, 1956),
 for the occurrence probability of 10\super{-2}, 10\super{-3}, and 10\super{-4}~yr\super{-1}, respectively.
The mean SEP flux is found as $\approx$~40 (cm\super{2}~s)\super{-1} in agreement with estimates from the lunar rocks.
On average, extreme SPEs contribute about 10\% to the total SEP fluence.

\subsection{Lunar and meteoritic rocks}
\label{sec:lunar}

Since energy spectra of SEP and GCR are dramatically different, one may think of a natural spectrometer
 to separate their effects and thus evaluate their fluxes independently.
A spectrometer that is able to separate cosmic rays is lunar (or meteoritic) rocks.
\epubtkImage{LR_lunar.png}{%
  \begin{figure}[htbp]
    \centerline{\includegraphics[width=8cm]{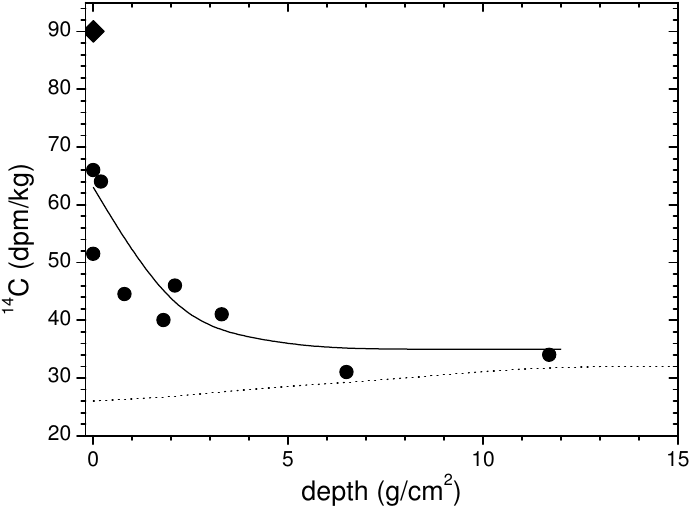}}
    \caption{Measured (dots) and calculated (curves) \super{14}C activity
    in a lunar sample 68815 \citep{jull98}. The big diamond implies
    contamination of a thin surface layer by \super{14}C implanted from
    solar wind. The dotted curve represents the expected production
    due to GCR, while the solid curve is the best fit SEP+GCR model
    production.}
    \label{Fig:lunar}
\end{figure}}

Figure~\ref{Fig:lunar} depicts an example of \super{14}C measured in a lunar sample \citep{jull98}.
The dotted line shows the expected production of radiocarbon by GCR.
The production increases with depth due to the development of a nucleonic cascade in the matter,
 initiated by energetic GCR particles, similar to the atmospheric cascade.
Less energetic particles of solar origin produce the isotope only in upper layers of
 the rock, since their low energy does not allow them to initiate a cascade.
On the other hand, thanks to their high flux in the lower energy range, the production
 of \super{14}C in the upper layers is much higher than that from GCR.
Thus, by first measuring the isotope activity in deep layers one can evaluate
 the average GCR flux, and then the measured excess in the
 upper level yields an estimate for the SEP flux in both integral intensity and spectral shape.
The result is based on model computations and therefore is slightly model dependent
 but makes it possible to give a robust estimate of the GCR and SEP in the past.

A disadvantage of this approach is that lunar samples are not stratified
 and do not allow for temporal separation.
The measured isotope activity is a balance between production and decay and, therefore,
 represents the production (and the ensuing flux) integrated over the life-time of
 the isotope before the sample has been measured.
However, using different isotopes with different life times, one can evaluate
 the cosmic-ray flux integrated over different timescales.

Estimates of the average SEP flux \f{30} on different
 timescales, as obtained from various isotopes measured in lunar samples, are
 collected in Table~\ref{tab:lunar}.
Based on isotopes with different life-times (see Table~\ref{tab:lunar}) one can
 evaluate the average flux of SEP on different time scale (see Figure~\ref{Fig:SEP_sch}).
The average \f{30} flux for the last five solar cycles (1954\,--\,2008)
 is consistent with the average flux estimated in the past for longer timescales
 from 10\super{3} to 10\super{7}~years \citep[cf.][]{reedy02,reedy12}.

\begin{table}[htbp]
\caption[Estimates of $4\pi$ omni-directional integral (above 30~MeV)
  flux.]{Estimates of $4\pi$ omni-directional integral (above 30~MeV)
  flux, \f{30} in [cm\super{2}~s]\super{-1}, of solar energetic particles,
  obtained from different sources.}
\label{tab:lunar}
\begin{center}
{\small
\begin{tabular}{lcccc}
\toprule
Timescale & Method & Source & Reference & \f{30} (\cms) \\
\midrule
1954\,--\,2008 & measurements & space-borne & \cite{reedy12} & 35\\
10\super{4} yr & \super{14}C & lunar rock & \cite{jull98} & 42 \\
10\super{5} yr & \super{41}Ca & lunar rock& \cite{fink98} & 56 \\
5~\texttimes~10\super{5} yr & \super{36}Cl & lunar rock& \cite{nishiizumi09} & 46\\
10\super{6} yr & \super{26}Al & lunar rock& \cite{kohl78} & 25\\
10\super{6} yr & \super{26}Al & lunar rock& \cite{grismore01} & 55\\
10\super{6} yr & \super{10}Be, \super{26}Al & lunar rock& \cite{michel96} & 24\\
10\super{6} yr & \super{10}Be, \super{26}Al & lunar rock& \cite{fink98} & 32\\
10\super{6} yr & \super{10}Be, \super{26}Al & lunar rock& \cite{nishiizumi09} & 24\\
2~\texttimes~10\super{6} yr & \super{10}Be, \super{26}Al & lunar rock& \cite{nishiizumi98} & $\sim$~35\\
5~\texttimes~10\super{6} yr & \super{53}Mn & lunar rock& \cite{kohl78} & 25\\
2~\texttimes~10\super{6} yr & \super{21}Ne, \super{22}Ne, \super{38}Ar & lunar rock& \cite{rao94} & 22\\
\bottomrule
\end{tabular}}
\end{center}
\end{table}

However, this method is not able to provide an estimate of the occurrence rate
 of extreme SEP events.
If one assumes that the entire average SEP flux is produced within one extreme event occurring at
 half of the isotope's life-time ago \citep{reedy96}, an upper limit for the occurrence of extreme SEP events
 can be placed.
This is an unrealistically extreme assumption, which may lead to an overestimate by many orders of magnitude,
  but it sets the very conservative upper limit which cannot be exceeded.

\subsection{Nitrates in polar ice}
\label{sec:nit}

It has been discussed until recently that another quantitative index of strong SEP events
 (with \F{30}~\texttimes~10\super{9}~cm\super{-2}) might be related to nitrate ($\mathrm{NO}_3^-$) records measured
 in polar ice cores.
The concentration of nitrates has been measured in polar ice from both the Southern \citep[South Pole, e.g.,][]{dreschhoff90}
 and Northern \citep[Greenland, e.g.,][]{zeller95, dreschhoff98} polar caps, depicting pronounced spikes
 associated with strong SEP events \citep{mccracken01}.
As a result of the analysis a list of large SEP events since 1560 and their fluences have been
 published \citep[see Table~1 in][]{mccracken01} and widely used.

However, as shown by several independent recent studies \citep{wolff12,usoskin_SEP_12} on the example
 of the Carrington event (September 1859), the nitrate spikes are not related to SEP events.
According to \citet{mccracken01}, the nitrate spike and the associated SEP event was the strongest in the
 entire record (\F{30}~$\approx$~2~\texttimes~10\super{10}~cm\super{-2}).
\citet{wolff12} have measured, with high resolution, nitrate content in 14 ice cores from Antarctic
 and Greenland for a few decades around 1859.
Only one Greenland series depicts a spike which can be associated with the event, all other series
 have no signatures.
Moreover, all similar spikes found in Greenland datasets are accompanied by chemical tracers (ammonium, formate,
 black carbon, etc.) clearly pointing to the anthropogenic source of nitrates -- biomass burning plumes.
No significant spikes have been found in the Antarctic records.
\citet{wolff12} concluded that ``Nitrate spikes cannot be used to derive the statistics of SEPs.''
Another confirmation of this conclusion was made by \citet{usoskin_SEP_12}, who calculated, from the \F{30} fluence
 proposed by \citet{mccracken01} for the Carrington event, the \super{10}Be production.
If the Carrington SEP event was so strong, it would have necessarily left its clear signature in the annually resolved
 \super{10}Be record, which however contradicts to the real data from NGRIP and Dye3 ice cores.

Thus, the nitrate record in polar ice cannot serve as an index of SEP events.
On the other hand, it may be used to study long-term variability of GCR (see Section~\ref{sec:nit_GCR}).

\subsection{Summary}

In this section, estimates of the averaged long-term flux of SEPs are
 discussed.

Measurements of cosmogenic isotopes with different life times in lunar and meteoritic rocks allow
 one to make rough estimates of the SEP flux over different timescales.
The directly space-borne-measured SEP flux for past decades is broadly consistent
 with estimates on longer timescales -- up to millions of years.
The same measurements can provide a very conservative upper estimate for the occurrence rate of extreme
 SEP events.
Terrestrial cosmogenic isotope data in dated archives (tree trunks, ice cores) give a possibility to
 assess the occurrence rate of strong SEP events on the time scales up to ten of millennia.
Measurements of nitrates in polar ice have been shown to be an invalid index of strong SEP events in the past.

Different estimates of the extreme (quantified as the fluence of SEP with energy above 10~MeV) SEP event occurrence
 probability are summarized in Figure~\ref{Fig:SEP_sch}).

An analysis of various kinds of data suggests that the distribution of the intensity
 of SEP events has a break, and the occurrence of extra-strong events (with the \F{30}
 fluence exceeding 5~\texttimes~10\super{10}~cm\super{-2}) is unlikely on the multi-millennial time scale.

\epubtkImage{LR_SEP_sch.png}{%
  \begin{figure}[hb!]
    \centerline{\includegraphics[width=7.3cm]{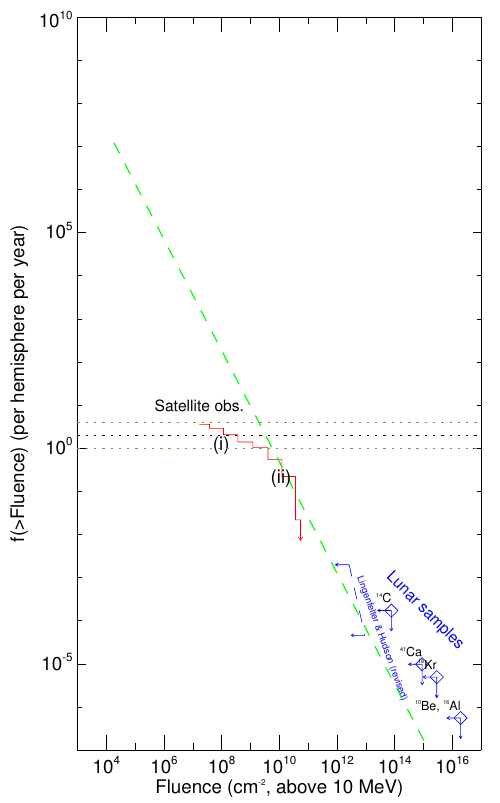}}
    \caption{Cumulative frequency distribution of SEP events with fluences greater than \F{10} (for particles with energies above 10~MeV). \emph{Red histogram:} satellite-based direct observations; \emph{Blue diamonds:} conservative upper limits derived from lunar isotopes (see Section~\ref{sec:lunar}); \emph{Blue dashed line:} upper limit based on \super{14}C record \citep{hudson10};
     Image reproduced by permission from~\cite{schrijver12}, copyright by AGU.}
    \label{Fig:SEP_sch}
\end{figure}}

\newpage

\section{Conclusions}
\label{sec:conc}

In this review the present knowledge of long-term solar activity
 on a multi-millennial timescale, as reconstructed using the indirect proxy method, is discussed.

Although the concept of solar activity is intuitively understandable as a deviation
 from the ``quiet'' sun concept, there is no clear definition for it, and
 different indices have been proposed to quantify different aspects of variable solar activity.
One of the most common and practical indices is sunspot number, which forms the longest
 available series of direct scientific observations.
While all other indices have a high correlation with sunspot numbers, dominated by the
 11-year cycle, the relationship between them at other timescales (short- and long-term trends)
 may vary to a great extent.

On longer timescales, quantitative information of past solar activity can only be obtained
 using the method based upon indirect proxy, i.e., quantitative parameters, which can be
 measured nowadays but represent the signatures, stored in natural archives, of the different effects of
 solar magnetic activity in the past.
Such traceable signatures can be related to nuclear or chemical effects caused by cosmic rays
 in the Earth's atmosphere, lunar rocks or meteorites.
The most common proxy of solar activity is formed by data from the cosmogenic radionuclides,
 \super{10}Be and \super{14}C, produced by cosmic rays in the Earth's atmosphere and stored in
 independently-dated stratified natural archives, such as tree rings or ice cores.
Using a recently-developed physics-based model it is now possible to reconstruct the temporal
 behavior of solar activity in the past, over many millennia.
The most robust results can be obtained for the Holocene epoch, which started more than 11,000~years ago,
 whose stable climate minimizes possible uncertainties in the reconstruction.
An indirect verification of long-term solar-activity reconstructions supports their veracity
 and confirms that variations of cosmogenic nuclides on the long-term scale (centuries to millennia) during
 the Holocene make a solid basis for studies of solar variability in the past.
However, such reconstructions may still contain systematic uncertainties related to unknown
 changes in the geomagnetic field or climate of the past, especially in the early part of the Holocene.

Measurements of the concentration of different cosmogenic isotopes in lunar and meteoritic rocks make it
 possible to estimate the SEP flux on different timescales.
Directly space-borne-measured SEP flux for recent decades is broadly consistent
 with estimates on longer timescales -- up to millions of years.
The occurrence of extra-strong events, with the fluence of SEP (with energy greater than 30~MeV)
 exceeding 5~\texttimes~10\super{10}~cm\super{-2} is unlikely on the multimillenial time scale.

In general, the following main features are observed in the long-term
evolution of solar magnetic activity.

\begin{itemize}

\item Solar activity is dominated by the 11-year Schwabe cycle on
 an interannual timescale.
Some additional longer characteristic times
 can be found, including the Gleissberg secular cycle, de~Vries/Suess
 cycle, and a quasi-cycle of 2000\,--\,2400~years.
However, all these longer cycles are intermittent and cannot be regarded as strict
 phase-locked periodicities.

\item One of the main features of long-term solar activity is that it
  contains an essential chaotic/ stochastic component, which leads to
  irregular variations and makes solar-activity predictions impossible
  for a scale exceeding one solar cycle.

\item The sun spends about 70\% of its time at moderate magnetic
  activity levels, about 15\,--\,20\% of its time in a
  grand minimum and about 10\,--\,15\% in a grand maximum.

\item Grand minima are a typical but rare phenomena in solar behavior.
Their occurrence appears not periodically, but rather as the
 result of a chaotic process within clusters separated by
 2000\,--\,2500~years.
Grand minima tend to be of two distinct types: short (Maunder-like) and longer (Sp\"orer-like).

\item The recent level of solar activity (after the 1940s) was very high,
 corresponding to a prolonged grand maximum, but it is ceasing now to the normal moderate level.
 Grand maxima are also rare and irregularly occurring events, though the exact rate of their
 occurrence is still a subject of debates.

\end{itemize}

These observational features of the long-term behavior of solar
activity have important implications, especially for the development of
theoretical solar-dynamo models and for solar-terrestrial studies.

\newpage


\section{Acknowledgements}
\label{section:acknowledgements}

I thank the editorial board of the \emph{Living Reviews in Solar Physics} for the invitation
 to prepare this review and organizational aid.
This work would be impossible without encouraging discussions, as well as direct and indirect help,
 from my colleagues Gennady Kovaltsov and Sami Solanki, whose invaluable support is acknowledged
 with my greatest gratitude.
I have also greatly benefited from lively and open discussions and debates
 with Rainer Arlt, Edouard Bard, J\"urg Beer, Narendra Bhandari, late Giuliana Cini Castagnoli, Paul Charbonneau, Ed Cliver, Vincent Courtillot, Cornelis de Jager, Gisella Dreschhoff,
 Erwin Fl\"uckiger, Peter Foukal, Yves Gallet, Agn\'es Genevey, David Hathaway, Monika Korte, Natalia Krivova,
 Bernd Kromer, Devendra Lal, Michael Lockwood, Ken McCracken, David Moss, Kalevi Mursula, Raimund Muscheler, Hiroko Miyahara,
 Heikki Nevanlinna, Keran O'Brien, Valery Ostryakov, Alexander Ruzmaikin, Dieter Schmitt, Manfred Sch\"ussler, Margaret Shea, Don Smart, Ian Snowball, Dmitry Sokoloff, Willie Soon,
 Leif Svalgaard, Carla Taricco, Rita Traversi, J\'ose Vaquero, and many others.
I thank Evguenia Usoskina for help with editing this text.
I am happy to acknowledge those colleagues who work in this field and make it living and vibrant.

\newpage



\bibliography{refs}

\begin{thebibliography}{401}
\expandafter\ifx\csname natexlab\endcsname\relax\def\natexlab#1{#1}\fi
\expandafter\ifx\csname url\endcsname\relax
  \def\url#1{{\tt #1}}\fi
\expandafter\ifx\csname urlprefix\endcsname\relax\def\urlprefix{URL }\fi
\providecommand{\eprint}[2][]{\url{#2}}
\catcode`\% 12

\bibitem[Abreu {\it et~al.\/}(2008)]{abreu08}
Abreu, J.A., Beer, J., Steinhilber, F., Tobias, S.M. and Weiss, N.O., 2008,
  ``For how long will the current grand maximum of solar activity persist?'',
  {\it Geophys. Res. Lett.\/}, {\bf 35}, L20109.
  {\small[\href{http://dx.doi.org/10.1029/2008GL035442}{DOI}]},
  {\small[\href{http://adsabs.harvard.edu/abs/2008GeoRL..3520109A}{ADS}]}

\bibitem[Abreu {\it et~al.\/}(2012)]{abreu12}
Abreu, J.A., Beer, J., Ferriz-Mas, A., McCracken, K.G. and Steinhilber, F.,
  2012, ``Is there a planetary influence on solar activity?'', {\it Astron.
  Astrophys.\/}, {\bf 548}, A88.
  {\small[\href{http://dx.doi.org/10.1051/0004-6361/201219997}{DOI}]},
  {\small[\href{http://adsabs.harvard.edu/abs/2012A&A...548A..88A}{ADS}]}

\bibitem[Argento {\it et~al.\/}(2013)]{argento12}
Argento, D.C., Reedy, R.C. and Stone, J.O., 2013, ``Modeling the earth's cosmic
  radiation'', {\it Nucl. Instrum. Methods B\/}, {\bf 294}, 464--469.
  {\small[\href{http://dx.doi.org/10.1016/j.nimb.2012.05.022}{DOI}]},
  {\small[\href{http://adsabs.harvard.edu/abs/2013NIMPB.294..464A}{ADS}]}

\bibitem[Arlt(2008)]{arlt08}
Arlt, R., 2008, ``Digitization of Sunspot Drawings by Staudacher in
  1749\,--\,1796'', {\it Solar Phys.\/}, {\bf 247}, 399--410.
  {\small[\href{http://dx.doi.org/10.1007/s11207-007-9113-4}{DOI}]},
  {\small[\href{http://adsabs.harvard.edu/abs/2008SoPh..247..399A}{ADS}]}

\bibitem[Arlt(2009)]{arlt09}
Arlt, R., 2009, ``The Butterfly Diagram in the Eighteenth Century'', {\it Solar
  Phys.\/}, {\bf 255}, 143--153.
  {\small[\href{http://dx.doi.org/10.1007/s11207-008-9306-5}{DOI}]},
  {\small[\href{http://adsabs.harvard.edu/abs/2009SoPh..255..143A}{ADS}]},
  {\small[\href{http://arxiv.org/abs/0812.2233}{{arXiv:0812.2233}}]}

\bibitem[Arlt and Abdolvand(2011)]{arlt11}
Arlt, R. and Abdolvand, A., 2011, ``First solar butterfly diagram from
  Schwabe's observations in 1825--1867'', in {\it Physics of Sun and Star
  Spots\/}, Proceedings of IAU Symposium 273 held in Ventura, California, USA,
  August 22\,--\,26, 2010, (Eds.) Choudhary, D.P., Strassmeier, K.G., IAU
  Symposium, S273, pp. 286--289, Cambridge University Press, Cambridge; New
  York. {\small[\href{http://dx.doi.org/10.1017/S1743921311015390}{DOI}]},
  {\small[\href{http://adsabs.harvard.edu/abs/2011IAUS..273..286A}{ADS}]},
  {\small[\href{http://arxiv.org/abs/1010.3131}{{arXiv:1010.3131
  {\small[astro-ph.SR]}}}]}

\bibitem[Arlt(2013)]{arlt13}
Arlt, R. et~al., 2013, {\it Mon. Not. R. Astron. Soc.\/}, submitted

\bibitem[Balmaceda {\it et~al.\/}(2005)]{balmaceda05}
Balmaceda, L.A., Solanki, S.K. and Krivova, N.A., 2005, ``A cross-calibrated
  sunspot areas time series since 1874'', {\it Mem. Soc. Astron. Ital.\/}, {\bf
  76}, 929--932.
  {\small[\href{http://adsabs.harvard.edu/abs/1983SoPh...87...23B}{ADS}]}

\bibitem[Baranyi {\it et~al.\/}(2001)]{baranyi01}
Baranyi, T., Gy{\~{o}}ri, L., Ludm{\'{a}}ny, A. and Coffey, H.E., 2001,
  ``Comparison of sunspot area data bases'', {\it Mon. Not. R. Astron. Soc.\/},
  {\bf 323}, 223--230.
  {\small[\href{http://dx.doi.org/10.1046/j.1365-8711.2001.04195.x}{DOI}]},
  {\small[\href{http://adsabs.harvard.edu/abs/2001MNRAS.323..223B}{ADS}]}

\bibitem[Bard {\it et~al.\/}(1987)]{bard87}
Bard, E., Arnold, M., Maurice, P. and Duplessy, J.-C., 1987, ``Measurements of
  bomb radiocarbon in the ocean by means of accelerator mass spectrometry:
  Technical aspects'', {\it Nucl. Instrum. Methods B\/}, {\bf 29}, 297--301.
  {\small[\href{http://dx.doi.org/10.1016/0168-583X(87)90253-9}{DOI}]},
  {\small[\href{http://adsabs.harvard.edu/abs/1987NIMPB..29..297B}{ADS}]}

\bibitem[Bard {\it et~al.\/}(1997)]{bard97}
Bard, E., Raisbeck, G.M., Yiou, F. and Jouzel, J., 1997, ``Solar modulation of
  cosmogenic nuclide production over the last millennium: comparison between
  $^{14}$C and $^{10}$Be records'', {\it Earth Planet. Sci. Lett.\/}, {\bf
  150}, 453--462.
  {\small[\href{http://dx.doi.org/10.1016/S0012-821X(97)00082-4}{DOI}]},
  {\small[\href{http://adsabs.harvard.edu/abs/1997E&PSL.150..453B}{ADS}]}

\bibitem[Barnard {\it et~al.\/}(2011)]{barnard11}
Barnard, L., Lockwood, M., Hapgood, M.A., Owens, M.J., Davis, C.J. and
  Steinhilber, F., 2011, ``Predicting space climate change'', {\it Geophys.
  Res. Lett.\/}, {\bf 38}, L16103.
  {\small[\href{http://dx.doi.org/10.1029/2011GL048489}{DOI}]},
  {\small[\href{http://adsabs.harvard.edu/abs/2011GeoRL..3816103B}{ADS}]}

\bibitem[Basurah(2004)]{basurah04}
Basurah, H.M., 2004, ``Auroral Evidence for Early High Solar Activities'', {\it
  Solar Phys.\/}, {\bf 225}, 209--212.
  {\small[\href{http://dx.doi.org/10.1007/s11207-004-1372-8}{DOI}]},
  {\small[\href{http://adsabs.harvard.edu/abs/2004SoPh..225..209B}{ADS}]}

\bibitem[Beer(2000)]{beer00}
Beer, J., 2000, ``Long-term indirect indices of solar variability'', {\it Space
  Sci. Rev.\/}, {\bf 94}, 53--66.
  {\small[\href{http://adsabs.harvard.edu/abs/2000SSRv...94...53B}{ADS}]}

\bibitem[Beer {\it et~al.\/}(1983)]{beer83}
Beer, J., Siegenthaler, U., Oeschger, H., Andr{\'{e}}e, M., Bonani, G., Suter,
  M., Wolfli, W., Finkel, R.C. and Langway, C.C., 1983, ``Temporal $^{10}$Be
  Variations'', in {\it 18th International Cosmic Ray Conference\/}, Bangalore,
  India, 22 August\,--\,3 September, 1983, (Ed.) Durgaprasad, D. et~al., 9, pp.
  317--320, TIFR, Bombay.
  {\small[\href{http://adsabs.harvard.edu/abs/1983ICRC....9..317B}{ADS}]}

\bibitem[Beer {\it et~al.\/}(1988)]{beer88}
Beer, J., Siegenthaler, U., Oeschger, H., Bonani, G. and Finkel, R.C., 1988,
  ``Information on past solar activity and geomagnetism from Be-10 in the Camp
  Century ice core'', {\it Nature\/}, {\bf 331}, 675--679.
  {\small[\href{http://dx.doi.org/10.1038/331675a0}{DOI}]},
  {\small[\href{http://adsabs.harvard.edu/abs/1988Natur.331..675B}{ADS}]}

\bibitem[Beer {\it et~al.\/}(1990)]{beer90}
Beer, J., Blinov, A., Bonani, G., Finkel, R.C., Hofmann, H.J., Lehmann, B.,
  Oeschger, H., Sigg, A., Schwander, J., Staffelbach, T., Stauffer, B., Suter,
  M. and W{\"{o}}tfli, W., 1990, ``Use of $^{10}$Be in polar ice to trace the
  11-year cycle of solar activity'', {\it Nature\/}, {\bf 347}, 164--166.
  {\small[\href{http://dx.doi.org/10.1038/347164a0}{DOI}]},
  {\small[\href{http://adsabs.harvard.edu/abs/1990Natur.347..164B}{ADS}]}

\bibitem[Beer {\it et~al.\/}(1998)]{beer98}
Beer, J., Tobias, S. and Weiss, N., 1998, ``An Active Sun Throughout the
  Maunder Minimum'', {\it Solar Phys.\/}, {\bf 181}, 237--249.
  {\small[\href{http://adsabs.harvard.edu/abs/1998SoPh..181..237B}{ADS}]}

\bibitem[Beer {\it et~al.\/}(2003)]{beer03}
Beer, J., Vonmoos, M.V., Muscheler, R., McCracken, K.G. and Mende, W., 2003,
  ``Heliospheric Modulation over the past 10,000 Years as derived from
  Cosmogenic Nuclides'', in {\it 28th International Cosmic Ray Conference\/},
  July 31\,--\,August 7, 2003, Trukuba, Japan, (Eds.) Kajita, T., Asaoka, Y.,
  Kawachi, A., Matsubara, Y., Sasaki, M., 7, pp. 4147--4150, Universal Academy
  Press, Tokyo, Japan.
  {\small[\href{http://adsabs.harvard.edu/abs/2003ICRC....7.4147B}{ADS}]}

\bibitem[Beer {\it et~al.\/}(2012)]{beer12}
Beer, J., McCracken, K.G. and von Steiger, R., 2012, {\it Cosmogenic
  Radionuclides: Theory and Applications in the Terrestrial and Space
  Environments\/}, Physics of Earth and Space Environments, Springer, Berlin;
  Heidelberg

\bibitem[Belmaker {\it et~al.\/}(2008)]{belmaker08}
Belmaker, R., Lazar, B., Tepelyakov, N., Stein, M. and Beer, J., 2008,
  ``$^{10}$Be in Lake Lisan sediments -- A proxy for production or climate?'',
  {\it Earth Planet. Sci. Lett.\/}, {\bf 269}, 447--456.
  {\small[\href{http://dx.doi.org/10.1016/j.epsl.2008.02.032}{DOI}]}

\bibitem[Belov {\it et~al.\/}(2006)]{belov06}
Belov, A.V., Gushchina, R.T., Obridko, V.N., Shelting, B.D. and Yanke, V.G.,
  2006, ``Long-term variations of galactic cosmic rays in the past and future
  from observations of various solar activity characteristics'', {\it J. Atmos.
  Terr. Phys.\/}, {\bf 68}, 1161--1166.
  {\small[\href{http://dx.doi.org/10.1016/j.jastp.2006.01.001}{DOI}]},
  {\small[\href{http://adsabs.harvard.edu/abs/2006JATP...68.1161B}{ADS}]}

\bibitem[Berggren {\it et~al.\/}(2009)]{berggren09}
Berggren, A.-M., Beer, J., Possnert, G., Aldahan, A., Kubik, P., Christl, M.,
  Johnsen, S.J., Abreu, J. and Vinther, B.M., 2009, ``A 600-year annual
  $^{10}$Be record from the NGRIP ice core, Greenland'', {\it Geophys. Res.
  Lett.\/}, {\bf 36}, L11801.
  {\small[\href{http://dx.doi.org/10.1029/2009GL038004}{DOI}]},
  {\small[\href{http://adsabs.harvard.edu/abs/2009GeoRL..3611801B}{ADS}]}

\bibitem[Bezuglov {\it et~al.\/}(2012)]{bezuglov12}
Bezuglov, M.V., Malyshevsky, V.S., Malykhina, T.V., Torgovkin, A.V., Fomin,
  G.V. and Shramenko, B.I., 2012, ``Photonuclear channel of $^{7}$Be production
  in the Earth's atmosphere'', {\it Phys. Atom. Nucl.\/}, {\bf 75}, 393--397.
  {\small[\href{http://dx.doi.org/10.1134/S1063778812030040}{DOI}]},
  {\small[\href{http://adsabs.harvard.edu/abs/2012PAN....75..393B}{ADS}]}

\bibitem[Bhandari {\it et~al.\/}(1966)]{bhandari66}
Bhandari, P.A., Lal, D. and Rama, 1966, ``Stratospheric circulation studies
  based on natural and artificail radioactive tracer elements'', {\it
  Tellus\/}, {\bf 18}, 391--406

\bibitem[Bonino {\it et~al.\/}(1995)]{bonino95}
Bonino, G., Castagnoli, G.C., Bhandari, N. and Taricco, C., 1995, ``Behavior of
  the Heliosphere over prolonged solar quiet periods by $^{44}$Ti measurements
  in meteorites'', {\it Science\/}, {\bf 270}, 1648--1650.
  {\small[\href{http://dx.doi.org/10.1126/science.270.5242.1648}{DOI}]},
  {\small[\href{http://adsabs.harvard.edu/abs/1995Sci...270.1648B}{ADS}]}

\bibitem[Bracewell(1986)]{bracewell86}
Bracewell, R.N., 1986, ``Simulating the sunspot cycle'', {\it Nature\/}, {\bf
  323}, 516--519. {\small[\href{http://dx.doi.org/10.1038/323516a0}{DOI}]},
  {\small[\href{http://adsabs.harvard.edu/abs/1986Natur.323..516B}{ADS}]}

\bibitem[Brandenburg and Sokoloff(2002)]{brandenburg02}
Brandenburg, A. and Sokoloff, D., 2002, ``Local and Nonlocal Magnetic Diffusion
  and Alpha-Effect Tensors in Shear Flow Turbulence'', {\it Geophys. Astrophys.
  Fluid Dyn.\/}, {\bf 96}, 319--344.
  {\small[\href{http://dx.doi.org/10.1080/03091920290032974}{DOI}]},
  {\small[\href{http://adsabs.harvard.edu/abs/2002GApFD..96..319B}{ADS}]},
  {\small[\href{http://arxiv.org/abs/arXiv:astro-ph/0111568}{{arXiv:astro-ph/0%
111568}}]}

\bibitem[Brandenburg and Spiegel(2008)]{brandenburg08}
Brandenburg, A. and Spiegel, E.A., 2008, ``Modeling a Maunder minimum'', {\it
  Astron. Nachr.\/}, {\bf 329}, 351--358.
  {\small[\href{http://adsabs.harvard.edu/abs/2008AN....329..351B}{ADS}]},
  {\small[\href{http://arxiv.org/abs/arXiv:0801.2156}{{arXiv:0801.2156}}]}

\bibitem[Brandenburg {\it et~al.\/}(1989)]{brandenburg89}
Brandenburg, A., Krause, F., Meinel, R., Moss, D. and Tuominen, I., 1989, ``The
  stability of nonlinear dynamos and the limited role of kinematic growth
  rates'', {\it Astron. Astrophys.\/}, {\bf 213}, 411--422.
  {\small[\href{http://adsabs.harvard.edu/abs/1989A&A...213..411B}{ADS}]}

\bibitem[Bray and Loughhead(1964)]{bray64}
Bray, R.J. and Loughhead, R.E., 1964, {\it Sunspots\/}, International
  Astrophysics Series, 7, Chapman \& Hall, London

\bibitem[Broeker and Peng(1986)]{broeker86}
Broeker, W.S. and Peng, T.-H., 1986, ``Carbon cycle: 1985 Glacial to
  interglacial changes in the operation of the global carbon cycle'', {\it
  Radiocarbon\/}, {\bf 28}, 309--327

\bibitem[Brooke and Moss(1994)]{brooke94}
Brooke, J.M. and Moss, D., 1994, ``Non-linear dynamos in torus geometry:
  transition to chaos'', {\it Mon. Not. R. Astron. Soc.\/}, {\bf 266},
  733--739.
  {\small[\href{http://adsabs.harvard.edu/abs/1994MNRAS.266..733B}{ADS}]}

\bibitem[Burger {\it et~al.\/}(2000)]{burger00}
Burger, R.A., Potgieter, M.S. and Heber, B., 2000, ``Rigidity dependence of
  cosmic ray proton latitudinal gradients measured by the Ulysses spacecraft:
  Implications for the diffusion tensor'', {\it J. Geophys. Res.\/}, {\bf 105},
  27,447--27,456. {\small[\href{http://dx.doi.org/10.1029/2000JA000153}{DOI}]},
  {\small[\href{http://adsabs.harvard.edu/abs/2000JGR...10527447B}{ADS}]}

\bibitem[Bushby and Tobias(2007)]{bushby07}
Bushby, P.J. and Tobias, S.M., 2007, ``On Predicting the Solar Cycle Using
  Mean-Field Models'', {\it Astrophys. J.\/}, {\bf 661}, 1289--1296.
  {\small[\href{http://dx.doi.org/10.1086/516628}{DOI}]},
  {\small[\href{http://adsabs.harvard.edu/abs/2007ApJ...661.1289B}{ADS}]},
  {\small[\href{http://arxiv.org/abs/0704.2345}{{arXiv:0704.2345
  {\small[astro-ph]}}}]}

\bibitem[Caballero-Lopez and Moraal(2004)]{caballero04}
Caballero-Lopez, R.A. and Moraal, H., 2004, ``Limitations of the force field
  equation to describe cosmic ray modulation'', {\it J. Geophys. Res.\/}, {\bf
  109}, A01101. {\small[\href{http://dx.doi.org/10.1029/2003JA010098}{DOI}]},
  {\small[\href{http://adsabs.harvard.edu/abs/2004JGRA..10901101C}{ADS}]}

\bibitem[Cane and Lario(2006)]{cane06}
Cane, H.V. and Lario, D., 2006, ``An Introduction to CMEs and Energetic
  Particles'', {\it Space Sci. Rev.\/}, {\bf 123}, 45--56.
  {\small[\href{http://dx.doi.org/10.1007/s11214-006-9011-3}{DOI}]},
  {\small[\href{http://adsabs.harvard.edu/abs/2006SSRv..123...45C}{ADS}]}

\bibitem[Carbonell {\it et~al.\/}(1993)]{carbonell93}
Carbonell, M., Oliver, R. and Ballester, J.L., 1993, ``On the asymmetry of
  solar activity'', {\it Astron. Astrophys.\/}, {\bf 274}, 497--504.
  {\small[\href{http://adsabs.harvard.edu/abs/1993A&A...274..497C}{ADS}]}

\bibitem[Carbonell {\it et~al.\/}(1994)]{carbonell94}
Carbonell, M., Oliver, R. and Ballester, J.L., 1994, ``A search for chaotic
  behaviour in solar activity'', {\it Astron. Astrophys.\/}, {\bf 290},
  983--994.
  {\small[\href{http://adsabs.harvard.edu/abs/1994A&A...290..983C}{ADS}]}

\bibitem[Casas {\it et~al.\/}(2006)]{casas06}
Casas, R., Vaquero, J.M. and V{\'{a}}zquez, M., 2006, ``Solar Rotation in the
  17th century'', {\it Solar Phys.\/}, {\bf 234}, 379--392.
  {\small[\href{http://dx.doi.org/10.1007/s11207-006-0036-2}{DOI}]},
  {\small[\href{http://adsabs.harvard.edu/abs/2006SoPh..234..379C}{ADS}]}

\bibitem[Castagnoli and Lal(1980)]{castagnoli80}
Castagnoli, G. and Lal, D., 1980, ``Solar modulation effects in terrestrial
  production of carbon-14'', {\it Radiocarbon\/}, {\bf 22}, 133--158

\bibitem[Charbonneau(2001)]{charbonneau01}
Charbonneau, P., 2001, ``Multiperiodicity, Chaos, and Intermittency in a
  Reduced Model of the Solar Cycle'', {\it Solar Phys.\/}, {\bf 199}, 385--404.
  {\small[\href{http://adsabs.harvard.edu/abs/2001SoPh..199..385C}{ADS}]}

\bibitem[Charbonneau(2010)]{charbonneau10}
Charbonneau, P., 2010, ``Dynamo Models of the Solar Cycle'', {\it Living Rev.
  Solar Phys.\/}, {\bf 7}, lrsp-2010-3.
  {\small[\href{http://adsabs.harvard.edu/abs/2010LRSP....7....3C}{ADS}]}. URL
  (accessed 25 August 2012):
  \newline\url{http://www.livingreviews.org/lrsp-2010-3}

\bibitem[Charbonneau and Dikpati(2000)]{charbonneau00}
Charbonneau, P. and Dikpati, M., 2000, ``Stochastic Fluctuations in a
  Babcock-Leighton Model of the Solar Cycle'', {\it Astrophys. J.\/}, {\bf
  543}, 1027--1043. {\small[\href{http://dx.doi.org/10.1086/317142}{DOI}]},
  {\small[\href{http://adsabs.harvard.edu/abs/2000ApJ...543.1027C}{ADS}]}

\bibitem[Charbonneau {\it et~al.\/}(2004)]{charbonneau04}
Charbonneau, P., Blais-Laurier, G. and St-Jean, C., 2004, ``Intermittency and
  Phase Persistence in a Babcock-Leighton Model of the Solar Cycle'', {\it
  Astrophys. J. Lett.\/}, {\bf 616}, L183--L186.
  {\small[\href{http://dx.doi.org/10.1086/426897}{DOI}]},
  {\small[\href{http://adsabs.harvard.edu/abs/2004ApJ...616L.183C}{ADS}]}

\bibitem[Choudhuri(1992)]{choudhuri92}
Choudhuri, A.R., 1992, ``Stochastic fluctuations of the solar dynamo'', {\it
  Astron. Astrophys.\/}, {\bf 253}, 277--285.
  {\small[\href{http://adsabs.harvard.edu/abs/1992A&A...253..277C}{ADS}]}

\bibitem[Choudhuri {\it et~al.\/}(2007)]{choudhuri07}
Choudhuri, A.R., Chatterjee, P. and Jiang, J., 2007, ``Predicting Solar Cycle
  24 With a Solar Dynamo Model'', {\it Phys. Rev. Lett.\/}, {\bf 98}, 131103.
  {\small[\href{http://dx.doi.org/10.1103/PhysRevLett.98.131103}{DOI}]},
  {\small[\href{http://adsabs.harvard.edu/abs/2007PhRvL..98m1103C}{ADS}]},
  {\small[\href{http://arxiv.org/abs/astro-ph/0701527}{{arXiv:astro-ph/0701527%
}}]}

\bibitem[Clark and Stephenson(1978)]{clark78}
Clark, D.H. and Stephenson, F.R., 1978, ``An Interpretation of the
  Pre-Telescopic Sunspot Records from the Orient'', {\it Quart. J. R. Astron.
  Soc.\/}, {\bf 19}, 387--410.
  {\small[\href{http://adsabs.harvard.edu/abs/1978QJRAS..19..387C}{ADS}]}

\bibitem[Clem and Dorman(2000)]{clem00}
Clem, J.M. and Dorman, L.I., 2000, ``Neutron Monitor Response Functions'', {\it
  Space Sci. Rev.\/}, {\bf 93}, 335--359.
  {\small[\href{http://dx.doi.org/10.1023/A:1026508915269}{DOI}]},
  {\small[\href{http://adsabs.harvard.edu/abs/2000SSRv...93..335C}{ADS}]}

\bibitem[Clette {\it et~al.\/}(2007)]{clette07}
Clette, F., Berghmans, D., Vanlommel, P., van~der Linden, R.A.M.,
  Koeckelenbergh, A. and Wauters, L., 2007, ``From the Wolf number to the
  International Sunspot Index: 25 years of SIDC'', {\it Adv. Space Res.\/},
  {\bf 40}, 919--928.
  {\small[\href{http://dx.doi.org/10.1016/j.asr.2006.12.045}{DOI}]},
  {\small[\href{http://adsabs.harvard.edu/abs/2007AdSpR..40..919C}{ADS}]}

\bibitem[Cliver {\it et~al.\/}(1998)]{cliver98}
Cliver, E.W., Boriakoff, V. and Bounar, K.H., 1998, ``Geomagnetic activity and
  the solar wind during the Maunder Minimum'', {\it Geophys. Res. Lett.\/},
  {\bf 25}, 897--900.
  {\small[\href{http://dx.doi.org/10.1029/98GL00500}{DOI}]},
  {\small[\href{http://adsabs.harvard.edu/abs/1998GeoRL..25..897C}{ADS}]}

\bibitem[Conway(1998)]{conway98}
Conway, A.J., 1998, ``Time series, neural networks and the future of the Sun'',
  {\it New Astron. Rev.\/}, {\bf 42}, 343--394.
  {\small[\href{http://dx.doi.org/10.1016/S1387-6473(98)00041-4}{DOI}]},
  {\small[\href{http://adsabs.harvard.edu/abs/1998NewAR..42..343C}{ADS}]}

\bibitem[Cooke {\it et~al.\/}(1991)]{cooke91}
Cooke, D.J., Humble, J.E., Shea, M.A., Smart, D.F., Lund, N., Rasmussen, I.L.,
  Byrnak, B., Goret, P. and Petrou, N., 1991, ``On cosmic-ray cut-off
  terminology'', {\it Nuovo Cimento C\/}, {\bf 14}, 213--234.
  {\small[\href{http://dx.doi.org/10.1007/BF02509357}{DOI}]},
  {\small[\href{http://adsabs.harvard.edu/abs/1991NCimC..14..213C}{ADS}]}

\bibitem[Covas {\it et~al.\/}(1998)]{covas98}
Covas, E., Tavakol, R., Tworkowski, A. and Brandenburg, A., 1998,
  ``Axisymmetric mean field dynamos with dynamic and algebraic
  $\alpha$-quenching'', {\it Astron. Astrophys.\/}, {\bf 329}, 350--360.
  {\small[\href{http://adsabs.harvard.edu/abs/1998A&A...329..350C}{ADS}]}

\bibitem[Damon and Sonett(1991)]{damon91}
Damon, P.E. and Sonett, C.P., 1991, ``Solar and terrestrial components of the
  atmospheric $^{14}$C variation spectrum'', in {\it The Sun in Time\/}, (Eds.)
  Sonett, C.P., Giampapa, M.S., Matthews, M.S., pp. 360--388, University of
  Arizona Press, Tucson.
  {\small[\href{http://adsabs.harvard.edu/abs/1991suti.conf..360D}{ADS}]}

\bibitem[Damon {\it et~al.\/}(1978)]{damon78}
Damon, P.E., Lerman, J.C. and Long, A., 1978, ``Temporal Fluctuations of
  Atmospheric $^{14}$C: Causal Factors and Implications'', {\it Annu. Rev.
  Earth Planet. Sci.\/}, {\bf 6}, 457--494.
  {\small[\href{http://dx.doi.org/10.1146/annurev.ea.06.050178.002325}{DOI}]},
  {\small[\href{http://adsabs.harvard.edu/abs/1978AREPS...6..457D}{ADS}]}

\bibitem[de~Carvalho and Prado(2000)]{carvalho00}
de~Carvalho, J.X. and Prado, C.P.C., 2000, ``Self-Organized Criticality in the
  Olami-Feder-Christensen Model'', {\it Phys. Rev. Lett.\/}, {\bf 84},
  4006--4009.
  {\small[\href{http://dx.doi.org/10.1103/PhysRevLett.84.4006}{DOI}]},
  {\small[\href{http://adsabs.harvard.edu/abs/2000PhRvL..84.4006D}{ADS}]}

\bibitem[de~Jager(2005)]{dejager05}
de~Jager, C., 2005, ``Solar Forcing of Climate. 1: Solar Variability'', {\it
  Space Sci. Rev.\/}, {\bf 120}, 197--241.
  {\small[\href{http://dx.doi.org/10.1007/s11214-005-7046-5}{DOI}]},
  {\small[\href{http://adsabs.harvard.edu/abs/2005SSRv..120..197D}{ADS}]}

\bibitem[de~Jong and Mook(1982)]{dejong82}
de~Jong, A.F.M. and Mook, W.G., 1982, ``An anomalous Suess effect above
  Europe'', {\it Nature\/}, {\bf 298}, 641--644.
  {\small[\href{http://dx.doi.org/10.1038/298641a0}{DOI}]},
  {\small[\href{http://adsabs.harvard.edu/abs/1982Natur.298..641D}{ADS}]}

\bibitem[de~Meyer(1998)]{demeyer98}
de~Meyer, F., 1998, ``Modulation of the Solar Magnetic Cycle'', {\it Solar
  Phys.\/}, {\bf 181}, 201--219.
  {\small[\href{http://dx.doi.org/10.1023/A:1005079132329}{DOI}]},
  {\small[\href{http://adsabs.harvard.edu/abs/1998SoPh..181..201D}{ADS}]}

\bibitem[Delaygue and Bard(2011)]{delaygue11}
Delaygue, G. and Bard, E., 2011, ``An Antarctic view of Beryllium-10 and solar
  activity for the past millennium'', {\it Climate Dyn.\/}, {\bf 36},
  2201--2218. {\small[\href{http://dx.doi.org/10.1007/s00382-010-0795-1}{DOI}]}

\bibitem[Dikpati and Gilman(2006)]{dikpati06}
Dikpati, M. and Gilman, P.A., 2006, ``Simulating and Predicting Solar Cycles
  Using a Flux-Transport Dynamo'', {\it Astrophys. J.\/}, {\bf 649}, 498--514.
  {\small[\href{http://dx.doi.org/10.1086/506314}{DOI}]},
  {\small[\href{http://adsabs.harvard.edu/abs/2006ApJ...649..498D}{ADS}]}

\bibitem[Dikpati {\it et~al.\/}(2008)]{dikpati08}
Dikpati, M., de~Toma, G. and Gilman, P.A., 2008, ``Polar Flux, Cross-equatorial
  Flux, and Dynamo-generated Tachocline Toroidal Flux as Predictors of Solar
  Cycles'', {\it Astrophys. J.\/}, {\bf 675}, 920--930.
  {\small[\href{http://dx.doi.org/10.1086/524656}{DOI}]},
  {\small[\href{http://adsabs.harvard.edu/abs/2008ApJ...675..920D}{ADS}]}

\bibitem[Donadini {\it et~al.\/}(2010)]{donadini10}
Donadini, F., Korte, M. and Constable, C., 2010, ``Millennial Variations of the
  Geomagnetic Field: from Data Recovery to Field Reconstruction'', {\it Space
  Sci. Rev.\/}, {\bf 155}, 219--246.
  {\small[\href{http://dx.doi.org/10.1007/s11214-010-9662-y}{DOI}]},
  {\small[\href{http://adsabs.harvard.edu/abs/2010SSRv..155..219D}{ADS}]}

\bibitem[Donnelly {\it et~al.\/}(1994)]{donnelly94}
Donnelly, R.F., White, O.R. and Livingston, W.C., 1994, ``The solar Ca II K
  index and the Mg II core-to-wing ratio'', {\it Solar Phys.\/}, {\bf 152},
  69--76. {\small[\href{http://dx.doi.org/10.1007/BF01473185}{DOI}]},
  {\small[\href{http://adsabs.harvard.edu/abs/1994SoPh..152...69D}{ADS}]}

\bibitem[Dorman(2004)]{dorman04}
Dorman, L.I., 2004, {\it Cosmic Rays in the Earth's Atmosphere and
  Underground\/}, Kluwer Academic Publishers, Dordrecht.
  {\small[\href{http://books.google.com/books?id=mKLv68WBu5kC}{Google Books}]}

\bibitem[Dreschhoff and Zeller(1990)]{dreschhoff90}
Dreschhoff, G.A.M. and Zeller, E.J., 1990, ``Evidence of individual solar
  proton events in Antarctic snow'', {\it Solar Phys.\/}, {\bf 127}, 333--346.
  {\small[\href{http://dx.doi.org/10.1007/BF00152172}{DOI}]},
  {\small[\href{http://adsabs.harvard.edu/abs/1990SoPh..127..333D}{ADS}]}

\bibitem[Dreschhoff and Zeller(1998)]{dreschhoff98}
Dreschhoff, G.A.M. and Zeller, E.J., 1998, ``Ultra-High Resolution Nitrate in
  Polar Ice as Indicator of Past Solar Activity'', {\it Solar Phys.\/}, {\bf
  177}, 365--374.
  {\small[\href{http://dx.doi.org/10.1023/A:1004932530313}{DOI}]},
  {\small[\href{http://adsabs.harvard.edu/abs/1998SoPh..177..365D}{ADS}]}

\bibitem[Eddy(1976)]{eddy76}
Eddy, J.A., 1976, ``The Maunder Minimum'', {\it Science\/}, {\bf 192},
  1189--1202.
  {\small[\href{http://dx.doi.org/10.1126/science.192.4245.1189}{DOI}]},
  {\small[\href{http://adsabs.harvard.edu/abs/1976Sci...192.1189E}{ADS}]}

\bibitem[Eddy(1977{\natexlab{a}})]{eddy77}
Eddy, J.A., 1977{\natexlab{a}}, ``The case of the missing sunspots'', {\it
  Scientific American\/}, {\bf 236}, 80--88.
  {\small[\href{http://adsabs.harvard.edu/abs/1977SciAm.236...80E}{ADS}]}

\bibitem[Eddy(1977{\natexlab{b}})]{eddy_climate_77}
Eddy, J.A., 1977{\natexlab{b}}, ``Climate and the changing sun'', {\it Climatic
  Change\/}, {\bf 1}, 173--190.
  {\small[\href{http://dx.doi.org/10.1007/BF01884410}{DOI}]}

\bibitem[Eddy(1983)]{eddy83}
Eddy, J.A., 1983, ``The Maunder Minimum: A reappraisal'', {\it Solar Phys.\/},
  {\bf 89}, 195--207.
  {\small[\href{http://dx.doi.org/10.1007/BF00211962}{DOI}]},
  {\small[\href{http://adsabs.harvard.edu/abs/1983SoPh...89..195E}{ADS}]}

\bibitem[Eddy {\it et~al.\/}(1989)]{eddy89}
Eddy, J.A., Stephenson, F.R. and Yau, K.K.C., 1989, ``On pre-telescopic sunspot
  records'', {\it Quart. J. R. Astron. Soc.\/}, {\bf 30}, 65--73.
  {\small[\href{http://adsabs.harvard.edu/abs/1989QJRAS..30...65E}{ADS}]}

\bibitem[Elsasser {\it et~al.\/}(1956)]{elsasser56}
Elsasser, W., Nay, E.P. and Winkler, J.R., 1956, ``Cosmic-ray intensity and
  geomagnetism'', {\it Nature\/}, {\bf 178}, 1226--1227.
  {\small[\href{http://dx.doi.org/10.1038/1781226a0}{DOI}]},
  {\small[\href{http://adsabs.harvard.edu/abs/1987RvGeo..25....1F}{ADS}]}

\bibitem[Feynman and Gabriel(1990)]{feynman90}
Feynman, J. and Gabriel, S.B., 1990, ``Period and phase of the 88-year solar
  cycle and the Maunder minimum: Evidence for a chaotic Sun'', {\it Solar
  Phys.\/}, {\bf 127}, 393--403.
  {\small[\href{http://dx.doi.org/10.1007/BF00152176}{DOI}]},
  {\small[\href{http://adsabs.harvard.edu/abs/1990SoPh..127..393F}{ADS}]}

\bibitem[Field {\it et~al.\/}(2006)]{field06}
Field, C.V., Schmidt, G.A., Koch, D. and Salyk, C., 2006, ``Modeling production
  and climate-related impacts on $^{10}$Be concentration in ice cores'', {\it
  J. Geophys. Res.\/}, {\bf 111}, D15107.
  {\small[\href{http://dx.doi.org/10.1029/2005JD006410}{DOI}]},
  {\small[\href{http://adsabs.harvard.edu/abs/2006JGRD..11115107F}{ADS}]}

\bibitem[Fink {\it et~al.\/}(1998)]{fink98}
Fink, D., Klein, J., Middleton, R., Vogt, S., Herzog, G.F. and Reedy, R.C.,
  1998, ``$^{41}$Ca, $^{26}$Al, and $^{10}$Be in lunar basalt 74275 and
  $^{10}$Be in the double drive tube 74002/74001'', {\it Geochim. Cosmochim.
  Acta\/}, {\bf 62}, 2389--2402.
  {\small[\href{http://dx.doi.org/10.1016/S0016-7037(98)00134-3}{DOI}]},
  {\small[\href{http://adsabs.harvard.edu/abs/1998GeCoA..62.2389F}{ADS}]}

\bibitem[Finkel and Nishiizumi(1997)]{finkel97}
Finkel, R.C. and Nishiizumi, K., 1997, ``Beryllium 10 concentrations in the
  Greenland Ice Sheet Project 2 ice core from 3--40 ka'', {\it J. Geophys.
  Res.\/}, {\bf 102}, 26,699--26,706.
  {\small[\href{http://dx.doi.org/10.1029/97JC01282}{DOI}]},
  {\small[\href{http://adsabs.harvard.edu/abs/1997JGR...10226699F}{ADS}]}

\bibitem[Foukal(1996)]{foukal96}
Foukal, P., 1996, ``The behavior of solar magnetic plages measured from Mt.
  Wilson observations between 1915-1984'', {\it Geophys. Res. Lett.\/}, {\bf
  23}, 2169--2172. {\small[\href{http://dx.doi.org/10.1029/96GL01356}{DOI}]},
  {\small[\href{http://adsabs.harvard.edu/abs/1996GeoRL..23.2169F}{ADS}]}

\bibitem[Foukal {\it et~al.\/}(2006)]{foukal06}
Foukal, P., Fr{\"{o}}hlich, C., Spruit, H. and Wigley, T.M.L., 2006,
  ``Variations in solar luminosity and their effect on the Earth's climate'',
  {\it Nature\/}, {\bf 443}, 161--166.
  {\small[\href{http://dx.doi.org/10.1038/nature05072}{DOI}]},
  {\small[\href{http://adsabs.harvard.edu/abs/2006Natur.443..161F}{ADS}]}

\bibitem[Fraser-Smith(1987)]{fraser87}
Fraser-Smith, A.C., 1987, ``Centered and eccentric geomagnetic dipoles and
  their poles, 1600--1985'', {\it Rev. Geophys.\/}, {\bf 25}, 1--16.
  {\small[\href{http://dx.doi.org/10.1029/RG025i001p00001}{DOI}]},
  {\small[\href{http://adsabs.harvard.edu/abs/1987RvGeo..25....1F}{ADS}]}

\bibitem[Frick {\it et~al.\/}(1997)]{frick97}
Frick, P., Galyagin, D., Hoyt, D.V., Nesme-Ribes, E., Schatten, K.H., Sokoloff,
  D. and Zakharov, V., 1997, ``Wavelet analysis of solar activity recorded by
  sunspot groups'', {\it Astron. Astrophys.\/}, {\bf 328}, 670--681.
  {\small[\href{http://adsabs.harvard.edu/abs/1997A&A...328..670F}{ADS}]}

\bibitem[Fr{\"{o}}hlich(2006)]{frohlich06}
Fr{\"{o}}hlich, C., 2006, ``Solar Irradiance Variability since 1978'', {\it
  Space Sci. Rev.\/}, {\bf 125}, 53--65.
  {\small[\href{http://adsabs.harvard.edu/abs/2006SSRv..125...53F}{ADS}]}

\bibitem[Fr{\"o}hlich(2012)]{frohlich12}
Fr{\"o}hlich, C., 2012, ``Total Solar Irradiance Observations'', {\it Surv.
  Geophys.\/}, {\bf 33}, 453--473.
  {\small[\href{http://dx.doi.org/10.1007/s10712-011-9168-5}{DOI}]},
  {\small[\href{http://adsabs.harvard.edu/abs/2012SGeo...33..453F}{ADS}]}

\bibitem[Garcia-Mu{\~{n}}oz {\it et~al.\/}(1975)]{garcia75}
Garcia-Mu{\~{n}}oz, M., Mason, G.M. and Simpson, J.A., 1975, ``The anomalous
  $^{4}$He component in the cosmic-ray spectrum at $\lesssim$50 MeV per nucleon
  during 1972--1974'', {\it Astrophys. J.\/}, {\bf 202}, 265--275.
  {\small[\href{http://dx.doi.org/10.1086/153973}{DOI}]},
  {\small[\href{http://adsabs.harvard.edu/abs/1975ApJ...202..265G}{ADS}]}

\bibitem[Genevey {\it et~al.\/}(2008)]{genevey08}
Genevey, A., Gallet, Y., Constable, C.G., Korte, M. and Hulot, G., 2008,
  ``ArcheoInt: An upgraded compilation of geomagnetic field intensity data for
  the past ten millennia'', {\it Geochem. Geophys. Geosyst.\/}, {\bf 9},
  Q04038. {\small[\href{http://dx.doi.org/10.1029/2007GC001881}{DOI}]}

\bibitem[Gerber {\it et~al.\/}(2002)]{gerber03}
Gerber, S., Joos, F., Br{\"{u}}gger, P., Stocker, T.F., Mann, M.E., Sitch, S.
  and Scholze, M., 2002, ``Constraining temperature variations over the last
  millennium by comparing simulated and observed atmospheric CO$_{2}$'', {\it
  Climate Dyn.\/}, {\bf 20}, 281--299.
  {\small[\href{http://adsabs.harvard.edu/abs/2002ClDy...20..281G}{ADS}]}

\bibitem[Gibson {\it et~al.\/}(2011)]{gibson11}
Gibson, S.E., de~Toma, G., Emery, B., Riley, P., Zhao, L., Elsworth, Y.,
  Leamon, R.J., Lei, J., McIntosh, S., Mewaldt, R.A., Thompson, B.J. and Webb,
  D., 2011, ``The Whole Heliosphere Interval in the Context of a Long and
  Structured Solar Minimum: An Overview from Sun to Earth'', {\it Solar
  Phys.\/}, {\bf 274}, 5--27.
  {\small[\href{http://dx.doi.org/10.1007/s11207-011-9921-4}{DOI}]},
  {\small[\href{http://adsabs.harvard.edu/abs/2011SoPh..274....5G}{ADS}]}

\bibitem[Gleeson and Axford(1968)]{gleeson68}
Gleeson, L.J. and Axford, W.I., 1968, ``Solar Modulation of Galactic Cosmic
  Rays'', {\it Astrophys. J.\/}, {\bf 154}, 1011--1026.
  {\small[\href{http://dx.doi.org/10.1086/149822}{DOI}]},
  {\small[\href{http://adsabs.harvard.edu/abs/1968ApJ...154.1011G}{ADS}]}

\bibitem[Gleissberg(1939)]{gleissberg39}
Gleissberg, W., 1939, ``A long-periodic fluctuation of the sun-spot numbers'',
  {\it Observatory\/}, {\bf 62}, 158--159.
  {\small[\href{http://adsabs.harvard.edu/abs/1939Obs....62..158G}{ADS}]}

\bibitem[Gleissberg(1948)]{gleissberg48}
Gleissberg, W., 1948, ``A preliminary forecast of solar activity'', {\it
  Popular Astron.\/}, {\bf 56}, 399.
  {\small[\href{http://adsabs.harvard.edu/abs/1948PA.....56..399G}{ADS}]}

\bibitem[Gleissberg(1952)]{gleissberg52}
Gleissberg, W., 1952, {\it Die H{\"a}ufigkeit der Sonnenflecken\/},
  Akademie-Verlag, Berlin

\bibitem[Gleissberg(1971)]{gleissberg71}
Gleissberg, W., 1971, ``The Probable Behaviour of Sunspot Cycle 21'', {\it
  Solar Phys.\/}, {\bf 21}, 240--245.
  {\small[\href{http://dx.doi.org/10.1007/BF00155794}{DOI}]},
  {\small[\href{http://adsabs.harvard.edu/abs/1971SoPh...21..240G}{ADS}]}

\bibitem[Goslar(2001)]{goslar01}
Goslar, T., 2001, ``Absolute production of radiocarbon and the long-term trend
  of atmospheric radiocarbon'', {\it Radiocarbon\/}, {\bf 43}, 743--749

\bibitem[Goslar(2003)]{goslar03}
Goslar, T., 2003, ``$^{14}$C as an Indicator of Solar Variability'', {\it PAGES
  News\/}, {\bf 11}(2/3), 12--14. URL (accessed 26 February 2008):
  \newline\url{http://www.pages-igbp.org/products/newsletters/nl2003_2a.pdf}

\bibitem[Gray {\it et~al.\/}(2010)]{gray10}
Gray, L.J., Beer, J., Geller, M., Haigh, J.D., Lockwood, M., Matthes, K.,
  Cubasch, U., Fleitmann, D., Harrison, G., Hood, L., Luterbacher, J., Meehl,
  G.A., Shindell, D., van Geel, B. and White, W., 2010, ``Solar Influences on
  Climate'', {\it Rev. Geophys.\/}, {\bf 48}, RG4001.
  {\small[\href{http://dx.doi.org/10.1029/2009RG000282}{DOI}]},
  {\small[\href{http://adsabs.harvard.edu/abs/2010RvGeo..48.4001G}{ADS}]}

\bibitem[Grieder(2001)]{grieder01}
Grieder, P.K.F., 2001, {\it Cosmic Rays at Earth: Researcher's Reference Manual
  and Data Book\/}, Elsevier Science, Amsterdam; New York

\bibitem[Grismore {\it et~al.\/}(2001)]{grismore01}
Grismore, R., Llewellyn, R.A., Brown, M.D., Dowson, S.T. and Cumblidge, K.,
  2001, ``Measurements of the concentrations of $^{26}$Al in lunar rocks 15555
  and 60025'', {\it Earth Planet. Sci. Lett.\/}, {\bf 187}, 163--171.
  {\small[\href{http://dx.doi.org/10.1016/S0012-821X(01)00271-0}{DOI}]},
  {\small[\href{http://adsabs.harvard.edu/abs/2001E&PSL.187..163G}{ADS}]}

\bibitem[Haigh(2007)]{haighLR}
Haigh, J.D., 2007, ``The Sun and the Earth's Climate'', {\it Living Rev. Solar
  Phys.\/}, {\bf 4}, lrsp-2007-2.
  {\small[\href{http://adsabs.harvard.edu/abs/2007LRSP....4....2H}{ADS}]}. URL
  (accessed 25 February 2008):
  \newline\url{http://www.livingreviews.org/lrsp-2007-2}

\bibitem[Haigh {\it et~al.\/}(2010)]{haigh10}
Haigh, J.D., Winning, A.R., Toumi, R. and Harder, J.W., 2010, ``An influence of
  solar spectral variations on radiative forcing of climate'', {\it Nature\/},
  {\bf 467}, 696--699.
  {\small[\href{http://dx.doi.org/10.1038/nature09426}{DOI}]},
  {\small[\href{http://adsabs.harvard.edu/abs/2010Natur.467..696H}{ADS}]}

\bibitem[Hale {\it et~al.\/}(1919)]{hale1919}
Hale, G.E., Ellerman, F., Nicholson, S.B. and Joy, A.H., 1919, ``The Magnetic
  Polarity of Sun-Spots'', {\it Astrophys. J.\/}, {\bf 49}, 153--178.
  {\small[\href{http://dx.doi.org/10.1086/142452}{DOI}]},
  {\small[\href{http://adsabs.harvard.edu/abs/1919ApJ....49..153H}{ADS}]}

\bibitem[Hanslmeier {\it et~al.\/}(2013)]{hanslmeier13}
Hanslmeier, A., Braj{\v{s}}a, R., {\v{C}}alogovi{\'{c}}, J., Vr{\v{s}}nak, B.,
  Ru{\v{z}}djak, D., Steinhilber, F., MacLeod, C.L., Ivezi{\'{c}}, {\v Z}. and
  Skoki{\'{c}}, I., 2013, ``The chaotic solar cycle. II. Analysis of cosmogenic
  $^{10}$Be data'', {\it Astron. Astrophys.\/}, {\bf 550}, A6.
  {\small[\href{http://dx.doi.org/10.1051/0004-6361/201015215}{DOI}]},
  {\small[\href{http://adsabs.harvard.edu/abs/2013A&A...550A...6H}{ADS}]}

\bibitem[Harvey and White(1999)]{harvey99}
Harvey, K.L. and White, O.R., 1999, ``What is solar cycle minimum?'', {\it J.
  Geophys. Res.\/}, {\bf 104}(A9), 19,759--19,764.
  {\small[\href{http://dx.doi.org/10.1029/1999JA900211}{DOI}]},
  {\small[\href{http://adsabs.harvard.edu/abs/1999JGR...10419759H}{ADS}]}

\bibitem[Hathaway(2009)]{hathaway09}
Hathaway, D.H., 2009, ``Solar Cycle Forecasting'', {\it Space Sci. Rev.\/},
  {\bf 144}, 401--412.
  {\small[\href{http://dx.doi.org/10.1007/s11214-008-9430-4}{DOI}]},
  {\small[\href{http://adsabs.harvard.edu/abs/2009SSRv..144..401H}{ADS}]}

\bibitem[Hathaway(2010)]{hathawayLR}
Hathaway, D.H., 2010, ``The Solar Cycle'', {\it Living Rev. Solar Phys.\/},
  {\bf 7}, lrsp-2010-1.
  {\small[\href{http://adsabs.harvard.edu/abs/2010LRSP....7....1H}{ADS}]}. URL
  (accessed 25 March 2010):
  \newline\url{http://www.livingreviews.org/lrsp-2010-1}

\bibitem[Hathaway and Wilson(2004)]{hathaway04}
Hathaway, D.H. and Wilson, R.M., 2004, ``What the Sunspot Record Tells Us About
  Space Climate'', {\it Solar Phys.\/}, {\bf 224}, 5--19.
  {\small[\href{http://dx.doi.org/10.1007/s11207-005-3996-8}{DOI}]}

\bibitem[Hathaway {\it et~al.\/}(1999)]{hathaway99}
Hathaway, D.H., Wilson, R.M. and Reichmann, E.J., 1999, ``A Synthesis of Solar
  Cycle Prediction Techniques'', {\it J. Geophys. Res.\/}, {\bf 104},
  22,375--22,388. {\small[\href{http://dx.doi.org/10.1029/1999JA900313}{DOI}]},
  {\small[\href{http://adsabs.harvard.edu/abs/1999JGR...10422375H}{ADS}]}

\bibitem[Heikkil{\"a} {\it et~al.\/}(2008)]{heikkila07}
Heikkil{\"a}, U., Beer, J. and Feichter, J., 2008, ``Modeling cosmogenic
  radionuclides $^{10}$Be and $^{7}$Be during the Maunder Minimum using the
  ECHAM5-HAM General Circulation Model'', {\it Atmos. Chem. Phys.\/}, {\bf 8},
  2797--2809. {\small[\href{http://dx.doi.org/10.5194/acp-8-2797-2008}{DOI}]},
  {\small[\href{http://adsabs.harvard.edu/abs/2008ACP.....8.2797H}{ADS}]}. URL
  (accessed 11 September 2012):
  \newline\url{http://www.atmos-chem-phys.net/8/2797/2008/}

\bibitem[Heikkil{\"a} {\it et~al.\/}(2009)]{heikkila09}
Heikkil{\"a}, U., Beer, J. and Feichter, J., 2009, ``Meridional transport and
  deposition of atmospheric $^{10}$Be'', {\it Atmos. Chem. Phys.\/}, {\bf 9},
  515--527. {\small[\href{http://dx.doi.org/10.5194/acp-9-515-2009}{DOI}]},
  {\small[\href{http://adsabs.harvard.edu/abs/2009ACP.....9..515H}{ADS}]}. URL
  (accessed 11 September 2012):
  \newline\url{http://www.atmos-chem-phys.net/9/515/2009/}

\bibitem[Herbst {\it et~al.\/}(2010)]{herbst10}
Herbst, K., Kopp, A., Heber, B., Steinhilber, F., Fichtner, H., Scherer, K. and
  Matthi{\"a}, D., 2010, ``On the importance of the local interstellar spectrum
  for the solar modulation parameter'', {\it J. Geophys. Res.\/}, {\bf 115},
  D00I20. {\small[\href{http://dx.doi.org/10.1029/2009JD012557}{DOI}]},
  {\small[\href{http://adsabs.harvard.edu/abs/2010JGRA..11500I20H}{ADS}]}

\bibitem[Hongre {\it et~al.\/}(1998)]{hongre98}
Hongre, L., Hulot, G. and Khokhlov, A., 1998, ``An analysis of the geomagnetic
  field over the past 2000 years'', {\it Phys. Earth Planet. Inter.\/}, {\bf
  106}, 311--335.
  {\small[\href{http://dx.doi.org/10.1016/S0031-9201(97)00115-5}{DOI}]},
  {\small[\href{http://adsabs.harvard.edu/abs/1998PEPI..106..311H}{ADS}]}

\bibitem[Horiuchi {\it et~al.\/}(1999)]{horiuchi99}
Horiuchi, K., Minoura, K., Kobayashi, K., Nakamura, T., Hatori, S., Matsuzaki,
  H. and Kawai, T., 1999, ``Last-glacial to post-glacial $^{10}$Be fluctuations
  in a sediment core from the Academician Ridge, Lake Baikal'', {\it Geophys.
  Res. Lett.\/}, {\bf 26}, 1047--1050.
  {\small[\href{http://dx.doi.org/10.1029/1999GL900163}{DOI}]},
  {\small[\href{http://adsabs.harvard.edu/abs/1999GeoRL..26.1047H}{ADS}]}

\bibitem[Horiuchi {\it et~al.\/}(2007)]{horiuchi07}
Horiuchi, K., Ohta, A., Uchida, T., Matsuzaki, H., Shibata, Y. and Motoyama,
  H., 2007, ``Concentration of $^{10}$Be in an ice core from the Dome Fuji
  station, Eastern Antarctica: Preliminary results from 1500 to 1810 yr AD'',
  {\it Nucl. Instrum. Methods B\/}, {\bf 259}, 584--587.
  {\small[\href{http://dx.doi.org/10.1016/j.nimb.2007.01.306}{DOI}]},
  {\small[\href{http://adsabs.harvard.edu/abs/2007NIMPB.259..584H}{ADS}]}

\bibitem[Horiuchi {\it et~al.\/}(2008)]{horiuchi08}
Horiuchi, K., Uchida, T., Sakamoto, Y., Ohta, A., Matsuzaki, H., Shibata, Y.
  and Motoyama, H., 2008, ``Ice core record of $^{10}$Be over the past
  millennium from Dome Fuji, Antarctica: A new proxy record of past solar
  activity and a powerful tool for stratigraphic dating'', {\it Quat.
  Geochronology\/}, {\bf 3}, 253--261.
  {\small[\href{http://dx.doi.org/10.1016/j.quageo.2008.01.003}{DOI}]}

\bibitem[Hoyng(1993)]{hoyng93}
Hoyng, P., 1993, ``Helicity fluctuations in mean field theory: an explanation
  for the variability of the solar cycle?'', {\it Astron. Astrophys.\/}, {\bf
  272}, 321--339.
  {\small[\href{http://adsabs.harvard.edu/abs/1993A&A...272..321H}{ADS}]}

\bibitem[Hoyt and Schatten(1998)]{hoyt98}
Hoyt, D.V. and Schatten, K., 1998, ``Group Sunspot Numbers: A New Solar
  Activity Reconstruction'', {\it Solar Phys.\/}, {\bf 179}, 189--219.
  {\small[\href{http://adsabs.harvard.edu/abs/1998SoPh..179..189H}{ADS}]}

\bibitem[Hoyt and Schatten(1996)]{hoyt96}
Hoyt, D.V. and Schatten, K.H., 1996, ``How Well Was the Sun Observed during the
  Maunder Minimum?'', {\it Solar Phys.\/}, {\bf 165}, 181--192.
  {\small[\href{http://dx.doi.org/10.1007/BF00149097}{DOI}]},
  {\small[\href{http://adsabs.harvard.edu/abs/1996SoPh..165..181H}{ADS}]}

\bibitem[Hoyt {\it et~al.\/}(1994)]{hoyt94}
Hoyt, D.V., Schatten, K.H. and Nesme-Ribes, E., 1994, ``The one hundredth year
  of Rudolf Wolf's death: Do we have the correct reconstruction of solar
  activity?'', {\it Geophys. Res. Lett.\/}, {\bf 21}, 2067--2070.
  {\small[\href{http://dx.doi.org/10.1029/94GL01698}{DOI}]},
  {\small[\href{http://adsabs.harvard.edu/abs/1994GeoRL..21.2067H}{ADS}]}

\bibitem[Hudson(2010)]{hudson10}
Hudson, H.~S., 2010, ``Solar flares add up'', {\it Nature Phys.\/}, {\bf 6},
  637--638. {\small[\href{http://dx.doi.org/10.1038/nphys1764}{DOI}]},
  {\small[\href{http://adsabs.harvard.edu/abs/2010NatPh...6..637H}{ADS}]}

\bibitem[Jennings and Weiss(1991)]{jennings91}
Jennings, R.L. and Weiss, N.O., 1991, ``Symmetry breaking in stellar dynamos'',
  {\it Mon. Not. R. Astron. Soc.\/}, {\bf 252}, 249--260.
  {\small[\href{http://adsabs.harvard.edu/abs/1991MNRAS.252..249J}{ADS}]}

\bibitem[Jiang {\it et~al.\/}(2007)]{jiang07}
Jiang, J., Chatterjee, P. and Choudhuri, A.R., 2007, ``Solar activity forecast
  with a dynamo model'', {\it Mon. Not. R. Astron. Soc.\/}, {\bf 381},
  1527--1542.
  {\small[\href{http://dx.doi.org/10.1111/j.1365-2966.2007.12267.x}{DOI}]},
  {\small[\href{http://adsabs.harvard.edu/abs/2007MNRAS.381.1527J}{ADS}]},
  {\small[\href{http://arxiv.org/abs/0707.2258}{{arXiv:0707.2258
  {\small[astro-ph]}}}]}

\bibitem[Joos(1994)]{joos94}
Joos, F., 1994, ``Imbalance in the budget'', {\it Nature\/}, {\bf 370},
  181--182. {\small[\href{http://dx.doi.org/10.1038/370181a0}{DOI}]}

\bibitem[Jull {\it et~al.\/}(1998)]{jull98}
Jull, A.J.T., Cloudt, S., Donahue, D.J., Sisterson, J.M., Reedy, R.C. and
  Masarik, J., 1998, ``$^{14}$C depth profiles in Apollo 15 and 17 cores and
  lunar rock 68815'', {\it Geochim. Cosmochim. Acta\/}, {\bf 62}, 3025--3036.
  {\small[\href{http://dx.doi.org/10.1016/S0016-7037(98)00193-8}{DOI}]},
  {\small[\href{http://adsabs.harvard.edu/abs/1998GeCoA..62.3025J}{ADS}]}

\bibitem[Kane(2007)]{kane07}
Kane, R.P., 2007, ``A Preliminary Estimate of the Size of the Coming Solar
  Cycle 24, based on Ohl's Precursor Method'', {\it Solar Phys.\/}, {\bf 243},
  205--217. {\small[\href{http://dx.doi.org/10.1007/s11207-007-0475-4}{DOI}]},
  {\small[\href{http://adsabs.harvard.edu/abs/2007SoPh..243..205K}{ADS}]}

\bibitem[K{\"a}pyl{\"a} {\it et~al.\/}(2012)]{kapyla12}
K{\"a}pyl{\"a}, P.J., Mantere, M.J. and Brandenburg, A., 2012, ``Cyclic
  Magnetic Activity due to Turbulent Convection in Spherical Wedge Geometry'',
  {\it Astrophys. J. Lett.\/}, {\bf 755}, L22.
  {\small[\href{http://dx.doi.org/10.1088/2041-8205/755/1/L22}{DOI}]},
  {\small[\href{http://adsabs.harvard.edu/abs/2012ApJ...755L..22K}{ADS}]}

\bibitem[Karak and Nandy(2012)]{karak12}
Karak, B.B. and Nandy, D., 2012, ``Turbulent Pumping of Magnetic Flux Reduces
  Solar Cycle Memory and thus Impacts Predictability of the Sun's Activity'',
  {\it Astrophys. J. Lett.\/}, {\bf 761}, L13.
  {\small[\href{http://dx.doi.org/10.1088/2041-8205/761/1/L13}{DOI}]},
  {\small[\href{http://adsabs.harvard.edu/abs/2012ApJ...761L..13K}{ADS}]},
  {\small[\href{http://arxiv.org/abs/1206.2106}{{arXiv:1206.2106
  {\small[astro-ph.SR]}}}]}

\bibitem[Kigoshi and Hasegawa(1966)]{kigoshi66}
Kigoshi, K. and Hasegawa, H., 1966, ``Secular variation of atmospheric
  radiocarbon concentration and its dependence on geomagnetism'', {\it J.
  Geophys. Res.\/}, {\bf 71}, 1065--1071.
  {\small[\href{http://adsabs.harvard.edu/abs/1966JGR....71.1065K}{ADS}]}

\bibitem[Klecker {\it et~al.\/}(2006)]{klecker06}
Klecker, B., Kunow, H., Cane, H.V., Dalla, S., Heber, B., Kecskemety, K.,
  Klein, K.-L., K{\'{o}}ta, J., Kucharek, H., Lario, D., Lee, M.A., Popecki,
  M.A., Posner, A., Rodriguez-Pacheco, J., Sanderson, T., Simnett, G.M. and
  Roelof, E.C., 2006, ``Energetic Particle Observations'', {\it Space Sci.
  Rev.\/}, {\bf 123}, 217--250.
  {\small[\href{http://dx.doi.org/10.1007/s11214-006-9018-9}{DOI}]},
  {\small[\href{http://adsabs.harvard.edu/abs/2006SSRv..123..217K}{ADS}]}

\bibitem[Kleczek(1952)]{kleczek52}
Kleczek, J., 1952, ``Solar flare index calculations'', {\it Publ. Centr.
  Astron. Inst. Czechoslovakia\/}, {\bf 22}

\bibitem[Koch {\it et~al.\/}(2006)]{koch06}
Koch, D., Schmidt, G.A. and Field, C.V., 2006, ``Sulfur, sea salt, and
  radionuclide aerosols in GISS ModelE'', {\it J. Geophys. Res.\/}, {\bf
  111}(D10), D06206.
  {\small[\href{http://dx.doi.org/10.1029/2004JD005550}{DOI}]},
  {\small[\href{http://adsabs.harvard.edu/abs/2006JGRD..11106206K}{ADS}]}

\bibitem[Kocharov {\it et~al.\/}(1995)]{kocharov95}
Kocharov, G.E., Ostryakov, V.M., Peristykh, A.N. and Vasil'ev, V.A., 1995,
  ``Radiocarbon Content Variations and Maunder Minimum of Solar Activity'',
  {\it Solar Phys.\/}, {\bf 159}, 381--391.
  {\small[\href{http://dx.doi.org/10.1007/BF00686539}{DOI}]},
  {\small[\href{http://adsabs.harvard.edu/abs/1995SoPh..159..381K}{ADS}]}

\bibitem[Kohl {\it et~al.\/}(1978)]{kohl78}
Kohl, C.P., Murrell, M.T., Russ~III, G.P. and Arnold, J.R., 1978, ``Evidence
  for the constancy of the solar cosmic ray flux over the past ten million
  years: $^{53}$Mn and $^{26}$Al measurements'', in {\it Lunar and Planetary
  Science IX\/}, Proceedings of the conference, Houston, TX, March 13\,--\,17,
  1978, Geochim. Cosmochim. Acta Suppl., 10, pp. 2299--2310, Pergamon Press,
  New York.
  {\small[\href{http://adsabs.harvard.edu/abs/1978LPSC....9.2299K}{ADS}]}

\bibitem[Korte and Constable(2005)]{korte05}
Korte, M. and Constable, C.G., 2005, ``The geomagnetic dipole moment over the
  last 7000 years -- new results from a global model'', {\it Earth Planet. Sci.
  Lett.\/}, {\bf 236}, 348--358.
  {\small[\href{http://dx.doi.org/10.1016/j.epsl.2004.12.031}{DOI}]},
  {\small[\href{http://adsabs.harvard.edu/abs/2005E&PSL.236..348K}{ADS}]}

\bibitem[Korte and Constable(2006)]{korte06}
Korte, M. and Constable, C.G., 2006, ``Centennial to millennial geomagnetic
  secular variation'', {\it Geophys. J. Int.\/}, {\bf 167}, 43--52.
  {\small[\href{http://dx.doi.org/10.1111/j.1365-246X.2006.03088.x}{DOI}]},
  {\small[\href{http://adsabs.harvard.edu/abs/2006GeoJI.167...43K}{ADS}]}

\bibitem[Korte and Constable(2008)]{korte08}
Korte, M. and Constable, C.G., 2008, ``Spatial and temporal resolution of
  millennial scale geomagnetic field models'', {\it Adv. Space Res.\/}, {\bf
  41}, 57--69.
  {\small[\href{http://dx.doi.org/10.1016/j.asr.2007.03.094}{DOI}]},
  {\small[\href{http://adsabs.harvard.edu/abs/2008AdSpR..41...57K}{ADS}]}

\bibitem[Korte {\it et~al.\/}(2011)]{korte11}
Korte, M., Constable, C., Donadini, F. and Holme, R., 2011, ``Reconstructing
  the Holocene geomagnetic field'', {\it Earth Planet. Sci. Lett.\/}, {\bf
  312}, 497--505.
  {\small[\href{http://dx.doi.org/10.1016/j.epsl.2011.10.031}{DOI}]},
  {\small[\href{http://adsabs.harvard.edu/abs/2011E&PSL.312..497K}{ADS}]}

\bibitem[Kovaltsov and Usoskin(2007)]{kovaltsov07}
Kovaltsov, G.A. and Usoskin, I.G., 2007, ``Regional cosmic ray induced
  ionization and geomagnetic field changes'', {\it Adv. Geosci.\/}, {\bf 13},
  31--35.
  {\small[\href{http://adsabs.harvard.edu/abs/2007AdG....13...31K}{ADS}]}. URL
  (accessed 14 October 2008):
  \newline\url{http://www.adv-geosci.net/13/31/2007/}

\bibitem[Kovaltsov and Usoskin(2010)]{kovaltsov10}
Kovaltsov, G.A. and Usoskin, I.G., 2010, ``A new 3D numerical model of
  cosmogenic nuclide $^{10}$Be production in the atmosphere'', {\it Earth
  Planet. Sci. Lett.\/}, {\bf 291}, 182--188.
  {\small[\href{http://dx.doi.org/10.1016/j.epsl.2010.01.011}{DOI}]},
  {\small[\href{http://adsabs.harvard.edu/abs/2010E&PSL.291..182K}{ADS}]}

\bibitem[Kovaltsov {\it et~al.\/}(2004)]{kovaltsov04}
Kovaltsov, G.A., Usoskin, I.G. and Mursula, K., 2004, ``An Upper Limit on
  Sunspot Activity During the Maunder Minimum'', {\it Solar Phys.\/}, {\bf
  224}, 95--101.
  {\small[\href{http://dx.doi.org/10.1007/s11207-005-4281-6}{DOI}]},
  {\small[\href{http://adsabs.harvard.edu/abs/2004SoPh..224...95K}{ADS}]}

\bibitem[Kovaltsov {\it et~al.\/}(2012)]{kovaltsov12}
Kovaltsov, G.A., Mishev, A. and Usoskin, I.G., 2012, ``A new model of
  cosmogenic production of radiocarbon $^{14}$C in the atmosphere'', {\it Earth
  Planet. Sci. Lett.\/}, {\bf 337}, 114--120.
  {\small[\href{http://dx.doi.org/10.1016/j.epsl.2012.05.036}{DOI}]},
  {\small[\href{http://adsabs.harvard.edu/abs/2012E&PSL.337..114K}{ADS}]},
  {\small[\href{http://arxiv.org/abs/1206.6974}{{arXiv:1206.6974
  {\small[physics.ao-ph]}}}]}

\bibitem[Kremliovsky(1994)]{kremliovsky94}
Kremliovsky, M.N., 1994, ``Can we understand time scales of solar activity?'',
  {\it Solar Phys.\/}, {\bf 151}, 351--370.
  {\small[\href{http://dx.doi.org/10.1007/BF00679081}{DOI}]},
  {\small[\href{http://adsabs.harvard.edu/abs/1994SoPh..151..351K}{ADS}]}

\bibitem[Kremliovsky(1995)]{kremliovsky95}
Kremliovsky, M.N., 1995, ``Limits of Predictability of Solar Activity'', {\it
  Solar Phys.\/}, {\bf 159}, 371--380.
  {\small[\href{http://dx.doi.org/10.1007/BF00686538}{DOI}]},
  {\small[\href{http://adsabs.harvard.edu/abs/1995SoPh..159..371K}{ADS}]}

\bibitem[Krivova {\it et~al.\/}(2007)]{krivova07}
Krivova, N.A., Balmaceda, L. and Solanki, S.K., 2007, ``Reconstruction of solar
  total irradiance since 1700 from the surface magnetic flux'', {\it Astron.
  Astrophys.\/}, {\bf 467}, 335--346.
  {\small[\href{http://dx.doi.org/10.1051/0004-6361:20066725}{DOI}]},
  {\small[\href{http://adsabs.harvard.edu/abs/2007A&A...467..335K}{ADS}]}

\bibitem[K{\v{r}}ivsk{\'{y}}(1984)]{krivsky84}
K{\v{r}}ivsk{\'{y}}, L., 1984, ``Long-term fluctuations of solar activity
  during the last thousand years'', {\it Solar Phys.\/}, {\bf 93}, 189--194.
  {\small[\href{http://adsabs.harvard.edu/abs/1984SoPh...93..189K}{ADS}]}

\bibitem[K{\v{r}}ivsk{\'{y}} and Pejml(1988)]{krivsky88}
K{\v{r}}ivsk{\'{y}}, L. and Pejml, K., 1988, ``Solar Activity, Aurorae and
  Climate in Central Europe in the Last 1000 Years'', {\it Publ. Astron. Inst.
  Czechoslovak Acad. Sci.\/}, {\bf 75}, 32

\bibitem[Kuklin(1976)]{kuklin76}
Kuklin, G.V., 1976, ``Cyclical and Secular Variations of Solar Activity'', in
  {\it Basic Mechanisms of Solar Activity\/}, Symposium no. 71 held in Prague,
  Czechoslovakia, 25\,--\,29 August 1975, (Eds.) Bumba, V., Kleczek, J., IAU
  Symposium, 71, pp. 147--148, D. Reidel, Dordrecht; Boston.
  {\small[\href{http://adsabs.harvard.edu/abs/1976IAUS...71..147K}{ADS}]}

\bibitem[Kurths and Ruzmaikin(1990)]{kurths90}
Kurths, J. and Ruzmaikin, A.A., 1990, ``On forecasting the sunspot numbers'',
  {\it Solar Phys.\/}, {\bf 126}, 407--410.
  {\small[\href{http://dx.doi.org/10.1007/BF00153060}{DOI}]},
  {\small[\href{http://adsabs.harvard.edu/abs/1990SoPh..126..407K}{ADS}]}

\bibitem[Lal(1987)]{lal87}
Lal, D., 1987, ``$^{10}$Be in polar ice: data reflect changes in cosmic ray
  flux or polar meteorology'', {\it Geophys. Res. Lett.\/}, {\bf 14}, 785--788.
  {\small[\href{http://dx.doi.org/10.1029/GL014i008p00785}{DOI}]},
  {\small[\href{http://adsabs.harvard.edu/abs/1987GeoRL..14..785L}{ADS}]}

\bibitem[Lal and Peters(1967)]{lal67}
Lal, D. and Peters, B., 1967, ``Cosmic Ray Produced Radioactivity on the
  Earth'', in {\it Kosmische Strahlung II / Cosmic Rays II\/}, (Ed.) Sittle,
  K., Handbuch der Physik / Encyclopedia of Physics, 9/46/2, pp. 551--612,
  Springer, Berlin.
  {\small[\href{http://dx.doi.org/10.1007/978-3-642-46079-1_7}{DOI}]},
  {\small[\href{http://adsabs.harvard.edu/abs/1967HDP....46..551L}{ADS}]}

\bibitem[Lal and Suess(1968)]{lal68}
Lal, D. and Suess, H.E., 1968, ``The radioactivity of the atmosphere and
  hydrosphere'', {\it Annu. Rev. Nucl. Sci.\/}, {\bf 18}, 407--434.
  {\small[\href{http://dx.doi.org/10.1146/annurev.ns.18.120168.002203}{DOI}]}

\bibitem[Lal {\it et~al.\/}(2005)]{lal05}
Lal, D., Jull, A.J.T., Pollard, D. and Vacher, L., 2005, ``Evidence for large
  century time-scale changes in solar activity in the past 32 Kyr, based on
  in-situ cosmogenic $^{14}$C in ice at Summit, Greenland'', {\it Earth Planet.
  Sci. Lett.\/}, {\bf 234}, 335--349.
  {\small[\href{http://dx.doi.org/10.1016/j.epsl.2005.02.011}{DOI}]},
  {\small[\href{http://adsabs.harvard.edu/abs/2005E&PSL.234..335L}{ADS}]}

\bibitem[Lawrence {\it et~al.\/}(1995)]{lawrence95}
Lawrence, J.K., Cadavid, A.C. and Ruzmaikin, A.A., 1995, ``Turbulent and
  Chaotic Dynamics Underlying Solar Magnetic Variability'', {\it Astrophys.
  J.\/}, {\bf 455}, 366.
  {\small[\href{http://dx.doi.org/10.1086/176583}{DOI}]},
  {\small[\href{http://adsabs.harvard.edu/abs/1995ApJ...455..366L}{ADS}]}

\bibitem[Lee {\it et~al.\/}(2004)]{lee04}
Lee, E.H., Ahn, Y.S., Yang, H.J. and Chen, K.Y., 2004, ``The Sunspot and
  Auroral Activity Cycle Derived from Korean Historical Records of the 11th
  18th Century'', {\it Solar Phys.\/}, {\bf 224}, 373--386.
  {\small[\href{http://dx.doi.org/10.1007/s11207-005-5199-8}{DOI}]},
  {\small[\href{http://adsabs.harvard.edu/abs/2004SoPh..224..373L}{ADS}]}

\bibitem[Letfus(1999)]{letfus99}
Letfus, V., 1999, ``Daily relative sunspot numbers 1749--1848: reconstruction
  of missing observations'', {\it Solar Phys.\/}, {\bf 184}, 201--211.
  {\small[\href{http://dx.doi.org/10.1023/A:1005086320594}{DOI}]},
  {\small[\href{http://adsabs.harvard.edu/abs/1999SoPh..184..201L}{ADS}]}

\bibitem[Letfus(2000)]{letfus00}
Letfus, V., 2000, ``Relative sunspot numbers in the first half of eighteenth
  century'', {\it Solar Phys.\/}, {\bf 194}, 175--184.
  {\small[\href{http://adsabs.harvard.edu/abs/2000SoPh..194..175L}{ADS}]}

\bibitem[Li {\it et~al.\/}(2001)]{li01}
Li, K.J., Yun, H.S. and Gu, X.M., 2001, ``Latitude Migration of Sunspot
  Groups'', {\it Astron. J.\/}, {\bf 122}, 2115--2117.
  {\small[\href{http://dx.doi.org/10.1086/323089}{DOI}]},
  {\small[\href{http://adsabs.harvard.edu/abs/2001AJ....122.2115L}{ADS}]}

\bibitem[Lingenfelter(1963)]{lingenfelter63}
Lingenfelter, R.E., 1963, ``Production of Carbon 14 by Cosmic-Ray Neutrons'',
  {\it Rev. Geophys. Space Phys.\/}, {\bf 1}, 35--55.
  {\small[\href{http://dx.doi.org/10.1029/RG001i001p00035}{DOI}]},
  {\small[\href{http://adsabs.harvard.edu/abs/1963RvGSP...1...35L}{ADS}]}

\bibitem[Lingenfelter and Hudson(1980)]{lingenfelter80}
Lingenfelter, R.E. and Hudson, H.S., 1980, ``Solar particle fluxes and the
  ancient sun'', in {\it The Ancient Sun: Fossil Record in the Earth, Moon and
  Meteorites\/}, Proceedings of the conference held at Boulder, Colorado,
  October 16\,--\,19, 1979, (Eds.) Pepin, R.O., Eddy, J.A., Merrill, R.B., pp.
  69--79, Pergamon Press, New York; Oxford.
  {\small[\href{http://adsabs.harvard.edu/abs/1980asfr.symp...69L}{ADS}]}

\bibitem[Lockwood {\it et~al.\/}(1999)]{lockwood99}
Lockwood, M., Stamper, R. and Wild, M.N., 1999, ``A doubling of the Sun's
  coronal magnetic field during the past 100 years'', {\it Nature\/}, {\bf
  399}, 437--439. {\small[\href{http://dx.doi.org/10.1038/20867}{DOI}]},
  {\small[\href{http://adsabs.harvard.edu/abs/1999Natur.399..437L}{ADS}]}

\bibitem[Lockwood {\it et~al.\/}(2011)]{lockwood11}
Lockwood, M., Owens, M.J., Barnard, L., Davis, C.J. and Steinhilber, F., 2011,
  ``The persistence of solar activity indicators and the descent of the Sun
  into Maunder Minimum conditions'', {\it Geophys. Res. Lett.\/}, {\bf 38},
  L22105. {\small[\href{http://dx.doi.org/10.1029/2011GL049811}{DOI}]},
  {\small[\href{http://adsabs.harvard.edu/abs/2011GeoRL..3822105L}{ADS}]}

\bibitem[Love(2011)]{love11}
Love, J.J., 2011, ``Secular trends in storm-level geomagnetic activity'', {\it
  Ann. Geophys.\/}, {\bf 29}, 251--262.
  {\small[\href{http://dx.doi.org/10.5194/angeo-29-251-2011}{DOI}]},
  {\small[\href{http://adsabs.harvard.edu/abs/2011AnGeo..29..251L}{ADS}]}

\bibitem[Lukianova {\it et~al.\/}(2009)]{lukianova09}
Lukianova, R., Alekseev, G. and Mursula, K., 2009, ``Effects of station
  relocation in the $aa$ index'', {\it J. Geophys. Res.\/}, {\bf 114}, A02105.
  {\small[\href{http://dx.doi.org/10.1029/2008JA013824}{DOI}]},
  {\small[\href{http://adsabs.harvard.edu/abs/2009JGRA..11402105L}{ADS}]}

\bibitem[Masarik and Beer(1999)]{masarik99}
Masarik, J. and Beer, J., 1999, ``Simulation of particle fluxes and cosmogenic
  nuclide production in the Earth's atmosphere'', {\it J. Geophys. Res.\/},
  {\bf 104}(D10), 12,099--12,111.
  {\small[\href{http://dx.doi.org/10.1029/1998JD200091}{DOI}]},
  {\small[\href{http://adsabs.harvard.edu/abs/1999JGR...10412099M}{ADS}]}

\bibitem[Masarik and Beer(2009)]{masarik09}
Masarik, J. and Beer, J., 2009, ``An updated simulation of particle fluxes and
  cosmogenic nuclide production in the Earth's atmosphere'', {\it J. Geophys.
  Res.\/}, {\bf 114}(D13), D11103.
  {\small[\href{http://dx.doi.org/10.1029/2008JD010557}{DOI}]},
  {\small[\href{http://adsabs.harvard.edu/abs/2009JGRD..11411103M}{ADS}]}

\bibitem[Matsumoto {\it et~al.\/}(2004)]{matsumoto04}
Matsumoto, K., Sarmiento, J.L., Key, R.M., Aumont, O., Bullister, J.L.,
  Caldeira, K., Campin, J.-M., Doney, S.C., Drange, H., Dutay, J.-C., Follows,
  M., Gao, Y., Gnanadesikan, A., Gruber, N., Ishida, A., Joos, F., Lindsay, K.,
  Maier-Reimer, E., Marshall, J.C., Matear, R.J., Monfray, P., Mouchet, A.,
  Najjar, R., Plattner, G.-K., Schlitzer, R., Slater, R., Swathi, P.S.,
  Totterdell, I.J., Weirig, M.-F., Yamanaka, Y., Yool, A. and Orr, J.C., 2004,
  ``Evaluation of ocean carbon cycle models with data-based metrics'', {\it
  Geophys. Res. Lett.\/}, {\bf 31}, L07303.
  {\small[\href{http://dx.doi.org/10.1029/2003GL018970}{DOI}]},
  {\small[\href{http://adsabs.harvard.edu/abs/2004GeoRL..3107303M}{ADS}]}

\bibitem[Mayaud(1972)]{mayaud72}
Mayaud, P.-N., 1972, ``The $aa$ indices: A 100-year series characterizing the
  magnetic activity'', {\it J. Geophys. Res.\/}, {\bf 77}, 6870--6874.
  {\small[\href{http://dx.doi.org/10.1029/JA077i034p06870}{DOI}]},
  {\small[\href{http://adsabs.harvard.edu/abs/1972JGR....77.6870M}{ADS}]}

\bibitem[McCracken(2004)]{mccracken_JGR_04}
McCracken, K.G., 2004, ``Geomagnetic and atmospheric effects upon the
  cosmogenic $^{10}$Be observed in polar ice'', {\it J. Geophys. Res.\/}, {\bf
  109}(A18), A04101.
  {\small[\href{http://dx.doi.org/10.1029/2003JA010060}{DOI}]},
  {\small[\href{http://adsabs.harvard.edu/abs/2004JGRA..10904101M}{ADS}]}

\bibitem[McCracken(2007)]{mccracken07}
McCracken, K.G., 2007, ``Heliomagnetic field near Earth, 1428--2005'', {\it J.
  Geophys. Res.\/}, {\bf 112}(A11), A09106.
  {\small[\href{http://dx.doi.org/10.1029/2006JA012119}{DOI}]},
  {\small[\href{http://adsabs.harvard.edu/abs/2007JGRA..11209106M}{ADS}]}

\bibitem[McCracken and Beer(2007)]{mccracken_beer_07}
McCracken, K.G. and Beer, J., 2007, ``Long-term changes in the cosmic ray
  intensity at Earth, 1428-2005'', {\it J. Geophys. Res.\/}, {\bf 112}(A11),
  A10101. {\small[\href{http://dx.doi.org/10.1029/2006JA012117}{DOI}]},
  {\small[\href{http://adsabs.harvard.edu/abs/2007JGRA..11210101M}{ADS}]}

\bibitem[McCracken {\it et~al.\/}(2001)]{mccracken01}
McCracken, K.G., Dreschhoff, G.A.M., Zeller, E.J., Smart, D.F. and Shea, M.A.,
  2001, ``Solar cosmic ray events for the period 1561--1994: 1. Identification
  in polar ice, 1561--1950'', {\it J. Geophys. Res.\/}, {\bf 106},
  21,585--21,598. {\small[\href{http://dx.doi.org/10.1029/2000JA000237}{DOI}]},
  {\small[\href{http://adsabs.harvard.edu/abs/2001JGR...10621585M}{ADS}]}

\bibitem[McCracken {\it et~al.\/}(2004)]{mccracken04}
McCracken, K.G., McDonald, F.B., Beer, J., Raisbeck, G.M. and Yiou, F., 2004,
  ``A phenomenological study of the long-term cosmic ray modulation, 850--1958
  AD'', {\it J. Geophys. Res.\/}, {\bf 109}(A18), A12103.
  {\small[\href{http://dx.doi.org/10.1029/2004JA010685}{DOI}]},
  {\small[\href{http://adsabs.harvard.edu/abs/2004JGRA..10912103M}{ADS}]}

\bibitem[McDonald {\it et~al.\/}(2010)]{mcdonald10}
McDonald, F.B., Webber, W.R. and Reames, D.V., 2010, ``Unusual time histories
  of galactic and anomalous cosmic rays at 1 AU over the deep solar minimum of
  cycle 23/24'', {\it Geophys. Res. Lett.\/}, {\bf 37}, L18101.
  {\small[\href{http://dx.doi.org/10.1029/2010GL044218}{DOI}]},
  {\small[\href{http://adsabs.harvard.edu/abs/2010GeoRL..3718101M}{ADS}]}

\bibitem[McHargue and Damon(1991)]{mchargue91}
McHargue, L.R. and Damon, P.E., 1991, ``The Global Beryllium 10 Cycle'', {\it
  Rev. Geophys.\/}, {\bf 29}, 141--158.
  {\small[\href{http://dx.doi.org/10.1029/91RG00072}{DOI}]},
  {\small[\href{http://adsabs.harvard.edu/abs/1991RvGeo..29..141M}{ADS}]}

\bibitem[Mega {\it et~al.\/}(2003)]{mega03}
Mega, M.S., Allegrini, P., Grigolini, P., Latora, V., Palatella, L., Rapisarda,
  A. and Vinciguerra, S., 2003, ``Power-Law Time Distribution of Large
  Earthquakes'', {\it Phys. Rev. Lett.\/}, {\bf 90}, 188501.
  {\small[\href{http://dx.doi.org/10.1103/PhysRevLett.90.188501}{DOI}]},
  {\small[\href{http://adsabs.harvard.edu/abs/2003PhRvL..90r8501M}{ADS}]},
  {\small[\href{http://arxiv.org/abs/arXiv:cond-mat/0212529}{{arXiv:cond-mat/0%
212529}}]}

\bibitem[Mendoza(1997)]{mendoza97}
Mendoza, B., 1997, ``Geomagnetic activity and wind velocity during the Maunder
  minimum'', {\it Ann. Geophys.\/}, {\bf 15}, 397--402.
  {\small[\href{http://dx.doi.org/10.1007/s00585-997-0397-3}{DOI}]},
  {\small[\href{http://adsabs.harvard.edu/abs/1997AnGeo..15..397M}{ADS}]}. URL
  (accessed 10 October 2008):
  \newline\url{http://www.ann-geophys.net/15/397/1997/}

\bibitem[Mewaldt(2006)]{mewaldt06}
Mewaldt, R.A., 2006, ``Solar energetic particle composition, energy spectra,
  and space weather'', {\it Space Sci. Rev.\/}, {\bf 124}, 303--316.
  {\small[\href{http://dx.doi.org/10.1007/s11214-006-9091-0}{DOI}]},
  {\small[\href{http://adsabs.harvard.edu/abs/2006SSRv..124..303M}{ADS}]}

\bibitem[Mewaldt {\it et~al.\/}(2007)]{mewaldt07}
Mewaldt, R.A., Cohen, C.M.S., Mason, G.M., Haggerty, D.K. and Desai, M.I.,
  2007, ``Long-Term Fluences of Solar Energetic Particles from H to Fe'', {\it
  Space Sci. Rev.\/}, {\bf 130}, 323--328.
  {\small[\href{http://dx.doi.org/10.1007/s11214-007-9200-8}{DOI}]},
  {\small[\href{http://adsabs.harvard.edu/abs/2007SSRv..130..323M}{ADS}]}

\bibitem[Mewaldt {\it et~al.\/}(2012)]{mewaldt12}
Mewaldt, R.A., Looper, M.D., Cohen, C.M.S., Haggerty, D.K., Labrador, A.W.,
  Leske, R.A., Mason, G.M., Mazur, J.E. and von Rosenvinge, T.T., 2012,
  ``Energy Spectra, Composition, and Other Properties of Ground-Level Events
  During Solar Cycle 23'', {\it Space Sci. Rev.\/}, {\bf 171}, 97--120.
  {\small[\href{http://dx.doi.org/10.1007/s11214-012-9884-2}{DOI}]},
  {\small[\href{http://adsabs.harvard.edu/abs/2012SSRv..171...97M}{ADS}]}

\bibitem[Michel and Neumann(1998)]{michel98}
Michel, R. and Neumann, S., 1998, ``Interpretation of cosmogenic nuclides in
  meteorites on the basis of accelerator experiments and physical model
  calculations'', {\it Proc. Indian Acad. Sci. (Earth Planet. Sci.)\/}, {\bf
  107}, 441--457.
  {\small[\href{http://adsabs.harvard.edu/abs/1998PIASE.107..441M}{ADS}]}

\bibitem[Michel {\it et~al.\/}(1996)]{michel96}
Michel, R., Leya, I. and Borges, L., 1996, ``Production of cosmogenic nuclides
  in meteoroids: accelerator experiments and model calculations to decipher the
  cosmic ray record in extraterrestrial matter'', {\it Nucl. Instrum. Methods
  B\/}, {\bf 113}, 434--444.
  {\small[\href{http://dx.doi.org/10.1016/0168-583X(95)01345-8}{DOI}]},
  {\small[\href{http://adsabs.harvard.edu/abs/1996NIMPB.113..434M}{ADS}]}

\bibitem[Mikaloff~Fletcher {\it et~al.\/}(2006)]{mikaloff06}
Mikaloff~Fletcher, S.E., Gruber, N., Jacobson, A.R., Doney, S.C., Dutkiewicz,
  S., Gerber, M., Follows, M., Joos, F., Lindsay, K., Menemenlis, D., Mouchet,
  A., M{\"{u}}ller, S.A. and Sarmiento, J.L., 2006, ``Inverse estimates of
  anthropogenic CO$_{2}$ uptake, transport, and storage by the ocean'', {\it
  Global Biogeochem. Cycles\/}, {\bf 20}, GB2002.
  {\small[\href{http://dx.doi.org/10.1029/2005GB002530}{DOI}]},
  {\small[\href{http://adsabs.harvard.edu/abs/2006GBioC..20B2002M}{ADS}]}

\bibitem[Mininni {\it et~al.\/}(2000)]{mininni00}
Mininni, P.D., G{\'{o}}mez, D.O. and Mindlin, G.B., 2000, ``Stochastic
  Relaxation Oscillator Model for the Solar Cycle'', {\it Phys. Rev. Lett.\/},
  {\bf 85}, 5476--5479.
  {\small[\href{http://dx.doi.org/10.1103/PhysRevLett.85.5476}{DOI}]},
  {\small[\href{http://adsabs.harvard.edu/abs/2000PhRvL..85.5476M}{ADS}]}

\bibitem[Mininni {\it et~al.\/}(2001)]{mininni01}
Mininni, P.D., G{\'{o}}mez, D.O. and Mindlin, G.B., 2001, ``Simple Model of a
  Stochastically Excited Solar Dynamo'', {\it Solar Phys.\/}, {\bf 201},
  203--223. {\small[\href{http://dx.doi.org/10.1023/A:1017515709106}{DOI}]},
  {\small[\href{http://adsabs.harvard.edu/abs/2001SoPh..201..203M}{ADS}]}

\bibitem[Mininni {\it et~al.\/}(2002)]{mininni02}
Mininni, P.D., G{\'{o}}mez, D.O. and Mindlin, G.B., 2002, ``Biorthogonal
  Decomposition Techniques Unveil the Nature of the Irregularities Observed in
  the Solar Cycle'', {\it Phys. Rev. Lett.\/}, {\bf 89}(6), 061101.
  {\small[\href{http://adsabs.harvard.edu/abs/2002PhRvL..89f1101M}{ADS}]}

\bibitem[Miyahara {\it et~al.\/}(2004)]{miyahara04}
Miyahara, H., Masuda, K., Muraki, Y., Furuzawa, H., Menjo, H. and Nakamura, T.,
  2004, ``Cyclicity of Solar Activity During the Maunder Minimum Deduced from
  Radiocarbon Content'', {\it Solar Phys.\/}, {\bf 224}, 317--322.
  {\small[\href{http://dx.doi.org/10.1007/s11207-005-6501-5}{DOI}]},
  {\small[\href{http://adsabs.harvard.edu/abs/2004SoPh..224..317M}{ADS}]}

\bibitem[Miyahara {\it et~al.\/}(2006{\natexlab{a}})]{miyahara_JGR_06}
Miyahara, H., Masuda, K., Muraki, Y., Kitagawa, H. and Nakamura, T.,
  2006{\natexlab{a}}, ``Variation of solar cyclicity during the Spoerer
  Minimum'', {\it J. Geophys. Res.\/}, {\bf 111}(A10), A03103.
  {\small[\href{http://dx.doi.org/10.1029/2005JA011016}{DOI}]},
  {\small[\href{http://adsabs.harvard.edu/abs/2006JGRA..11103103M}{ADS}]}

\bibitem[Miyahara {\it et~al.\/}(2006{\natexlab{b}})]{miyahara06}
Miyahara, H., Sokoloff, D. and Usoskin, I.G., 2006{\natexlab{b}}, ``The Solar
  Cycle at the Maunder Minimum Epoch'', in {\it Advances in Geosciences, Vol.
  2: Solar Terrestrial (ST)\/}, (Eds.) Ip, W.-H., Duldig, M., pp. 1--20, World
  Scientific, Singapore; Hackensack, NJ.
  {\small[\href{http://adsabs.harvard.edu/abs/2006aogs....2....1M}{ADS}]}

\bibitem[Miyahara {\it et~al.\/}(2010)]{miyahara10}
Miyahara, H., Kitazawa, K., Nagaya, K., Yokoyama, Y., Matsuzaki, H., Masuda,
  K., Nakamura, T. and Muraki, Y., 2010, ``Is the Sun heading for another
  Maunder Minimum? Precursors of the grand solar minima'', {\it J. Cosmol.\/},
  {\bf 8}, 1970--1982.
  {\small[\href{http://adsabs.harvard.edu/abs/2010JCos....8.1970M}{ADS}]}

\bibitem[Miyake {\it et~al.\/}(2012)]{miyake12}
Miyake, F., Nagaya, K., Masuda, K. and Nakamura, T., 2012, ``A signature of
  cosmic-ray increase in ad 774--775 from tree rings in Japan'', {\it
  Nature\/}, {\bf 486}, 240--242.
  {\small[\href{http://dx.doi.org/10.1038/nature11123}{DOI}]},
  {\small[\href{http://adsabs.harvard.edu/abs/2011ACPD...1114003M}{ADS}]}

\bibitem[Moraal and Stoker(2010)]{moraal10}
Moraal, H. and Stoker, P.H., 2010, ``Long-term neutron monitor observations and
  the 2009 cosmic ray maximum'', {\it J. Geophys. Res.\/}, {\bf 115}, A12109.
  {\small[\href{http://dx.doi.org/10.1029/2010JA015413}{DOI}]},
  {\small[\href{http://adsabs.harvard.edu/abs/2010JGRA..11512109M}{ADS}]}

\bibitem[Morfill {\it et~al.\/}(1991)]{morfill91}
Morfill, G.E., Scheingraber, H., Voges, W. and Sonett, C.P., 1991, ``Sunspot
  number variations - Stochastic or chaotic'', in {\it The Sun in Time\/},
  (Eds.) Sonett, C.P., Giampapa, M.S., Matthews, M.S., pp. 30--58, University
  of Arizona Press, Tucson.
  {\small[\href{http://adsabs.harvard.edu/abs/1991suti.conf...30M}{ADS}]}

\bibitem[Moss {\it et~al.\/}(1992)]{moss92}
Moss, D., Brandenburg, A., Tavakol, R. and Tuominen, I., 1992, ``Stochastic
  effects in mean-field dynamos'', {\it Astron. Astrophys.\/}, {\bf 265},
  843--849.
  {\small[\href{http://adsabs.harvard.edu/abs/1992A&A...265..843M}{ADS}]}

\bibitem[Moss {\it et~al.\/}(2008)]{moss08}
Moss, D., Sokoloff, D., Usoskin, I.G. and Tutubalin, V., 2008, ``Solar Grand
  Minima and Random Fluctuations in Dynamo Parameters'', {\it Solar Phys.\/},
  {\bf 250}, 221--234.
  {\small[\href{http://dx.doi.org/10.1007/s11207-008-9202-z}{DOI}]},
  {\small[\href{http://adsabs.harvard.edu/abs/2008SoPh..250..221M}{ADS}]},
  {\small[\href{http://arxiv.org/abs/0806.3331}{{arXiv:0806.3331}}]}

\bibitem[Mossman(1989)]{mossman89}
Mossman, J.E., 1989, ``A comprehensive search for sunspots without the aid of a
  telescope, 1981--1982'', {\it Quart. J. R. Astron. Soc.\/}, {\bf 30}, 59--64.
  {\small[\href{http://adsabs.harvard.edu/abs/1989QJRAS..30...59M}{ADS}]}

\bibitem[Mundt {\it et~al.\/}(1991)]{mundt91}
Mundt, M.D., Maguire~II, W.B. and Chase, R.R.P., 1991, ``Chaos in the Sunspot
  Cycle: Analysis and Prediction'', {\it J. Geophys. Res.\/}, {\bf 96},
  1705--1716. {\small[\href{http://dx.doi.org/10.1029/90JA02150}{DOI}]},
  {\small[\href{http://adsabs.harvard.edu/abs/1991JGR....96.1705M}{ADS}]}

\bibitem[Mursula {\it et~al.\/}(2001)]{mursula01}
Mursula, K., Usoskin, I.G. and Kovaltsov, G.A., 2001, ``Persistent 22-year
  cycle in sunspot activity: Evidence for a relic solar magnetic field'', {\it
  Solar Phys.\/}, {\bf 198}, 51--56.
  {\small[\href{http://dx.doi.org/10.1023/A:1005218414790}{DOI}]},
  {\small[\href{http://adsabs.harvard.edu/abs/2001SoPh..198...51M}{ADS}]}

\bibitem[Mursula {\it et~al.\/}(2003)]{mursula03}
Mursula, K., Usoskin, I.G. and Kovaltsov, G.A., 2003, ``Reconstructing the
  long-term cosmic ray intensity: linear relations do not work'', {\it Ann.
  Geophys.\/}, {\bf 21}, 863--867.
  {\small[\href{http://adsabs.harvard.edu/abs/2003AnGeo..21..863M}{ADS}]}. URL
  (accessed 10 October 2008):
  \newline\url{http://www.ann-geophys.net/21/863/2003/}

\bibitem[Muscheler {\it et~al.\/}(2005)]{muscheler_Nat_05}
Muscheler, R., Joos, F., M{\"{u}}ller, S.A. and Snowball, I., 2005, ``Climate:
  How unusual is today's solar activity?'', {\it Nature\/}, {\bf 436}, E3--E4.
  {\small[\href{http://dx.doi.org/10.1038/nature04045}{DOI}]},
  {\small[\href{http://adsabs.harvard.edu/abs/2005Natur.436E...3M}{ADS}]}

\bibitem[Muscheler {\it et~al.\/}(2007)]{muscheler07}
Muscheler, R., Joos, F., Beer, J., M{\"{u}}ller, S.A., Vonmoos, M. and
  Snowball, I., 2007, ``Solar activity during the last 1000 yr inferred from
  radionuclide records'', {\it Quat. Sci. Rev.\/}, {\bf 26}, 82--97.
  {\small[\href{http://dx.doi.org/10.1016/j.quascirev.2006.07.012}{DOI}]},
  {\small[\href{http://adsabs.harvard.edu/abs/2007QSRv...26...82M}{ADS}]}

\bibitem[Nagaya {\it et~al.\/}(2012)]{nagaya12}
Nagaya, K., Kitazawa, K., Miyake, F., Masuda, K., Muraki, Y., Nakamura, T.,
  Miyahara, H. and Matsuzaki, H., 2012, ``Variation of the Schwabe Cycle Length
  During the Grand Solar Minimum in the 4th Century BC Deduced from Radiocarbon
  Content in Tree Rings'', {\it Solar Phys.\/}, {\bf 280}, 223--236.
  {\small[\href{http://dx.doi.org/10.1007/s11207-012-0045-2}{DOI}]},
  {\small[\href{http://adsabs.harvard.edu/abs/2012SoPh..280..223N}{ADS}]}

\bibitem[Nagovitsyn(1997)]{nagovistyn97}
Nagovitsyn, Y.A., 1997, ``A nonlinear mathematical model for the solar
  cyclicity and prospects for reconstructing the solar activity in the past'',
  {\it Astron. Lett.\/}, {\bf 23}, 742--748.
  {\small[\href{http://adsabs.harvard.edu/abs/1997AstL...23..742N}{ADS}]}

\bibitem[Nevanlinna(1995)]{nevanlinna95}
Nevanlinna, H., 1995, ``Auroral observations in Finland -- visual sightings
  during the 18th and 19th centuries'', {\it J. Geomag. Geoelectr.\/}, {\bf
  47}, 953--960

\bibitem[Nevanlinna(2004{\natexlab{a}})]{nevanlinna04}
Nevanlinna, H., 2004{\natexlab{a}}, ``Historical Space Climate Data from
  Finland: Compilation and Analysis'', {\it Solar Phys.\/}, {\bf 224},
  395--405. {\small[\href{http://dx.doi.org/10.1007/s11207-005-3749-8}{DOI}]},
  {\small[\href{http://adsabs.harvard.edu/abs/2004SoPh..224..395N}{ADS}]}

\bibitem[Nevanlinna(2004{\natexlab{b}})]{nevanlinna04a}
Nevanlinna, H., 2004{\natexlab{b}}, ``Results of the Helsinki magnetic
  observatory 1844--1912'', {\it Ann. Geophys.\/}, {\bf 22}, 1691--1704.
  {\small[\href{http://adsabs.harvard.edu/abs/2004AnGeo..22.1691N}{ADS}]}. URL
  (accessed 14 October 2008):
  \newline\url{http://www.ann-geophys.net/22/1691/2004/}

\bibitem[Newton(1928)]{newton928}
Newton, H.W., 1928, ``The Sun's cycle of activity'', {\it Quart. J. R.
  Meteorol. Soc.\/}, {\bf 54}(227), 161--174

\bibitem[Ney(1959)]{ney59}
Ney, E.P., 1959, ``Cosmic Radiation and the Weather'', {\it Nature\/}, {\bf
  183}, 451--452. {\small[\href{http://dx.doi.org/10.1038/183451a0}{DOI}]}

\bibitem[Nishiizumi {\it et~al.\/}(2009)]{nishiizumi09}
Nishiizumi, K., Arnold, J.R., Kohl, C.P., Caffee, M.W., Masarik, J. and Reedy,
  R.C., 2009, ``Solar cosmic ray records in lunar rock 64455'', {\it Geochim.
  Cosmochim. Acta\/}, {\bf 73}, 2163--2176.
  {\small[\href{http://adsabs.harvard.edu/abs/2009GeCoA..73.2163N}{ADS}]}

\bibitem[Nishizumi {\it et~al.\/}(1997)]{nishiizumi98}
Nishizumi, K., Caffee, M.W. and Arnold, J.R., 1997, ``$^{10}$Be from the active
  Sun'', in {\it Lunar and Planetary Science XXVIII\/}, Proceedings of the
  conference, March 17\,--\,21, 1997, 1027, Lunar and Planetary Institute,
  Houston.
  {\small[\href{http://adsabs.harvard.edu/abs/1997LPI....28.1027N}{ADS}]}

\bibitem[O'Brien(1979)]{obrien79}
O'Brien, K., 1979, ``Secular variations in the production of cosmogenic
  isotopes in the earth's atmosphere'', {\it J. Geophys. Res.\/}, {\bf 84},
  423--431. {\small[\href{http://dx.doi.org/10.1029/JA084iA02p00423}{DOI}]},
  {\small[\href{http://adsabs.harvard.edu/abs/1979JGR....84..423O}{ADS}]}

\bibitem[O'Brien and Burke(1973)]{obrien73}
O'Brien, K. and Burke, G.D.P., 1973, ``Calculated cosmic ray neutron monitor
  response to solar modulation of galactic cosmic rays'', {\it J. Geophys.
  Res.\/}, {\bf 78}, 3013--3019.
  {\small[\href{http://dx.doi.org/10.1029/JA078i016p03013}{DOI}]},
  {\small[\href{http://adsabs.harvard.edu/abs/1973JGR....78.3013O}{ADS}]}

\bibitem[Oeschger {\it et~al.\/}(1974)]{oeschger74}
Oeschger, H., Siegenthaler, U., Schotterer, U. and Gugelmann, A., 1974, ``A box
  diffusion model to study the carbon dioxide exchange in nature'', {\it
  Tellus\/}, {\bf 27}, 168--192

\bibitem[Ogurtsov(2004)]{ogurtsov04}
Ogurtsov, M.G., 2004, ``New Evidence for Long-Term Persistence in the Sun's
  Activity'', {\it Solar Phys.\/}, {\bf 220}, 93--105.
  {\small[\href{http://dx.doi.org/10.1023/B:sola.0000023439.59453.e5}{DOI}]},
  {\small[\href{http://adsabs.harvard.edu/abs/2004SoPh..220...93O}{ADS}]}

\bibitem[Ogurtsov {\it et~al.\/}(2002)]{ogurtsov02}
Ogurtsov, M.G., Nagovitsyn, Y.A., Kocharov, G.E. and Jungner, H., 2002,
  ``Long-period cycles of the Sun's activity recorded in direct solar data and
  proxies'', {\it Solar Phys.\/}, {\bf 211}, 371--394.
  {\small[\href{http://dx.doi.org/10.1023/A:1022411209257}{DOI}]},
  {\small[\href{http://adsabs.harvard.edu/abs/2002SoPh..211..371O}{ADS}]}

\bibitem[Oguti and Egeland(1995)]{oguti95}
Oguti, T. and Egeland, A., 1995, ``Auroral occurrence in Norwegian archives'',
  {\it J. Geomag. Geoelectr.\/}, {\bf 47}, 353--359

\bibitem[Oliver and Ballester(1996)]{oliver96}
Oliver, R. and Ballester, J.L., 1996, ``Rescaled Range Analysis of the
  Asymmetry of Solar Activity'', {\it Solar Phys.\/}, {\bf 169}, 215--224.
  {\small[\href{http://dx.doi.org/10.1007/BF00153842}{DOI}]},
  {\small[\href{http://adsabs.harvard.edu/abs/1996SoPh..169..215O}{ADS}]}

\bibitem[Oliver and Ballester(1998)]{oliver98}
Oliver, R. and Ballester, J.L., 1998, ``Is there memory in solar activity?'',
  {\it Phys. Rev. E\/}, {\bf 58}, 5650--5654.
  {\small[\href{http://dx.doi.org/10.1103/PhysRevE.58.5650}{DOI}]},
  {\small[\href{http://adsabs.harvard.edu/abs/1998PhRvE..58.5650O}{ADS}]}

\bibitem[Ossendrijver(2003)]{ossendrijver03}
Ossendrijver, M., 2003, ``The solar dynamo'', {\it Astron. Astrophys. Rev.\/},
  {\bf 11}, 287--367.
  {\small[\href{http://dx.doi.org/10.1007/s00159-003-0019-3}{DOI}]},
  {\small[\href{http://adsabs.harvard.edu/abs/2003A&ARv..11..287O}{ADS}]}

\bibitem[Ossendrijver(2000)]{ossendrijver00}
Ossendrijver, M.A.J.H., 2000, ``The dynamo effect of magnetic flux tubes'',
  {\it Astron. Astrophys.\/}, {\bf 359}, 1205--1210.
  {\small[\href{http://adsabs.harvard.edu/abs/2000A&A...359.1205O}{ADS}]}

\bibitem[Ossendrijver {\it et~al.\/}(1996)]{ossendrijver96}
Ossendrijver, M.A.J.H., Hoyng, P. and Schmitt, D., 1996, ``Stochastic
  excitation and memory of the solar dynamo'', {\it Astron. Astrophys.\/}, {\bf
  313}, 938--948.
  {\small[\href{http://adsabs.harvard.edu/abs/1996A&A...313..938O}{ADS}]}

\bibitem[Ostriakov and Usoskin(1990)]{ostryakov90}
Ostriakov, V.M. and Usoskin, I.G., 1990, ``On the dimension of solar
  attractor'', {\it Solar Phys.\/}, {\bf 127}, 405--412.
  {\small[\href{http://dx.doi.org/10.1007/BF00152177}{DOI}]},
  {\small[\href{http://adsabs.harvard.edu/abs/1990SoPh..127..405O}{ADS}]}

\bibitem[Ostryakov and Usoskin(1990)]{ostryakov90b}
Ostryakov, V.M. and Usoskin, I.G., 1990, ``Correlation dimensions of structured
  signals'', {\it Sov. Tech. Phys. Lett.\/}, {\bf 16}, 658--659

\bibitem[Owens {\it et~al.\/}(2012)]{owens12}
Owens, M.J., Usoskin, I.G. and Lockwood, M., 2012, ``Heliospheric modulation of
  galactic cosmic rays during grand solar minima: Past and future variations'',
  {\it Geophys. Res. Lett.\/}, {\bf 39}, L19102.
  {\small[\href{http://dx.doi.org/10.1029/2012GL053151}{DOI}]},
  {\small[\href{http://adsabs.harvard.edu/abs/2012GeoRL..3919102O}{ADS}]}

\bibitem[{\"{O}}zg{\"{u}}{\c{c}} {\it et~al.\/}(2003)]{ozguc03}
{\"{O}}zg{\"{u}}{\c{c}}, A., Ata{\c{c}}, T. and Ryb{\'{a}}k, J., 2003,
  ``Temporal variability of the flare index (1966--2001)'', {\it Solar
  Phys.\/}, {\bf 214}, 375--396.
  {\small[\href{http://dx.doi.org/10.1023/A:1024225802080}{DOI}]},
  {\small[\href{http://adsabs.harvard.edu/abs/2003SoPh..214..375O}{ADS}]}

\bibitem[Panchev and Tsekov(2007)]{panchev07}
Panchev, S. and Tsekov, M., 2007, ``Empirical evidences of persistence and
  dynamical chaos in solar terrestrial phenomena'', {\it J. Atmos. Sol.-Terr.
  Phys.\/}, {\bf 69}, 2391--2404.
  {\small[\href{http://dx.doi.org/10.1016/j.jastp.2007.07.011}{DOI}]},
  {\small[\href{http://adsabs.harvard.edu/abs/2007JASTP..69.2391P}{ADS}]}

\bibitem[Parker(1955)]{parker55}
Parker, E.N., 1955, ``Hydromagnetic Dynamo Models'', {\it Astrophys. J.\/},
  {\bf 122}, 293--314. {\small[\href{http://dx.doi.org/10.1086/146087}{DOI}]},
  {\small[\href{http://adsabs.harvard.edu/abs/1955ApJ...122..293P}{ADS}]}

\bibitem[Parker(1965)]{parker65}
Parker, E.N., 1965, ``The passage of energetic charged particles through
  interplanetary space'', {\it Planet. Space Sci.\/}, {\bf 13}, 9--49.
  {\small[\href{http://dx.doi.org/10.1016/0032-0633(65)90131-5}{DOI}]},
  {\small[\href{http://adsabs.harvard.edu/abs/1965P&SS...13....9P}{ADS}]}

\bibitem[Pedro {\it et~al.\/}(2011)]{pedro10}
Pedro, J.B., Heikkil{\"a}, U.E., Klekociuk, A., Smith, A.M., van Ommen, T.D.
  and Curran, M.A.J., 2011, ``Beryllium-10 transport to Antarctica: Results
  from seasonally resolved observations and modeling'', {\it J. Geophys.
  Res.\/}, {\bf 116}, D23120.
  {\small[\href{http://dx.doi.org/10.1029/2011JD016530}{DOI}]},
  {\small[\href{http://adsabs.harvard.edu/abs/2011JGRD..11623120P}{ADS}]}

\bibitem[Peristykh and Damon(1998)]{peristykh98}
Peristykh, A.N. and Damon, P.E., 1998, ``Modulation of atmospheric $^{14}$C
  concentration by the solar wind and irradiance components of the Hale and
  Schwabe solar cycles'', {\it Solar Phys.\/}, {\bf 177}, 343--343.
  {\small[\href{http://dx.doi.org/10.1023/A:1004982321191}{DOI}]},
  {\small[\href{http://adsabs.harvard.edu/abs/1998SoPh..177..343P}{ADS}]}

\bibitem[Peristykh and Damon(2003)]{peristykh03}
Peristykh, A.N. and Damon, P.E., 2003, ``Persistence of the Gleissberg 88-year
  solar cycle over the last $\sim$12,000 years: Evidence from cosmogenic
  isotopes'', {\it J. Geophys. Res.\/}, {\bf 108}, 1003.
  {\small[\href{http://dx.doi.org/10.1029/2002JA009390}{DOI}]},
  {\small[\href{http://adsabs.harvard.edu/abs/2003JGRA..108.1003P}{ADS}]}

\bibitem[Pesnell(2012)]{pesnell12}
Pesnell, W.D., 2012, ``Solar Cycle Predictions (Invited Review)'', {\it Solar
  Phys.\/}, {\bf 281}, 507--532.
  {\small[\href{http://dx.doi.org/10.1007/s11207-012-9997-5}{DOI}]},
  {\small[\href{http://adsabs.harvard.edu/abs/2012SoPh..281..507P}{ADS}]}

\bibitem[Petrovay(2010)]{petrovay10}
Petrovay, K., 2010, ``Solar Cycle Prediction'', {\it Living Rev. Solar
  Phys.\/}, {\bf 7}, 6.
  {\small[\href{http://adsabs.harvard.edu/abs/2010LRSP....7....6P}{ADS}]},
  {\small[\href{http://arxiv.org/abs/1012.5513}{{arXiv:1012.5513
  {\small[astro-ph.SR]}}}]}

\bibitem[Pipin {\it et~al.\/}(2012)]{pipin12}
Pipin, V.V., Sokoloff, D.D. and Usoskin, I.G., 2012, ``Variations of the solar
  cycle profile in a solar dynamo with fluctuating dynamo governing
  parameters'', {\it Astron. Astrophys.\/}, {\bf 542}, A26.
  {\small[\href{http://dx.doi.org/10.1051/0004-6361/201118733}{DOI}]},
  {\small[\href{http://adsabs.harvard.edu/abs/2012A&A...542A..26P}{ADS}]},
  {\small[\href{http://arxiv.org/abs/1112.6218}{{arXiv:1112.6218
  {\small[astro-ph.SR]}}}]}

\bibitem[Polygiannakis {\it et~al.\/}(2003)]{polygiannakis03}
Polygiannakis, J., Preka-Papadema, P. and Moussas, X., 2003, ``On signal-noise
  decomposition of time-series using the continuous wavelet transform:
  application to sunspot index'', {\it Mon. Not. R. Astron. Soc.\/}, {\bf 343},
  725--734.
  {\small[\href{http://dx.doi.org/10.1046/j.1365-8711.2003.06705.x}{DOI}]},
  {\small[\href{http://adsabs.harvard.edu/abs/2003MNRAS.343..725P}{ADS}]},
  {\small[\href{http://arxiv.org/abs/arXiv:physics/0301030}{{arXiv:physics/030%
1030}}]}

\bibitem[Price {\it et~al.\/}(1992)]{price92}
Price, C.P., Prichard, D. and Hogenson, E.A., 1992, ``Do the sunspot numbers
  form a `chaotic' set?'', {\it J. Geophys. Res.\/}, {\bf 97}, 19,113--19,120.
  {\small[\href{http://dx.doi.org/10.1029/92JA01459}{DOI}]},
  {\small[\href{http://adsabs.harvard.edu/abs/1992JGR....9719113P}{ADS}]}

\bibitem[Pulkkinen(2007)]{pulkkinenLR}
Pulkkinen, T., 2007, ``Space Weather: Terrestrial Perspective'', {\it Living
  Rev. Solar Phys.\/}, {\bf 4}, lrsp-2007-1. URL (accessed 25 February 2008):
  \newline\url{http://www.livingreviews.org/lrsp-2007-1}

\bibitem[Raisbeck {\it et~al.\/}(1981)]{raisbeck81}
Raisbeck, G.M., Yiou, F., Fruneau, M., Loiseaux, J.M., Lieuvin, M. and Ravel,
  J.C., 1981, ``Cosmogenic $^{10}$Be/$^{7}$Be as a probe of atmospheric
  transport processes'', {\it Geophys. Res. Lett.\/}, {\bf 8}, 1015--1018.
  {\small[\href{http://dx.doi.org/10.1029/GL008i009p01015}{DOI}]},
  {\small[\href{http://adsabs.harvard.edu/abs/1981GeoRL...8.1015R}{ADS}]}

\bibitem[Raisbeck {\it et~al.\/}(1990)]{raisbeck90}
Raisbeck, G.M., Yiou, F., Jouzel, J. and Petit, J.R., 1990, ``$^{10}$Be and
  $\delta$ $^{2}$H in Polar Ice Cores as a Probe of the Solar Variability's
  Influence on Climate'', {\it Philos. Trans. R. Soc. London, Ser. A\/}, {\bf
  330}, 463--469.
  {\small[\href{http://adsabs.harvard.edu/abs/1990RSPTA.330..463R}{ADS}]}

\bibitem[Rao {\it et~al.\/}(1994)]{rao94}
Rao, M.N., Garrison, D.H., Bogard, D.D. and Reedy, R.C., 1994, ``Determination
  of the flux and energy distribution of energetic solar protons in the past 2
  Myr using lunar rock 68815'', {\it Geochim. Cosmochim. Acta\/}, {\bf 58},
  4231--4245.
  {\small[\href{http://dx.doi.org/10.1016/0016-7037(94)90275-5}{DOI}]},
  {\small[\href{http://adsabs.harvard.edu/abs/1994GeCoA..58.4231R}{ADS}]}

\bibitem[Reedy(1996)]{reedy96}
Reedy, R.C., 1996, ``Constraints on Solar Particle Events from Comparisons of
  Recent Events and Million-Year Averages'', in {\it Solar Drivers of the
  Interplanetary and Terrestrial Disturbances\/}, Proceedings of the 16th
  International Workshop, National Solar Observatory/Sacramento Peak, Sunspot,
  New Mexico, USA, 16\,--\,20 October 1995, (Eds.) Balasubramaniam, K.S., Keil,
  S.L., Smartt, R.N., ASP Conference Series, 95, pp. 429--436, Astronomical
  Society of the Pacific, San Francisco.
  {\small[\href{http://adsabs.harvard.edu/abs/1996ASPC...95..429R}{ADS}]}

\bibitem[Reedy(2002)]{reedy02}
Reedy, R.C., 2002, ``Recent Solar Energetic Particles: Updates and Trends'', in
  {\it 33rd Lunar and Planetary Science Conference\/}, League City, TX, March
  11\,--\,15, 2002, 1938, Lunar and Planetary Institute, Houston.
  {\small[\href{http://adsabs.harvard.edu/abs/2002LPI....33.1938R}{ADS}]}. URL
  (accessed 5 March 2013):
  \newline\url{http://www.lpi.usra.edu/meetings/lpsc2002/pdf/1938.pdf}

\bibitem[Reedy(2012)]{reedy12}
Reedy, R.C., 2012, ``Update on Solar-Proton Fluxes During the Last Five Solar
  Activity Cycles'', in {\it 43rd Lunar and Planetary Science Conference\/},
  The Woodlands, TX, March 19\,--\,23, 2012, 43, 1285, Lunar and Planetary
  Institute, Houston.
  {\small[\href{http://adsabs.harvard.edu/abs/2012LPI....43.1285R}{ADS}]}. URL
  (accessed 5 March 2013):
  \newline\url{http://www.lpi.usra.edu/meetings/lpsc2012/pdf/1285.pdf}

\bibitem[Reeves {\it et~al.\/}(1992)]{reeves92}
Reeves, G.D., Cayton, T.E., Gary, S.P. and Belian, R.D., 1992, ``The great
  solar energetic particle events of 1989 observed from geosynchronous orbit'',
  {\it J. Geophys. Res.\/}, {\bf 97}, 6219--6226.
  {\small[\href{http://dx.doi.org/10.1029/91JA03102}{DOI}]},
  {\small[\href{http://adsabs.harvard.edu/abs/1992JGR....97.6219R}{ADS}]}

\bibitem[Reimer {\it et~al.\/}(2004)]{reimer04}
Reimer, P.J., Baillie, M.G.L., Bard, E., Bayliss, A., Beck, J.W., Bertrand,
  C.J.H., Blackwell, P.G., Buck, C.E., Burr, G.S., Cutler, K.B., Damon, P.E.,
  Edwards, R.L., Fairbanks, R.G., Friedrich, M., Guilderson, T.P., Hogg, A.G.,
  Hughen, K.A., Kromer, B., McCormac, G., Manning, S., Ramsey, C.B., Reimer,
  R.W., Remmele, S., Southon, J.R., Stuiver, M., Talamo, S., Taylor, F.W.,
  van~der Plicht, J. and Weyhenmeyer, C.E., 2004, ``IntCal04 terrestrial
  radiocarbon age calibration, 0--26 cal kyr BP'', {\it Radiocarbon\/}, {\bf
  46}(3), 1029--1058. URL (accessed 11 September 2012):
  \newline\url{https://journals.uair.arizona.edu/index.php/radiocarbon/article%
/view/4167/3592}

\bibitem[Reimer {\it et~al.\/}(2009)]{reimer09}
Reimer, P.J., Baillie, M.G.L., Bard, E., Bayliss, A., Beck, J.W., Blackwell,
  P.G., Ramsey, C.~Bronk, Buck, C.E., Burr, G.S., Edwards, R.L., Friedrich, M.,
  Grootes, P.M., Guilderson, T.P., Hajdas, I., Heaton, T.J., Hogg, A.G.,
  Hughen, K.A., Kaiser, K.F., Kromer, B., McCormac, F.G., Manning, S.W.,
  Reimer, R.W., Richards, D.A., Southon, J.R., Talamo, S., Turney, C.S.M.,
  van~der Plicht, J. and Weyhenmeye, C.E., 2009, ``IntCal09 and Marine09
  Radiocarbon Age Calibration Curves, 0--50,000 Years cal BP'', {\it
  Radiocarbon\/}, {\bf 51}(4), 1111--1150. URL (accessed 11 September 2012):
  \newline\url{https://journals.uair.arizona.edu/index.php/radiocarbon/article%
/view/3569}

\bibitem[Ribes and Nesme-Ribes(1993)]{ribes93}
Ribes, J.C. and Nesme-Ribes, E., 1993, ``The solar sunspot cycle in the Maunder
  minimum AD1645 to AD1715'', {\it Astron. Astrophys.\/}, {\bf 276}, 549--563.
  {\small[\href{http://adsabs.harvard.edu/abs/1993A&A...276..549R}{ADS}]}

\bibitem[Rigozo {\it et~al.\/}(2001)]{rigozo01}
Rigozo, N.R., Echer, E., Vieira, L.E.A. and Nordemann, D.J.R., 2001,
  ``Reconstruction of Wolf Sunspot Numbers on the Basis of Spectral
  Characteristics and Estimates of Associated Radio Flux and Solar Wind
  Parameters for the Last Millennium'', {\it Solar Phys.\/}, {\bf 203},
  179--191. {\small[\href{http://dx.doi.org/10.1023/A:1012745612022}{DOI}]},
  {\small[\href{http://adsabs.harvard.edu/abs/2001SoPh..203..179R}{ADS}]}

\bibitem[Rozanov {\it et~al.\/}(2012)]{rozanov12}
Rozanov, E., Calisto, M., Egorova, T., Peter, T. and Schmutz, W., 2012,
  ``Influence of the Precipitating Energetic Particles on Atmospheric Chemistry
  and Climate'', {\it Surv. Geophys.\/}, {\bf 33}, 483--501.
  {\small[\href{http://dx.doi.org/10.1007/s10712-012-9192-0}{DOI}]},
  {\small[\href{http://adsabs.harvard.edu/abs/2012SGeo...33..483R}{ADS}]}

\bibitem[Rozelot(1994)]{rozelot94}
Rozelot, J.P., 1994, ``On the stability of the 11-year solar cycle period (and
  a few others)'', {\it Solar Phys.\/}, {\bf 149}, 149--154.
  {\small[\href{http://dx.doi.org/10.1007/BF00645186}{DOI}]},
  {\small[\href{http://adsabs.harvard.edu/abs/1994SoPh..149..149R}{ADS}]}

\bibitem[Rozelot(1995)]{rozelot95}
Rozelot, J.P., 1995, ``On the chaotic behaviour of the solar activity'', {\it
  Astron. Astrophys.\/}, {\bf 297}, L45--L48.
  {\small[\href{http://adsabs.harvard.edu/abs/1995A&A...297L..45R}{ADS}]}

\bibitem[Ruth {\it et~al.\/}(2007)]{ruth07}
Ruth, U., Barnola, J.-M., Beer, J., Bigler, M., Blunier, T., Castellano, E.,
  Fischer, H., Fundel, F., Huybrechts, P., Kaufmann, P., Kipfstuhl, S.,
  Lambrecht, A., Morganti, A., Oerter, H., Parrenin, F., Rybak, O., Severi, M.,
  Udisti, R., Wilhelms, F. and Wolff, E.W., 2007, ```EDML1': a chronology for
  the EPICA deep ice core from Dronning Maud Land, Antarctica, over the last
  150 000 years'', {\it Clim. Past\/}, {\bf 3}, 475--484.
  {\small[\href{http://dx.doi.org/10.5194/cp-3-475-2007}{DOI}]},
  {\small[\href{http://adsabs.harvard.edu/abs/2007CliPa...3..475R}{ADS}]}. URL
  (accessed 11 September 2012):
  \newline\url{http://www.clim-past.net/3/475/2007/}

\bibitem[Ruzmaikin {\it et~al.\/}(1994)]{ruzmaikin94}
Ruzmaikin, A., Feynman, J. and Robinson, P., 1994, ``Long-term persistence of
  solar activity'', {\it Solar Phys.\/}, {\bf 149}, 395--403.
  {\small[\href{http://dx.doi.org/10.1007/BF00690625}{DOI}]},
  {\small[\href{http://adsabs.harvard.edu/abs/1994SoPh..149..395R}{ADS}]}

\bibitem[Ruzmaikin(1981)]{ruzmaikin81}
Ruzmaikin, A.A., 1981, ``The solar cycle as a strange attractor'', {\it
  Comments Astrophys.\/}, {\bf 9}, 85--93.
  {\small[\href{http://adsabs.harvard.edu/abs/1981ComAp...9...85R}{ADS}]}

\bibitem[Rybansk{\'{y}} {\it et~al.\/}(2005)]{rybansky05}
Rybansk{\'{y}}, M., Ru{\v{s}}in, V., Minarovjech, M., Klocok, L. and Cliver,
  E.W., 2005, ``Reexamination of the coronal index of solar activity'', {\it J.
  Geophys. Res.\/}, {\bf 110}(A9), 8106.
  {\small[\href{http://dx.doi.org/10.1029/2005JA011146}{DOI}]},
  {\small[\href{http://adsabs.harvard.edu/abs/2005JGRA..11008106R}{ADS}]}

\bibitem[Sabine {\it et~al.\/}(2004)]{sabine04}
Sabine, C.L., Feely, R.A., Gruber, N., Key, R.M., Lee, K., Bullister, J.L.,
  Wanninkhof, R., Wong, C.S., Wallace, D.W.R., Tilbrook, B., Millero, F.J.,
  Peng, T.-H., Kozyr, A., Ono, T. and Rios, A.F., 2004, ``The Oceanic Sink for
  Anthropogenic CO$_{2}$'', {\it Science\/}, {\bf 305}, 367--371.
  {\small[\href{http://dx.doi.org/10.1126/science.1097403}{DOI}]},
  {\small[\href{http://adsabs.harvard.edu/abs/2004Sci...305..367S}{ADS}]}

\bibitem[Salakhutdinova(1999)]{salakhutdinova99}
Salakhutdinova, I.I., 1999, ``Identifying the quasi-regular and stochastic
  components of solar cyclicity and their properties'', {\it Solar Phys.\/},
  {\bf 188}, 377--396.
  {\small[\href{http://dx.doi.org/10.1023/A:1005265229175}{DOI}]},
  {\small[\href{http://adsabs.harvard.edu/abs/1999SoPh..188..377S}{ADS}]}

\bibitem[Scherer and Fichtner(2004)]{scherer_AA04}
Scherer, K. and Fichtner, H., 2004, ``Constraints on the heliospheric magnetic
  field variation during the Maunder Minimum from cosmic ray modulation
  modelling'', {\it Astron. Astrophys.\/}, {\bf 413}, L11--L14.
  {\small[\href{http://dx.doi.org/10.1051/0004-6361:20034636}{DOI}]},
  {\small[\href{http://adsabs.harvard.edu/abs/2004A&A...413L..11S}{ADS}]}

\bibitem[Scherer {\it et~al.\/}(2004)]{scherer04}
Scherer, K., Fahr, H.-J., Fichtner, H. and Heber, B., 2004, ``Long-Term
  Modulation of Cosmic Rays in the Heliosphere and its Influence at Earth'',
  {\it Solar Phys.\/}, {\bf 224}, 305--316.
  {\small[\href{http://dx.doi.org/10.1007/s11207-005-5687-x}{DOI}]},
  {\small[\href{http://adsabs.harvard.edu/abs/2004SoPh..224..305S}{ADS}]}

\bibitem[Schlamminger(1990)]{schlamminger90}
Schlamminger, L., 1990, ``Aurora borealis during the Maunder minimum'', {\it
  Mon. Not. R. Astron. Soc.\/}, {\bf 247}, 67--69.
  {\small[\href{http://adsabs.harvard.edu/abs/1990MNRAS.247...67S}{ADS}]}

\bibitem[Schmalz and Stix(1991)]{schmaltz91}
Schmalz, S. and Stix, M., 1991, ``An $\alpha\Omega$ dynamo with order and
  chaos'', {\it Astron. Astrophys.\/}, {\bf 245}, 654--661.
  {\small[\href{http://adsabs.harvard.edu/abs/1991A&A...245..654S}{ADS}]}

\bibitem[Schmidt {\it et~al.\/}(2012)]{schmidt12}
Schmidt, G.A., Jungclaus, J.H., Ammann, C.M., Bard, E., Braconnot, P., Crowley,
  T.J., Delaygue, G., Joos, F., Krivova, N.A., Muscheler, R., Otto-Bliesner,
  B.L., Pongratz, J., Shindell, D.T., Solanki, S.K., Steinhilber, F. and
  Vieira, L.E.A., 2012, ``Climate forcing reconstructions for use in PMIP
  simulations of the Last Millennium (v1.1)'', {\it Geosci. Model Dev.\/}, {\bf
  5}, 185--191. {\small[\href{http://dx.doi.org/10.5194/gmd-5-185-2012}{DOI}]},
  {\small[\href{http://adsabs.harvard.edu/abs/2012GMD.....5..185S}{ADS}]}. URL
  (accessed 11 September 2012):
  \newline\url{http://www.geosci-model-dev.net/5/185/2012/}

\bibitem[Schmitt {\it et~al.\/}(1996)]{schmitt96}
Schmitt, D., Sch{\"{u}}ssler, M. and Ferriz-Mas, A., 1996, ``Intermittent solar
  activity by an on-off dynamo'', {\it Astron. Astrophys.\/}, {\bf 311},
  L1--L4.
  {\small[\href{http://adsabs.harvard.edu/abs/1996A&A...311L...1S}{ADS}]}

\bibitem[Schove(1955)]{schove55}
Schove, D.J., 1955, ``The Sunspot Cycle, 649 B.C. to A.D. 2000'', {\it J.
  Geophys. Res.\/}, {\bf 60}, 127--146.
  {\small[\href{http://dx.doi.org/10.1029/JZ060i002p00127}{DOI}]},
  {\small[\href{http://adsabs.harvard.edu/abs/1955JGR....60..127S}{ADS}]}

\bibitem[Schove(1979)]{schove79}
Schove, D.J., 1979, ``Sunspot Turning-Points and Aurorae Since A.D. 1510'',
  {\it Solar Phys.\/}, {\bf 63}, 423--432.
  {\small[\href{http://dx.doi.org/10.1007/BF00174546}{DOI}]},
  {\small[\href{http://adsabs.harvard.edu/abs/1979SoPh...63..423S}{ADS}]}

\bibitem[Schove(1983)]{schove83}
Schove, D.J., 1983, ``Sunspot, auroral, radiocarbon and climatic fluctuations
  since 7000 BC'', {\it Ann. Geophys.\/}, {\bf 1}, 391--396.
  {\small[\href{http://adsabs.harvard.edu/abs/1983AnGeo...1..391S}{ADS}]}

\bibitem[Schrijver {\it et~al.\/}(2012)]{schrijver12}
Schrijver, C.J., Beer, J., Baltensperger, U., Cliver, E.W., G\"udel, M, Hudson,
  H.S., McCracken, K.G., Osten, R.A., Peter, T., Soderblom, D.R., Usoskin, I.G.
  and Wolff, E.W., 2012, ``Estimating the frequency of extremely energetic
  solar events, based on solar, stellar, lunar, and terrestrial records'', {\it
  J. Geophys. Res.\/}, {\bf 117}, A08103.
  {\small[\href{http://dx.doi.org/10.1029/2012JA017706}{DOI}]},
  {\small[\href{http://adsabs.harvard.edu/abs/2012JGRA..11708103S}{ADS}]}

\bibitem[Schr{\"{o}}der(1992)]{schroder92}
Schr{\"{o}}der, W., 1992, ``On the existence of the 11-year cycle in solar and
  auroral activity before and after the so-called Maunder minimum'', {\it J.
  Geomag. Geoelectr.\/}, {\bf 44}, 119--128

\bibitem[Sch{\"u}pbach {\it et~al.\/}(2011)]{schupbach11}
Sch{\"u}pbach, S., Federer, U., Bigler, M., Fischer, H. and Stocker, T.F.,
  2011, ``A refined TALDICE-1a age scale from 55 to 112 ka before present for
  the Talos Dome ice core based on high-resolution methane measurements'', {\it
  Clim. Past\/}, {\bf 7}, 1001--1009.
  {\small[\href{http://dx.doi.org/10.5194/cp-7-1001-2011}{DOI}]},
  {\small[\href{http://adsabs.harvard.edu/abs/2011CliPa...7.1001S}{ADS}]}. URL
  (accessed 11 September 2012):
  \newline\url{http://www.clim-past.net/7/1001/2011/}

\bibitem[Sch{\"{u}}ssler {\it et~al.\/}(1994)]{schussler94}
Sch{\"{u}}ssler, M., Caligari, P., Ferriz-Mas, A. and Moreno-Insertis, F.,
  1994, ``Instability and eruption of magnetic flux tubes in the solar
  convection zone'', {\it Astron. Astrophys.\/}, {\bf 281}, L69--L72.
  {\small[\href{http://adsabs.harvard.edu/abs/1994A&A...281L..69S}{ADS}]}

\bibitem[Sch{\"{u}}ssler {\it et~al.\/}(1997)]{schussler97}
Sch{\"{u}}ssler, M., Schmitt, D. and Ferriz-Mas, A., 1997, ``Long-term
  Variation of Solar Activity by a Dynamo Based on Magnetic Flux Tubes'', in
  {\it 1st Advances in Solar Physics Euroconference: Advances in the Physics of
  Sunspots\/}, (Eds.) Schmieder, B., del Toro~Iniesta, J.C., V{\'{a}}zquez, M.,
  ASP Conference Series, 118, pp. 39--44, Astronomical Society of the Pacific,
  San Francisco.
  {\small[\href{http://adsabs.harvard.edu/abs/1997ASPC..118...39S}{ADS}]}

\bibitem[Scuderi(1990)]{scuderi90}
Scuderi, L.A., 1990, ``Oriental sunspot observations and volcanism'', {\it
  Quart. J. R. Astron. Soc.\/}, {\bf 31}, 109--120.
  {\small[\href{http://adsabs.harvard.edu/abs/1990QJRAS..31..109S}{ADS}]}

\bibitem[Sello(2000)]{sello00}
Sello, S., 2000, ``Wavelet entropy as a measure of solar cycle complexity'',
  {\it Astron. Astrophys.\/}, {\bf 363}, 311--315.
  {\small[\href{http://adsabs.harvard.edu/abs/2000A&A...363..311S}{ADS}]},
  {\small[\href{http://arxiv.org/abs/arXiv:astro-ph/0005334}{{arXiv:astro-ph/0%
005334}}]}

\bibitem[Semeniuk {\it et~al.\/}(2011)]{semeniuk11}
Semeniuk, K., Fomichev, V.I., McConnell, J.C., Fu, C., Melo, S.M.L. and
  Usoskin, I.G., 2011, ``Middle atmosphere response to the solar cycle in
  irradiance and ionizing particle precipitation'', {\it Atmos. Chem. Phys.\/},
  {\bf 11}, 5045--5077.
  {\small[\href{http://dx.doi.org/10.5194/acp-11-5045-2011}{DOI}]},
  {\small[\href{http://adsabs.harvard.edu/abs/2011ACP....11.5045S}{ADS}]}. URL
  (accessed 11 September 2012):
  \newline\url{http://www.atmos-chem-phys.net/11/5045/2011/}

\bibitem[Serre and Nesme-Ribes(2000)]{serre00}
Serre, T. and Nesme-Ribes, E., 2000, ``Nonlinear analysis of solar cycles'',
  {\it Astron. Astrophys.\/}, {\bf 360}, 319--330.
  {\small[\href{http://adsabs.harvard.edu/abs/2000A&A...360..319S}{ADS}]}

\bibitem[Shapiro {\it et~al.\/}(2011)]{shapiro11}
Shapiro, A.I., Schmutz, W., Rozanov, E., Schoell, M., Haberreiter, M., Shapiro,
  A.V. and Nyeki, S., 2011, ``A new approach to the long-term reconstruction of
  the solar irradiance leads to large historical solar forcing'', {\it Astron.
  Astrophys.\/}, {\bf 529}, A67.
  {\small[\href{http://dx.doi.org/10.1051/0004-6361/201016173}{DOI}]},
  {\small[\href{http://adsabs.harvard.edu/abs/2011A&A...529A..67S}{ADS}]},
  {\small[\href{http://arxiv.org/abs/1102.4763}{{arXiv:1102.4763
  {\small[astro-ph.SR]}}}]}

\bibitem[Shea and Smart(1990)]{shea90}
Shea, M.A. and Smart, D.F., 1990, ``A summary of major solar proton events'',
  {\it Solar Phys.\/}, {\bf 127}, 297--320.
  {\small[\href{http://dx.doi.org/10.1007/BF00152170}{DOI}]},
  {\small[\href{http://adsabs.harvard.edu/abs/1990SoPh..127..297S}{ADS}]}

\bibitem[Shea and Smart(2002)]{shea02}
Shea, M.A. and Smart, D.F., 2002, ``Solar proton event patterns: the rising
  portion of five solar cycles'', {\it Adv. Space Res.\/}, {\bf 29}, 325--330.
  {\small[\href{http://dx.doi.org/10.1016/S0273-1177(01)00592-0}{DOI}]},
  {\small[\href{http://adsabs.harvard.edu/abs/2002AdSpR..29..325S}{ADS}]}

\bibitem[Shea {\it et~al.\/}(2006)]{shea06}
Shea, M.A., Smart, D.F., McCracken, K.G., Dreschhoff, G.A.M. and Spence, H.E.,
  2006, ``Solar proton events for 450 years: The Carrington event in
  perspective'', {\it Adv. Space Res.\/}, {\bf 38}, 232--238.
  {\small[\href{http://dx.doi.org/10.1016/j.asr.2005.02.100}{DOI}]},
  {\small[\href{http://adsabs.harvard.edu/abs/2006AdSpR..38..232S}{ADS}]}

\bibitem[Shindell {\it et~al.\/}(1999)]{shindell99}
Shindell, D., Rind, D., Balachandran, N., Lean, J.L. and Lonergan, P., 1999,
  ``Solar Cycle Variability, Ozone, and Climate'', {\it Science\/}, {\bf 284},
  305--308.
  {\small[\href{http://dx.doi.org/10.1126/science.284.5412.305}{DOI}]},
  {\small[\href{http://adsabs.harvard.edu/abs/1999Sci...284..305S}{ADS}]}

\bibitem[Siegenthaler {\it et~al.\/}(1980)]{siegenthaler80}
Siegenthaler, U., Heimann, M. and Oeschger, H., 1980, ``$^{14}$C variations
  caused by changes in the global carbon cycle'', {\it Radiocarbon\/}, {\bf
  22}, 177--191

\bibitem[Silverman(1992)]{silverman92}
Silverman, S.M., 1992, ``Secular variation of the aurora for the past 500
  years'', {\it Rev. Geophys.\/}, {\bf 30}, 333--351.
  {\small[\href{http://dx.doi.org/10.1029/92RG01571}{DOI}]}

\bibitem[Silverman(2006)]{silverman06}
Silverman, S.M., 2006, ``Comparison of the aurora of September 1/2, 1859 with
  other great auroras'', {\it Adv. Space Res.\/}, {\bf 38}, 136--144.
  {\small[\href{http://dx.doi.org/10.1016/j.asr.2005.03.157}{DOI}]}

\bibitem[Siscoe(1980)]{siscoe80}
Siscoe, G.L., 1980, ``Evidence in the auroral record for secular solar
  variability'', {\it Rev. Geophys. Space Phys.\/}, {\bf 18}, 647--658.
  {\small[\href{http://dx.doi.org/10.1029/RG018i003p00647}{DOI}]},
  {\small[\href{http://adsabs.harvard.edu/abs/1980RvGSP..18..647S}{ADS}]}

\bibitem[Siscoe and Verosub(1983)]{siscoe83}
Siscoe, G.L. and Verosub, K.L., 1983, ``High medieval auroral incidence over
  China and Japan: implications for the medieval site of the geomagnetic
  pole'', {\it Geophys. Res. Lett.\/}, {\bf 10}, 345--348.
  {\small[\href{http://dx.doi.org/10.1029/GL010i004p00345}{DOI}]},
  {\small[\href{http://adsabs.harvard.edu/abs/1983GeoRL..10..345S}{ADS}]}

\bibitem[Smart and Shea(2002)]{smart02}
Smart, D.F. and Shea, M.A., 2002, ``A review of solar proton events during the
  22nd solar cycle'', {\it Adv. Space Res.\/}, {\bf 30}, 1033--1044.
  {\small[\href{http://dx.doi.org/10.1016/S0273-1177(02)00497-0}{DOI}]},
  {\small[\href{http://adsabs.harvard.edu/abs/2002AdSpR..30.1033S}{ADS}]}

\bibitem[Snow {\it et~al.\/}(2005)]{snow05}
Snow, M., McClintock, W.E., Woods, T.N., White, O.R., Harder, J.W. and Rottman,
  G., 2005, ``The Mg II Index from SORCE'', {\it Solar Phys.\/}, {\bf 230},
  325--344. {\small[\href{http://dx.doi.org/10.1007/s11207-005-6879-0}{DOI}]},
  {\small[\href{http://adsabs.harvard.edu/abs/2005SoPh..230..325S}{ADS}]}

\bibitem[Snowball and Muscheler(2007)]{snowball07}
Snowball, I. and Muscheler, R., 2007, ``Palaeomagnetic intensity data: an
  Achilles heel of solar activity reconstructions'', {\it Holocene\/}, {\bf
  17}, 851--859.
  {\small[\href{http://dx.doi.org/10.1177/0959683607080531}{DOI}]}

\bibitem[Sokoloff(2004)]{sokoloff04}
Sokoloff, D., 2004, ``The Maunder Minimum and the Solar Dynamo'', {\it Solar
  Phys.\/}, {\bf 224}, 145--152.
  {\small[\href{http://dx.doi.org/10.1007/s11207-005-4176-6}{DOI}]},
  {\small[\href{http://adsabs.harvard.edu/abs/2004SoPh..224..145S}{ADS}]}

\bibitem[Sokoloff and Nesme-Ribes(1994)]{sokoloff94}
Sokoloff, D. and Nesme-Ribes, E., 1994, ``The Maunder minimum: A mixed-parity
  dynamo mode?'', {\it Astron. Astrophys.\/}, {\bf 288}, 293--298.
  {\small[\href{http://adsabs.harvard.edu/abs/1994A&A...288..293S}{ADS}]}

\bibitem[Solanki and Krivova(2004)]{solanki04a}
Solanki, S.K. and Krivova, N.A., 2004, ``Solar Irradiance Variations: From
  Current Measurements to Long-Term Estimates'', {\it Solar Phys.\/}, {\bf
  224}, 197--208.
  {\small[\href{http://dx.doi.org/10.1007/s11207-005-6499-8}{DOI}]},
  {\small[\href{http://adsabs.harvard.edu/abs/2004SoPh..224..197S}{ADS}]}

\bibitem[Solanki {\it et~al.\/}(2000)]{solanki00}
Solanki, S.K., Sch{\"{u}}ssler, M. and Fligge, M., 2000, ``Evolution of the
  Sun's large-scale magnetic field since the Maunder minimum'', {\it Nature\/},
  {\bf 408}, 445--447.
  {\small[\href{http://dx.doi.org/10.1038/35044027}{DOI}]},
  {\small[\href{http://adsabs.harvard.edu/abs/2000Natur.408..445S}{ADS}]}

\bibitem[Solanki {\it et~al.\/}(2002)]{solanki02}
Solanki, S.K., Sch{\"{u}}ssler, M. and Fligge, M., 2002, ``Secular variation of
  the Sun's magnetic flux'', {\it Astron. Astrophys.\/}, {\bf 383}, 706--712.
  {\small[\href{http://dx.doi.org/10.1051/0004-6361:20011790}{DOI}]},
  {\small[\href{http://adsabs.harvard.edu/abs/2002A&A...383..706S}{ADS}]}

\bibitem[Solanki {\it et~al.\/}(2004)]{solanki_Nat_04}
Solanki, S.K., Usoskin, I.G., Kromer, B., Sch{\"{u}}ssler, M. and Beer, J.,
  2004, ``Unusual activity of the Sun during recent decades compared to the
  previous 11,000 years'', {\it Nature\/}, {\bf 431}, 1084--1087.
  {\small[\href{http://dx.doi.org/10.1038/nature02995}{DOI}]},
  {\small[\href{http://adsabs.harvard.edu/abs/2004Natur.431.1084S}{ADS}]}

\bibitem[Solanki {\it et~al.\/}(2005)]{solanki_Nat_05}
Solanki, S.K., Usoskin, I.G., Kromer, B., Sch{\"{u}}ssler, M. and Beer, J.,
  2005, ``Climate: How unusual is today's solar activity? Reply'', {\it
  Nature\/}, {\bf 436}, E4--E5.
  {\small[\href{http://dx.doi.org/10.1038/nature04046}{DOI}]},
  {\small[\href{http://adsabs.harvard.edu/abs/2005Natur.436E...4S}{ADS}]}

\bibitem[Sonett(1983)]{sonett83}
Sonett, C.P., 1983, ``The great solar anomaly ca. 1780-1800: An error in
  compiling the record?'', {\it J. Geophys. Res.\/}, {\bf 88}, 3225--3228.
  {\small[\href{http://dx.doi.org/10.1029/JA088iA04p03225}{DOI}]},
  {\small[\href{http://adsabs.harvard.edu/abs/1983JGR....88.3225S}{ADS}]}

\bibitem[Sonett and Finney(1990)]{sonett90}
Sonett, C.P. and Finney, S.A., 1990, ``The Spectrum of Radiocarbon'', {\it
  Philos. Trans. R. Soc. London, Ser. A\/}, {\bf 330}, 413--425.
  {\small[\href{http://adsabs.harvard.edu/abs/1990RSPTA.330..413S}{ADS}]}

\bibitem[Soon and Yaskell(2003)]{soon03}
Soon, W.W.-H. and Yaskell, S.H., 2003, {\it The Maunder Minimum and the
  Variable Sun-Earth Connection\/}, World Scientific, Singapore; River Edge,
  NJ. {\small[\href{http://books.google.com/books?id=HfdG-HPiBdMC}{Google
  Books}]}

\bibitem[Steig {\it et~al.\/}(1996)]{steig96}
Steig, E.J., Polissar, P.J., Stuiver, M., Grootes, P.M. and Finkel, R.C., 1996,
  ``Large amplitude solar modulation cycles of $^{10}$Be in Antarctica:
  Implications for atmospheric mixing processes and interpretation of the ice
  core record'', {\it Geophys. Res. Lett.\/}, {\bf 23}, 523--526.
  {\small[\href{http://dx.doi.org/10.1029/96GL00255}{DOI}]},
  {\small[\href{http://adsabs.harvard.edu/abs/1996GeoRL..23..523S}{ADS}]}

\bibitem[Steinhilber {\it et~al.\/}(2008)]{steinhilber08}
Steinhilber, F., Abreu, J.A. and Beer, J., 2008, ``Solar modulation during the
  Holocene'', {\it Astrophys. Space Sci. Trans.\/}, {\bf 4}, 1--6.
  {\small[\href{http://adsabs.harvard.edu/abs/2008ASTRA...4....1S}{ADS}]}

\bibitem[Steinhilber {\it et~al.\/}(2009)]{steinhilber09}
Steinhilber, F., Beer, J. and Fr{\"o}hlich, C., 2009, ``Total solar irradiance
  during the Holocene'', {\it Geophys. Res. Lett.\/}, {\bf 36}, L19704.
  {\small[\href{http://dx.doi.org/10.1029/2009GL040142}{DOI}]},
  {\small[\href{http://adsabs.harvard.edu/abs/2009GeoRL..3619704S}{ADS}]}

\bibitem[Steinhilber {\it et~al.\/}(2010)]{steinhilber10}
Steinhilber, F., Abreu, J.A., Beer, J. and McCracken, K.G., 2010,
  ``Interplanetary magnetic field during the past 9300 years inferred from
  cosmogenic radionuclides'', {\it J. Geophys. Res.\/}, {\bf 115}, A01104.
  {\small[\href{http://dx.doi.org/10.1029/2009JA014193}{DOI}]},
  {\small[\href{http://adsabs.harvard.edu/abs/2010JGRA..11501104S}{ADS}]}

\bibitem[Steinhilber {\it et~al.\/}(2012)]{steinhilber12}
Steinhilber, F., Abreu, J.A., Beer, J., Brunner, I., Christl, M., Fischer, H.,
  Heikkil{\"{a}}, U., Kubik, P.W., Mann, M., McCracken, K.G., Miller, H.,
  Miyahara, H., Oerter, H. and Wilhelms, F., 2012, ``9,400 years of cosmic
  radiation and solar activity from ice cores and tree rings'', {\it Proc.
  Natl. Acad. Sci. USA\/}, {\bf 109}(16), 5967--5971.
  {\small[\href{http://dx.doi.org/10.1073/pnas.1118965109}{DOI}]},
  {\small[\href{http://adsabs.harvard.edu/abs/2012PNAS..109.5967S}{ADS}]}

\bibitem[Stephenson {\it et~al.\/}(2004)]{stephenson04}
Stephenson, F.R., Willis, D.M. and Hallinan, T.J., 2004, ``Aurorae: The
  earliest datable observation of the aurora borealis'', {\it Astron.
  Geophys.\/}, {\bf 45}, 6.15--6.17.
  {\small[\href{http://dx.doi.org/10.1046/j.1468-4004.2003.45615.x}{DOI}]},
  {\small[\href{http://adsabs.harvard.edu/abs/2004A&G....45f..15S}{ADS}]}

\bibitem[Stuiver(1961)]{stuiver61}
Stuiver, M., 1961, ``Variations in Radiocarbon Concentration and Sunspot
  Activity'', {\it J. Geophys. Res.\/}, {\bf 66}, 273--276.
  {\small[\href{http://dx.doi.org/10.1029/JZ066i001p00273}{DOI}]},
  {\small[\href{http://adsabs.harvard.edu/abs/1961JGR....66..273S}{ADS}]}

\bibitem[Stuiver and Braziunas(1989)]{stuiver89}
Stuiver, M. and Braziunas, T.F., 1989, ``Atmospheric $^{14}$C and century-scale
  solar oscillations'', {\it Nature\/}, {\bf 338}, 405--408.
  {\small[\href{http://dx.doi.org/10.1038/338405a0}{DOI}]},
  {\small[\href{http://adsabs.harvard.edu/abs/1989Natur.338..405S}{ADS}]}

\bibitem[Stuiver and Braziunas(1993)]{stuiver93}
Stuiver, M. and Braziunas, T.F., 1993, ``Sun, ocean, climate and atmospheric
  $^{14}$CO$_2$: an evaluation of casual and spectral relationships'', {\it
  Holocene\/}, {\bf 3}(4), 289--305.
  {\small[\href{http://dx.doi.org/10.1177/095968369300300401}{DOI}]}

\bibitem[Stuiver and Pollach(1977)]{stuiver77}
Stuiver, M. and Pollach, H., 1977, ``Discussion: Reporting of $^{14}$C data'',
  {\it Radiocarbon\/}, {\bf 19}, 355--363

\bibitem[Stuiver and Quay(1980)]{stuiver80}
Stuiver, M. and Quay, P.D., 1980, ``Changes in Atmospheric Carbon-14 Attributed
  to a Variable Sun'', {\it Science\/}, {\bf 207}, 11--19.
  {\small[\href{http://dx.doi.org/10.1126/science.207.4426.11}{DOI}]},
  {\small[\href{http://adsabs.harvard.edu/abs/1980Sci...207...11S}{ADS}]}

\bibitem[Stuiver {\it et~al.\/}(1991)]{stuiver91}
Stuiver, M., Braziunas, T.F., Becker, B. and Kromer, B., 1991, ``Climatic,
  Solar, Oceanic, and Geomagnetic Influences on Late-Glacial and Holocene
  Atmospheric $^{14}$C/$^{12}$C Change'', {\it Quat. Res.\/}, {\bf 35}, 1--24.
  {\small[\href{http://dx.doi.org/10.1016/0033-5894(91)90091-I}{DOI}]}

\bibitem[Stuiver {\it et~al.\/}(1998)]{stuiver98}
Stuiver, M., Reimer, P.J., Bard, E., Burr, G.S., Hughen, K.A., Kromer, B.,
  McCormac, G., van~der Plicht, J. and Spurk, M., 1998, ``INTCAL98 radiocarbon
  age calibration, 24,000--0 cal BP'', {\it Radiocarbon\/}, {\bf 40}(3),
  1041--1083. URL (accessed 11 September 2012):
  \newline\url{https://journals.uair.arizona.edu/index.php/radiocarbon/article%
/view/3781}

\bibitem[Suess(1955)]{suess55}
Suess, H.E., 1955, ``Radiocarbon concentration in modern wood'', {\it
  Science\/}, {\bf 122}, 415--417.
  {\small[\href{http://dx.doi.org/10.1126/science.122.3166.415-a}{DOI}]}

\bibitem[Suess(1965)]{suess65}
Suess, H.E., 1965, ``Secular Variations of the Cosmic-Ray-Produced Carbon 14 in
  the Atmosphere and Their Interpretations'', {\it J. Geophys. Res.\/}, {\bf
  70}, 5937. {\small[\href{http://dx.doi.org/10.1029/JZ070i023p05937}{DOI}]},
  {\small[\href{http://adsabs.harvard.edu/abs/1965JGR....70.5937S}{ADS}]}

\bibitem[Suess(1980)]{suess80}
Suess, H.E., 1980, ``The radiocarbon record in tree rings of the last 8000
  years'', {\it Radiocarbon\/}, {\bf 22}, 200--209

\bibitem[Svalgaard(2012)]{svalgaard12}
Svalgaard, L., 2012, ``How well do we know the sunspot number?'', {\it Proc.
  IAU\/}, {\bf 286}, 27--33.
  {\small[\href{http://dx.doi.org/10.1017/S1743921312004590}{DOI}]},
  {\small[\href{http://adsabs.harvard.edu/abs/2012IAUS..286...27S}{ADS}]}

\bibitem[Svensmark(1998)]{svensmark98}
Svensmark, H., 1998, ``Influence of Cosmic Rays on Earth's Climate'', {\it
  Phys. Rev. Lett.\/}, {\bf 81}, 5027--5030.
  {\small[\href{http://dx.doi.org/10.1103/PhysRevLett.81.5027}{DOI}]},
  {\small[\href{http://adsabs.harvard.edu/abs/1998PhRvL..81.5027S}{ADS}]}

\bibitem[Sweeney {\it et~al.\/}(2007)]{sweeney07}
Sweeney, C., Gloor, E., Jacobson, A.R., Key, R.M., McKinley, G., Sarmiento,
  J.L. and Wanninkhof, R., 2007, ``Constraining global air-sea gas exchange for
  CO$_{2}$ with recent bomb $^{14}$C measurements'', {\it Global Biogeochem.
  Cycles\/}, {\bf 21}, GB2015.
  {\small[\href{http://dx.doi.org/10.1029/2006GB002784}{DOI}]},
  {\small[\href{http://adsabs.harvard.edu/abs/2007GBioC..21B2015S}{ADS}]}

\bibitem[Tans {\it et~al.\/}(1979)]{tans79}
Tans, P.P., de~Jong, A.F.M. and Mook, W.G., 1979, ``Natural atmospheric
  $^{14}$C variation and the Suess effect'', {\it Nature\/}, {\bf 280},
  826--828. {\small[\href{http://dx.doi.org/10.1038/280826a0}{DOI}]}

\bibitem[Tapping(1987)]{tapping87}
Tapping, K.F., 1987, ``Recent solar radio astronomy at centimeter wavelengths:
  The temporal variability of the 10.7-cm flux'', {\it J. Geophys. Res.\/},
  {\bf 92}, 829--838.
  {\small[\href{http://dx.doi.org/10.1029/JD092iD01p00829}{DOI}]},
  {\small[\href{http://adsabs.harvard.edu/abs/1987JGR....92..829T}{ADS}]}

\bibitem[Tapping and Charrois(1994)]{tapping94}
Tapping, K.F. and Charrois, D.P., 1994, ``Limits to the Accuracy of the 10.7 cm
  Flux'', {\it Solar Phys.\/}, {\bf 150}, 305--315.
  {\small[\href{http://dx.doi.org/10.1007/BF00712892}{DOI}]},
  {\small[\href{http://adsabs.harvard.edu/abs/1994SoPh..150..305T}{ADS}]}

\bibitem[Taricco {\it et~al.\/}(2006)]{taricco06}
Taricco, C., Bhandari, N., Cane, D., Colombetti, P. and Verma, N., 2006,
  ``Galactic cosmic ray flux decline and periodicities in the interplanetary
  space during the last 3 centuries revealed by $^{44}$Ti in meteorites'', {\it
  J. Geophys. Res.\/}, {\bf 111}(A10), A08102.
  {\small[\href{http://dx.doi.org/10.1029/2005JA011459}{DOI}]},
  {\small[\href{http://adsabs.harvard.edu/abs/2006JGRA..11108102T}{ADS}]}

\bibitem[Temmer {\it et~al.\/}(2002)]{temmer02}
Temmer, M., Veronig, A. and Hanslmeier, A., 2002, ``Hemispheric Sunspot Numbers
  $R_{n}$ and $R_{s}$: Catalogue and N-S asymmetry analysis'', {\it Astron.
  Astrophys.\/}, {\bf 390}, 707--715.
  {\small[\href{http://dx.doi.org/10.1051/0004-6361:20020758}{DOI}]},
  {\small[\href{http://adsabs.harvard.edu/abs/2002A&A...390..707T}{ADS}]},
  {\small[\href{http://arxiv.org/abs/arXiv:astro-ph/0208436}{{arXiv:astro-ph/0%
208436}}]}

\bibitem[Tinsley and Zhou(2006)]{tinsley06}
Tinsley, B.A. and Zhou, L., 2006, ``Initial results of a global circuit model
  with variable stratospheric and tropospheric aerosols'', {\it J. Geophys.
  Res.\/}, {\bf 111}(D10), D16205.
  {\small[\href{http://dx.doi.org/10.1029/2005JD006988}{DOI}]},
  {\small[\href{http://adsabs.harvard.edu/abs/2006JGRD..11116205T}{ADS}]}

\bibitem[Tobias {\it et~al.\/}(2006)]{tobias06}
Tobias, S., Hughes, D. and Weiss, N., 2006, ``Unpredictable Sun leaves
  researchers in the dark'', {\it Nature\/}, {\bf 442}, 26.
  {\small[\href{http://dx.doi.org/10.1038/442026c}{DOI}]},
  {\small[\href{http://adsabs.harvard.edu/abs/2006Natur.442...26T}{ADS}]}

\bibitem[Tobias {\it et~al.\/}(1995)]{tobias95}
Tobias, S.M., Weiss, N.O. and Kirk, V., 1995, ``Chaotically modulated stellar
  dynamos'', {\it Mon. Not. R. Astron. Soc.\/}, {\bf 273}, 1150--1166.
  {\small[\href{http://adsabs.harvard.edu/abs/1995MNRAS.273.1150T}{ADS}]}

\bibitem[Toptygin(1985)]{toptygin85}
Toptygin, I.N., 1985, {\it Cosmic Rays in Interplanetary Magnetic Fields\/},
  Geophysics and Astrophysics Monographs, Kluwer Academic Publishers, Dordrecht

\bibitem[Traversi {\it et~al.\/}(2012)]{traversi12}
Traversi, R., Usoskin, I.G., Solanki, S.K., Becagli, S., Frezzotti, M., Severi,
  M., Stenni, B. and Udisti, R., 2012, ``Nitrate in Polar Ice: A New Tracer of
  Solar Variability'', {\it Solar Phys.\/}, {\bf 280}, 237--254.
  {\small[\href{http://dx.doi.org/10.1007/s11207-012-0060-3}{DOI}]},
  {\small[\href{http://adsabs.harvard.edu/abs/2012SoPh..280..237T}{ADS}]}

\bibitem[Tylka and Dietrich(2009)]{tylka09}
Tylka, A. and Dietrich, W., 2009, ``A new and comprehensive analysis of proton
  spectra in ground-level enhanced (GLE) solar particle events'', in {\it 31th
  International Cosmic Ray Conference\/}, 31st ICRC, held on July 7\,--\,15,
  2009, Lod\'z, Poland, ICRC0273, Universal Academy Press, Lod\'z

\bibitem[Tylka {\it et~al.\/}(1997)]{tylka97}
Tylka, A.J., Dietrich, W.F. and Boberg, P.R., 1997, ``Probability distribution
  of high-energy solar-heavy-ion fluxes from IMP-8: 1973-1996'', {\it IEEE
  Trans. Nucl. Sci.\/}, {\bf 44}, 2140--2149.
  {\small[\href{http://dx.doi.org/10.1109/23.659029}{DOI}]},
  {\small[\href{http://adsabs.harvard.edu/abs/1997ITNS...44.2140T}{ADS}]}

\bibitem[Usoskin(2011)]{usoskin_MSI_11}
Usoskin, I.G., 2011, ``Cosmic rays and climate forcing'', {\it Mem. Soc.
  Astron. Ital.\/}, {\bf 82}, 937--942.
  {\small[\href{http://adsabs.harvard.edu/abs/2011MmSAI..82..937U}{ADS}]}

\bibitem[Usoskin and Kovaltsov(2004)]{usoskin_Rev_04}
Usoskin, I.G. and Kovaltsov, G.A., 2004, ``Long-Term Solar Activity: Direct and
  Indirect Study'', {\it Solar Phys.\/}, {\bf 224}, 37--47.
  {\small[\href{http://dx.doi.org/10.1007/s11207-005-3997-7}{DOI}]},
  {\small[\href{http://adsabs.harvard.edu/abs/2004SoPh..224...37U}{ADS}]}

\bibitem[Usoskin and Kovaltsov(2006)]{usoskin_JGR_06}
Usoskin, I.G. and Kovaltsov, G.A., 2006, ``Cosmic ray induced ionization in the
  atmosphere: Full modeling and practical applications'', {\it J. Geophys.
  Res.\/}, {\bf 111}, D21206.
  {\small[\href{http://dx.doi.org/10.1029/2006JD007150}{DOI}]},
  {\small[\href{http://adsabs.harvard.edu/abs/2006JGRD..11121206U}{ADS}]}

\bibitem[Usoskin and Kovaltsov(2008{\natexlab{a}})]{usoskin_7Be_08}
Usoskin, I.G. and Kovaltsov, G.A., 2008{\natexlab{a}}, ``Production of
  cosmogenic $^{7}$Be isotope in the atmosphere: Full 3D modelling'', {\it J.
  Geophys. Res.\/}, {\bf 113}, D12107.
  {\small[\href{http://dx.doi.org/10.1029/2007JD009725}{DOI}]}

\bibitem[Usoskin and Kovaltsov(2008{\natexlab{b}})]{usoskin_CR_08}
Usoskin, I.G. and Kovaltsov, G.A., 2008{\natexlab{b}}, ``Cosmic rays and
  climate of the Earth: Possible connection'', {\it C. R. Geosci.\/}, {\bf
  340}, 441--450.
  {\small[\href{http://dx.doi.org/10.1016/j.crte.2007.11.001}{DOI}]},
  {\small[\href{http://adsabs.harvard.edu/abs/2008CRGeo.340..441U}{ADS}]}

\bibitem[Usoskin and Kovaltsov(2012)]{usoskin_SEP_12}
Usoskin, I.G. and Kovaltsov, G.A., 2012, ``Occurrence of extreme solar particle
  events: Assessment from historical proxy data'', {\it Astrophys. J.\/}, {\bf
  757}, 92. {\small[\href{http://dx.doi.org/10.1088/0004-637X/757/1/92}{DOI}]},
  {\small[\href{http://adsabs.harvard.edu/abs/2012ApJ...757...92U}{ADS}]},
  {\small[\href{http://arxiv.org/abs/1207.5932}{{arXiv:1207.5932
  {\small[astro-ph.SR]}}}]}

\bibitem[Usoskin and Kromer(2005)]{usoskin_14C_05}
Usoskin, I.G. and Kromer, B., 2005, ``Reconstruction of the $^{14}$C production
  rate from measured relative abundance'', {\it Radiocarbon\/}, {\bf 47},
  31--37

\bibitem[Usoskin and Mursula(2003)]{usoskin_SP03}
Usoskin, I.G. and Mursula, K., 2003, ``Long-Term Solar Cycle Evolution: Review
  of Recent Developments'', {\it Solar Phys.\/}, {\bf 218}, 319--343.
  {\small[\href{http://dx.doi.org/10.1023/B:SOLA.0000013049.27106.07}{DOI}]},
  {\small[\href{http://adsabs.harvard.edu/abs/2003SoPh..218..319U}{ADS}]}

\bibitem[Usoskin {\it et~al.\/}(1998)]{usoskin_JGR_98}
Usoskin, I.G., Kananen, H., Mursula, K., Tanskanen, P. and Kovaltsov, G.A.,
  1998, ``Correlative study of solar activity and cosmic ray intensity'', {\it
  J. Geophys. Res.\/}, {\bf 103}(A5), 9567--9574.
  {\small[\href{http://dx.doi.org/10.1029/97JA03782}{DOI}]},
  {\small[\href{http://adsabs.harvard.edu/abs/1998JGR...103.9567U}{ADS}]}

\bibitem[Usoskin {\it et~al.\/}(2000)]{usoskin_MM_AA_00}
Usoskin, I.G., Mursula, K. and Kovaltsov, G.A., 2000, ``Cyclic behaviour of
  sunspot activity during the Maunder minimum'', {\it Astron. Astrophys.\/},
  {\bf 354}, L33--L36.
  {\small[\href{http://adsabs.harvard.edu/abs/2000A&A...354L..33U}{ADS}]}

\bibitem[Usoskin {\it et~al.\/}(2001{\natexlab{a}})]{usoskin_JGR_MM_01}
Usoskin, I.G., Mursula, K. and Kovaltsov, G.A., 2001{\natexlab{a}},
  ``Heliospheric modulation of cosmic rays and solar activity during the
  Maunder minimum'', {\it J. Geophys. Res.\/}, {\bf 106}, 16,039--16,046.
  {\small[\href{http://dx.doi.org/10.1029/2000JA000105}{DOI}]},
  {\small[\href{http://adsabs.harvard.edu/abs/2001JGR...10616039U}{ADS}]}

\bibitem[Usoskin {\it et~al.\/}(2001{\natexlab{b}})]{usoskin_lost_AA_01}
Usoskin, I.G., Mursula, K. and Kovaltsov, G.A., 2001{\natexlab{b}}, ``Was one
  sunspot cycle lost in late XVIII century?'', {\it Astron. Astrophys.\/}, {\bf
  370}, L31--L34.
  {\small[\href{http://dx.doi.org/10.1051/0004-6361:20010319}{DOI}]},
  {\small[\href{http://adsabs.harvard.edu/abs/2001A&A...370L..31U}{ADS}]}

\bibitem[Usoskin {\it et~al.\/}(2002{\natexlab{a}})]{usoskin_lost_GRL_02}
Usoskin, I.G., Mursula, K. and Kovaltsov, G.A., 2002{\natexlab{a}}, ``Lost
  sunspot cycle in the beginning of Dalton minimum: New evidence and
  consequences'', {\it Geophys. Res. Lett.\/}, {\bf 29}, 36--1.
  {\small[\href{http://dx.doi.org/10.1029/2002GL015640}{DOI}]},
  {\small[\href{http://adsabs.harvard.edu/abs/2002GeoRL..29x..36U}{ADS}]}

\bibitem[Usoskin {\it et~al.\/}(2002{\natexlab{b}})]{usoskin_JGR_02}
Usoskin, I.G., Mursula, K., Solanki, S.K., Sch{\"{u}}ssler, M. and Kovaltsov,
  G.A., 2002{\natexlab{b}}, ``A physical reconstruction of cosmic ray intensity
  since 1610'', {\it J. Geophys. Res.\/}, {\bf 107}, 1374.
  {\small[\href{http://dx.doi.org/10.1029/2002JA009343}{DOI}]},
  {\small[\href{http://adsabs.harvard.edu/abs/2002JGRA..107.1374U}{ADS}]}

\bibitem[Usoskin {\it et~al.\/}(2003{\natexlab{a}})]{usoskin_SP_daily03}
Usoskin, I.G., Mursula, K. and Kovaltsov, G.A., 2003{\natexlab{a}},
  ``Reconstruction of monthly and yearly group sunspot numbers from sparse
  daily observations'', {\it Solar Phys.\/}, {\bf 218}, 295--305.
  {\small[\href{http://dx.doi.org/10.1023/B:SOLA.0000013029.99907.97}{DOI}]}

\bibitem[Usoskin {\it et~al.\/}(2003{\natexlab{b}})]{usoskin_lost_03}
Usoskin, I.G., Mursula, K. and Kovaltsov, G.A., 2003{\natexlab{b}}, ``The lost
  sunspot cycle: Reanalysis of sunspot statistics'', {\it Astron.
  Astrophys.\/}, {\bf 403}, 743--748.
  {\small[\href{http://dx.doi.org/10.1051/0004-6361:20030398}{DOI}]},
  {\small[\href{http://adsabs.harvard.edu/abs/2003A&A...403..743U}{ADS}]}

\bibitem[Usoskin {\it et~al.\/}(2003{\natexlab{c}})]{usoskin_PRL_03}
Usoskin, I.G., Solanki, S.K., Sch{\"{u}}ssler, M., Mursula, K. and Alanko, K.,
  2003{\natexlab{c}}, ``Millennium-Scale Sunspot Number Reconstruction:
  Evidence for an Unusually Active Sun since the 1940s'', {\it Phys. Rev.
  Lett.\/}, {\bf 91}, 211101.
  {\small[\href{http://dx.doi.org/10.1103/PhysRevLett.91.211101}{DOI}]},
  {\small[\href{http://adsabs.harvard.edu/abs/2003PhRvL..91u1101U}{ADS}]},
  {\small[\href{http://arxiv.org/abs/arXiv:astro-ph/0310823}{{arXiv:astro-ph/0%
310823}}]}

\bibitem[Usoskin {\it et~al.\/}(2004)]{usoskin_AA_04}
Usoskin, I.G., Mursula, K., Solanki, S.K., Sch{\"{u}}ssler, M. and Alanko, K.,
  2004, ``Reconstruction of solar activity for the last millennium using
  $^{10}$Be data'', {\it Astron. Astrophys.\/}, {\bf 413}, 745--751.
  {\small[\href{http://dx.doi.org/10.1051/0004-6361:20031533}{DOI}]},
  {\small[\href{http://adsabs.harvard.edu/abs/2004A&A...413..745U}{ADS}]},
  {\small[\href{http://arxiv.org/abs/arXiv:astro-ph/0309556}{{arXiv:astro-ph/0%
309556}}]}

\bibitem[Usoskin {\it et~al.\/}(2005)]{usoskin_Phi_05}
Usoskin, I.G., Alanko-Huotari, K., Kovaltsov, G.A. and Mursula, K., 2005,
  ``Heliospheric modulation of cosmic rays: Monthly reconstruction for
  1951--2004'', {\it J. Geophys. Res.\/}, {\bf 110}(A9), A12108.
  {\small[\href{http://dx.doi.org/10.1029/2005JA011250}{DOI}]},
  {\small[\href{http://adsabs.harvard.edu/abs/2005JGRA..11012108U}{ADS}]}

\bibitem[Usoskin {\it et~al.\/}(2006{\natexlab{a}})]{usoskin_GRL_06}
Usoskin, I.G., Solanki, S.K. and Korte, M., 2006{\natexlab{a}}, ``Solar
  activity reconstructed over the last 7000 years: The influence of geomagnetic
  field changes'', {\it Geophys. Res. Lett.\/}, {\bf 33}, 8103.
  {\small[\href{http://dx.doi.org/10.1029/2006GL025921}{DOI}]},
  {\small[\href{http://adsabs.harvard.edu/abs/2006GeoRL..3308103U}{ADS}]}

\bibitem[Usoskin {\it et~al.\/}(2006{\natexlab{b}})]{usoskin_GRL_SCR06}
Usoskin, I.G., Solanki, S.K., Kovaltsov, G.A., Beer, J. and Kromer, B.,
  2006{\natexlab{b}}, ``Solar proton events in cosmogenic isotope data'', {\it
  Geophys. Res. Lett.\/}, {\bf 33}, L08\,107.
  {\small[\href{http://dx.doi.org/10.1029/2006GL026059}{DOI}]},
  {\small[\href{http://adsabs.harvard.edu/abs/2006GeoRL..3308107U}{ADS}]}

\bibitem[Usoskin {\it et~al.\/}(2006{\natexlab{c}})]{usoskin_Ti_06}
Usoskin, I.G., Solanki, S.K., Taricco, C., Bhandari, N. and Kovaltsov, G.A.,
  2006{\natexlab{c}}, ``Long-term solar activity reconstructions: direct test
  by cosmogenic $^{44}$Ti in meteorites'', {\it Astron. Astrophys.\/}, {\bf
  457}, L25--L28.
  {\small[\href{http://dx.doi.org/10.1051/0004-6361:20065803}{DOI}]},
  {\small[\href{http://adsabs.harvard.edu/abs/2006A&A...457L..25U}{ADS}]}

\bibitem[Usoskin {\it et~al.\/}(2007)]{usoskin_AA_07}
Usoskin, I.G., Solanki, S.K. and Kovaltsov, G.A., 2007, ``Grand minima and
  maxima of solar activity: new observational constraints'', {\it Astron.
  Astrophys.\/}, {\bf 471}, 301--309.
  {\small[\href{http://dx.doi.org/10.1051/0004-6361:20077704}{DOI}]},
  {\small[\href{http://adsabs.harvard.edu/abs/2007A&A...471..301U}{ADS}]},
  {\small[\href{http://arxiv.org/abs/0706.0385}{{arXiv:0706.0385}}]}

\bibitem[Usoskin {\it et~al.\/}(2008)]{usoskin_CRII_08}
Usoskin, I.G., Korte, M. and Kovaltsov, G.A., 2008, ``Role of centennial
  geomagnetic changes in local atmospheric ionization'', {\it Geophys. Res.
  Lett.\/}, {\bf 35}, L05811.
  {\small[\href{http://dx.doi.org/10.1029/2007GL033040}{DOI}]},
  {\small[\href{http://adsabs.harvard.edu/abs/2008GeoRL..3505811U}{ADS}]}

\bibitem[Usoskin {\it et~al.\/}(2009{\natexlab{a}})]{usoskin_7Be_09}
Usoskin, I.G., Field, C.V., Schmidt, G.A., Lepp{\"{a}}nen, A.-P., Aldahan, A.,
  Kovaltsov, G.A., Possnert, G. and Ungar, R.K., 2009{\natexlab{a}},
  ``Short-term production and synoptic influences on atmospheric $^{7}$Be
  concentrations'', {\it J. Geophys. Res.\/}, {\bf 114}(D13), D06108.
  {\small[\href{http://dx.doi.org/10.1029/2008JD011333}{DOI}]},
  {\small[\href{http://adsabs.harvard.edu/abs/2009JGRD..11406108U}{ADS}]}

\bibitem[Usoskin {\it et~al.\/}(2009{\natexlab{b}})]{usoskin_10Be_09}
Usoskin, I.G., Horiuchi, K., Solanki, S.K., Kovaltsov, G.A. and Bard, E.,
  2009{\natexlab{b}}, ``On the common solar signal in different cosmogenic
  isotope data sets'', {\it J. Geophys. Res.\/}, {\bf 114}(A13), A03112.
  {\small[\href{http://dx.doi.org/10.1029/2008JA013888}{DOI}]},
  {\small[\href{http://adsabs.harvard.edu/abs/2009JGRA..11403112U}{ADS}]}

\bibitem[Usoskin {\it et~al.\/}(2009{\natexlab{c}})]{usoskin_lost_09}
Usoskin, I.G., Mursula, K., Arlt, R. and Kovaltsov, G.A., 2009{\natexlab{c}},
  ``A Solar Cycle Lost in 1793--1800: Early Sunspot Observations Resolve the
  Old Mystery'', {\it Astrophys. J. Lett.\/}, {\bf 700}, L154--L157.
  {\small[\href{http://dx.doi.org/10.1088/0004-637X/700/2/L154}{DOI}]},
  {\small[\href{http://adsabs.harvard.edu/abs/2009ApJ...700L.154U}{ADS}]},
  {\small[\href{http://arxiv.org/abs/0907.0063}{{arXiv:0907.0063
  {\small[astro-ph.SR]}}}]}

\bibitem[Usoskin {\it et~al.\/}(2009{\natexlab{d}})]{usoskin_sok_09}
Usoskin, I.G., Sokoloff, D. and Moss, D., 2009{\natexlab{d}}, ``Grand Minima of
  Solar Activity and the Mean-Field Dynamo'', {\it Solar Phys.\/}, {\bf 254},
  345--355. {\small[\href{http://dx.doi.org/10.1007/s11207-008-9293-6}{DOI}]},
  {\small[\href{http://adsabs.harvard.edu/abs/2009SoPh..254..345U}{ADS}]}

\bibitem[Usoskin {\it et~al.\/}(2010)]{usoskin_Geo_10}
Usoskin, I.G., Mironova, I.A., Korte, M. and Kovaltsov, G.A., 2010, ``Regional
  millennial trend in the cosmic ray induced ionization of the troposphere'',
  {\it J. Atmos. Sol.-Terr. Phys.\/}, {\bf 72}, 19--25.
  {\small[\href{http://dx.doi.org/10.1016/j.jastp.2009.10.003}{DOI}]},
  {\small[\href{http://adsabs.harvard.edu/abs/2010JASTP..72...19U}{ADS}]}

\bibitem[Usoskin {\it et~al.\/}(2011)]{usoskin_bazi_11}
Usoskin, I.G., Bazilevskaya, G.A. and Kovaltsov, G.A., 2011, ``Solar modulation
  parameter for cosmic rays since 1936 reconstructed from ground-based neutron
  monitors and ionization chambers'', {\it J. Geophys. Res.\/}, {\bf 116},
  A02104. {\small[\href{http://dx.doi.org/10.1029/2010JA016105}{DOI}]},
  {\small[\href{http://adsabs.harvard.edu/abs/2011JGRA..11602104U}{ADS}]}

\bibitem[Usoskin {\it et~al.\/}(2013)]{usoskin_775_13}
Usoskin, I.G., Kromer, B., Ludlow, F., Beer, J., Friedrich, M., Kovaltsov,
  G.A., Solanki, S.K. and Wacker, L., 2013, ``The AD775 cosmic event revisited:
  the Sun is to blame'', {\it Astron. Astrophys.\/}, {\bf 552}, L3.
  {\small[\href{http://dx.doi.org/10.1051/0004-6361/201321080}{DOI}]},
  {\small[\href{http://adsabs.harvard.edu/abs/2013A&A...552L...3U}{ADS}]},
  {\small[\href{http://arxiv.org/abs/1302.6897}{{arXiv:1302.6897
  {\small[astro-ph.SR]}}}]}

\bibitem[Vaquero(2007)]{vaquero_rev07}
Vaquero, J.M., 2007, ``Historical sunspot observations: A review'', {\it Adv.
  Space Res.\/}, {\bf 40}, 929--941.
  {\small[\href{http://dx.doi.org/10.1016/j.asr.2007.01.087}{DOI}]},
  {\small[\href{http://adsabs.harvard.edu/abs/2007AdSpR..40..929V}{ADS}]},
  {\small[\href{http://arxiv.org/abs/arXiv:astro-ph/0702068}{{arXiv:astro-ph/0%
702068}}]}

\bibitem[Vaquero and V{\'{a}}zquez(2009)]{vaquero09}
Vaquero, J.M. and V{\'{a}}zquez, M., 2009, {\it The Sun Recorded Through
  History: Scientific Data Extracted from Historical Documents\/}, Astrophysics
  and Space Science Library, 361, Springer, Berlin; New York.
  {\small[\href{http://books.google.com/books?id=iWkyiEeiyksC}{Google Books}]}

\bibitem[Vaquero {\it et~al.\/}(2002)]{vaquero02}
Vaquero, J.M., Gallego, M.C. and Garc{\'{\i}}a, J.A., 2002, ``A 250-year cycle
  in naked-eye observations of sunspots'', {\it Geophys. Res. Lett.\/}, {\bf
  29}, 58--1. {\small[\href{http://dx.doi.org/10.1029/2002GL014782}{DOI}]},
  {\small[\href{http://adsabs.harvard.edu/abs/2002GeoRL..29t..58V}{ADS}]}

\bibitem[Vaquero {\it et~al.\/}(2004)]{vaquero04}
Vaquero, J.M., Gallego, M.C. and S{\'{a}}nchez-Bajo, F., 2004, ``Reconstruction
  of a Monthly Homogeneous Sunspot Area Series Since 1832'', {\it Solar
  Phys.\/}, {\bf 221}, 179--189.
  {\small[\href{http://dx.doi.org/10.1023/B:SOLA.0000033360.67976.bd}{DOI}]},
  {\small[\href{http://adsabs.harvard.edu/abs/2004SoPh..221..179V}{ADS}]}

\bibitem[Vaquero {\it et~al.\/}(2005)]{vaquero05}
Vaquero, J.M., Trigo, R.M. and Gallego, M.C., 2005, ``A `lost' sunspot
  observation in 1785'', {\it Astron. Nachr.\/}, {\bf 326}, 112--114.
  {\small[\href{http://dx.doi.org/10.1002/ansa.200410343}{DOI}]},
  {\small[\href{http://adsabs.harvard.edu/abs/2005AN....326..112V}{ADS}]}

\bibitem[Vaquero {\it et~al.\/}(2007)]{vaquero07}
Vaquero, J.M., Trigo, R.M., Gallego, M.C. and Moreno-Corral, M.A., 2007, ``Two
  Early Sunspots Observers: Teodoro de Almeida and Jos{\'e} Antonio Alzate'',
  {\it Solar Phys.\/}, {\bf 240}, 165--175.
  {\small[\href{http://dx.doi.org/10.1007/s11207-006-0264-5}{DOI}]},
  {\small[\href{http://adsabs.harvard.edu/abs/2007SoPh..240..165V}{ADS}]}

\bibitem[Vaquero {\it et~al.\/}(2011)]{vaquero11}
Vaquero, J.M., Gallego, M.C., Usoskin, I.G. and Kovaltsov, G.A., 2011,
  ``Revisited Sunspot Data: A New Scenario for the Onset of the Maunder
  Minimum'', {\it Astrophys. J. Lett.\/}, {\bf 731}, L24.
  {\small[\href{http://dx.doi.org/10.1088/2041-8205/731/2/L24}{DOI}]},
  {\small[\href{http://adsabs.harvard.edu/abs/2011ApJ...731L..24V}{ADS}]},
  {\small[\href{http://arxiv.org/abs/1103.1520}{{arXiv:1103.1520
  {\small[astro-ph.SR]}}}]}

\bibitem[Vaquero {\it et~al.\/}(2012)]{vaquero12}
Vaquero, J.M., Trigo, R.M. and Gallego, M.C., 2012, ``A Simple Method to Check
  the Reliability of Annual Sunspot Number in the Historical Period
  1610--1847'', {\it Solar Phys.\/}, {\bf 277}, 389--395.
  {\small[\href{http://dx.doi.org/10.1007/s11207-011-9901-8}{DOI}]},
  {\small[\href{http://adsabs.harvard.edu/abs/2012SoPh..277..389V}{ADS}]},
  {\small[\href{http://arxiv.org/abs/1111.2633}{{arXiv:1111.2633
  {\small[astro-ph.SR]}}}]}

\bibitem[Vasiliev and Dergachev(2002)]{vassiliev02}
Vasiliev, S.S. and Dergachev, V.A., 2002, ``The $\sim$2400-year cycle in
  atmospheric radiocarbon concentration: bispectrum of $^{14}$C data over the
  last 8000 years'', {\it Ann. Geophys.\/}, {\bf 20}, 115--120.
  {\small[\href{http://adsabs.harvard.edu/abs/2002AnGeo..20..115V}{ADS}]}. URL
  (accessed 14 October 2008):
  \newline\url{http://www.ann-geophys.net/20/115/2002/}

\bibitem[V{\'a}zquez and Vaquero(2010)]{vaquero10}
V{\'a}zquez, M. and Vaquero, J.M., 2010, ``Aurorae Observed at the Canary
  Islands'', {\it Solar Phys.\/}, {\bf 267}, 431--444.
  {\small[\href{http://dx.doi.org/10.1007/s11207-010-9650-0}{DOI}]},
  {\small[\href{http://adsabs.harvard.edu/abs/2010SoPh..267..431V}{ADS}]}

\bibitem[Versteegh(2005)]{versteegh05}
Versteegh, G.J.M., 2005, ``Solar Forcing of Climate. 2: Evidence from the
  Past'', {\it Space Sci. Rev.\/}, {\bf 120}, 243--286.
  {\small[\href{http://dx.doi.org/10.1007/s11214-005-7047-4}{DOI}]},
  {\small[\href{http://adsabs.harvard.edu/abs/2005SSRv..120..243V}{ADS}]}

\bibitem[Vieira {\it et~al.\/}(2011)]{vieira11}
Vieira, L.E.A., Solanki, S.K., Krivova, N.A. and Usoskin, I.G., 2011,
  ``Evolution of the solar irradiance during the Holocene'', {\it Astron.
  Astrophys.\/}, {\bf 531}, A6.
  {\small[\href{http://dx.doi.org/10.1051/0004-6361/201015843}{DOI}]},
  {\small[\href{http://adsabs.harvard.edu/abs/2011A&A...531A...6V}{ADS}]},
  {\small[\href{http://arxiv.org/abs/1103.4958}{{arXiv:1103.4958
  {\small[astro-ph.SR]}}}]}

\bibitem[Viereck and Puga(1999)]{viereck99}
Viereck, R.A. and Puga, L.C., 1999, ``The NOAA Mg II core-to-wing solar index:
  Construction of a 20-year time series of chromospheric variability from
  multiple satellites'', {\it J. Geophys. Res.\/}, {\bf 104}, 9995--10\,006.
  {\small[\href{http://dx.doi.org/10.1029/1998JA900163}{DOI}]},
  {\small[\href{http://adsabs.harvard.edu/abs/1999JGR...104.9995V}{ADS}]}

\bibitem[Vitinsky(1965)]{vitinsky65}
Vitinsky, Y.I., 1965, {\it Solar Activity Forecasting\/}, Israel Program for
  Scientific Translations, Jerusalem

\bibitem[Vitinsky {\it et~al.\/}(1986)]{vitinsky86}
Vitinsky, Y.I., Kopecky, M. and Kuklin, G.V., 1986, {\it Statistics of Sunspot
  Activity (in Russian)\/}, Nauka, Moscow

\bibitem[Vogt {\it et~al.\/}(1990)]{vogt90}
Vogt, S., Herzog, G.F. and Reedy, R.C., 1990, ``Cosmogenic nuclides in
  extraterrestrial materials'', {\it Rev. Geophys.\/}, {\bf 28}, 253--275.
  {\small[\href{http://dx.doi.org/10.1029/RG028i003p00253}{DOI}]},
  {\small[\href{http://adsabs.harvard.edu/abs/1990RvGeo..28..253V}{ADS}]}

\bibitem[Vonmoos {\it et~al.\/}(2006)]{vonmoos06}
Vonmoos, M.V., Beer, J. and Muscheler, R., 2006, ``Large variations in Holocene
  solar activity: Constraints from $^{10}$Be in the Greenland Ice Core Project
  ice core'', {\it J. Geophys. Res.\/}, {\bf 111}(A10), A10105.
  {\small[\href{http://dx.doi.org/10.1029/2005JA011500}{DOI}]},
  {\small[\href{http://adsabs.harvard.edu/abs/2006JGRA..11110105V}{ADS}]}

\bibitem[Voss {\it et~al.\/}(1996)]{voss96}
Voss, H., Kurths, J. and Schwarz, U., 1996, ``Reconstruction of grand minima of
  solar activity from $\Delta^{14}$C data: Linear and nonlinear signal
  analysis'', {\it J. Geophys. Res.\/}, {\bf 101}, 15,637--15,644.
  {\small[\href{http://dx.doi.org/10.1029/96JA00542}{DOI}]},
  {\small[\href{http://adsabs.harvard.edu/abs/1996JGR...10115637V}{ADS}]}

\bibitem[Waldmeier(1961)]{waldmeier61}
Waldmeier, M., 1961, {\it The Sunspot-Activity in the Years 1610--1960\/},
  Schulthess u. Co. / Swiss Federal Observatory, Z{\"u}rich.
  {\small[\href{http://adsabs.harvard.edu/abs/1961QB525.W34......}{ADS}]}

\bibitem[Wang and Sheeley~Jr(2013)]{wang13}
Wang, Y.-M. and Sheeley~Jr, N.R., 2013, ``The Solar Wind and Interplanetary
  Field during Very Low Amplitude Sunspot Cycles'', {\it Astrophys. J.\/}, {\bf
  764}, 90. {\small[\href{http://dx.doi.org/10.1088/0004-637X/764/1/90}{DOI}]}

\bibitem[Wang {\it et~al.\/}(2005)]{wang05}
Wang, Y.-M., Lean, J.L. and Sheeley~Jr, N.R., 2005, ``Modeling the Sun's
  Magnetic Field and Irradiance since 1713'', {\it Astrophys. J.\/}, {\bf 625},
  522--538. {\small[\href{http://dx.doi.org/10.1086/429689}{DOI}]},
  {\small[\href{http://adsabs.harvard.edu/abs/2005ApJ...625..522W}{ADS}]}

\bibitem[Webber and Higbie(2003)]{webber03}
Webber, W.R. and Higbie, P.R., 2003, ``Production of cosmogenic Be nuclei in
  the Earth's atmosphere by cosmic rays: Its dependence on solar modulation and
  the interstellar cosmic ray spectrum'', {\it J. Geophys. Res.\/}, {\bf 108},
  1355. {\small[\href{http://dx.doi.org/10.1029/2003JA009863}{DOI}]},
  {\small[\href{http://adsabs.harvard.edu/abs/2003JGRA..108.1355W}{ADS}]}

\bibitem[Webber and Higbie(2009)]{webber09}
Webber, W.R. and Higbie, P.R., 2009, ``Galactic propagation of cosmic ray
  nuclei in a model with an increasing diffusion coefficient at low rigidities:
  A comparison of the new interstellar spectra with Voyager data in the outer
  heliosphere'', {\it J. Geophys. Res.\/}, {\bf 114}(A13), A02103.
  {\small[\href{http://dx.doi.org/10.1029/2008JA013689}{DOI}]},
  {\small[\href{http://adsabs.harvard.edu/abs/2009JGRA..11402103W}{ADS}]}

\bibitem[Webber {\it et~al.\/}(2007)]{webber07}
Webber, W.R., Higbie, P.R. and McCracken, K.G., 2007, ``Production of the
  cosmogenic isotopes $^{3}$H, $^{7}$Be, $^{10}$Be, and $^{36}$Cl in the
  Earth's atmosphere by solar and galactic cosmic rays'', {\it J. Geophys.
  Res.\/}, {\bf 112}, A10106.
  {\small[\href{http://dx.doi.org/10.1029/2007JA012499}{DOI}]},
  {\small[\href{http://adsabs.harvard.edu/abs/2007JGRA..11210106W}{ADS}]}

\bibitem[Weiss and Tobias(2000)]{weiss00}
Weiss, N.O. and Tobias, S.M., 2000, ``Physical Causes of Solar Activity'', {\it
  Space Sci. Rev.\/}, {\bf 94}, 99--112.
  {\small[\href{http://dx.doi.org/10.1023/A:1026790416627}{DOI}]},
  {\small[\href{http://adsabs.harvard.edu/abs/2000SSRv...94...99W}{ADS}]}

\bibitem[Weiss {\it et~al.\/}(1984)]{weiss84}
Weiss, N.O., Cattaneo, F. and Jones, C.A., 1984, ``Periodic and aperiodic
  dynamo waves'', {\it Geophys. Astrophys. Fluid Dyn.\/}, {\bf 30}, 305--341.
  {\small[\href{http://dx.doi.org/10.1080/03091928408219262}{DOI}]},
  {\small[\href{http://adsabs.harvard.edu/abs/1984GApFD..30..305W}{ADS}]}

\bibitem[Wheatland(2003)]{wheatland03}
Wheatland, M.S., 2003, ``The Coronal Mass Ejection Waiting-Time Distribution'',
  {\it Solar Phys.\/}, {\bf 214}, 361--373.
  {\small[\href{http://dx.doi.org/10.1023/A:1024222511574}{DOI}]},
  {\small[\href{http://adsabs.harvard.edu/abs/2003SoPh..214..361W}{ADS}]},
  {\small[\href{http://arxiv.org/abs/arXiv:astro-ph/0303019}{{arXiv:astro-ph/0%
303019}}]}

\bibitem[Willis and Stephenson(2001)]{willis01}
Willis, D.M. and Stephenson, F.R., 2001, ``Solar and auroral evidence for an
  intense recurrent geomagnetic storm during December in AD 1128'', {\it Ann.
  Geophys.\/}, {\bf 19}, 289--302.
  {\small[\href{http://adsabs.harvard.edu/abs/2001AnGeo..19..289W}{ADS}]}. URL
  (accessed 14 October 2008):
  \newline\url{http://www.ann-geophys.net/19/289/2001/}

\bibitem[Willis {\it et~al.\/}(1980)]{willis80}
Willis, D.M., Easterbrook, M.G. and Stephenson, F.R., 1980, ``Seasonal
  variation of oriental sunspot sightings'', {\it Nature\/}, {\bf 287},
  617--619. {\small[\href{http://dx.doi.org/10.1038/287617a0}{DOI}]},
  {\small[\href{http://adsabs.harvard.edu/abs/1980Natur.287..617W}{ADS}]}

\bibitem[Willis {\it et~al.\/}(1996)]{willis96}
Willis, D.M., Davda, V.N. and Stephenson, F.R., 1996, ``Comparison between
  Oriental and Occidental Sunspot Observations'', {\it Quart. J. R. Astron.
  Soc.\/}, {\bf 37}, 189--229.
  {\small[\href{http://adsabs.harvard.edu/abs/1996QJRAS..37..189W}{ADS}]}

\bibitem[Wilson(1994)]{wilson94}
Wilson, P.R., 1994, {\it Solar and Stellar Activity Cycles\/}, Cambridge
  Astrophysics Series, 24, Cambridge University Press, Cambridge; New York

\bibitem[Wilson(1988{\natexlab{a}})]{wilson88}
Wilson, R.M., 1988{\natexlab{a}}, ``Bimodality and the Hale cycle'', {\it Solar
  Phys.\/}, {\bf 117}, 269--278.
  {\small[\href{http://dx.doi.org/10.1007/BF00147248}{DOI}]},
  {\small[\href{http://adsabs.harvard.edu/abs/1988SoPh..117..269W}{ADS}]}

\bibitem[Wilson(1988{\natexlab{b}})]{wilson88a}
Wilson, R.M., 1988{\natexlab{b}}, ``On the long-term secular increase in
  sunspot number'', {\it Solar Phys.\/}, {\bf 115}, 397--408.
  {\small[\href{http://dx.doi.org/10.1007/BF00148736}{DOI}]},
  {\small[\href{http://adsabs.harvard.edu/abs/1988SoPh..115..397W}{ADS}]}

\bibitem[Wilson(1998)]{wilson98}
Wilson, R.M., 1998, ``A Comparison of Wolf's Reconstructed Record of Annual
  Sunspot Number with Schwabe's Observed Record of `Clusters of Spots' for the
  Interval of 1826--1868'', {\it Solar Phys.\/}, {\bf 182}, 217--230.
  {\small[\href{http://dx.doi.org/10.1023/A:1005046820210}{DOI}]},
  {\small[\href{http://adsabs.harvard.edu/abs/1998SoPh..182..217W}{ADS}]}

\bibitem[Wittmann and Xu(1987)]{wittman_xu87}
Wittmann, A.D. and Xu, Z.T., 1987, ``A catalogue of sunspot observations from
  165 BC to AD 1684'', {\it Astron. Astrophys. Suppl.\/}, {\bf 70}, 83--94.
  {\small[\href{http://adsabs.harvard.edu/abs/1987A&AS...70...83W}{ADS}]}

\bibitem[Wolff {\it et~al.\/}(2012)]{wolff12}
Wolff, E.W., Bigler, M., Curran, M.A.J., Dibb, J.E., Frey, M.M., Legrand, M.
  and McConnell, J.R., 2012, ``The Carrington event not observed in most ice
  core nitrate records'', {\it Geophys. Res. Lett.\/}, {\bf 39}, L08503.
  {\small[\href{http://dx.doi.org/10.1029/2012GL051603}{DOI}]},
  {\small[\href{http://adsabs.harvard.edu/abs/2012GeoRL..3908503W}{ADS}]}

\bibitem[Yang {\it et~al.\/}(2000)]{yang00}
Yang, S., Odah, H. and Shaw, J., 2000, ``Variations in the geomagnetic dipole
  moment over the last 12000 years'', {\it Geophys. J. Int.\/}, {\bf 140},
  158--162.
  {\small[\href{http://dx.doi.org/10.1046/j.1365-246x.2000.00011.x}{DOI}]},
  {\small[\href{http://adsabs.harvard.edu/abs/2000GeoJI.140..158Y}{ADS}]}

\bibitem[Yau and Stephenson(1988)]{yau_steph88}
Yau, K.K.C. and Stephenson, F.R., 1988, ``A Revised Catalog of Far Eastern
  Observations of Sunspots (165 BC to AD 1918)'', {\it Quart. J. R. Astron.
  Soc.\/}, {\bf 29}, 175--197.
  {\small[\href{http://adsabs.harvard.edu/abs/1988QJRAS..29..175Y}{ADS}]}

\bibitem[Yiou {\it et~al.\/}(1997)]{yiou97}
Yiou, F., Raisbeck, G.M., Baumgartner, S., Beer, J., Hammer, C., Johnsen, S.,
  Jouzel, J., Kubik, P.W., Lestringuez, J., Sti{\'{e}}venard, M., Suter, M. and
  Yiou, P., 1997, ``Beryllium 10 in the greenland ice core project ice core at
  Summit, Greenland'', {\it J. Geophys. Res.\/}, {\bf 102}, 26,783--26,794.
  {\small[\href{http://dx.doi.org/10.1029/97JC01265}{DOI}]},
  {\small[\href{http://adsabs.harvard.edu/abs/1997JGR...10226783Y}{ADS}]}

\bibitem[Yule(1927)]{yule927}
Yule, G.U., 1927, ``On a Method of Investigating Periodicities in Disturbed
  Series, with Special Reference to Wolfer's Sunspot Numbers'', {\it Philos.
  Trans. R. Soc. London, Ser. A\/}, {\bf 226}, 267--298. Online version
  (accessed 25 February 2008):
  \newline\url{http://visualiseur.bnf.fr/Visualiseur?Destination=Gallica&O=NUM%
M-56031}

\bibitem[Zeller and Dreschhoff(1995)]{zeller95}
Zeller, E.J. and Dreschhoff, G.A.M., 1995, ``Anomalous Nitrate Concentrations
  in Polar Ice Cores -- do they Result from Solar Particle Injections into the
  Polar Atmosphere?'', {\it Geophys. Res. Lett.\/}, {\bf 22}, 2521--2524.
  {\small[\href{http://dx.doi.org/10.1029/95GL02560}{DOI}]},
  {\small[\href{http://adsabs.harvard.edu/abs/1995GeoRL..22.2521Z}{ADS}]}

\bibitem[Zhentao(1990)]{zhentao90}
Zhentao, X., 1990, ``Solar Observations in Ancient China and Solar
  Variability'', {\it Philos. Trans. R. Soc. London, Ser. A\/}, {\bf 330},
  513--516.
  {\small[\href{http://adsabs.harvard.edu/abs/1990RSPTA.330..513Z}{ADS}]}

\bibitem[Zolotova and Ponyavin(2007)]{zolotova07}
Zolotova, N.V. and Ponyavin, D.I., 2007, ``Was the unusual solar cycle at the
  end of the XVIII century a result of phase asynchronization?'', {\it Astron.
  Astrophys.\/}, {\bf 470}, L17--L20.
  {\small[\href{http://dx.doi.org/10.1051/0004-6361:20077681}{DOI}]},
  {\small[\href{http://adsabs.harvard.edu/abs/2007A&A...470L..17Z}{ADS}]}

\end{thebibliography}

\end{document}